%% file: scet_paper.tex
\documentclass[a4paper,11pt]{article} 
\pdfoutput=1
\usepackage{jheppub}
\usepackage{slashed}
\usepackage[dvipsnames,table]{xcolor}
\usepackage{hyperref}
\usepackage{xspace}
\usepackage[tight]{subfigure}
\usepackage{amsmath}
\usepackage{mathtools}
\usepackage{amssymb}
\usepackage{amsfonts}
\usepackage{mathrsfs}
\usepackage{comment}
\usepackage{makecell}
\setcellgapes{4pt}
\usepackage{afterpage}
\usepackage{verbatim}
\usepackage{booktabs}
\usepackage{bbold}
\usepackage{array}
\usepackage[title,titletoc]{appendix}
\usepackage{scalerel}

\input{macros.tex}

\makeatletter
\def\paragraph{\@startsection{paragraph}{4}{\z@}{-2.00ex plus
 +1ex minus +.2ex}{0.5ex plus .2ex}{\it\normalsize}}
\makeatother

\newcommand{\eewwllll}{\text{e}^+\text{e}^-\rightarrow\text{W}^+\text{W}^-\rightarrow \mu^+\nu_\mu\bar{\nu}_\tau\tau^-}
\newcommand{\eezzllll}{\text{e}^+\text{e}^-\rightarrow\text{ZZ}\rightarrow\mu^+\mu^-\tau^+\tau^-}
\newcommand{\ppwwllll}{\text{p}\text{p}\rightarrow\text{W}^+\text{W}^-\rightarrow
  \mu^+\nu_\mu \bar{\nu}_\text{e}\text{e}^-}
\newcommand{\ppwzllll}{\text{p}\text{p}\rightarrow\text{W}^+\text{Z}\rightarrow
  \text{e}^+\nu_\text{e} \mu^+\mu^-}
\newcommand{\ppzzllll}{\text{p}\text{p}\rightarrow\text{ZZ}\rightarrow\text{e}^+\text{e}^-\mu^+\mu^-}
\newcommand{\aawwllll}{\gamma\gamma\rightarrow\text{W}^+\text{W}^-\rightarrow
  \mu^+\nu_\mu \bar{\nu}_\text{e}\text{e}^-}

\usepackage{tikz}
\tikzstyle{terminator} = [rectangle, draw, text centered, rounded corners, minimum height=2em]
\tikzstyle{flow} = [ draw, -latex']
\usepgflibrary{arrows}

\usepackage{pifont}

\title{Automated resummation of electroweak Sudakov logarithms in diboson production at future colliders}
\author{Ansgar Denner$^{a}$ and }
\author{Stefan Rode$^{a}$}
\affiliation{$^{a}$Universit\"at W\"urzburg, Institut f\"ur Theoretische Physik und Astrophysik, 97074 W\"urzburg, Germany}
\emailAdd{ansgar.denner@uni-wuerzburg.de}
\emailAdd{stefan.rode@uni-wuerzburg.de}
\date{\draftdate}
\abstract{At energies that are large with respect to the
  electroweak scale, the electroweak corrections to scattering
  processes involve large logarithms that have to be
  resummed to obtain decent predictions. Soft--collinear effective theory (SCET) has been proposed as a
  suitable framework to allow for this resummation, while retaining
  non-logarithmic corrections in a consistent way. In this paper, we
  investigate the approximations needed to use this approach for the
  calculation of electroweak corrections to off-shell diboson production
  at high-energy colliders. Upon implementing the method into a
  Monte-Carlo integration code, we provide resummed predictions for cross
  sections and distributions at a $3\TeV$ lepton collider and a
  $100\TeV$ proton collider.}

\begin{document}
\maketitle
\section{Introduction} 
The experimental precision at the LHC necessitates the inclusion of electroweak (EW) radiative corrections to achieve an adequate precision in the theoretical predictions. At each order in the coupling constant $\alpha$, EW radiative corrections contain Sudakov logarithms of the form
\begin{align}
\alpha^n\log^{k}\left(\frac{s_{ij}}{M^2_\text{W}}\right),\quad k\le 2n\label{SudakovLog}
\end{align}
with $s_{ij}=(p_i+p_j)^2$ denoting a kinematic invariant of two external momenta $p_i$ and $p_j$. If $|s_{ij}|\gg M_\text{W}^2$ the convergence of the perturbative series is spoiled. 
To some extent this is the case already at the LHC, where in
high-energy tails of distributions EW corrections of several ten percent have been obtained. However, future colliders 
such as the hadron--hadron option of the Future Circular Collider (FCC--hh) \cite{FCC_hh_lumi} or the Compact Linear Collider (CLIC) \cite{CLIC_proposal,CLIC_machine,CLIC_Lumi} will operate at even higher energies,
where this problem will be more severe, making the all-order resummation of the logarithmically enhanced corrections inevitable.

The Sudakov logarithms are particularly large for processes involving external gauge bosons. Diboson production therefore provides a natural playground to study the effects of resummation. 
The production of a pair of massive EW gauge bosons has extensively been analysed at the LHC:
WW production \cite{WW_CMS_7,WW_CMS_8,WW_ATLAS_8,WW_ATLAS_13}, WZ
production \cite{WZ_ATLAS_2, WZ_CMS, WZ_CMS_2,WZ_CMS_3, WZ_ATLAS,WZ_CMS_pol}, and ZZ production \cite{ZZ_CMS_8, ZZ_ATLAS_8,ZZ_ATLAS_13, ZZ_CMS_13}
have been studied with the particular goal to achieve a precise
measurement of the Standard Model (SM) triple-gauge-boson couplings.  
In turn, on the fixed-order-computation side QCD corrections up to
NNLO accuracy \cite{ZZ_NNLO_QCD_Zurich,WpWm_NNLO_QCD_on, ZZ_NNLO_QCD_differential, WpWm_NNLO_QCD_off, ZZ_NNLO_QCD_Munich, WpWm_NNLO_QCD_ps,ZZ_NNLO}, EW corrections to NLO \cite{eeWW_RACOONWW,WpWm_EW_Bierweiler,Baglio:2013toa,Z0Z0_EW_Higgs,WpWm_NLOEW_offshell,Z0Z0_EW_off,WmZ0_NLO_ew_off}, and the
combination of both \cite{Diboson_NNLO_QCD_NLO_EW} have been obtained for both on-shell and off-shell production. For the case of on-shell $\text{W}\text{W}$ production predictions have been presented in Ref.~\cite{Evolution_qqWW_NNLL} using
infrared (IR) evolution equations to achieve the resummation of large
EW logarithms.

We note that the dominant EW Sudakov corrections have
previously been incorporated into widely 
used event generators such as {\sc{MadGraph}} \cite{Pagani_Zaro, Pagani_Zaro_2} and {\sc{Sherpa}} \cite{Sherpa_Sudakov}. The latter code also embeds an approximate resummation formula. 

Using fixed-order methods, it has been found that the origin of the leading logarithmic corrections stems from the diagrams involving the exchange of a vector boson between pairs of external legs \cite{DennerPozz1},
and in particular from the regions where the loop momentum of the
virtual gauge boson is soft and/or collinear to the momentum of
either external particle. In logarithmic 
approximation these integrals can be calculated using the strategy of regions \cite{Strategy_of_regions}, in which the loop momentum is consistently expanded into hard, collinear, and soft modes. Within Soft-Collinear Effective Theory (SCET) \cite{First_SCET_pap,Second_SCET_pap, Third_SCET_pap, InvariantOperators} this calculational trick is promoted to the level of fields and the Lagrangian of an effective theory.  

Originally, SCET has been constructed to resum large logarithms in
radiative corrections in QCD. The generalisation of SCET to the EW
theory ($\text{SCET}_\text{EW}$) has been
presented in Refs.~\cite{CGKM1, CGKM2, SCET_4f, SCET_SU2,
 SCET_SM}. Within this framework the renormalisation group equation
(RGE) is used to resum the logarithmically enhanced corrections
(\ref{SudakovLog}). In addition $\text{SCET}_\text{EW}$ allows for a
systematic inclusion of all $\mathcal{O}(\alpha)$ corrections, which
is the main advantage compared to the resummation approach of IR
evolution equations
\cite{Evolution_Original,Evolution_4f_NNLL,Evolution_qqWW_NNLL}.
Moreover, $\text{SCET}_\text{EW}$ facilitates the systematic inclusion of power corrections. 

$\text{SCET}_\text{EW}$ has previously been applied to several processes. In the existing literature (see e.g.\ Refs.~\cite{SCET_Single_V, qqVV_HSM, SCET_VBF_Method, SCET_VBF_Results}), 
the focus has been on the analytic computation of simple processes, such as four-fermion processes, Higgs production in vector-boson fusion, and vector-boson production without decays. 

Within this work, in contrast, we aim at the computation of more complicated processes at the fully differential level
 involving also the decays of unstable particles. 
Furthermore, we want to incorporate the effects of phase-space cuts.
The occurring phase-space integration must therefore be performed
numerically, and a certain grade of automation is desirable. We
therefore incorporate the results of $\text{SCET}_\text{EW}$ into a
Monte Carlo (MC) integrator
in order to obtain fully differential cross sections. 
Using this tool, we study the quality of the $\text{SCET}_\text{EW}$ approximation as well as the impact of several contributions of 
the RGE-improved matrix elements for diboson production processes at CLIC and FCC--hh. These
colliders will operate at energies, which definitely
necessitate the resummation of EW Sudakov logarithms,
while it is a priori not clear, to which extent the assumptions
necessary for applying $\text{SCET}_\text{EW}$ are justified in these cases. 

This work is organised as follows: In Sec.~\ref{Sec:SCETEW} we introduce some aspects of the $\text{SCET}_\text{EW}$ framework from Refs.~\cite{CGKM1,CGKM2, SCET_4f, SCET_SU2,
  SCET_SM} along with some notation and conventions.  
In Sec.~\ref{Sec:Details} we give more specific computational and technical details on our approach. In Sec.~\ref{Sec:Results} we present numerical results for diboson production at the FCC--hh and CLIC.

\section{\texorpdfstring{$\text{SCET}_\text{EW}$}{SCET EW} and RGE-improved matrix elements}
\label{Sec:SCETEW}
In this section, we review a few basic facts about the $\text{SCET}_\text{EW}$ formalism along with our conventions.

The key idea of $\text{SCET}_\text{EW}$ is the expansion of the SM Lagrangian in powers of
\begin{align}
\lambda^2=\frac{M_\text{W}^2}{Q^2}
\end{align}
with the W-boson mass\footnote{Throughout this work we identify the EW scale with the W-boson mass. However, the Z, Higgs, and top-quark masses are considered to be of the same order of magnitude and would also be a reasonable choice for the EW scale. All other particles are assumed to be massless.} $M_\text{W}$ and some energy scale $\sqrt{Q^2}\gg M_\text{W}$. 
All propagators with a  squared momentum of order $Q^2$ are integrated out, 
leaving soft and collinear interactions to one or more given directions as dynamic degrees of freedom in the effective theory.   
For a process with \textit{all} kinematic invariants of order $Q^2$, logarithms of the form (\ref{SudakovLog}) can then consistently be resummed
by means of the RGE taking into account the running between a high
scale $\mu_\text{h}\sim\sqrt{Q^2}$ and a low scale $\mu_\text{l}\sim\MW$.

Throughout we work only at leading power, i.e.\ to
$\mathcal{O}(\lambda^0)$. All power-suppressed terms are thus
neglected.%
\footnote{In principle, SCET allows to consistently calculate
    also power-suppressed contributions, and also corrections to the
    GBET can be calculated systematically. This becomes, however,
    technically challenging. Moreover, if $Q$ and $\MW$ are close
    enough for power corrections to become phenomenologically relevant,
    the logarithmically enhanced corrections can be reliably treated
    using fixed-order perturbation theory.}  In the following, we
first introduce some basic notation (Sec.~\ref{Sec:Notation}).
Afterwards we describe the operator basis we used
(Sec.~\ref{Sec:OperatorBasis}) and the emerging RGE-improved matrix
elements (Sec.~\ref{Sec:MatchRun}).  Finally we demonstrate the
handling of longitudinal gauge bosons by means of the Goldstone-boson
equivalence theorem (GBET) (Sec.~\ref{Sec:GBET}).
\subsection{\texorpdfstring{$\text{SCET}_\text{EW}$}{SCET EW}: conventions and notation}
\label{Sec:Notation}
In the following, we consider all momentum four vectors in the Sudakov parametrisation
\begin{align}
p^\mu = (n\cdot p)\frac{\bar{n}^\mu}{2}+(\bar{n}\cdot p)\frac{n^\mu}{2}+p_\perp^\mu 
\end{align}
with two light-like reference vectors $n$, $\bar{n}$ satisfying
\begin{align}
\bar{n}\cdot n=2, \qquad n^2=\bar{n}^2=0.
\end{align}
Denoting the light-cone components of a momentum four-vector according to
\begin{align}
p=(n\cdot p,\bar{n}\cdot p,|{\bf p}_\perp|)= (p^+,p^-,p_\perp),
\end{align}
we define a momentum to have $n$-collinear, $\bar{n}$-collinear, and soft scaling, respectively, according to
\begin{align}
p_\text{c}^\mu\sim (\lambda^2,1,\lambda)Q,\qquad p_{\bar{\text{c}}}^\mu\sim (1,\lambda^2,\lambda)Q,\qquad p_\text{s}^\mu\sim (\lambda^2,\lambda^2,\lambda^2)Q
\label{MomentumScaling}
\end{align}
with $\lambda\ll 1$. An $n$-collinear four-momentum can thus be considered ``almost parallel'' to $n$. 
Note that the term ``soft'' is used for the scaling, which is sometimes called ``ultrasoft'', in particular if a mode with the scaling $(\lambda,\lambda,\lambda)$ is present. 

When considering a hard scattering process with $n$ distinct directions every external momentum $p_i$ defines a pair of reference vectors $n_i$, $\bar{n}_i$.
At leading power, all external fields are collinear, and interactions
involving external soft fields are power suppressed. 
Gauge-invariant interactions involving collinear scalars, fermions, and gauge bosons can be constructed using the combinations \cite{ CGKM2,SCET_Diboson,SCET_Intro}
\begin{align}
\Xi_{n_i,p_i}(x)&=W_{n_i}^\dagger(x)\Phi_{n_i,p_i}(x),\nonumber\\
\chi_{n_i,p_i}(x)&=W_{n_i}^\dagger(x)\xi_{n_i,p_i}(x),\nonumber\\
\mathcal{A}_{\perp,n_i,p_i}^\mu(x)&=\frac{1}{g}W_{n_i}^\dagger(x)\left(\text{i}D^\mu_{\perp,n_i,p_i} W_{n_i}(x)\right),\label{chidef}
\end{align}
with $\Phi$ denoting the Higgs doublet, and $\xi_{n_i,p_i}$ is the leading-power component of the Dirac spinor $\psi_{p_i}$ given by
\begin{equation}
\xi_{n_i,p_i}=\frac{\slashed{n}_i\slashed{\bar{n}}_i}{4}\psi_{p_i}.
\end{equation}
The collinear Wilson lines are defined as
\begin{align}
W_{n_i}(x)&=\hat{\text{P}}\exp\left(\text{i}\sum_{k=1}^2g_k\int_{-\infty}^0\text{d}s\, \bar{n}_i\cdot A^{(k),a}_{n_i,p_i}(x+s\bar{n})T^a\right),
\end{align}
with $\hat{\text{P}}$ denoting path ordering and $T^a$ generic
generators of the symmetry group. The quantity $D_{n_i,p_i,\perp}$ in (\ref{chidef}) is the perpendicular component of the $n_i$-collinear covariant derivative
\begin{equation}
D^\mu_{n_i,p_i}=\partial^\mu-\sum_{k=1,2}\text{i}g_kA^{(k),\mu,a}_{n_i,p_i}(x)T^a
\end{equation} 
with the collinear gauge fields $A_{n_i,p_i}^{(k),\mu,a}$ of the U(1) and SU(2) group with gauge couplings $g_1$ and $g_2$, respectively.

All field operators carry momentum labels indicating their hard momentum.  
For more details on the label formalism see Refs.~\cite{Second_SCET_pap, Third_SCET_pap}. For our purpose a field with momentum label $p_i$ 
can be viewed as a momentum eigenstate, the difference is only relevant for the consistent inclusion of power corrections.
An $n$-particle operator can then be written as\footnote{In some publications, such as Ref.~\cite{SCET_SU2}, each field is additionally dressed with a soft Wilson line
to decouple the soft and collinear interactions from each other. 
If the soft Wilson lines are omitted, the soft--collinear interactions are kept explicitly in the SCET Lagrangian.}
\begin{align}
\mathcal{O}=\phi_{n_1,p_1}(x)\ldots\phi_{n_n,p_n}(x)\label{OlabelForm}
\end{align}
with each $\phi_{n_i,p_i}$ denoting one of the operator products in (\ref{chidef}).

Radiative corrections to operators of the form (\ref{OlabelForm}) are calculated using the $\text{SCET}_\text{EW}$ Lagrangian. As far as gauge bosons and
fermions are concerned, it has the same form as the SCET Lagrangian for QCD, which can, for instance, be found in Refs.~\cite{First_SCET_pap, Second_SCET_pap, Third_SCET_pap}.
The treatment of scalars in the SM is described in more detail in Refs.~\cite{CGKM1, CGKM2}. 
If the W and Z masses can be neglected, as for the high-scale matching and anomalous-dimension computations, the occurring loop integrals have very similar properties as in the QCD case.

For the computation of low-scale corrections, $\text{SCET}_\text{EW}$ integrals with finite masses have to be considered, which
suffer from the collinear or factorisation anomaly and require additional regularisation
schemes on top of dimensional regularisation \cite{SCET_wo_regulator, AnalyticRegularisation}.
\subsection{Operator basis}
\label{Sec:OperatorBasis}
In the $\text{SCET}_\text{EW}$ literature \cite{CGKM1, CGKM2, SCET_4f,SCET_SU2,SCET_SM, qqVV_HSM} the high-scale Wilson coefficients 
and the anomalous dimension have always been expressed in the space of SU(2)-gauge-covariant operators. At the low scale, the operators are then matched
onto another set of operators in the physical basis, and the low-scale
corrections are calculated in this basis.

This choice is particularly convenient for an analytic computation and in principle
any SM matrix element can be expressed in terms of the matching coefficients of these operators. 
However, the fact that we would like to use the fixed-order automation apparatus motivates 
us to rewrite all occurring expressions in terms of scattering amplitudes, which can be associated with partonic processes.\footnote{Here and in the following a \textit{process} always refers to the scattering of elementary particles such as quarks and leptons, and not, for instance, hadrons.}

To this end we break up 
the fermion and scalar doublets as well as the gauge-boson triplets and consider operators which are monomials of fields corresponding to physical particles.
In the following we discuss the particular choices for fermions, gauge bosons, and scalars.   

\paragraph{Fermions}
For fermions we use the flavour and charge eigenstates with
left-handed (L) and right-handed (R) chiralities of each field. As an example consider the four-fermion operator 
\begin{align}
\mathcal{O}&=\bar{u}_{\text{L},n_1,p_1}W_{n_1}\gamma_\mu W^\dagger_{n_2}u_{\text{L},n_2,p_2}\bar{e}_{\text{L},n_3,p_3}W_{n_3}\gamma^\mu  W^\dagger_{n_4} e_{\text{L},n_4,p_4}.\label{uuee}
\end{align}
Each fermion field is one component of an SU(2) doublet. Equation~(\ref{uuee}) can be related to the scattering process $\bar{\text{u}}_\text{L}\text{u}_\text{L}\rightarrow\text{e}^+_\text{L}\text{e}^-_\text{L}$, which we make use of for the automation procedure.
\paragraph{Gauge bosons}
For processes with external gauge bosons we use a mixture of the symmetric and the physical basis in the different contributions.
For charged gauge bosons we employ the charge eigenstates $W^\pm$ rather than the SU(2)-covariant $W^{1/2}$ fields used in Ref.~\cite{SCET_SU2}.

For neutral gauge bosons the operators are constructed in the
symmetric basis $W^3/B$, which simplifies the anomalous dimension a
lot. For the low-scale corrections one has to apply the
back-transformation, because they depend on the masses of the external
particles. This is discussed in more detail in Sec.~\ref{se:low-scale mixing}.
\paragraph{Scalars}
The scalars are treated in close analogy to the neutral gauge bosons: If we denote the Higgs doublet by
\begin{align}
\Phi=\begin{pmatrix}\phi^+\\\phi_2 \end{pmatrix}=\begin{pmatrix}\phi^+\\\frac{1}{\sqrt{2}}(v+\eta+\text{i}\chi) \end{pmatrix},\qquad \phi^-=(\phi^+)^\dagger\label{Higgsdef},
\end{align}
we construct the operators from the fields $\phi^\pm$, $\phi_2$, and $\phi_2^*$. At the low energy scale the lower components have to be rotated into the mass eigenstates $\chi$, $\eta$. 
Here $\eta$ denotes the physical Higgs field and $\phi^\pm$, $\chi$ the would-be Goldstone-boson fields. 

\subsection{Matching and running}
\label{Sec:MatchRun}
To extract physical predictions in $\text{SCET}_\text{EW}$, the operators (\ref{OlabelForm}) have to be matched against the full SM.
For each operator the difference is absorbed into a Wilson coefficient $C(\mu)$.  
Because $\text{SCET}_\text{EW}$ reproduces the dependence on the EW scales
of $\mathcal{O}(M_\text{W})$ by construction, the Wilson coefficients can depend only on the high scales. 
The matching is, thus, most easily calculated with the low scales set
to zero \cite{SCET_SU2}. In practice this implies that the matching computation
is done in the symmetric phase of the Standard Model (SySM).
  
The low-scale corrections on the other hand have to be computed keeping the full mass dependence. Owing to the simplified structure of the loop corrections in $\text{SCET}_\text{EW}$ they do, 
however, not depend on the internal structure of the process. Instead they are obtained only from the quantum numbers and momenta of the external particles and can be computed once and for all
for each SM particle.
\paragraph{Tree-level matching}
In the basis described above the tree-level matching condition for a single process can be phrased as \cite{SCET_Intro}
\begin{equation}
\mathcal{M}_{\text{SySM}}=\sum_{k} C^{(k)}(\mu)\left<0\right|\mathcal{O}^{(k)}\left|p_1,\ldots,p_n\right>_\text{SCET},\label{spinpolvec}
\end{equation}
with  $k$ running over the different operators. The expressions $\left<0|\mathcal{O}^{(k)}|p_1,\ldots,p_n\right>$
contain the Dirac matrices, spinors, and polarisation vectors, whereas the non-trivial dependence on the kinematics is incorporated in the Wilson coefficients $C^{(k)}$. 
\paragraph{One-loop matching}
When the low scales are put to zero all loop corrections in $\text{SCET}_\text{EW}$ vanish, because they involve only scaleless integrals. The SySM matrix element has both UV and IR divergences.  
While the UV poles in the full and the effective theory can in general be different, the IR poles have to agree \cite{qqVV_HSM} and hence cancel in the difference. The Wilson coefficients are thus calculated from the IR-finite part of the SySM matrix element, and after renormalisation one obtains
\begin{align}
\left.\mathcal{M}_{\text{SySM}}^\text{1-loop}(\mu_\text{h})\right|_\text{IR-finite}&=\sum_kC^{(k),\text{1-loop}}(\mu_\text{h})\left<0|\mathcal{O}^{(k)}|p_1,\ldots,p_n\right>_\text{SCET},\label{spinpolvec_1loop}
\end{align}
which is the version of (\ref{spinpolvec}) at one loop. 

\paragraph{Running}
Since the non-trivial SU(2)-charge flow induces mixing
between operators of different weak isospin structure, the anomalous dimension mixes matrix elements
associated with different processes into each other. The anomalous dimension can be written in terms of gauge-group operators~\cite{SCET_SU2},
\begin{align}
\boldsymbol{\gamma}&=\sum_{\left<ij\right>}\left(\mathbf{t}_i\cdot\mathbf{t}_j\Gamma_{\text{cusp},2}+\frac{Y_iY_j}{4}\Gamma_{\text{cusp},1}\mathbb{1}\right)\log\left(\frac{\mu^2}{-s_{ij}-\text{i}0}\right)+\sum_i\gamma_i\mathbb{1}
\label{ADfirst}
\end{align}
with $\left<ij\right>$ denoting the sum over pairs of external
particles (without double counting) and with the $\text{SU}(2)$ 
operators $\boldsymbol{t}$ being analogous to the QCD colour operators introduced in Ref.~\cite{CataniSeymour}. Their action on an external field $\psi_i$ with 
gauge index $\alpha$ is given by \cite{SCET_SU2}
\begin{align}
\mathbf{t}_i\psi_{j,\alpha}=\sum_{a=1}^3t^a_{i,\alpha\beta}\psi_{i,\beta}\delta_{ij}\label{tSU2}
\end{align}
with $t^a_{i,\alpha\beta}$ denoting components of the SU(2) generators in the representation of
$\psi$. We use the linear combinations 
\begin{align}
t^\pm_i=\frac{1}{\sqrt{2}}\left(t^1_i\mp \text{i}t^2_i\right) \qquad\rightarrow\qquad t^1_i\cdot t^1_j+t^2_i\cdot t^2_j=t^+_i\cdot t^-_j+t^-_i\cdot t^+_j\label{tSU2_pm}
\end{align} 
instead of the $t^{1/2}$. This basis has also been employed in Ref.~\cite{SCET_SM}, with a different normalisation convention. 
It is convenient, because the $t^\pm$ operators rotate matrix elements in the basis described in Sec.~\ref{Sec:OperatorBasis} into other matrix elements associated to scattering
processes. Thus, we can express all parts of the calculation in terms of matrix elements
and never have to evaluate either the Wilson coefficients or operator
expectation values separately. The symbol $\text{i}0$ appearing in Eq.~(\ref{ADfirst}) is an infinitesimal imaginary part arising from the Feynman prescription (Feynman i$\varepsilon$).
\paragraph{RGE-improved matrix elements}
To obtain the RGE-improved matrix elements, the Wilson coefficients are matched at the high scale and run down to the low scale. Their scale dependence is governed by 
the respective anomalous dimension and the RGE, whose solution achieves the exponentiation of the large logarithms.
This procedure results in the formula 
\begin{align}
\mathcal{M}^\text{SCET}=\sum_{jl}\boldsymbol{D}_{1l}(\mu_\text{l})\left[\hat{\text{P}}\exp\left(-\int_{\mu_\text{l}}^{\mu_\text{h}}\frac{\text{d}\mu}{\mu}\boldsymbol{\gamma}(\mu)\right)\right]_{lj}\mathcal{M}_{\text{SySM},j}(\mu_\text{h}),
\label{Basisbase}
\end{align}
where $j$ and $l$ run over all processes that arise, when pairs of fields of the respective process are rotated into their SU(2) partners and $\mathcal{M}_{\text{SySM},j}$ is the matrix element corresponding to process $j$. 
The processes are assumed to be ordered in a way that $j=1$
corresponds to the original untransformed process. The index
\text{SySM} is omitted in the following.

Beyond that, (\ref{Basisbase}) contains the following ingredients:
\begin{itemize}
\item The $\text{SCET}_\text{EW}$ anomalous dimension $\boldsymbol{\gamma}$: The matrix exponential describes the RG running from $\mu_\text{h}$ to $\mu_\text{l}$. The path-ordering symbol $\hat{\text{P}}$ is defined according to
\begin{align}
\hat{\text{P}}\exp\left(-\int_{\mu_\text{l}}^{\mu_\text{h}}\frac{\text{d}\mu}{\mu}\,\boldsymbol{\gamma}(\mu)\right)&=\mathbb{1}-\int_{\mu_\text{l}}^{\mu_\text{h}}\frac{\text{d}\mu}{\mu}\boldsymbol{\gamma}(\mu)+\frac 12\int_{\mu_\text{l}}^{\mu_\text{h}}\frac{\text{d}\mu}{\mu}\int_{\mu_\text{l}}^{\mu}\frac{\text{d}\mu'}{\mu'}\boldsymbol{\gamma}(\mu')\boldsymbol{\gamma}(\mu)+\ldots\,,\label{PDef}
\end{align}
thus sorting matrices with smaller arguments to the left. 
\item The low-scale mixing matrix $\boldsymbol{D}$: It takes the
explicit $\text{SCET}_\text{EW}$ low-scale corrections into account and can be defined via $\boldsymbol{D}=\mathbb{1}+\boldsymbol{D}^{(1)}+\mathcal{O}(\alpha^2)$ with
\begin{align}
\left<0|\mathcal{O}^{\text{1-loop}}_i|p_1,\ldots,p_n\right>_{\text{SCET}_\text{EW}}=\boldsymbol{D}^{(1)}_{ij}\left<0|\mathcal{O}^{\text{tree}}_j|p_1,\ldots,p_n\right>_{\text{SCET}_\text{EW}}.\label{deltaMD}
\end{align}
The expression on the l.h.s.\ arises from one-loop corrections using the $\text{SCET}_\text{EW}$ Lagrangian. Decomposing them into the basis of the respective tree-level expectation values 
implicitly defines $\boldsymbol{D}^{(1)}$. 
\end{itemize}
Both $\boldsymbol{\gamma}$ and $\boldsymbol{D}^{(1)}$ are universal quantities that can be constructed in a process-independent manner, for more details see Sec.~\ref{Sec:Ingredients}.
\subsection{Treatment of longitudinally polarised gauge bosons}
\label{Sec:GBET}
Special care is required when dealing with the longitudinal
polarisation states of the massive gauge bosons.

In the unbroken phase of the theory, where the Wilson coefficients and the anomalous dimension are computed, gauge bosons are massless and hence always transversely polarised, while in the low-energy theory with massive gauge bosons longitudinal degrees of freedom are present. These degrees of freedom can, however, be identified with those of the scalar would-be Goldstone bosons in the symmetric phase, reflected in the GBET \cite{GBET1974,GBET1985,GBET1986, GBET1988},
 which in terms of on-shell matrix elements reads \cite{BDJ}
\begin{align}
        \mathcal{M}^{\phi_1\ldots\hat{W}^\pm_\text{L}\ldots\phi_n}&{}=\pm\left(1-\frac{\Sigma^{\text{WW}}_\text{L}(M_\text{W}^2)}{M_\text{W}^2}-\frac{\Sigma^{\text{W}\phi}(M_\text{W}^2)}{M_\text{W}}+\frac 12 \delta Z_\text{W}+\frac{\delta M_\text{W}}{M_\text{W}}\right)\mathcal{M}^{\phi_1\ldots\phi^\pm\ldots\phi_n}\nonumber\\
&\qquad+\mathcal{O}(\alpha^2)+\mathcal{O}\left(\frac{M_\text{W}}{E}\right),\nonumber\\
        \mathcal{M}^{\phi_1\ldots\hat{Z}_\text{L}\ldots\phi_n}&{}=\text{i}\left(1-\frac{\Sigma_\text{L}^{\text{ZZ}}(M_\text{Z}^2)}{M_\text{Z}^2}+\text{i}\frac{\Sigma^{\text{Z}\chi}(M_\text{Z}^2)}{M_\text{Z}}
+\frac 12\delta Z_\text{ZZ} +\frac{\delta M_\text{Z}}{M_\text{Z}}\right)\mathcal{M}^{\phi_1\ldots\chi\ldots\phi_n}\nonumber\\
&\qquad+\mathcal{O}(\alpha^2)+\mathcal{O}\left(\frac{M_\text{Z}}{E}\right),\label{GBET}
\end{align}
with $E$ being the gauge boson's energy. As indicated by the hats, we assume the gauge-boson field on the l.h.s.\ to be renormalised, which introduces the field-renormalisation constants on the r.h.s.
The Goldstone-boson fields are kept unrenormalised ($\delta Z_{\phi/\chi}=0$) by convention. 

By means of the GBET the Wilson coefficients and the anomalous dimension for a process involving longitudinally polarised gauge bosons can be obtained 
from the respective quantities with the gauge bosons replaced by the corresponding Goldstone bosons.

\section{Details of the calculation}
\label{Sec:Details}
We present an implementation of RGE-improved $\text{SCET}_\text{EW}$ results into the MC integrator \mocanlo. 
We aim for the inclusion of all $\mathcal{O}(\alpha)$ effects in diboson-production processes, including decay effects, real hard-photon
radiation, as well as the resummation of the dominant Sudakov logarithms. This requires the usage of both fixed-order methods
and $\text{SCET}_\text{EW}$.

In this section, we discuss the interplay between the fixed-order techniques and the $\text{SCET}_\text{EW}$ resummation (Sec.~\ref{Sec:Strategy}), 
followed by a detailed discussion of the ingredients of the RGE-improved master formula (\ref{base}) (Sec.~\ref{Sec:Ingredients}). 
In Sec.~\ref{Sec:LogCount} we discuss the logarithm-counting scheme. We conclude by a brief discussion of the technical setup (Sec.~\ref{Sec:Technic}).
\subsection{Strategy: Application of
  \texorpdfstring{$\text{SCET}_\text{EW}$}{SCET EW} to diboson production}
\label{Sec:Strategy}
The main problem when applying (\ref{Basisbase}) to processes with many external particles is to ensure the validity of the Sudakov condition
\begin{align}
|s_{ij}|\gg M^2_\text{W},\quad i,j\in\{1,\ldots,n\},\label{SCETcond}
\end{align}
for \textit{all} pairs of external particles. 
In the following we discuss the necessary steps to achieve these conditions.  

\subsubsection{Double-pole approximation}
\begin{figure}
\centering
\includegraphics[page=2,scale=0.99]{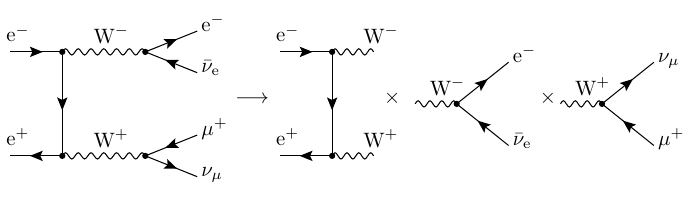}
\caption{Factorisation of the matrix element in the DPA in
  $\text{W}^+\text{W}^-$ production and $\text{W}$ decays.}
\label{Fig:DPA}
\end{figure}

We consider pair production of W and Z bosons in pp and $\text{e}^+\text{e}^-$ collisions, i.~e. processes of the form
\begin{align}
\text{pp}/\text{e}^+\text{e}^-&\rightarrow VV'\rightarrow 4\ell, \quad V,V'\in\{\text{W}^\pm,\PZ\}.\label{BBprime}
\end{align}
As discussed at the beginning of Sec.~\ref{Sec:SCETEW}, applying $\text{SCET}_\text{EW}$ requires all external 
kinematic invariants of a given process to be large compared to the EW scale. This is obviously not the case 
in the processes (\ref{BBprime}), because the invariants of the virtual
gauge bosons are of the order of their masses in the 
resonant regions, which dominate the cross section. We therefore use the double-pole approximation (DPA) \cite{Aeppli_DPA,Aeppli_DPA_2,eeWW_RACOONWW}
in order to factorise the process into subprocesses associated with
gauge-boson production and decay (see Fig.~\ref{Fig:DPA}). 
To achieve this
in a gauge-invariant manner one has to project the momenta of the bosons' decay products such that the gauge bosons are on shell. There is some 
freedom how to do this exactly, and different on-shell projections modify the result only up to $\mathcal{O}(\Gamma_V/M_V)$ \cite{eeWW_RACOONWW}.
We choose an on-shell projection, which preserves the spatial directions of the decay products in appropriate frames as described in Ref.~\cite{Pol_ZZ_Denner}.

On the level of LO and virtual NLO matrix elements the factorisation
in the DPA can be schematically written as
\begin{align}
\mathcal{M}_\text{LO}&\rightarrow \mathcal{M}^\text{prod}_\text{LO}\times\mathcal{M}^{\text{dec},1}_\text{LO}\times \mathcal{M}^{\text{dec},2}_\text{LO},\nonumber\\
\mathcal{M}_\text{NLO}&\rightarrow \left(\frac{\mathcal{M}^\text{prod}_\text{NLO}}{\mathcal{M}^\text{prod}_\text{LO}}+\frac{\mathcal{M}^\text{dec,1}_\text{NLO}}{\mathcal{M}^\text{dec,1}_\text{LO}}
+\frac{\mathcal{M}^\text{dec,2}_\text{NLO}}{\mathcal{M}^\text{dec,2}_\text{LO}}+\delta_\text{nfact}\right)\mathcal{M}_\text{LO},\label{DPAfacti}
\end{align}
with $\delta_\text{nfact}$ comprising the non-factorisable corrections, defined as the part of the difference between the corrections 
to the full process and the production/decay cascade that is non-analytic in the off-shell behaviour \cite{Aeppli_DPA_2,Denner_nfact,Schwan_nfact}.
The key points of (\ref{DPAfacti}) with respect to the isolation of logarithmic corrections are: 
\begin{itemize}
\item The decay processes can not depend on the large invariants and are therefore free of large logarithms. 
The same holds for $\delta_\text{nfact}$ (see, for instance, Ref.~\cite{AccomandoKaiser}).
\item If the particles of the production process are well separated, the process fulfils Eq.~(\ref{SCETcond}) in the high-energy limit.
\end{itemize}
After applying the DPA, we can treat the production matrix element with $\text{SCET}_\text{EW}$, achieving a resummation of all logarithmically enhanced corrections.
The decay and non-factorisable corrections are computed in the full SM. 

\subsubsection{Polarisation definitions}
\label{Sec:PolDef}
The matching and running within $\text{SCET}_\text{EW}$ takes place on the level of helicity amplitudes: 
both the matching contributions and the logarithmic corrections depend on the external spins and helicities of both fermions and gauge bosons. 
This requires a notion for the polarisation state of the virtual gauge bosons,
which are treated as the external states of the production subprocesses. We employ the polarisation definition introduced in Ref.~\cite{PolGiov_LO}: 
For each involved gauge boson $V$ with momentum $k$ the matrix element is decomposed according to \cite{Pol_WW_Denner}
\begin{equation}
\mathcal{M}=\sum_{\lambda=\pm1,0}\frac{\mathcal{M}^\text{prod}_\lambda(\tilde
  k)\mathcal{M}^\text{dec}_\lambda(\tilde k)}{k^2-M_V^2+\text{i}\Gamma_V M_V}\label{Mlambda}
\end{equation}
with $\mathcal{M}^\text{prod/dec}_\lambda(\tilde k)$ denoting the production and decay matrix element with an external polarised gauge boson after 
the momenta have been projected on shell ($k\to\tilde k$).
The polarisation of a massive gauge boson is a frame-dependent quantity with the frame dependence entering via the polarisation vectors, 
\begin{align}
\varepsilon^\mu_{\lambda=-1}&=\frac{1}{\sqrt{2}}(0,\cos\theta\cos\phi+\text{i}\sin\phi,\cos\theta\sin\phi-\text{i}\cos\phi,-\sin\theta),\nonumber\\
\varepsilon^\mu_{\lambda=+1}&=\frac{1}{\sqrt{2}}(0,-\cos\theta\cos\phi+\text{i}\sin\phi,-\cos\theta\sin\phi-\text{i}\cos\phi,\sin\theta),\nonumber\\
\varepsilon^\mu_{\lambda=0}&=\frac{1}{M}(|\boldsymbol{\tilde k}|,E\sin\theta\cos\phi,E\sin\theta\sin\phi,E\cos\theta),
\end{align}
with $\theta$ and $\phi$ denoting the polar and azimuthal angles of
the boson's three-momentum $\boldsymbol{\tilde k}$ in a certain frame. 
In the following we always choose polarisation vectors defined in the two-boson centre-of-mass frame.
While this is the preferred frame for diboson production,
    the calculation can as well be performed in other frames. However,
    one has to make sure that all longitudinal gauge bosons have
large energies in order not to spoil the validity of the GBET used to
treat the longitudinal polarisations.

When the matrix element in (\ref{Mlambda}) is squared, we sum the polarised matrix elements incoherently,
\begin{equation}
|\mathcal{M}|^2\approx\sum_{\lambda}\frac{|\mathcal{M}^\text{prod}_\lambda(\tilde
  k)|^2|\mathcal{M}^\text{dec}_\lambda(\tilde k)|^2}{(k^2-M_V^2+\text{i}\Gamma_V M_V)^2},
\end{equation}
thereby neglecting the interference contributions between the
different polarisation states $+1,-1,0$. While these contributions can be sizable on the level of
phase-space points, they are expected to integrate to zero in sufficiently inclusive cut setups \cite{PolGiov_LO,Pol_WW_Denner,Pol_ZZ_Denner}.
In particular, the observable under consideration has to be inclusive with respect to the angles of the gauge-bosons'
decay products.

More details on the calculation of polarisation effects in diboson processes and NLO results can be found in Refs.~\cite{Pol_WW_Denner, Pol_WZ_NLO, Pol_ZZ_Denner,Pol_Semilep,Pol_Semilep_WW, Pol_WZ_Thi_1,Pol_WZ_Thi_2, Pol_WZ_Thi_3,Pol_WW_Thi}.
\subsubsection{Infrared subtraction}
IR singularities owing to photon exchange and radiation are treated with the Catani--Seymour dipole subtraction \cite{CataniSeymour,DittmaierFermions,CataniSeymour_Massive,Kabelschacht}.
After applying the Catani--Seymour dipole-sub\-trac\-tion scheme any cross section is obtained in the form \cite{CataniSeymour}
\begin{align}
\sigma=\int_n\text{d}\sigma_\text{Born}+\int_n\left(\text{d}\sigma_\text{virt}+\int_1\text{d}\sigma_\text{sub}\right)+\int_{n+1}\left(\text{d}\sigma_\text{real}-\text{d}\sigma_\text{sub}\right)\label{VirtRep}
\end{align}  
with each of the three terms being integrated separately. The index below
the integral sign denotes the number of particles of the respective phase space. 
Both the second and the third term are IR finite: While the subtraction terms cancel the real corrections
in the singular phase-space regions, the integrated subtraction terms
exhibit explicit regulated poles which cancel those in the virtual contribution
$\text{d}\sigma_\text{virt}$. 

Both the pole approximation and $\text{SCET}_\text{EW}$ are only applied to the virtual contributions, since the dominant Sudakov logarithms arise there.
Therefore we substitute
\begin{align}
\int_n\left(\text{d}\sigma_\text{virt}+\int_1\text{d}\sigma_\text{sub}\right)\rightarrow \int_n\left(\text{d}\sigma^\text{SCET}_\text{virt}+\int_1\text{d}\sigma_\text{sub}\right)\label{EFTsubst}
\end{align}
with $\text{d}\sigma_\text{virt}^\text{SCET}$ obtained by subtracting the tree-level result from the squared $\text{SCET}_\text{EW}$ matrix element:
\begin{equation}
\text{d}\sigma_\text{virt}^\text{SCET}=\text{d}\Phi^{(m)}\left(\left|\mathcal{M}^\text{SCET}\right|^2-\left|\mathcal{M}_0\right|^2\right).\label{dPhi}
\end{equation}
The tree-level matrix element $\left|\mathcal{M}_0\right|^2$ has to be subtracted because it is already contained in $\text{d}\sigma_\text{Born}$.
This procedure is necessary because the matrix element (\ref{Basisbase}) does not naturally decompose into tree-level and loop-level contributions. 
Instead, the amplitude obtained from matching a set of operators contains the Born contribution and dominant virtual corrections to all orders.   
\subsubsection{Modifications of the factorisation formula}
It is, however, not possible to use (\ref{Basisbase}) as virtual matrix element in (\ref{dPhi}): The IR divergences contained in $\boldsymbol{D}$ are multiplied with the resummed matrix element instead of the Born matrix element. Because the subtraction contributions remain unmodified, this would lead to a non-cancellation of IR divergences on the r.h.s.\ of (\ref{EFTsubst}). 

After substituting 
\begin{align}
\sum_l\boldsymbol{D}_{1l}(\mu_\text{l})&\left[\hat{\text{P}}\exp\left(-\int_{\mu_\text{l}}^{\mu_\text{h}}\text{d}\log\mu\,\boldsymbol{\gamma}(\mu)\right)\right]_{lj}=\notag\\
&{}=\sum_l\left(\delta_{1l}+\boldsymbol{D}^{(1)}_{1l}(\mu_\text{l})\right)\left[\hat{\text{P}}\exp\left(-\int_{\mu_\text{l}}^{\mu_\text{h}}\text{d}\log\mu\,\boldsymbol{\gamma}(\mu)\right)\right]_{lj}\nonumber\\
&{}\rightarrow\boldsymbol{D}^{(1)}_{1j}(\mu_\text{l})+\left[\hat{\text{P}}\exp\left(-\int_{\mu_\text{l}}^{\mu_\text{h}}\text{d}\log\mu\,\boldsymbol{\gamma}(\mu)\right)\right]_{1j},\label{Dsubst}
\end{align}
the IR divergences in $\boldsymbol{D}^{(1)}$ are multiplied with the Born matrix element. 
It should clearly be said that (\ref{Basisbase}) is the consistent way
of including all terms of $\mathcal{O}\bigl(\alpha^2\log^2(M_\text{W}^2/s)\bigr)$ 
and the substitution (\ref{Dsubst}) misses some of these
contributions: The product of the non-logarithmic parts related to virtual-photon exchange
[of $\mathcal{O}(\alpha)$] and the terms in the matrix exponential [the dominant ones are of $\mathcal{O}(\alpha\mathsf{L}^2)$]
are discarded in (\ref{Dsubst}) for the sake of IR
finiteness. Since we include the real photonic corrections only in
fixed order, we have to discard IR-singular higher-order virtual
corrections to ensure IR finiteness.%
\footnote{We only consider real-photon radiation but do not
    include radiation of massive EW gauge bosons or $\Pq\bar\Pq$ pairs.}

In addition, we split off the parameter-renormalisation (PR) contributions into a separate contribution: 
\begin{align}
\mathcal{M}_j\rightarrow\mathcal{M}_j+\delta_\text{PR}\mathcal{M}_j.
\end{align}
If the running of the EW coupling constants is not considered, $\delta_\text{PR}$ contains logarithmically enhanced and finite parts of the renormalisation constants $\delta e$, $\delta s_\text{w}$, and $\delta c_\text{w}$.
If the running of the coupling constants is taken into account, the
logarithmically enhanced terms are contained in the anomalous
dimension, while the finite remainders are still part of $\delta_\text{PR}$.  

We therefore use 
\begin{align}
\mathcal{M}_\text{SCET}=\sum_{j}\left[\boldsymbol{D}^{(1)}(\mu_\text{l})+\hat{\text{P}}\exp\left(-\int_{\mu_\text{l}}^{\mu_\text{h}}\text{d}\log\mu\,\boldsymbol{\gamma}(\mu)\right)+\delta_\text{PR}\mathbb{1}\right]_{1j}\left.\mathcal{M}_{j}(\mu_\text{h})\right|_{\{M\}=0,\ \text{IR-finite}}\label{base}
\end{align}
as a master formula for the MC code. Remember that the basic ingredients are 
\begin{itemize}
\item the IR-finite parts of the SySM matrix elements $\mathcal{M}_j(\mu_\text{h})$,
\item the PR contributions $\delta_\text{PR}$,
\item the anomalous-dimension matrix $\boldsymbol{\gamma}$,
\item the low-scale $\text{SCET}_\text{EW}$ corrections $\boldsymbol{D}(\mu_\text{l})$. 
\end{itemize}
We choose the high and the low scale according to
\begin{align}
\mu_\text{h}=\sqrt{\hat{s}},\qquad \mu_\text{l}=M_\text{W},\label{partonic}
\end{align}
with $\sqrt{\hat{s}}$ denoting the centre-of-mass energy of the partonic subprocess. Note that if no resummation is applied, the result is completely independent of the scale choices: Changing $\mu_\text{h}$ merely shifts contributions from the high-scale matching into the anomalous dimension and vice versa, while changing $\mu_\text{l}$ reshuffles contributions between the low-scale corrections and the anomalous dimension. The dependence of the final result on the precise choice of $\mu_\text{h}$ and $\mu_\text{l}$ is expected to be small. 
\subsection{Ingredients of the virtual matrix element}
\label{Sec:Ingredients}
In this section, we discuss the ingredients of the RGE-improved matrix element (\ref{base}) and their implementation.
\subsubsection{Construction of the relevant operators}
From the form of the anomalous dimension (\ref{ADfirst}) entering (\ref{base}) we see that the set of contributing operators (and therefore matrix elements) to a given process
\begin{align}
\phi_1\phi_2\to\phi_3\ldots\phi_n
\label{4to0}
\end{align}
is obtained by applying the SU(2) operators $\boldsymbol{t}_i\cdot\boldsymbol{t}_j$ arbitrarily often to any pair of particles. A single transformation can be written as 
\begin{align}
\boldsymbol{t}_i\cdot\boldsymbol{t}_j[\phi_1\ldots\phi_i\ldots\phi_j\ldots\phi_n]\quad\Longrightarrow\quad\phi_1\ldots\varphi_i\ldots\varphi_j\ldots\phi_n
\label{4tildeto0}
\end{align}
with $\varphi_{i}$ denoting SU(2) partners of $\phi_i$ (for an external $W^3$ both $W^+$ and $W^-$ have to be considered).  
We refer to (\ref{4tildeto0}) as a two-particle transformation in the following.
An algorithm to compute all possible processes connected to (\ref{4to0}) is implemented as follows. Starting with a list of processes that contains the initial one as its only element:
\begin{itemize}
\item Apply all possible two-particle transformations to the external fields. Check whether the resulting process violates charge conservation.
If not, append it to the list.
\item Go to the next process in the list and apply all possible two-particle transformations. 
Check whether the resulting process violates charge conservation. 
If not, check whether the process is already in the list. If not, add it.
\item Repeat the above until the end of the list is reached (that is, the iteration of the last process did not produce any new ones).
\end{itemize}
These steps provide a way to obtain all relevant processes contributing to (\ref{base}).   
\subsubsection{Tree-level matrix elements}
With the considerations of the previous section the automated
computation of expectation values of $\text{SCET}_\text{EW}$ operators
can be performed in a straightforward manner, once a program is at
hand that can calculate on-shell amplitudes in a generic gauge
theory. We use \recolatwo \cite{Recola2} equipped with a model file of
the SM in the symmetric phase to evaluate the amplitudes
numerically. The model file is generated using {\sc{FeynRules}}
\cite{FeynRules} and renormalised using the in-house software
{\sc{Rept1l}} (for more details see Ref.~\cite{Rept1l}). 

Given this technical toolkit, all we have to do is to express every part of (\ref{base}) in a basis of matrix elements representing physical processes in the SySM.

The generalisation to any SM process is rather obvious: The SU(2)-transformed fields are constructed in a suited basis of the respective representation
and the matrix element reads
\begin{equation}
\mathcal{M}^{\phi_1\ldots\phi_n}_\text{SCET}(\{p\})=\sum_j\boldsymbol{A}_{1j}(\mu_\text{h},\mu_\text{l})\mathcal{M}^{\tilde{\phi}^{(j)}_{1}\ldots\tilde{\phi}^{(j)}_{n}}\label{phi1j}
\end{equation}
with $j$ running over the number of processes that can be generated by
applying SU(2) transformations to any number of pairs of external
fields of the considered process.
Thus, $\tilde{\phi}^{(j)}_{i}$ is either equal to $\phi^{(j)}_{i}$ or to $\varphi^{(j)}_{i}$ as defined in (\ref{4tildeto0}).
The matrix $\boldsymbol{A}$ incorporates the path-ordered exponential in (\ref{base}).
With a suited model file the transformed matrix elements can be
evaluated for all possible combinations of fields using \recolatwo. 

\subsubsection{High-scale matching contributions}
The high-scale matching part is particularly desirable to be
automated, as it requires process-specific loop computations. For
four-fermion processes the respective analytical calculation has been
performed in Ref.~\cite{SCET_4f}, while the results for diboson
production can be found in Ref.~\cite{qqVV_HSM}.

In the operator basis we have chosen, the one-loop matching
contributions are included simply by promoting the matrix elements in
(\ref{phi1j}) from tree level to one loop (in the SySM):
\begin{align}
\mathcal{M}^{\phi_1\ldots\phi_n}(\{p\})=\sum_j\boldsymbol{A}_{1j}(\mu_\text{h},\mu_\text{l})\left(\mathcal{M}^{\text{tree},\tilde{\phi}_{1j}\ldots\tilde{\phi}_{1n}}+\mathcal{M}^{\text{1-loop},\tilde{\phi}_{1j}\ldots\tilde{\phi}_{1n}}_\text{IR-finite}\right)
.\label{C11}
\end{align}
The quantities on the r.h.s.\ of (\ref{C11}) can again directly be calculated with \recolatwo.
It should, however, be noted that for a consistent one-loop matching all transformed processes have to be evaluated at one loop. For processes which many contributing SySM processes such as $\gamma\gamma\rightarrow \text{W}^+\text{W}^-$ with
a total of 36 processes 
this procedure is rather time consuming. 

Apart from that, the SySM parameters and fields are
$\overline{\text{MS}}$ renormalised, and the high-scale matching corrections induce a UV-scale dependence. We choose  
\begin{align}
\mu_\text{UV}=\mu_\text{h}=\sqrt{\hat{s}}.
\end{align}
\subsubsection{Anomalous dimension} 
The most important quantity in the factorisation formula (\ref{base}) is the anomalous dimension, which governs the RGE running.
The form of the SCET anomalous dimension of the SM has been analysed
in Refs.~\cite{SCET_4f, SCET_SU2} and is  at one-loop order given by Eq.~(\ref{ADfirst}).
Using charge conservation, the decomposition
$s_{ij}=\eta_{ij}(\bar{n}_i\cdot p_i)(\bar{n}_j\cdot
p_j)(n_i\cdot n_j)/2$, which holds in the high-energy limit, and the
one-loop result for the cusp anomalous dimension,
$$\Gamma_{\text{cusp},k}=\frac{\alpha_k}{\pi},$$ 
where 
\begin{equation}
\alpha_1 = \frac{\alpha}{c_\text{w}^2}, \qquad \alpha_2 = \frac{\alpha}{s_\text{w}^2},
\end{equation}
the anomalous dimension can be decomposed into a collinear (C) and a soft (S) part as
\begin{align}
\boldsymbol{\gamma}&=\gamma_\text{C}\mathbb{1} + \boldsymbol{\gamma}_\text{S},
\nonumber\\
\gamma_\text{C}&{}=\frac{\alpha}{\pi}\sum_i
  C_i\log\left(\frac{\bar{n}_i\cdot
      p_i}{\mu}\right)+\sum_i\gamma_i, \nonumber\\
\boldsymbol{\gamma}_\text{S}&{}=-\frac{\alpha}{\pi}\sum_{\left<ij\right>}
\left(\frac{1}{s_\text{w}^2}\mathbf{t}_i\cdot\mathbf{t}_j
+\frac{1}{4c_\text{w}^2}\sigma_iY_i\sigma_jY_j\mathbb{1}\right)
\log\left(-\eta_{ij}\frac{n_i\cdot n_j}{2}-\text{i}0\right).
\label{ADSM}
\end{align}
The factor $\eta_{ij}$ is defined via
\begin{align}
\eta_{ij}= \begin{cases}
\text{$\phantom{-}1$ \quad if $i$,~$j$ are both incoming or both outgoing,}\\
\text{$-1$ \quad else.}
\end{cases}\label{etaijdef}
\end{align}
This distinction is important to get the correct $\text{i}\pi$ contributions. 
From now on, the hypercharges $Y_i$, isospin quantum numbers, and electric charges $Q_i$ always
refer to the particles. The sign factors $\sigma_i$ take the values
$\sigma_i=+1$ for incoming particles and outgoing antiparticles and
$\sigma_i=-1$ for incoming antiparticles and outgoing particles. In \refeq{ADSM},
$C_i$ refers to the EW Casimir invariant of particle $i$ defined via \cite{DennerPozz1} 
\begin{align}
C_i=\sum_{V_a=A,Z,W^\pm}I^{V_a}_iI^{\bar{V}_a}_i=\frac{1}{c_\text{w}^2}\frac{Y_i^2}{4}+\frac{C^\text{SU(2)}_i}{s_\text{w}^2}.
\end{align}

\paragraph{Collinear anomalous dimension}
 The collinear part, defined in \refeq{ADSM}, is given by the sum of one-particle contributions
  \begin{align}
\gamma_\text{C}(\mu)=\sum_{i}\gamma_{\text{C},i}(\mu)
 \end{align}
and contains the leading-logarithmic contribution
completely. For a general gauge theory the collinear anomalous
dimensions for gauge bosons and fermions with label momentum $p_i$ are
given by \cite{SCET_SM}
\begin{align}
\gamma_{\text{C},A_k}=\frac{\alpha_k}{4\pi}\left(4C_{\text{A},k}\log\left(\frac{\bar{n}_i\cdot p_i}{\mu}\right)-\beta_{0,k}\right),\qquad\gamma_\text{C,f}=\sum_{k=1}^2\frac{\alpha_kC_{\text{F},k}}{4\pi}\left(4\log\left(\frac{\bar{n}_i\cdot p_i}{\mu}\right)-3\right)
\end{align}
with the Casimir invariants $C_{\text{A},k}$ and
$C_{\text{F},k}$ of the U(1) and SU(2) gauge groups, respectively. In the SM the
Casimir operators are obtained from those corresponding to the
respective SU(2) representation and
hypercharges. The values for all SM particles are collected in App.~B
of Ref.~\cite{DennerPozz1}. 
The $\beta$-function coefficients of the SU(2) and U(1) subgroup are given by \cite{DennerPozz1,SCET_SM} 
\begin{equation}
\beta_{0,1}=-\frac{41}{6},\qquad\beta_{0,2}=\frac{19}{6}.\label{beta0Def}
\end{equation}
Just like the left-handed fermions, the SM scalars are in the fundamental representation of SU(2). Their anomalous dimension has two modifications with respect to the fermion one: 
\begin{itemize}
\item There is a different collinear factor for bosons \cite{DennerPozz2}, and 
\item there is an additional contribution related to the top-quark Yukawa coupling $g_\text{t}$.
\end{itemize}
This leads to the result:
\begin{align}
\gamma_{\text{C}\Phi}=\sum_{k=1}^2\frac{\alpha_k C_{\text{F},k}}{4\pi}\left(4\log\left(\frac{\bar{n}_i\cdot p_i}{\mu}\right)-4\right)+\frac{N_\text{C}g_\text{t}^2}{16\pi^2}\label{gammaCPhi}
\end{align}
with the top-quark Yukawa coupling $g_\text{t}$. 
\paragraph{Soft anomalous dimension}
The angular-dependent logarithmic corrections to the matrix element arise from the soft anomalous dimension $\boldsymbol{\gamma}_{\text{S}}$. It is given by the non-trivial sum over pairs of external particles in (\ref{ADSM}). 
The contribution of the soft anomalous dimension to the $\text{SCET}_\text{EW}$ amplitude (\ref{base}) is obtained by forming a vector of all contributing matrix elements
\begin{align}
\begin{pmatrix}
\mathcal{M}^{\phi^{(1)}_1\ldots\phi^{(1)}_n}\\
\vdots\\
\mathcal{M}^{\phi^{(m)}_1\ldots\phi^{(m)}_n}
\end{pmatrix}\label{matvec}
\end{align}
and performing a matrix multiplication. 
The neutral gauge-boson contribution to the soft anomalous dimension is a diagonal matrix 
\begin{align}
\boldsymbol{\gamma}_\text{S,N}=-\frac{\alpha}{\pi}\text{Diag}\,(N_1,\ldots,N_m) \label{SoftNeutral}
\end{align}
with 
\begin{align}
 N_k=\sum_{\left<ij\right>}\sigma_i\sigma_j\left(\frac{Y^{(k)}_iY^{(k)}_j}{4c^2_\text{w}}+\frac{I^{3(k)}_iI^{3(k)}_j}{s_\text{w}^2}\right)\log\left(-\eta_{ij}\frac{n_i\cdot n_j}{2}-\text{i}0\right),\label{NkDef}
\end{align}
with $i,j$ running over pairs of external particles and $k$ indicating the respective transformed process.

The off-diagonal elements of $\boldsymbol{\gamma}_\text{S}$ are
obtained from the $\text{W}^\pm$ couplings of all external particles
$\phi^{(j)}_k$ in (\ref{matvec}) using (\ref{tSU2}) and (\ref{tSU2_pm}). In closed
form the off-diagonal entries can be written as
\begin{align}
\left(\boldsymbol{\gamma}_\text{S}\right)_{kl}=-\frac{\alpha}{\pi s_\text{w}^2}
\sum_{\left<ij\right>}\prod_{n\neq i,j}\delta_{\phi^{(k)}_n\phi^{(l)}_n}\sum_{a=\pm}
t^a_{\beta_i^{(k)} \beta_i^{(l)}}t^{\bar{a}}_{\beta_j^{(k)}\beta_j^{(l)}}
\log\left(-\eta_{ij}\frac{n_i\cdot n_j}{2}-\text{i}0\right)
\label{gammaSoftOff}
\end{align}
for $k\neq l$. The indices $\beta_i^{(k)}$ denote the SU(2) indices of
particle at position $i$ of the process labelled by $k$. The symbol
$\bar{a}$ denotes the sign opposite to $a$.
The expression is non-zero if W-boson couplings connect particles at positions $i$ and $j$ in the processes $k$ and $l$. The Kronecker deltas assure that the entry is 
zero, if more than one particle-pair transformation is needed to
relate the processes $k$ and $l$. Equation (\ref{gammaSoftOff}) is best understood by means of an example, which we give in the following.
\paragraph{Construction of the soft-anomalous-dimension matrix for $\bar{\text{u}}_\text{L}\text{u}_\text{L}\rightarrow\text{e}_\text{L}^+\text{e}_\text{L}^-$}
As an example, we consider the process $\bar{\text{u}}_{\text{L}} \text{u}_{\text{L}}\rightarrow \text{e}^+_{\text{L}}\text{e}^-_{{\text{L}}}$. The corrections arising from the soft anomalous
dimension can be written as
\begin{align}
\delta\mathcal{M}_{\boldsymbol{\gamma}_\text{S}}^{\bar{\text{u}}_{\text{L}} \text{u}_{{\text{L}}}\text{e}_{\text{L}}^+\text{e}_{{\text{L}}}^-}=
\exp\left(-\int_{\mu_\text{l}}^{\mu_\text{h}}\text{d}\log\mu\,\boldsymbol{\gamma}(\mu)\right)
\cdot\mathcal{M}_j
\end{align}
with
\begin{equation}
\boldsymbol{\mathcal{M}}=
\begin{pmatrix}
\mathcal{M}^{\bar{\text{u}}_{\text{L}} \text{u}_{{\text{L}}} \text{e}_{\text{L}}^+\text{e}_{{\text{L}}}^-}\\\mathcal{M}^{\bar{\text{d}}_{\text{L}} \text{d}_{{\text{L}}} \text{e}_{\text{L}}^+\text{e}_{{\text{L}}}^-}\\\mathcal{M}^{\bar{\text{d}}_{\text{L}} \text{u}_{{\text{L}}} \bar{\nu}_{\text{L}} \text{e}_{{\text{L}}}^-}\\\mathcal{M}^{\bar{\text{u}}_{\text{L}} \text{d}_{{\text{L}}} \text{e}_{\text{L}}^+\nu_{{\text{L}}}}\\
\mathcal{M}^{\bar{\text{u}}_{\text{L}} \text{u}_{{\text{L}}} \bar{\nu}_{\text{L}}\nu_{{\text{L}}}}\\\mathcal{M}^{\bar{\text{d}}_{\text{L}} \text{d}_{{\text{L}}} \bar{\nu}_{\text{L}}\nu_{{\text{L}}}}\label{vecM}
\end{pmatrix}.
\end{equation}
Note that the last entry is not obtained from the first one via a
single external-pair transformation, instead two of them are
needed. Therefore, the last matrix element does not contribute at one
loop. It does nevertheless contribute to the exponential, because it
is connected to the other operators via external-pair transformations. 

Using the Mandelstam variables, defined via 
\begin{align}
s=(p_1+p_2)^2>0,\qquad t=(p_1-p_3)^2<0,\qquad u=(p_1-p_4)^2<0,
\end{align}
with $p_1$, $p_2$ being the incoming momenta and $p_3$, $p_4$ the outgoing momenta, we define the quantities \cite{qqVV_HSM} 
\begin{align}
\mathsf{L}_s=\log\left(\frac{s}{\mu^2_\text{h}}\right)-\text{i}\pi,\qquad
\mathsf{L}_t=\log\left(\frac{-t}{\mu^2_\text{h}}\right),\qquad
\mathsf{L}_u=\log\left(\frac{-u}{\mu^2_\text{h}}\right).
\end{align}
The $\text{i}\pi$ takes care of the correct analytical continuation of the complex logarithm according to the signs of $s$, $t$, and $u$. These logarithms are used as building blocks for the  off-diagonal elements of $\boldsymbol{\gamma}_\text{S}$. Because in the given example the W couplings are all identical, the matrix that is to be exponentiated has the following form:
\begin{align}
\boldsymbol{\gamma}_\text{S}=-\frac{\alpha}{4\pi s_\text{w}^2}
\begin{pmatrix}
s_\text{w}^2N_1&\mathsf{L}_s&\mathsf{L}_t&\mathsf{L}_t&\mathsf{L}_s&0\\
\mathsf{L}_s&s_\text{w}^2N_2&\mathsf{L}_u&\mathsf{L}_u&0&\mathsf{L}_s\\
\mathsf{L}_t&\mathsf{L}_u&s_\text{w}^2N_3&0&\mathsf{L}_u&\mathsf{L}_t\\
\mathsf{L}_t&\mathsf{L}_u&0&s_\text{w}^2N_4&\mathsf{L}_u&\mathsf{L}_t\\
\mathsf{L}_s&0&\mathsf{L}_u&\mathsf{L}_u&s_\text{w}^2N_5&\mathsf{L}_s\\
0&\mathsf{L}_s&\mathsf{L}_t&\mathsf{L}_t&\mathsf{L}_s&s_\text{w}^2N_6
\end{pmatrix}
\label{SoftMat}
\end{align}
with the $N_k$ defined according to (\ref{NkDef}). The off-diagonal
entries can be read off by identifying the respective pairs of
particles that are transformed with respect to one another in the list
of processes in (\ref{vecM}). If more than two particles have to be transformed, the entry is 0.
\paragraph{Path-ordered matrix exponential}
As argued in Appendix A of Ref.~\cite{SCET_SM} the path-ordering symbol can be ignored when exponentiating, because $\gamma(\mu)$ commutes with itself for different values of $\mu$.
The key arguments are the following:
\begin{itemize}
\item The collinear anomalous dimension is proportional to the unit matrix and commutes with itself for different values of the scale.
\item The soft anomalous dimension does not depend on $\mu$ at all.  
\end{itemize}
We can therefore replace the path-ordered matrix exponential by a normal one. 
The matrix parts of the anomalous dimensions such as (\ref{SoftMat}) have to be exponentiated using
\begin{align}
\exp{\boldsymbol{A}}=\sum_{k=0}^\infty\frac{\boldsymbol{A}^k}{k!}.\label{MatExp}
\end{align}
To evaluate (\ref{MatExp}) numerically we cut off the sum at some finite order, which we choose to be $k=6$. We have checked that the impact of including the $k=6$ terms is already of the order of $10^{-5}$ with respect to the Born matrix element on the level of single phase-space points.
\paragraph{Integration over the scale}
If the running of the gauge couplings is neglected, the anomalous
dimension is a linear function of $\log\mu$ and thus easy to
integrate.  

The one-loop running can also be taken into account as in QCD with a sum over the running couplings.
Decomposing the anomalous dimension as
\begin{align}
\boldsymbol{\gamma}(\alpha_1(\mu),\alpha_2(\mu),\mu)=\frac{\alpha_1(\mu)}{\pi}
\sum_i\frac{Y_i^2}{4}\log\frac{\mu_\text{h}}{\mu}\mathbb{1}
+\frac{\alpha_2(\mu)}{\pi}\sum_i{C_i^{\text{SU(2)}}}\log\frac{\mu_\text{h}}{\mu}\mathbb{1}+\boldsymbol{\gamma_\text{non-cusp}},
\end{align}
one can solve the integral including the one-loop running \cite{gammaInt, SCET_SM},
\begin{align}
\int_{\mu_\text{l}}^{\mu_\text{h}}\text{d}\log\mu\,\boldsymbol{\gamma}(\alpha_1(\mu),\alpha_2(\mu),\mu)&=-f_0^\text{EWSM}(\alpha_1,\alpha_2)+\boldsymbol{\gamma}_\text{non-cusp}\log\left(\frac{\mu_\text{h}}{\mu_\text{l}}\right)\nonumber\\
        &=-\frac{f^\text{SU(2)}_0(z_2)}{\alpha_2(\mu_\text{h})}-\frac{f^\text{U(1)}_0(z_1)}{\alpha_1(\mu_\text{h})}+\boldsymbol{\gamma}_\text{non-cusp}\log\left(\frac{\mu_\text{h}}{\mu_\text{l}}\right),\label{intmuewsm}
\end{align}
with 
\begin{align}
z_{1/2}&=\frac{\alpha_{1/2}(\mu_\text{l})}{\alpha_{1/2}(\mu_\text{h})},\nonumber\\
f^\text{U(1)}_0(z)&=\sum_i\frac{\pi Y_i^2}{\beta_{0,1/2}^2}\left(\log z +\frac{1}{z}-1\right),\nonumber\\
f^\text{SU(2)}_0(z)&=\sum_i\frac{4\pi C_i^{\text{SU(2)}}}{\beta_{0,1/2}^2}\left(\log z +\frac{1}{z}-1\right)\label{f12},
\end{align}
with the one-loop $\beta$-function coefficients $\beta_{0,1/2}$. 
From the two-loop running on, however, the RGEs for $\alpha_1$ and $\alpha_2$ are coupled and one can not analytically perform the integration in (\ref{intmuewsm}).

\subsubsection{Low-scale corrections}
\begin{figure}
\centering
\includegraphics[page=1,scale=0.85]{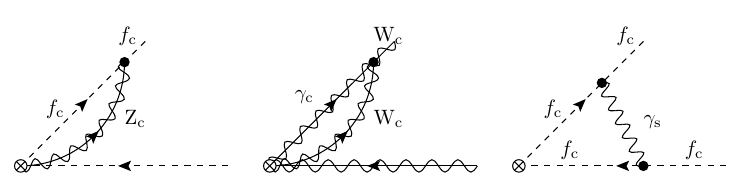}
\caption{Sample diagrams contributing to the low-scale corrections. Collinear fermions are represented by dashed arrows, collinear gauge bosons
by wiggly lines with an arrow line, and soft gauge bosons by wiggly lines.}
\label{Fig:LowScaleDiags}
\end{figure}
The low-scale corrections are obtained from the one-loop operator
matrix elements in $\text{SCET}_\text{EW}$. The respective Feynman
graphs are of the type depicted in Fig.~\ref{Fig:LowScaleDiags} with
non-zero W-, Z-, Higgs-boson and top-quark masses. The regularisation
techniques required for their calculation are described in
Refs.~\cite{SCET_4f, SCET_wo_regulator}. The W-boson exchange graphs induce
mixing between the operators, even though we are going to argue that
at one-loop level the non-trivial matrix structure can be avoided.

In a given operator basis we define the one-loop matrix structure according to (\ref{deltaMD}):
\begin{align}
\left<\mathcal{O}^{\text{1-loop}}_i\right>=\boldsymbol{D}^{(1)}_{ij}\left<\mathcal{O}^{\text{tree}}_j\right>.
\end{align}
with $\boldsymbol{D}^{(1)}$ being composed of one-particle (``collinear'') and two-particle (``soft'') contributions. 
In analogy to the collinear and soft anomalous dimension we define the collinear and soft low-scale corrections $D_\text{C}$ and $\boldsymbol{D}_\text{S}$, respectively:
\begin{align}
\boldsymbol{D}^{(1)}(\mu_\text{l})&=\sum_iD_{\text{C},i}(\mu_\text{l})\mathbb{1}+\boldsymbol{D}_\text{S},\nonumber\\
\boldsymbol{D}_\text{S}&=\sum_{\left<ij\right>}\left[\sum_{a=\pm}
\frac{\alpha}{2\pi s_\text{w}^2}\log\left(\frac{\mu^2_\text{l}}{M_\text{W}^2}\right)\boldsymbol{t}^{a}_i\cdot\boldsymbol{t}^{\bar{a}}_j 
+\frac{\alpha}{2\pi}\log\left(\frac{\mu_\text{l}^2}{M_\text{Z}^2}\right)\sigma_iI^Z_i\sigma_jI^Z_j\right.\nonumber\\
&\hspace{20pt}\left.{}+\frac{\alpha}{2\pi}\sigma_iQ_i\sigma_jQ_j\left(-\frac{1}{\varepsilon_\text{IR}}
+\log\left(\frac{\mu_\text{l}^2}{\mu_\text{IR}^2}\right)\right)
\mathbb{1}\right]\log\left(-\eta_{ij}\frac{n_i\cdot n_j}{2}-\text{i}0\right)\label{DForm}
\end{align}
with $I_i^Z$ being the Z-boson coupling of particle $i$ as defined for
fermions in (\ref{I^Z}) and more generally in Ref.~\cite{DennerPozz1}. 
Note that the collinear matching contains both loop and field-renormalisation contributions.
The fact that the low-scale $\text{SCET}_\text{EW}$ corrections have the structure (\ref{DForm}) has been shown in App.~C of Ref.~\cite{SCET_SU2}. 
In our approach, the results of Refs.~\cite{SCET_SU2, SCET_SM} are supplemented by the photon contributions proportional to $Q_iQ_j$, which are
not present in Ref.~\cite{SCET_SM}. The reason is that these results are based on a formalism, in which the $\text{SCET}_\text{EW}$ results are matched
onto a version of SCET with W, Z, Higgs bosons and top quark integrated out. The theory is called $\text{SCET}_\gamma$ and can be used to resum large logarithms
between the EW scale and a possibly much smaller factorisation scale. In $\text{SCET}_\gamma$ both the top quark and the W boson are treated as
boosted heavy quarks coupling to soft photons and gluons. The $\text{SCET}_\gamma$ loop graphs contain the IR poles in (\ref{DForm}), but no additional finite corrections. 

Because we identify the IR scale with the EW scale ($\mu_\text{IR}=\mu_\text{l}=M_\text{W}$), we 
have no need to consider running effects below $\mu_\text{l}$. Instead we can simply add the UV-finite loop corrections from $\text{SCET}_\gamma$ to 
the IR-finite $\text{SCET}_\text{EW}$ low-scale corrections to obtain
both the IR-finite ($\text{SCET}_\text{EW}$) and IR-divergent
($\text{SCET}_\gamma$) contributions.

While the soft low-scale corrections have a universal form,
the collinear low-scale corrections $D_{\text{C},i}$ depend on the spin of particle $i$. 
The results for all SM particles are collected in App.~\ref{Sec:SCFunctions}. For all external particles the correction factors contain logarithmic contributions of the form
\begin{align}
D^\text{weak}_{\text{C},i}\sim\log\frac{\bar{n}_i\cdot p_i}{\mu_\text{l}}\log\frac{M^2_\text{W/Z}}{\mu_\text{l}^2}+\ldots\label{Dweak}
\end{align}
in the weak part and of the form

\begin{align}
D^{\gamma}_{\text{C},i}\sim\log\frac{\bar{n}_i\cdot p_i}{\mu_\text{l}}\log\frac{\mu^2_\text{IR}}{\mu_\text{l}^2}+\ldots\label{Dphot}
\end{align}
in the photonic part. Recalling that $\bar{n}_i\cdot p_i$ is the large component of $p_i$, the expressions above are logarithmically enhanced in the high-energy limit. 
To obtain a maximally simple form for the low-scale corrections, we observe that, if the W and Z mass were equal, setting $\mu_\text{l}$ and $\mu_\text{IR}$ to their mass would remove all two-particle corrections. Although this is not possible, we can choose
\begin{align}
\mu_\text{IR}=\mu_\text{l}=M_\text{W}\label{Scalechoice}.
\end{align}
With this choice the logarithms in (\ref{Dphot}) are removed and the only remaining logarithms in (\ref{Dweak}) are suppressed by a factor of $\log c_\text{w}^2\approx0.25$. 

Besides this, the scale choice (\ref{Scalechoice}) has the advantage that the soft matching has no non-trivial matrix structure, as this again arises only because of the W-boson contributions in (\ref{DForm}). The one-loop soft matching $\boldsymbol{D}_\text{S}$ is then only related to Z-boson exchange and becomes a unit matrix:
\begin{align}
\boldsymbol{D}_\text{S}=D_\text{S}\mathbb{1}=\frac{\alpha}{2\pi}\sum_{\left<ij\right>}\log \left(c_\text{w}^2\right)\sigma_iI^Z_i\sigma_jI^Z_j\log\left(-\eta_{ij}\frac{n_i\cdot n_j}{2}-\text{i}0\right)\mathbb{1}.
\end{align}
The functions $D_{\text{C},i}$ can be computed once for all SM particles, and the two-particle contributions can be constructed for each process in a similar way as the soft anomalous dimension. 

Eventually the low-scale corrections (\ref{deltaMD}) are implemented as
\begin{align}
\mathcal{M}\rightarrow
\mathcal{M}+\sum_{i}D_{\text{C},i}\mathcal{M}+D_\text{S}\mathcal{M}
\label{gammaSDS}
\end{align}
on the level of the matrix element in (\ref{base}).
\paragraph{Field renormalisation and radiative corrections to the GBET}
We already mentioned that the low-scale corrections contain also field-renormalisation contributions. For fermions, transverse gauge bosons, and the Higgs boson this simply means 
that the corresponding field-renormalisation constant $\delta Z_i/2$ is added to the loop contributions:
\begin{equation}
D_{\text{C},i}=\frac 12\delta Z_i\left.\right|_{\mu_\text{UV}=\mu_\text{l}}+D_{\text{C},i}^\text{loop}.
\label{DC_FRC}
\end{equation}
In the case of longitudinal gauge bosons, the $\delta Z_i$
contribution is not present, since the fields of the unphysical
Goldstone bosons are not renormalised in our convention. Instead, one
has to account for the fact that the GBET, i.e.\ the relations
(\ref{GBET}), get perturbative corrections. Therefore, the collinear matching for longitudinal gauge bosons contains the radiative correction factors [see Eq.~(\ref{GBET})]
\begin{align}
        D_{\text{C},\text{W}_\text{L}}&=\left[-\frac{\Sigma^{WW}_\text{L}(M_\text{W}^2)}{M_\text{W}^2}-\frac{\Sigma^{W\phi}(M_\text{W}^2)}{M_\text{W}}+\frac{\delta M_\text{W}}{M_\text{W}}+\frac 12 \delta Z_\text{W}\right]_{\mu_\text{UV}=\mu_\text{l}} + D^\text{loop}_{\text{C},\text{W}_\text{L}},\nonumber\\
        D_{\text{C},\text{Z}_\text{L}}&=\left[-\frac{\Sigma^{ZZ}_\text{L}(M_\text{Z}^2)}{M_\text{Z}^2}+\text{i}\frac{\Sigma^{Z\chi}(M_\text{Z}^2)}{M_\text{Z}}+\frac{\delta M_\text{Z}}{M_\text{Z}}+\frac 12 \delta Z_\text{ZZ}\right]_{\mu_\text{UV}=\mu_\text{l}} + D^\text{loop}_{\text{C},\text{Z}_\text{L}}.
\label{GBET1LOOP}
\end{align}

\subsubsection{Mixing at the low scale}
\label{se:low-scale mixing}
Since we work in a basis of operators involving fields that are charge eigenstates, there is a one-to-one correspondence of fields in the physical basis at the low scale and fields in the symmetric basis at the high scale for all fermions and for the $\text{W}^\pm$ bosons. 
 For processes with external photons, Z bosons, or Higgs bosons this is not true. A mixing transformation has to be introduced, which we discuss in this section.
\paragraph{$\gamma$/Z mixing}
If we want to compute a process involving a photon or Z boson, the anomalous dimension and high-scale matching are conveniently expressed in terms of processes involving $\text{W}^3$ and B bosons. For these processes all aforementioned steps can be applied using the formulae in the previous sections. However, the low-scale contributions $\boldsymbol{D}$ have to be calculated for mass eigenstates. One has to apply a forth-and-back transformation:
\begin{itemize}
\item Before (\ref{base}) can be applied, a matrix element associated with external photons or Z bosons has to be decomposed 
into subamplitudes containing the SU(2) eigenstates $\text{W}^3/\text{B}$. 
In the description of (\ref{chidef}) we already stated that the $\text{SCET}_\text{EW}$ 
operators and Wilson lines are given in terms of gauge eigenstates.
\item The anomalous dimension, the running, and the high-scale
  matching are calculated in this basis. Each subamplitude obtains its
  own correction factor, which depends only on the weak isospin.
\item Finally, the subamplitudes are transformed back into mass
  eigenstates and the low-scale corrections are calculated. They
  depend both on the weak isospin and the mass eigenstate. 
\end{itemize} 
We start from a physical matrix element with photons and/or Z bosons. Repeatedly applying the transformation
\begin{align}
\begin{pmatrix}
A_\mu\\Z_\mu
\end{pmatrix}
=
\begin{pmatrix}
c_\text{w}&-s_\text{w}\\
s_\text{w}&c_\text{w}
\end{pmatrix}
\begin{pmatrix}
B_\mu\\W^3_\mu
\end{pmatrix},
\end{align}
one obtains for a matrix element involving $n_\gamma$ photons and $n_\text{Z}$ Z bosons:
\begin{align}
\mathcal{M}^{n_\gamma n_\text{Z}}=
\sum_{n_\text{W}^{(\gamma)}+n_\text{B}^{(\gamma)}=n_\gamma }\sum_{n_\text{W}^{(\text{Z})}+n_\text{B}^{(\text{Z})}
=n_\text{Z}}(-s_\text{w})^{n_\text{W}^{(\gamma)}}c_\text{w}^{n_\text{B}^{(\gamma)}}s_\text{w}^{n_\text{B}^{(\text{Z})}}c_\text{w}^{n_\text{W}^{(\text{Z})}}\mathcal{M}^{n_\text{W}^{(\gamma)}n_\text{B}^{(\gamma)}n^{(\text{Z})}_\text{B}n^{(\text{Z})}_\text{W}}.
\end{align}
The high-scale matching and the running via the anomalous dimensions are computed for each
contribution $\mathcal{M}^{n_\text{W}^{(\gamma)}n_\text{B}^{(\gamma)}n^{(\text{Z})}_\text{B}n^{(\text{Z})}_\text{W}}$ separately. In particular, there is no cross talk between the different $\mathcal{M}^{n_\text{W}^{(\gamma)}n_\text{B}^{(\gamma)}n^{(\text{Z})}_\text{B}n^{(\text{Z})}_\text{W}}$ due to the soft anomalous dimension and soft matching, since the respective matrices are block diagonal.

The low-scale $\text{SCET}_\text{EW}$ corrections depend both on the SU(2) structure, which is determined by the $W^3/B$ field in the operator, and on the external mass, which is determined by the external momentum \cite{SCET_SM}. This leads to the following expression for the one-particle low-scale corrections associated with the photons and Z bosons:
\begin{align}
\delta \mathcal{M}^{n_\text{W}^{(\gamma)}n_\text{B}^{(\gamma)}n^{(\text{Z})}_\text{B}n^{(\text{Z})}_\text{W}}=\hspace{300pt}\nonumber\\
=\left(n_\text{W}^{(\text{Z})}D_\text{C}^{W^3\rightarrow Z}+n_\text{W}^{(\gamma)}D_\text{C}^{W^3\rightarrow \gamma}+n_\text{B}^{(\text{Z})}D_\text{C}^{B\rightarrow Z}+n_\text{B}^{(\gamma)}D_\text{C}^{B\rightarrow \gamma}\right)\mathcal{M}^{n_\text{W}^{(\gamma)}n_\text{B}^{(\gamma)}n^{(\text{Z})}_\text{B}n^{(\text{Z})}_\text{W}}
\end{align}   
with the various $D_\text{C}$ factors collected in App.~\ref{Sec:SCFunctions}.

The formula for the two-particle contributions in (\ref{DForm}) can be
applied in a straightforward manner on each subamplitude. Of course
only the external $W^3$ fields in each operator get transformed but
not the external $B$ fields. 
\paragraph{Z/Higgs mixing}
The same strategy is applied for processes involving Higgs bosons or longitudinal Z-boson modes. The latter are represented by the neutral would-be Goldstone boson $\chi$, which is the imaginary part of the lower component field of the Higgs doublet $\phi_2$. This is an $I_3$  and hypercharge eigenstate and therefore the natural choice for the construction of operators in the SySM. The transformation reads  
\begin{align}
\begin{pmatrix}
\eta\\\chi
\end{pmatrix}
=\begin{pmatrix}
\frac{1}{\sqrt{2}}&\frac{1}{\sqrt{2}}\\
-\frac{\text{i}}{\sqrt{2}}&\frac{\text{i}}{\sqrt{2}}
\end{pmatrix}
\begin{pmatrix}
\phi_2\\\phi_2^*
\end{pmatrix}.
\end{align}
Thus, a matrix element with $n_\eta$ Higgs bosons and $n_\chi$ longitudinally polarised Z bosons has the hypercharge eigenstate decomposition 
\begin{align}
\mathcal{M}^{n_\eta n_{\chi}}=\hspace{350pt}\nonumber\\
=\sum_{n_{\phi_2}^{(\eta)}+n_{\phi_2^*}^{(\eta)}=n_\eta}\sum_{n_{\phi_2}^{(\chi)}+n_{\phi_2^*}^{(\chi)}=n_{\chi}}\left(\frac{1}{\sqrt{2}} \right)^{n_{\phi_2}^{(\eta)}} \left(\frac{1}{\sqrt{2}} \right)^{n_{\phi_2^*}^{(\eta)}} \left(\frac{\text{i}}{\sqrt{2}} \right)^{n_{\phi_2}^{(\chi)}} \left(\frac {-\text{i}}{\sqrt{2}} \right)^{n_{\phi_2^*}^{(\chi)}} \mathcal{M}^{n_{\phi_{2}}^{(\eta)}n_{\phi_2^*}^{(\eta)}n^{(\chi)}_{\phi_2^*}n^{(\chi)}_{\phi_2}}.
\end{align} 
The situation is simplified by the fact that $\phi_2$ and $\phi_2^*$ do not get different low-scale corrections. Thus each subamplitude receives the correction
\begin{align}   
\delta\mathcal{M}^{n_{\phi_{2}}^{(\eta)}n_{\phi_2^*}^{(\eta)}n^{(\chi)}_{\phi_2^*}n^{(\chi)}_{\phi_2}}=\left(n_\eta D_\text{C}^{\phi\rightarrow\eta}+n_\chi D_\text{C}^{\phi\rightarrow\text{Z}_\text{L}}\right)\mathcal{M}^{n_{\phi_{2}}^{(\eta)}n_{\phi_2^*}^{(\eta)}n^{(\chi)}_{\phi_2^*}n^{(\chi)}_{\phi_2}}.
\end{align}
\subsubsection{Coupling renormalisation constants}
The last missing contribution needed to match the $\text{SCET}_\text{EW}$ matrix element against the one in the full theory is the contribution associated with the coupling-constant renormalisation.
For the renormalisation of the coupling constants as well as the weak mixing angle we adopt the on-shell scheme. This is in contrast to the approach in Refs.~\cite{SCET_4f, SCET_SU2}, in which a scheme with a running electromagnetic coupling is employed. The pros and cons are:
\begin{itemize}
\item The logarithmic corrections associated with the running are not resummed if the on-shell scheme is used. Strictly speaking, an RGE-improved result beyond LL accuracy is not possible in the on-shell scheme. 
\item Within the on-shell scheme the $G_\text{F}$ input scheme can be employed in order to use the decay constant of the muon as an input value. This is one of the most precisely measured quantities in particle physics.  
\end{itemize} 
If not stated otherwise we stick to the on-shell scheme and include the logarithmically enhanced corrections perturbatively. 
Because the matching is calculated in the SySM, we introduce renormalisation constants associated with the U(1) and SU(2) coupling constants. They can be related to the usual SM renormalisation constants in an elementary way,
\begin{align}
g_1&=\frac{e}{c_\text{w}}\quad\rightarrow\quad\frac{\delta g_1}{g_1}=\frac{\delta e}{e}-\frac{\delta c_\text{w}}{c_\text{w}},\nonumber\\
g_2&=\frac{e}{s_\text{w}}\quad\rightarrow\quad\frac{\delta g_2}{g_2}=\frac{\delta e}{e}-\frac{\delta s_\text{w}}{s_\text{w}},
\end{align}
with $\delta s_\text{w}$ and $\delta c_\text{w}$ being the on-shell
renormalisation constants associated with the sine and cosine of the
weak mixing angle. 
In the following we divide the calculation into two separate contributions: The logarithmically enhanced corrections and the finite remainder. The coefficients for the logarithms 
are related to the UV poles and can therefore be obtained from the RGE. They arise, because the UV scale identified with $\mu_\text{h}$ is much larger
than the EW scale.
After the logarithmic part is split off, the finite part is simply obtained by setting $\mu_\text{UV}=\mu_\text{l}$ in the analytic expressions for the counterterms.  
Of course these quantities have to be computed only once for each coupling. 
For some distributions we do study the influence of the running coupling, in which case the logarithmic part of $\delta_\text{PR}$ has to be set to zero.
\paragraph{Logarithmic part}
The coefficients of the logarithmically enhanced corrections can be
determined by the $\beta$-function coefficients in Eq.~(\ref{beta0Def}) which are calculated from the self energies of the associated gauge bosons according to
\begin{align}
\beta_{0,1}&=-\frac13\sum_{\varphi=\phi^{1/2},\, f^{1/2}_\text{L},\,  f_\text{R}}\frac{\eta_\varphi Y_\varphi^2}{4},\nonumber\\
\beta_{0,2}&=\frac{11}{3}C_A-\frac13\sum_{\varphi=\Phi,f_\text{L}}\frac{\eta_\varphi}{2},
\end{align}
with $\Phi$ denoting the Higgs doublet with components $\phi^{1/2}$, $f_\text{L}$ all left-handed doublets with components $f_\text{L}^{1/2}$, and the $f_\text{R}$ the right-handed singlets.  The $\eta$~factors read $\eta_\Phi=1$ and $\eta_{f_\text{L}}=\eta_{f_\text{R}}=2$. Summation over the quark colours is implied. 

Taking the renormalisation of the coupling constant and the weak mixing angle into account in a consistent way requires the decomposition of any SySM amplitude according to their respective power in the couplings. Using $\alpha_i=g_i^2/(4\pi)$, any subamplitude $\mathcal{M}_{n_1n_2}$ proportional to $g_1^{n_1}g_2^{n_2}$ thus receives logarithmic corrections   
\begin{align}
\delta\mathcal{M}^\text{log}_{n_1n_2}=\delta^{\text{log}}_{\text{PR},n_1n_2}\mathcal{M}_{n_1n_2}=
-\left(n_1\frac{\alpha_1}{4\pi}\beta_0^\text{U(1)}+n_2\frac{\alpha_2}{4\pi}\beta_0^\text{SU(2)}\right)\log\frac{\mu_\text{h}^2}{\mu_\text{l}^2}\mathcal{M}_{n_1n_2}
\end{align}
from the respective counterterm contributions. 
Both the finite part of the field-renormalisation constants (\ref{DC_FRC}) and the radiative corrections to the GBET (\ref{GBET1LOOP}) are calculated using the one-loop library \collier~\cite{CollierMan}.  
\paragraph{Finite part}
In addition, there are finite remainders 
\begin{align}
\delta\mathcal{M}^\text{fin}_{n_1n_2}=\delta_{\text{PR},n_1n_2}^{\text{fin}}\mathcal{M}_{n_1n_2}=\left(\left(n_1+n_2\right)\frac{\delta
    e}{e}-n_1\left.\frac{\delta
      c_\text{w}}{c_\text{w}}-n_2\frac{\delta
      s_\text{w}}{s_\text{w}}\right)\right|_{\mu_\text{UV}=\mu_\text{l}}\mathcal{M}_{n_1n_2}.
\label{n1n2finite}
\end{align}
These as well as the charge renormalisation
constant $\delta e$ have to be calculated in the broken phase of the SM.

Finally, the contribution $\delta_\text{PR}$ in (\ref{base}) is obtained as the sum of the logarithmically enhanced and the finite corrections. 
The same methods can be applied for processes involving the top-quark Yukawa coupling
or the quartic Higgs coupling at tree level. 

\subsubsection{Decay corrections}
So far, we have discussed the 
contributions to the production process that are treated with $\text{SCET}_\text{EW}$. The corrections to the full processes (\ref{BBprime}) require also the NLO corrections to the decay of the bosons $V,\,V'$. They do not contain large logarithms, but have to be evaluated using the full mass dependence (at least as far as $M_\text{W/Z/H/t}$ are concerned). To this end we use a second instance of \recolaone that works within the SM. It is, however, not required to evaluate the decay processes at NLO for every phase-space point.

For a given set of momenta $\{p\}$ in the full (production and
decay) process the corrections to the squared matrix element read (remember
that interference contributions are neglected, and we suppress the
non-factorisable corrections for simplicity here)
\begin{align}
\delta|\mathcal{M}_\lambda(\{p\})|^2=\delta|\mathcal{M}_{\lambda,\text{prod}}(\{p\})|^2|\mathcal{M}_{\lambda,\text{LO,\,Dec}}(\{p\})|^2+|\mathcal{M}_{\lambda,\text{LO,\,prod}}(\{p\})|^2\delta|\mathcal{M}_{\lambda,\text{Dec}}(\{p\})|^2.\label{deltaMProdDec}
\end{align}
In the following, we argue that $\delta|\mathcal{M}_{\lambda,\text{Dec}}(\{p\})|^2$ can be constructed using $\{p\}$-independent building blocks, which have to be evaluated at NLO and some $\{p\}$-dependent tree-level quantities.
Writing the square of the decay matrix element of a boson of spin $\lambda$ into two massless fermions with helicities $s,s'$ as
\begin{align}
|\mathcal{M}_\lambda^{ss'}|^2=|\mathcal{M}_{\lambda,\text{LO}}^{ss'}|^2\bigl(1+2\delta_\lambda^{ss'}+\mathcal{O}(\alpha^2)\bigr),
\end{align}
we can use the following observations:
\begin{itemize}
\item The correction factor $\delta_\lambda^{ss'}=\delta^{ss'}$ does not depend on $\lambda$, because the polarisation definitions are ambiguous in the rest frame.
\item Moreover, $\delta^{ss'}$ does not depend on momenta and can be calculated in the rest frame of the boson once and for all.
\end{itemize}
While the first point is obvious, the second one requires justification. We demonstrate it by means of the Z-boson decay.
Introducing the chiral projection operators
\begin{align}
\omega_\pm=\frac{\mathbb{1}\pm\gamma_5}{2},
\end{align}
we can write the LO matrix element for the decay of a Z boson with momentum $k$ into two massless left-handed ($-$) or right-handed ($+$) fermions  with momenta $p$ and $q$,
\begin{align}
\mathcal{M}^\pm_0=\bar{u}(p)\omega_\mp\slashed{\epsilon}(k)(g_f^-\omega_-+g_f^+\omega_+)\omega_\pm v(q),
\end{align}
with the left-handed and right-handed form factors $g^-_f$ and
$g^+_f$. At one-loop $g^+_f$ and $g^-_f$ receive different correction
factors $\delta{g^+_f}$ and $\delta{g^-_f}$. 
Using $\omega_\pm\slashed{\epsilon}=\slashed{\epsilon}\omega_\mp$, we obtain for the ratio between the one-loop and tree-level matrix element: 
\begin{align}
\delta^{\pm\mp}=\frac{\mathcal{M}^\pm_1}{\mathcal{M}^\pm_0}=\frac{\bar{u}(p)\slashed{\epsilon}(k)\omega_\pm(\delta g_f^-\omega_-+\delta g_f^+\omega_+)\omega_\pm v(q)}
{\bar{u}(p)\slashed{\epsilon}(k)\omega_\pm(g_f^-\omega_-+g_f^+\omega_+)\omega_\pm v(q)}=\frac{\delta g_f^\pm}{g_f^\pm}.
\end{align}
The last expression depends only on masses but not on any angles and can be calculated only once for each external spin configuration $ss'=\pm\mp$. For massless fermions the other configurations do not contribute.  

Using this, the $\delta|\mathcal{M}_{\lambda,\text{Dec}}(\{p\})|^2$ in (\ref{deltaMProdDec}) are obtained as
\begin{align}
\delta|\mathcal{M}_{\lambda,\text{Dec}}(\{p\})|^2=\frac{\sum_{ss'}2\delta^{ss'}|\mathcal{M}_{\lambda,\text{LO}}^{ss'}(\{p\})|^2}{\sum_{ss'}|\mathcal{M}_{\lambda,\text{LO}}^{ss'}(\{p\})|^2}
\label{deltaMDec}
\end{align}
and thus depend on the external momenta only via tree-level matrix
elements. Therefore, we calculate $\delta^{ss'}$ before starting the actual integration and simply add the second term in (\ref{deltaMProdDec}) with $\delta|\mathcal{M}_{\lambda,\text{Dec}}(\{p\})|^2$ obtained as above for each phase-space point. 

We note that our method is independent of this particular feature and can also be
applied for processes with more general decays. Then, however, the decay
corrections need to be calculated for each phase-space point anew.

\subsection{Logarithm counting}\label{Sec:LogCount}
In this section, we describe the counting of large logarithms
within $\text{SCET}_\text{EW}$ and fixed-order computations and
specify the sets of terms that we include in our calculations. It is important to be aware of the rather disparate conventions in the $\text{SCET}_\text{EW}$ and the fixed-order literature. An extensive discussion on different logarithm-counting schemes in QCD and SCET can be found in Ref.~\cite{LogCount}. 
\paragraph{Which terms are present?}
The occurring contributions in any SM scattering amplitude computed in fixed-order perturbation theory can schematically be arranged as \cite{SCET_4f}  
\begin{align}
\mathcal{M}=
\begin{pmatrix}
1&&&&\\
\alpha\mathsf{L}^2 &\alpha\mathsf{L} &\alpha&&\\
\alpha^2\mathsf{L}^4 &\alpha^2\mathsf{L}^3 &\alpha^2\mathsf{L}^2&\alpha^2\mathsf{L}&\alpha^2\\
\alpha^3\mathsf{L}^6 &\alpha^3\mathsf{L}^5&\ldots&&&\\
\vdots&&&&
\end{pmatrix}\label{FOLogCount}
\end{align}
with 
\begin{equation}
\mathsf{L}=\log\left(\frac{s}{M^2_\text{W}}\right).
\end{equation}
In fixed-order computations, the first column of (\ref{FOLogCount}) is  commonly referred to as the leading-logarithmic ($\text{LL}_\text{FO}$), the second one as the next-to-leading logarithmic ($\text{NLL}_\text{FO}$), and the $n$-th column as the $\text{N}^{(n-1)}\text{LL}_\text{FO}$ contribution. 
If the $\text{SCET}_\text{EW}$ approach is applied, the scattering amplitude is obtained as an exponential. Because it can completely be decomposed into sum-over-pair contributions, the expansion for its logarithm is the same as the result for the Sudakov form factor obtained in Ref.~\cite{CollinsSudakov}:
\begin{align}
\log\mathcal{M}=
\begin{pmatrix}
\alpha\mathsf{L}^2 &\alpha\mathsf{L} &\alpha&\\
\alpha^2\mathsf{L}^3 &\alpha^2\mathsf{L}^2 &\alpha^2\mathsf{L}^1&\alpha^2\\
\alpha^3\mathsf{L}^4 &\alpha^3\mathsf{L}^3&\ldots&&\\
\vdots&&&&
\end{pmatrix}\label{EFTLogCount}
\end{align}
with the first column(s) again being defined as the $\text{LL}_\text{SCET}$, $\text{NLL}_\text{SCET}$, \ldots, $\text{N}^{(n-1)}\text{LL}_\text{SCET}$ contribution. These two logarithm-counting schemes differ by subleading contributions: Exponentiating the first column of (\ref{EFTLogCount}) does not only reproduce the first column of (\ref{FOLogCount}) but additional subleading terms that are related to the running of the coupling constants $\alpha_1$, $\alpha_2$. 

Furthermore, one should note that in order to fix precisely which terms to include at which order in the calculation one needs to know the hierarchy between $\alpha$ and $\mathsf{L}$. In Ref.~\cite{SCET_4f} this is sketched for two cases: 
The relevant one for EW corrections in the TeV range is the $\text{LL}^2$ regime, where $\alpha\mathsf{L}^2=\mathcal{O}(1)$,\footnote{The relation $\alpha\mathsf{L}^2=\mathcal{O}(1)$ can of course only provide a rough estimate, since $\mathsf{L}$
is a phase-space dependent quantity.} naively corresponding to 
\begin{align}
\sqrt{s}\approx27\,\text{TeV}.
\end{align}
One has, however, to keep in mind that finite prefactors in front of $\mathsf{L}$ can push this value down to energy scales in the range of a few TeV. It is therefore to be expected that for instance the CLIC collider accesses this regime. 
When $\alpha\mathsf{L}^2=\mathcal{O}(1)$, the terms in (\ref{FOLogCount}), (\ref{EFTLogCount}) are of the orders of magnitude
\begin{align}
\mathcal{M}=\begin{pmatrix}
                1&&&&\\
                1&\alpha^{1/2}&\alpha&&\\
                1&\alpha^{1/2}&\alpha&\alpha^{3/2}&\alpha^2\\
                1&&&\ldots&&\\
                \vdots &&&&
                \end{pmatrix}
,\qquad
\log\mathcal{M}=\begin{pmatrix}
                1&\alpha^{1/2}&\alpha&&\\
                \alpha^{1/2}&\alpha&\alpha^{3/2}&\alpha^2&\\
                \alpha&\alpha^{3/2}&\alpha^2&\alpha^{5/2}&\alpha^3\\
                \alpha^{3/2}&&\ldots&&\\
                \vdots &&&&
                \end{pmatrix}
.
\end{align}
Here, the first column in $\mathcal{M}$ has to be resummed, while
terms of $\mathcal{O}(\alpha^{1/2})$ and $\mathcal{O}(\alpha)$ have to be included at least perturbatively. A resummation of the $\alpha^{1/2}$ terms may also be necessary to achieve high accuracy. Note also that these numbers provide merely a vague order of magnitude (the actual corrections depend heavily on the finite prefactors such as $s_\text{w}$, $4\pi$, the Casimir operators and more). 
\paragraph{Which terms do we include?}
To investigate the impact of the respective grades of resummation we
would like to define a LL resummation scheme, which includes the
single-logarithmic terms [$\mathcal{O}(\alpha^{1/2})$ in the
$\text{LL}^2$ regime] perturbatively. This is ambiguous,
because it depends on whether these terms are included via
(\ref{FOLogCount}) or (\ref{EFTLogCount}). To make this difference
more explicit, consider the exponentiated form of the first row of (\ref{EFTLogCount}): 
\begin{align}\label{eq:Mexp}
\mathcal{M}=\exp\bigl(\alpha\mathsf{L}^2+\alpha\mathsf{L}+\alpha+\mathcal{O}(\alpha^2)\bigr)=\exp(\alpha\mathsf{L}^2)\exp\bigl(\alpha\mathsf{L}+\alpha+\mathcal{O}(\alpha^2)\bigr).
\end{align}
Consistently expanding the whole expression in $\alpha$ reproduces the terms in (\ref{FOLogCount}) order by order. There are two possibilities to resum the leading term while including the $\alpha\mathsf{L}$  and $\alpha$ terms perturbatively:
\begin{itemize}
\item Expand the second factor on the r.h.s.\ in (\ref{eq:Mexp}) to $\mathcal{O}(\alpha)$: This results in 
\begin{align}
\mathcal{M}_{\text{NLL}_\text{FO}}=\exp(\alpha\mathsf{L}^2)(1+\alpha\mathsf{L}+\alpha).\label{LL1}
\end{align}
Note that this includes the first two columns of the matrix in (\ref{FOLogCount}) and is therefore referred to as $\text{NLL}_\text{FO}$. 
\item Set the second exponential to 1 and add the $\alpha\mathsf{L}$ and $\alpha$-terms directly from (\ref{FOLogCount}): 
\begin{align}
\mathcal{M}_\text{LL}=\exp(\alpha\mathsf{L}^2)+\alpha\mathsf{L}+\alpha.\label{LL2}
\end{align}
\end{itemize}
The difference between the two formulae is subleading [of $\mathcal{O}(\alpha^2\mathsf{L}^3)$], but may still be sizeable. While (\ref{LL1}) can be expected to yield more precise predictions, (\ref{LL2}) can be used to study the impact
of the LL resummation, because it differs from the fixed-order NLO results only by means of the double-log resummation.\footnote{This is not entirely true: 
Another source of deviation between fixed order and LL+NLO in 
our implementation are the NLO contributions
from the transformed processes in the factorisation formula. This effect is, however, 
of $\mathcal{O}(\alpha^2\mathsf{L})$ compared to $\mathcal{O}(\alpha^2\mathsf{L}^4)$ for the double-log resummation.}
We therefore investigate the following combinations of contributions:
\begin{align}
\text{LL}+\text{NLO}&{}=\exp(\alpha\mathsf{L^2})+\alpha\mathsf{L}+\alpha\nonumber,\\
\text{NLL}_\text{FO}+\text{NLO}&{}=\exp(\alpha\mathsf{L^2})(1+\alpha\mathsf{L})+\alpha\nonumber,\\
\text{NLL}+\text{NLO}&{}=\exp(\alpha\mathsf{L^2}+\alpha\mathsf{L})+\alpha.
\label{LogCountings}
\end{align}
The supplement ``+NLO" refers to the included $\mathcal{O}(\alpha)$ terms.  
In the last case we take into account the first row of
(\ref{EFTLogCount}), i.e.\ the most important neglected terms are the $\alpha^2\mathsf{L}^3$, $\alpha^3\mathsf{L}^4$, and $\alpha^2\mathsf{L}^2$ term in (\ref{EFTLogCount}).
The former two are associated with the running of the EW couplings and are potentially sizable, which is why we define
\begin{align}
\text{LL}+\text{NLO}+\text{running}&=\exp\left(f^\text{EWSM}_0(\alpha_1,\alpha_2)\right)+\alpha\mathsf{L}+\alpha,\label{RunningDef}
\end{align}
with $f^\text{EWSM}_0(\alpha_1,\alpha_2)$ defined in (\ref{intmuewsm}). This differs from the LL\,+\,NLO case only by resumming the PR logarithms. 

The $\alpha^2\mathsf{L}^2$ term is associated with 
the two-loop anomalous dimension, which is rather involved due to the mixing of the several coupling constants of the SM. Its impact has been estimated to be $<1\%$ at $\sqrt{s}=4\,\text{TeV}$ in Ref.~\cite{SCET_4f}. 
We neglect it in the following.

\subsection{Technical setup}
\label{Sec:Technic}
\begin{figure}
\centering
\includegraphics[page=3,scale=0.75]{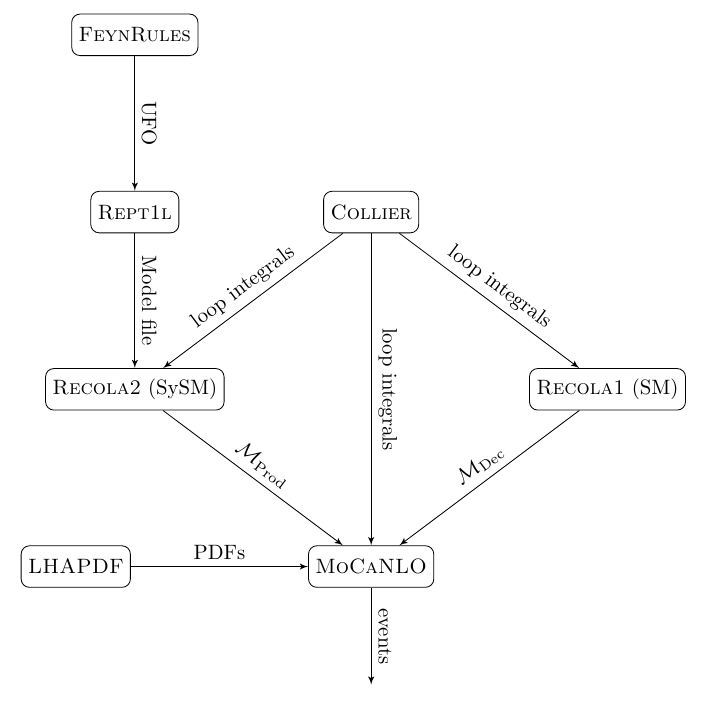}
\caption{Used software and dependencies in the setup.}
\label{Fig:Flowchart}
\end{figure}
In the previous section we described in detail the implementation of all ingredients of the $\text{SCET}_\text{EW}$ computation. To obtain numerical predictions for collider observables they have been implemented in the integrator \mocanlo.
\mocanlo is an in-house multichannel MC integration program that can calculate NLO QCD+EW cross sections on the level of weighted events to (in principle) arbitrarily complicated SM processes.
It provides both off-shell and pole-approximated results and has
been used for the computation of full NLO corrections to processes such as
vector-boson scattering \cite{VBS_offshell_ZZ} or $\text{t}\bar{\text{t}}\PW$ production~\cite{ttW}.

The programs used for the different parts of the computation and their dependencies are collected in the flowchart Fig.~\ref{Fig:Flowchart}.  {\sc{LHAPDF}} \cite{LHAPDF} is of course only used if protons in the initial state are considered.
The most delicate issue from the interface point of view is the usage of \collier: It is used for the decay correction by \recolaone, for the one-loop high-scale matching by \recolatwo and finally explicitly by \mocanlo to evaluate the loop-integrals in the low-scale matching matrix. 
Since we can calculate the relative decay corrections
  independently of the phase-space point, we can use  \recolaone,
  \recolatwo, and  \collier out of the box.%
  \footnote{We note that for processes that require the
      calculation of decay corrections for each phase-space point, the
      technical setup has to be extended. Either the decay
      corrections are calculated by a code other than \recola and
      \collier, or one uses more than one instance of \collier in
      parallel, or one needs to tune the cache system of \collier.}

This approach requires some care, but the given resources are exploited in an efficient manner.

\section{Results}
\label{Sec:Results}
\subsection{Numerical input}
We use the SM input parameters
\begin{align}
M_\text{W}^\text{OS}&=80.385\,\text{GeV},
&\Gamma^\text{OS}_\text{W}&=2.085\,\text{GeV},
\nonumber\\
M^\text{OS}_\text{Z}&=91.1876\,\text{GeV},
&\Gamma^\text{OS}_\text{Z}&=2.4952\,\text{GeV},
\nonumber\\
M_\text{H}&=125\,\text{GeV},
&\Gamma_\text{H}&=4.07\cdot10^{-3}\,\text{GeV},\label{numint}
\nonumber\\
m_\text{t}&=173.21\,\text{GeV},
&\Gamma_\text{t}&=0\,\text{GeV},\notag\\
G_\text{F}&=1.166138\cdot10^{-5}\,\text{GeV}^{-2}.
\end{align}
Note that the pole masses and widths that are employed as input parameters within the complex-mass scheme are related to the on-shell quantities via \cite{BardinMasses}
\begin{align}
M_V^\text{pole}=\frac{\left(M^\text{OS}_V\right)^2}
{\sqrt{\left({M_V^\text{OS}}\right)^2+\left({\Gamma_V^\text{OS}}\right)^2}},\qquad
\Gamma_V^\text{pole}=\frac{\Gamma^\text{OS}_VM^\text{OS}_V}
{\sqrt{\left({M_V^\text{OS}}\right)^2+\left({\Gamma_V^\text{OS}}\right)^2}}.\label{MassDefs}
\end{align}
As we use the Fermi constant as an input parameter, the EW coupling constant $\alpha$ is a derived quantity and is expressed in terms of the input as follows:
\begin{align}
\alpha=\frac{\sqrt{2}G_\text{F}M_\text{W}^2}{\pi}\left(1-\left(\frac{M^\text{pole}_\text{W}}{M^\text{pole}_\text{Z}}\right)^2\right).\label{AlphaDef}
\end{align}
For calculations involving initial-state protons we rely on the NNPDF3.1QED Parton-Distribu\-tion-Function (PDF) set \cite{NNPDF31} within the {\sc{LHAPDF}}
framework \cite{LHAPDF}. For the strong coupling constant, the number
of flavours, and the factorisation scale, which enter only via the
PDFs, we employ
\begin{align}
\alpha_\text{s}\left(M^\text{pole}_\text{Z}\right)=0.118, \qquad N_\text{f}=5, \qquad \mu_\text{F}=M_\text{W}^\text{pole}.
\end{align}
Throughout all calculations we neglect flavour-mixing effects and set
the quark mixing matrix to unity. This allows us to directly sum production channels involving quarks of different generations via PDF merging.

\subsection{Results for the CLIC collider}
The CLIC project aims at a new level of experimental precision in high-energy $\text{e}^+\text{e}^-$ collisions \cite{CLIC_proposal, CLIC_machine, CLIC_Lumi}. In several stages it is planned to operate at collision energies up to 
\begin{equation}
\sqrt{s}=3\,\text{TeV},
\end{equation}
which we assume in the following.
The fact that the bulk of interactions takes place at very high energies makes a high-energy lepton collider a particularly well-suited case of application for $\text{SCET}_\text{EW}$. There is, however, a number of questions related to how observables have to be defined when leptons with such a large energy interact. In order to demonstrate the effect of $\text{SCET}_\text{EW}$ we have to make some assumptions, of which a detailed discussion is beyond the scope of this work.
\begin{itemize}
\item The effects of initial-state radiation, which appear to be challenging for future lepton colliders (see for instance Ref.~\cite{ISR_epem} for details) are treated perturbatively. Thus, we do not use lepton PDFs but assume that only $\text{e}^+\text{e}^-$ pairs of exactly $3\,\text{TeV}$ contribute at LO. The occurring collinear singularities are regulated by the electron mass, leading to logarithmic contributions of the form
\begin{align}
\delta_\text{coll}=\frac{\alpha}{\pi}\left(\log\frac{s}{m^2_\text{e}}-1\right)\approx 7\%, \label{deltacoll}
\end{align}
assuming an electron mass of
\begin{align}
m_\text{e}=5.11\cdot10^{-4}\,\text{GeV}.
\end{align}
These logarithms remain unresummed within our framework. However,
resummation techniques for these collinear logarithms have been available for a long time \cite{ISR_resum_Kuraev, ISR_resum_review, ISR_resum_Cacciari, EW_review}. For precise predictions in lepton collisions at very high energies the inclusion of lepton-PDF effects is necessary. Recently, results for high-energy lepton PDFs including initial-state radiation of all SM particles have been published \cite{LeptonPDFs_SM}.
\item We assume all leptons to be distinguishable if their pair invariant mass is above $10\,\text{GeV}$ (the numerical value is inspired by LHC analyses). In particular we do not include corrections associated with real emission of massive gauge bosons or their decay products.
\end{itemize}
We consider two relevant special cases of (\ref{BBprime}) for $\text{e}^+\text{e}^-$ collisions: 
\begin{align}
&\eewwllll,
\label{eeWW}\\
&\eezzllll
\label{eeZZ}.
\end{align}
Processes involving $\tau$~leptons in the final state are not the phenomenologically most interesting ones:
\begin{itemize}
\item The process (\ref{eeWW}) is usually not considered in
  experimental analyses of $\text{W}^+\text{W}^-$ production, because
  the $\tau$ lepton has to be reconstructed via its decay products, which involve another W boson.
\item The process (\ref{eeZZ}) suffers from low statistics: As the branching fraction of a Z boson into two charged leptons is about 10\% \cite{PDG}, final states similar to (\ref{eeZZ}) account for only 1\% of all ZZ events. Without the overwhelming QCD background of a hadron collider experimental analyses will very likely be dominated by (semi-)hadronic and invisible decay channels.
\end{itemize}
We stick to the choice (\ref{eeWW}), (\ref{eeZZ}) for the following reasons:
\begin{itemize}
\item We avoid final-state electrons in order to suppress non-doubly-resonant background contributions. We choose different lepton flavours to minimise interference contributions, which can not be calculated in DPA.
\item The processes do not receive QCD corrections at NLO. This is merely a matter of simplicity, as we are only considering EW corrections within this work.
\item When decaying into quarks, the gauge bosons are very likely to produce single (fat) jets, complicating the signal/background ratio even more. In particular, assuming fully hadronic final states, the two processes develop a very similar signal and can only be distinguished by the respective jet invariant masses. However, if an efficient tagging of these jets can be achieved, the gauge bosons are basically detected directly and one can simply apply $\text{SCET}_\text{EW}$ to the production process. Since this is, however, speculative, we merely consider the fully leptonic final states given above.
\end{itemize}
However, we stress the fact that our approach is not limited to these
processes and can be generalised to all diboson processes and more
complicated processes involving resonant vector bosons. 

In the following, after a discussion of the event selection and kinematics in Sec.~\ref{Sec:EventSelection}, we work our way 
through the assumptions presented in order to check the applicability of $\text{SCET}_\text{EW}$.
In Sec.~\ref{Sec:DPARes} we investigate the quality of the DPA and in Sec.~\ref{Sec:PolRes} we 
collect some results for the production of polarised bosons in order
to estimate the error owing to the use of an incoherent polarisation sum. In Sec.~\ref{Sec:ValiRes} we check the validity of the assumption (\ref{SCETcond}) before 
presenting the $\text{SCET}_\text{EW}$ results broken down to individual contributions in Sec.~\ref{Sec:IndiRes} and the resummed results in Sec.~\ref{Sec:ResumRes}.

\subsubsection{Event selection and kinematics}
\label{Sec:EventSelection}

Photons are recombined with leptons if 
\begin{align}
\Delta
R_{\ell\gamma}=\sqrt{(y_\ell-y_\gamma)^2+(\Delta\phi_{\ell\gamma})^2}<0.1,
\label{recoco}
\end{align}
where $\Delta\phi_{\ell\gamma}$ denotes the azimuthal distance of the lepton and the photon 
and $y_\ell$, $y_\gamma$ their rapidities. 
We use the following charged-lepton acceptance cuts
\begin{align}
p_{\text{T},\ell}&>20\,\text{GeV},&10^\circ<\theta_\ell&<170^\circ,&
M_{\text{inv},\ell\ell'}&>10\,\text{GeV},&\label{standardcuts_CLIC}
\end{align}
with $\theta_\ell$ denoting the angle of the lepton with respect to
the the positron beam. 
In the ZZ case we impose an additional invariant-mass cut around the Z mass:
\begin{align}
81\,\text{GeV}<M_{\mu^+\mu^-}<101 \,\text{GeV},\qquad
81\,\text{GeV}<M_{\tau^+\tau^-}<101 \,\text{GeV}.\label{ZWindow_CLIC}
\end{align}
For the $\text{SCET}_\text{EW}$ calculations the condition
\begin{align}
s,|t|,|u|>M_\text{W}^2,\label{stu}
\end{align}
with $s,t,u$ being the usual Mandelstam variables in the production process
\begin{align}
\text{e}^+\text{e}^- \rightarrow V V'.
\end{align}
is enforced by means of an additional technical cut: If $\text{SCET}_\text{EW}$ is applied, the event is discarded if (\ref{stu}) is not fulfilled. This effectively restricts the fiducial phase space, and we define the High-Energy (HE) phase space to be:
\begin{align}
\text{d}\Pi_\text{HE}=\text{d}\Pi_{s,|t|,|u|>M_\text{W}^2}\label{SudPS}.
\end{align}

As stated above, we consistently define the scattering angle
with respect to the $z$~axis, which we choose along the positron beam direction.
According to the charge flow we define the \textit{forward region} for $\text{W}^+\text{W}^-$ production 
such that an outgoing $\text{W}^+$ boson 
(or its decay product) travels in positive $z$~direction and a $\text{W}^-$ travels in negative $z$~direction.
Because of the asymmetry of the weak interaction this region
has the largest cross section. Note that this definition implies that a $\mu^+$ with small scattering angle as well
as a $\tau^-$ with a large scattering angle are radiated in the forward region. 
Accordingly, a $\mu^+$ with large scattering angle or a $\tau^-$ with a small scattering angle are said
to be radiated in the backward region.

\subsubsection{Double-pole approximation}
\label{Sec:DPARes}
The application of $\text{SCET}_\text{EW}$ relies on the factorisation of a complicated process into a production and a decay part. The first validation step is thus to justify the application of the DPA.
To this end we calculate all considered processes both in DPA and fully off shell. The quality of the approximation is estimated using the quantity
\begin{align}
\Delta_\text{DPA}=1-\frac{\text{d}\sigma^\text{DPA}/\text{d}\mathcal{O}}{\text{d}\sigma^\text{full}/\text{d}\mathcal{O}},\label{dDPA}
\end{align}
where $\mathcal{O}$ denotes a generic kinematic variable.
If the DPA works properly, $\Delta_\text{DPA}$ is of the order of $\Gamma_V/M_V$, i.e.\ a few percent,
\begin{align}
\frac{\Gamma_\text{W}}{M_\text{W}}=2.6\%,\qquad\frac{\Gamma_\text{Z}}{M_\text{Z}}=2.7\%.
\end{align}
When the virtual corrections are computed, the respective relative corrections are defined as\footnote{Of course virtual corrections alone are not well-defined owing to IR singularities. We define them via their IR-finite part:
\begin{equation}
\text{d}\sigma^\text{virt}= \left.\text{d}\sigma^\text{virt}\right|_{1/\varepsilon_\text{IR}^2=1/\varepsilon_\text{IR}=0,\,\mu_\text{IR}=M_\text{W}}\label{muIR}.
\end{equation}}
\begin{equation}
\delta^\text{virt}_\text{DPA/full}=\frac{\text{d}\sigma^\text{virt,DPA/full}/\text{d}\mathcal{O}}{\text{d}\sigma^\text{born,DPA/full}/\text{d}\mathcal{O}}.
\end{equation}
One convenient way of applying the DPA is to compute only the virtual corrections in DPA, rescale them via 
\begin{align}
\frac{\text{d}\sigma^\text{virt,\,full}}{\text{d}\mathcal{O}}\rightarrow\frac{\text{d}\sigma^\text{virt,\,DPA}}{\text{d}\mathcal{O}}\times \frac{\text{d}\sigma^\text{born, full}}{\text{d}\mathcal{O}}\left(\frac{\text{d}\sigma^\text{born, DPA}}{\text{d}\mathcal{O}}\right)^{-1},\label{virtrescale}
\end{align}
and compute all other ingredients off shell. In this case the error owing to the DPA is given by the difference between the relative virtual corrections,
\begin{align}
\delta^\text{virt}_\text{full}-\delta^\text{virt}_\text{DPA}\label{deltadelta}.
\end{align}
We thus calculate and plot (\ref{deltadelta}) for the processes (\ref{eeWW}) and (\ref{eeZZ}). 
Note that in the differences \refeq{dDPA} and \refeq{deltadelta}, as
in the DPA, no interferences are neglected, as long as the unpolarised
cross section is understood.

For $\text{W}^+\text{W}^-$ pair production we obtain the fiducial cross section
\begin{align}
\sigma_\text{full}=1.760(7)\,\text{fb},\qquad\sigma_\text{DPA}=1.456(3)\,\text{fb},\qquad\Delta_\text{DPA}=17.3(4)\%.
\end{align}
The numbers in parentheses denote MC integration errors.
The relative virtual corrections read
\begin{align}
\delta^\text{virt}_\text{full}=-33.9(2)\%,\qquad\delta^\text{virt}_\text{DPA}=-34.6(1)\%,
\qquad\delta^\text{virt}_\text{full}-\delta^\text{virt}_\text{DPA}=0.7(2)\%.\label{DPAsubpercent}
\end{align}
While the Born cross section calculated in the DPA does not accurately
reproduce the full result,
$\delta^\text{virt}_\text{full}-\delta^\text{virt}_\text{DPA}$ is
smaller than one percent, indicating that calculating the relative
virtual corrections within the DPA provides a good approximation.

\begin{figure}
\centering
\includegraphics[width=0.45\textwidth]{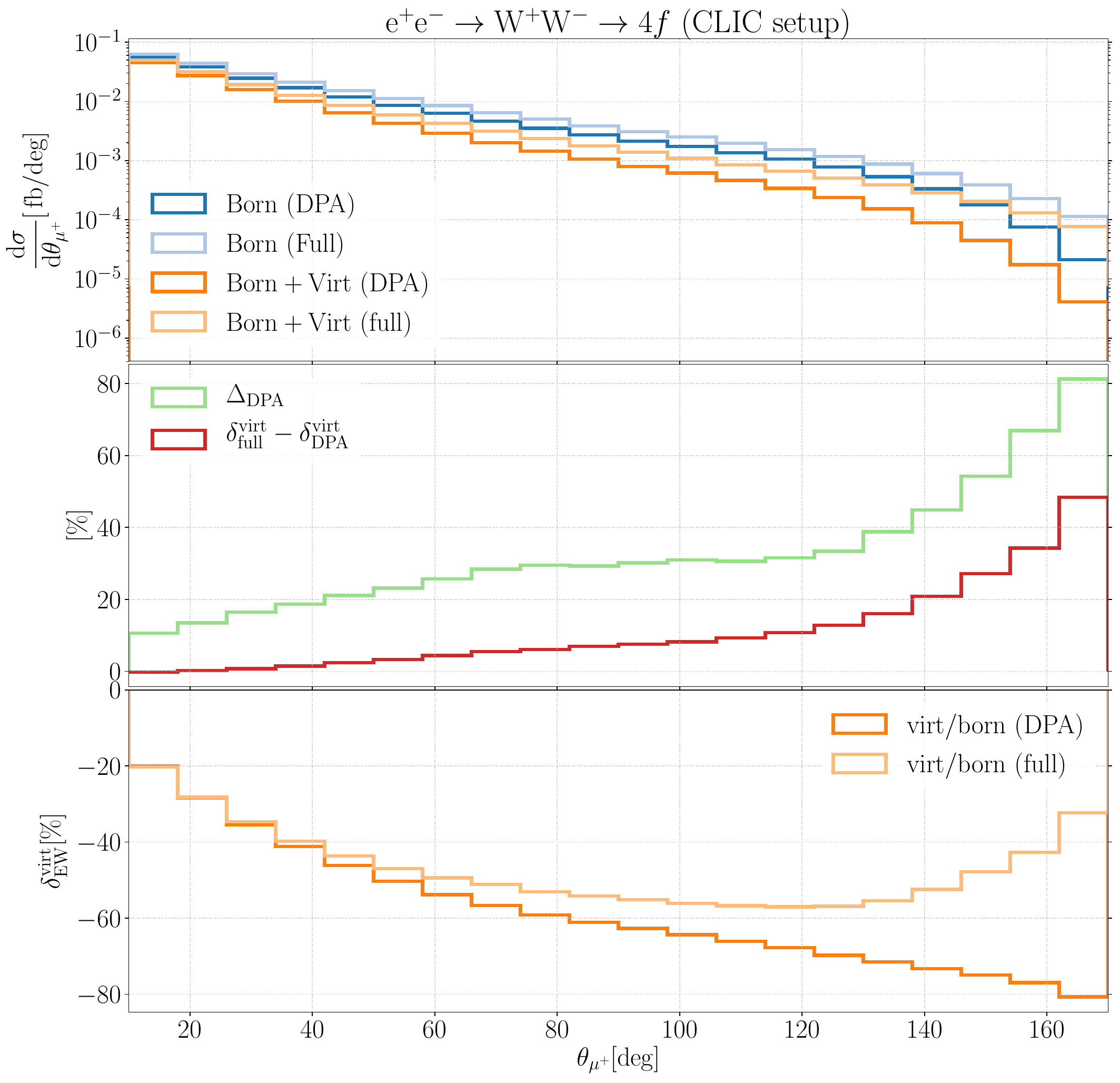}
\includegraphics[width=0.45\textwidth]{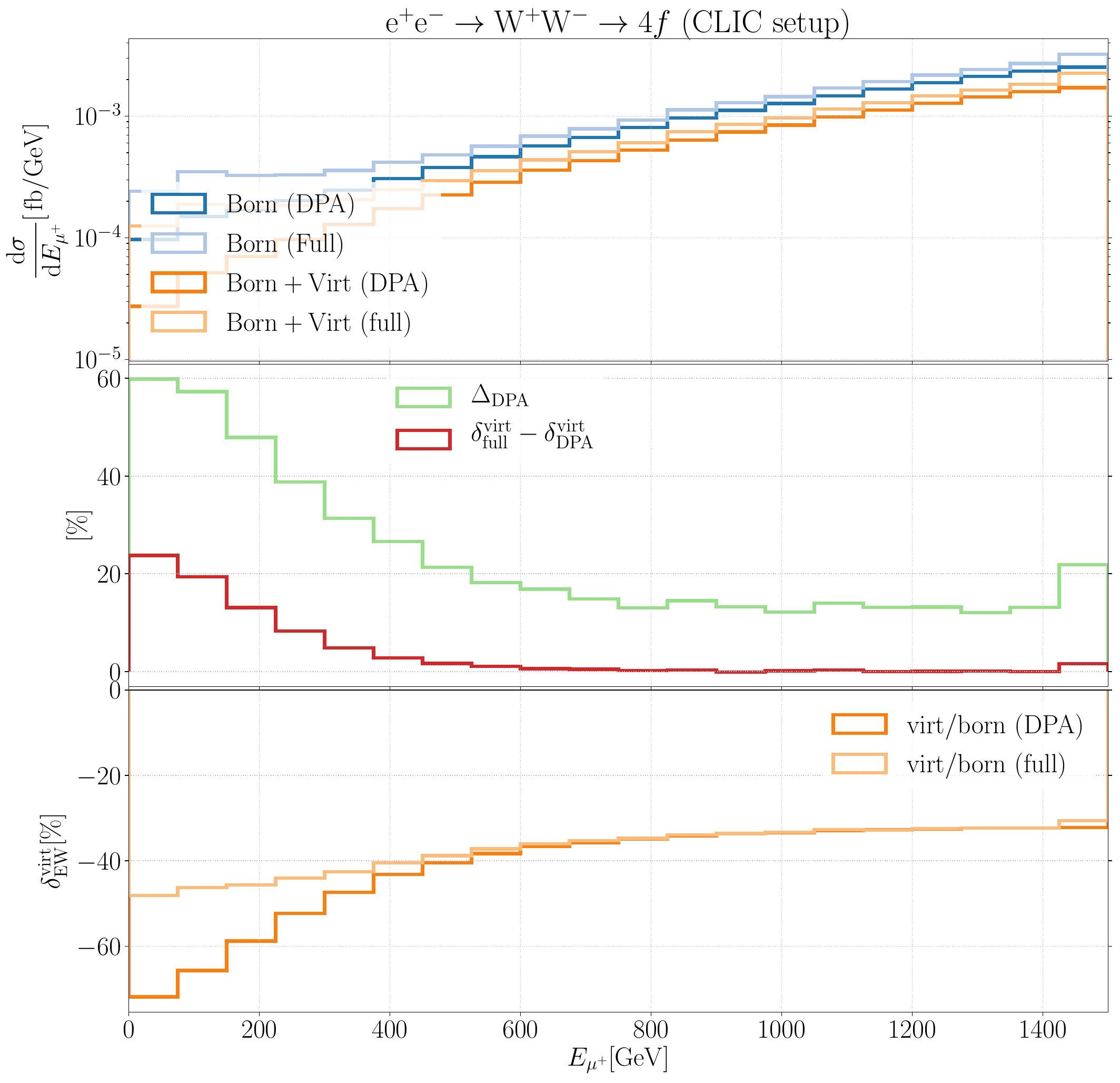}
\includegraphics[width=0.45\textwidth]{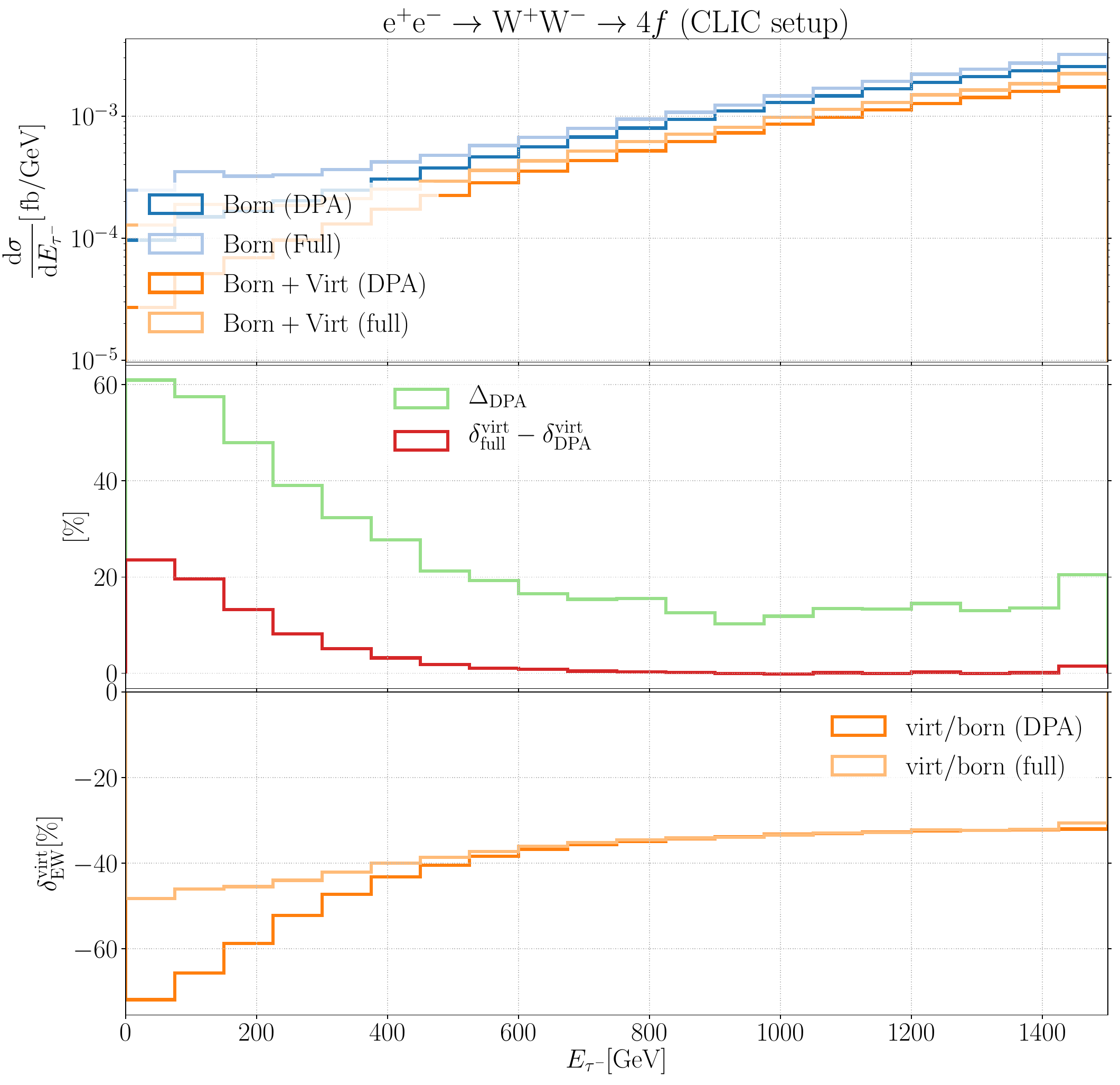}
\includegraphics[width=0.45\textwidth]{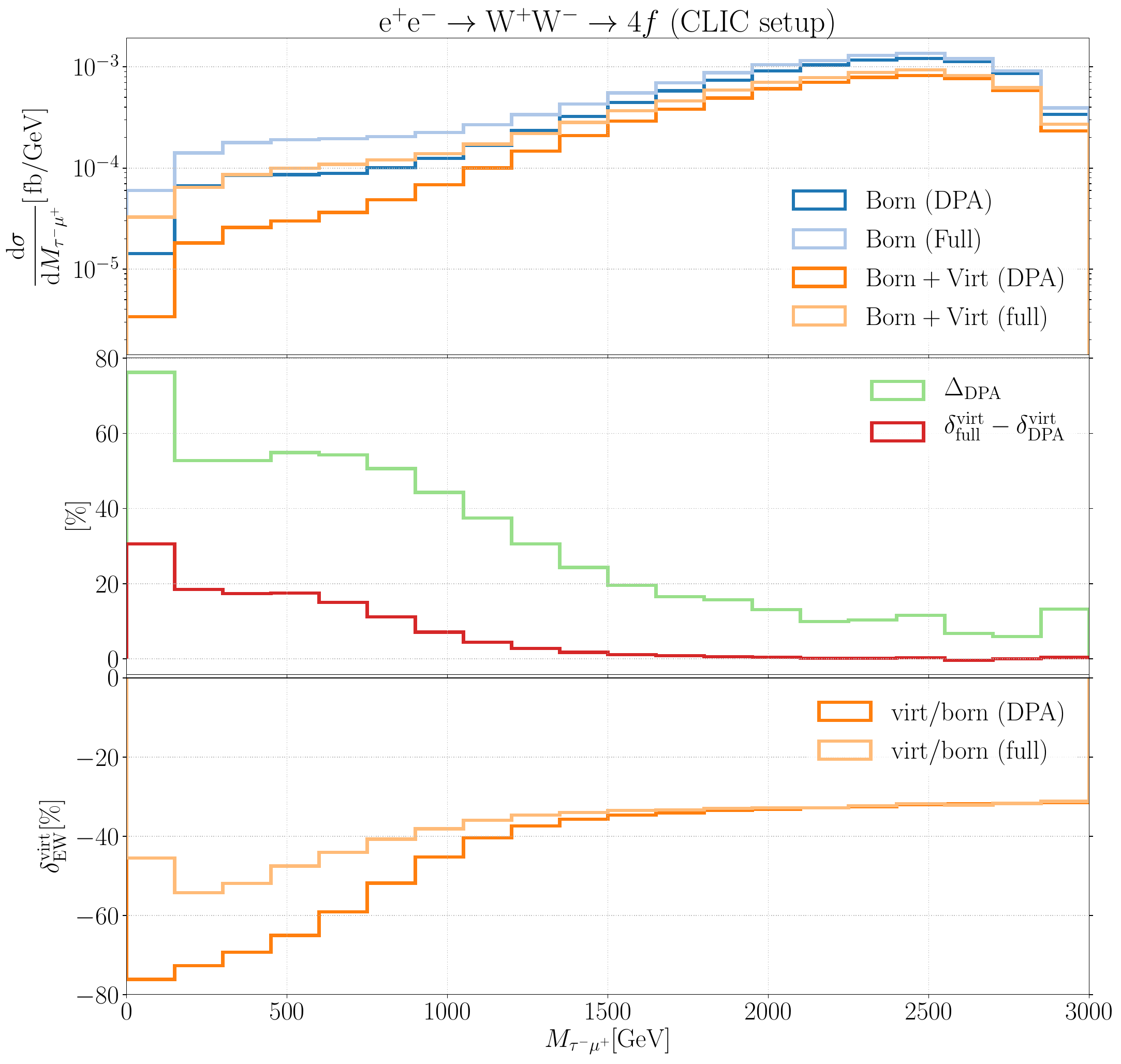}
\caption{Comparison between DPA and fully off-shell calculation for
  $\eewwllll$: Differential distributions in the $\mu^+$ production angle, the $\mu^+$ and $\tau^-$ energy, and the $\tau^-\mu^+$ invariant mass. The upper panels show the differential cross sections, the middle ones the deviations owing to the DPA with $\Delta_\text{DPA}$ as defined in (\ref{dDPA}) and $\delta^\text{virt}_\text{full}-\delta^\text{virt}_\text{DPA}$, while the lower ones show the IR-finite relative virtual EW corrections.}
\label{Fig:DPA_WW_ee}
\end{figure}%
Figure~\ref{Fig:DPA_WW_ee} shows a comparison between fully off-shell
results and the DPA differential in the muon production angle, the lepton energies, and the $\tau^-\mu^+$ invariant mass. 
The angular distribution is dominated by the forward region (see the definition at the end of Sec.~\ref{Sec:EventSelection}) owing to the
dominant contribution from the $t$-channel diagram. Towards the
backward region the cross section decreases by up to four orders of
magnitude. The off-shell virtual EW corrections are between $-20\%$
and $-40\%$ in the forward region, grow towards the central region and
decrease again to $-30\%$ in the backward region. Note that the
virtual corrections in DPA reach $-80\%$ in the backward region.
The difference $\Delta_\text{DPA}$ between full calculation and DPA increases from 10\% in
the forward direction to 80\% in the backward direction. The
corresponding difference of the relative corrections remains below 5\%
in the forward hemisphere and increases where the cross section is small.

The distributions in the lepton energies are peaked at high charged-lepton energies for helicity-conservation reasons:
Because in the forward region, where the cross section is large, the
$\text{W}^+\text{W}^-$ pair has a preferred polarisation configuration
of ($-,+$) and preferably decays into high-energy leptons and
low-energy neutrinos. In the high-energy tails the quality of the DPA
is satisfactory ($\Delta_\text{DPA}$ is about 15\%, but the difference
of the relative corrections is $<1\%$), while towards the low-energy
regime $\delta^\text{virt}_\text{full}-\delta^\text{virt}_\text{DPA}$
grows up to 25\%. 

The dilepton invariant-mass distribution has a maximum at
$M_{\tau^-\mu^+}\approx2500\,\text{GeV}$, which is consistent with the
other distributions, since configurations with back-to-back leptons
with high energies are preferred. These, in turn have a high dilepton
invariant mass. Again, in the region that dominates the cross section
($M_{\tau^-\mu^+}\gtrsim 1500\,\text{GeV}$), the relative corrections in
DPA and off shell differ only by subpercent effects.

All in all the DPA result at LO is never really appropriate,
which implies that the description of a process as a product of
a production and several decay processes has its limitations
at high energies. 
One particular mechanism that causes deviations is discussed in
Sec.~\ref{Sec:DPAFCC}.%
\footnote{A possible path to include EW high-energy logarithms in
  processes with resonances beyond the DPA has recently been proposed 
in \citere{Lindert:2023fcu}.}
As discussed above that does not imply that the DPA is worthless,
because it is still possible to compute only the relative virtual 
corrections in DPA and use (\ref{deltadelta}) as a measure of accuracy,
which is formally of order
$\mathcal{O}\left(\alpha{\Gamma_\text{W/Z}}/{M_\text{W/Z}}\right)$.
In this respect the DPA works best in the regions of phase space,
where the cross section is largest: In the forward region,
in the region of large dilepton invariant masses, and for large lepton
energies,
$\delta^\text{virt}_\text{full}-\delta^\text{virt}_\text{DPA}$ is at
the subpercent level. Because these regions dominate the cross section
the small value in (\ref{DPAsubpercent}) is obtained. In some regions
with small cross sections the DPA happens to fail completely. For instance, 
$\delta^\text{virt}_\text{full}-\delta^\text{virt}_\text{DPA}$ grows up to $\sim 50\%$ in the backward region,
where the cross section is dominated by singly-resonant contributions.

The fiducial cross section for ZZ production reads:
\begin{align}
\sigma_\text{full}=0.013047(3)\,\text{fb},\qquad\sigma_\text{DPA}=0.012274(3)\,\text{fb},\qquad\Delta_\text{DPA}=5.93(2)\%.
\end{align}
For the relative virtual corrections, defined as in (\ref{muIR}), we obtain
\begin{align}
\delta^\text{virt}_\text{full}=-44.58(2)\%,\qquad\delta^\text{virt}_\text{DPA}=-46.21(4)\%,\qquad\delta^\text{virt}_\text{full}-\delta^\text{virt}_\text{DPA}=1.63(4)\%,
\end{align}
Compared to
the $\text{W}^+\text{W}^-$-production results, $\Delta_\text{DPA}$ is
smaller, because the Z-window cuts isolate the doubly-resonant
contributions. The difference
$\delta^\text{virt}_\text{full}-\delta^\text{virt}_\text{DPA}$
is within the expected uncertainty of the DPA.
\begin{figure}
\centering
\includegraphics[width=0.45\textwidth]{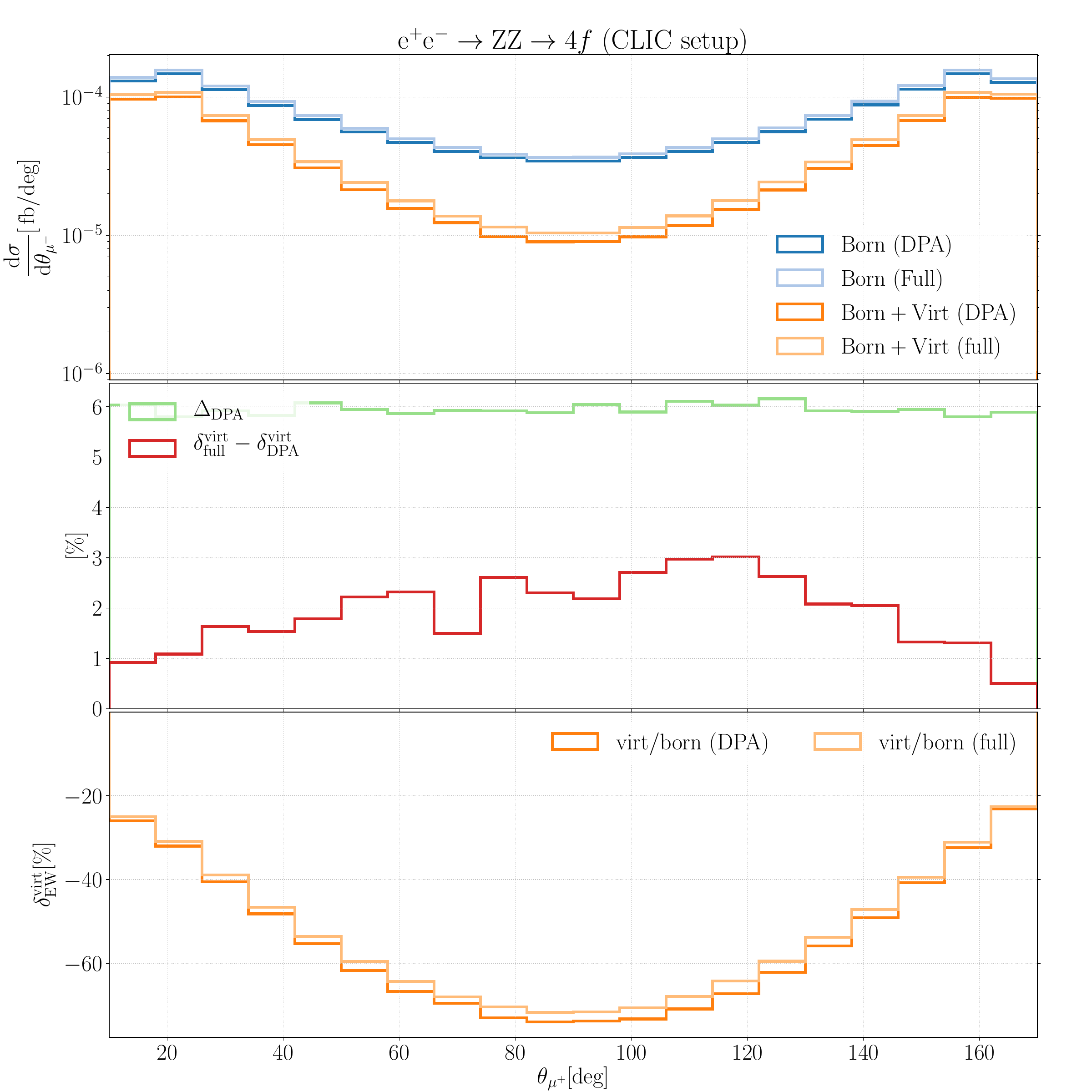}
\includegraphics[width=0.45\textwidth]{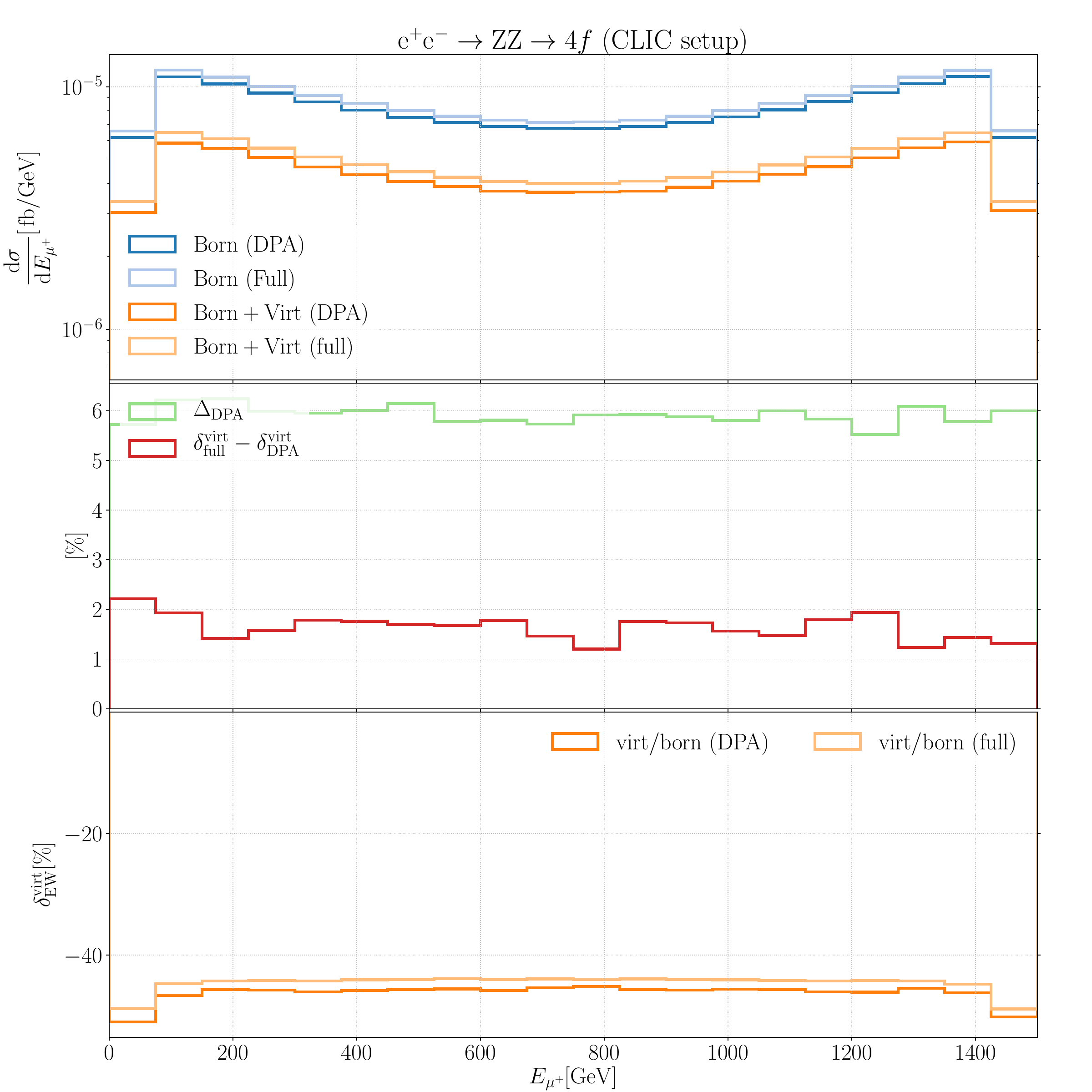}\\
\includegraphics[width=0.45\textwidth]{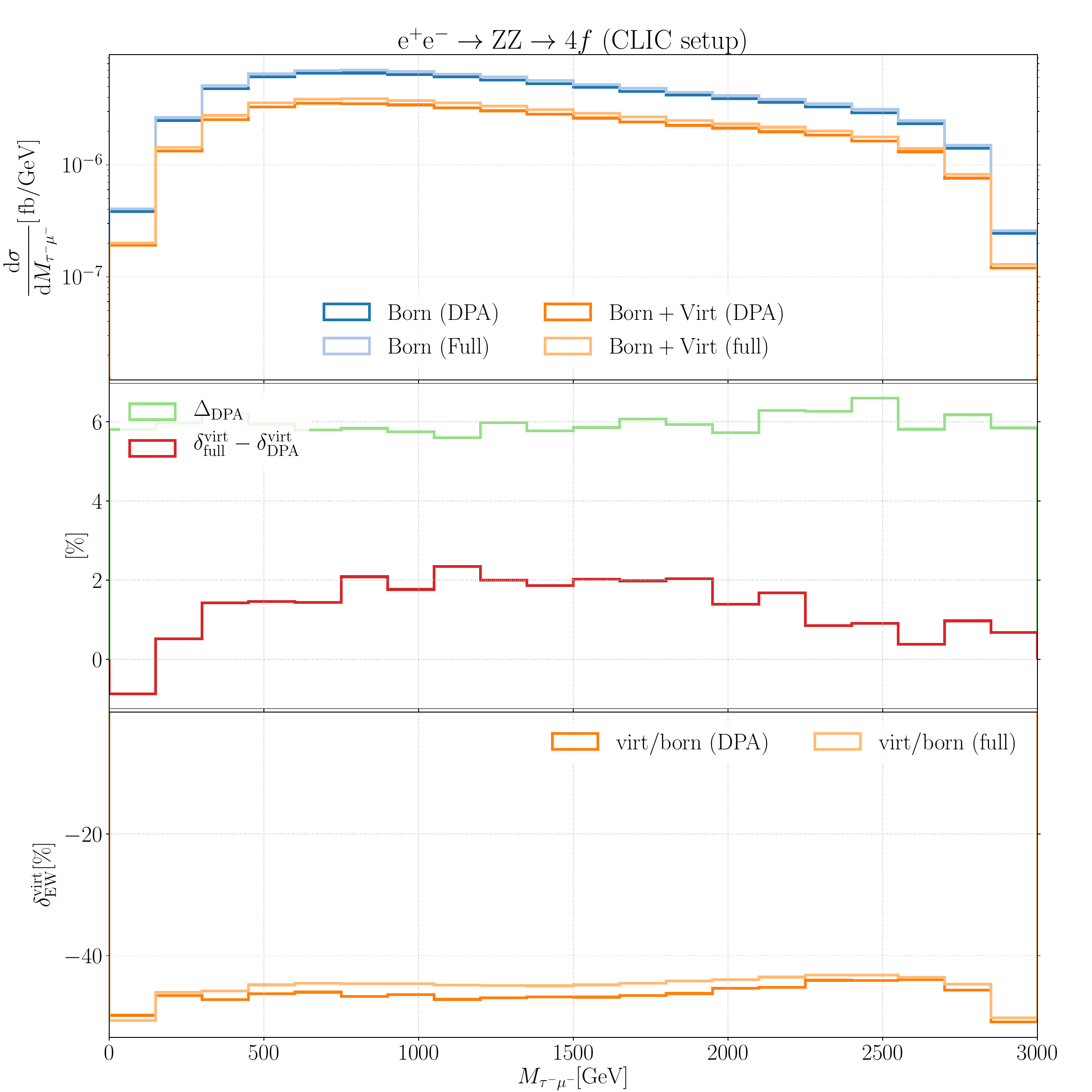}
\includegraphics[width=0.45\textwidth]{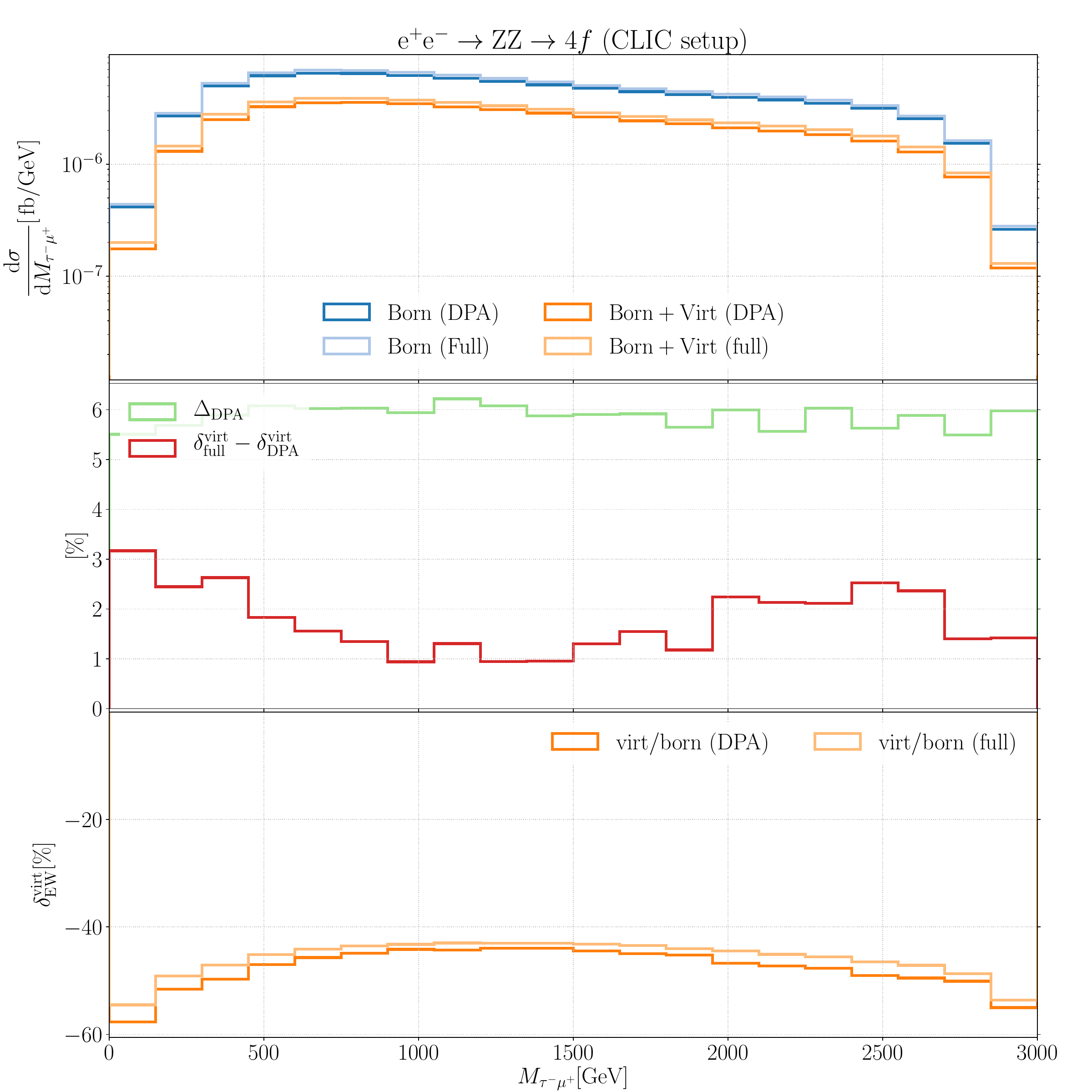}
\caption{Comparison between DPA and fully off-shell calculation for
  $\eezzllll$:
  Differential distributions in the $\mu^+$ production angle, the
  $\mu^+$ energy and the $\tau^-\mu^-$ and $\tau^-\mu^+$ invariant masses. The plots are organised as described in the caption of Fig.~\ref{Fig:DPA_WW_ee}.}
\label{Fig:DPA_ZZ_ee}
\end{figure}%
Differential results in the $\mu^+$ energy, the $\mu^+$ production angle, and the dilepton invariant masses of the $\tau^-\mu^-$ and the $\tau^-\mu^+$ system can be found in Fig.~\ref{Fig:DPA_ZZ_ee}. 
The distribution in the antimuon energy is peaked at low and high energies
with the first and last bin ($E<75\,\text{GeV}$, $E>1425\,\text{GeV}$)
being suppressed owing to the phase-space cuts. The virtual
corrections are approximately constant over energy both in DPA and
off shell. The same holds for $\Delta_\text{DPA}$ and
$\delta^\text{virt}_\text{full}-\delta^\text{virt}_\text{DPA}$, which
vary slightly around 6\% and 2\%, respectively.  

The scattering-angle distribution has a double-peak structure with
maxima at $\theta_{\mu}\approx 25^\circ$ and
$\theta_{\mu}\approx 155^\circ$, corresponding to the $t$- and
$u$-channel enhancements, respectively. In the central region the
cross section is suppressed by one order of magnitude. The virtual
corrections vary between $-75\%$ in the central region and $-20\%$ in
the tails. This distribution is the only one for ZZ~production in which a phase-space
dependence of the quality of the DPA can be observed: While
$\Delta_\text{DPA}$ does not vary over $\theta_\mu$,
$\delta^\text{virt}_\text{full}-\delta^\text{virt}_\text{DPA}$ is
slightly enhanced in the central region, reaching at most $\sim3\%$. 

The invariant-mass distributions have a maximum at $M_{\tau^-\mu^-}\approx M_{\tau^-\mu^+}\approx700\,\text{GeV}$. Similar to the energy distribution the quality of the DPA does not show a systematic difference between high and low values of $M_{\tau^-\mu^+}$.  

All in all the quality of the DPA for this process is also satisfactory. In contrast to the $\text{W}^+\text{W}^-$-production case the deviation is constant over many phase-space variables. This can be explained by the fact that 
it is mainly caused by the irreducible photon background, which is expected to be distributed similar to the resonant contributions. On the other hand, the singly-resonant contributions, which 
dominate the large deviations in some phase-space regions for
$\text{W}^+\text{W}^-$ production, are removed by the invariant-mass cuts~(\ref{ZWindow_CLIC}).

\subsubsection{Polarised cross sections}
\label{Sec:PolRes}
As explained in Sec.~\ref{Sec:PolDef}, we neglect 
interference terms between
purely transverse and longitudinal polarisation configurations throughout our calculations.
Furthermore we point out that the power-suppressed contributions from mixed longitudinal/transverse polarisation configurations
are not present in the SySM and thus also not included in the $\text{SCET}_\text{EW}$ computation.  
In close analogy to the DPA analysis we define the quantity
\begin{align}\Delta_\text{pol}=1-
\left(\frac{\text{d}\sigma_\text{TT}}{\text{d}\mathcal{O}}
+\frac{\text{d}\sigma_\text{LL}}{\text{d}\mathcal{O}}\right)\left(\frac{\text{d}\sigma^\text{DPA}_\text{UU}}{\text{d}\mathcal{O}}\right)^{-1},\label{Deltapol}
\end{align}
which estimates the error made by neglecting the mixed polarisation
contributions as well as the interference terms. Here and in the
following UU denotes the unpolarised cross section, while T and L
denote transversely and longitudinally polarised bosons, respectively. 
The error due to the interference terms alone is computed via
\begin{align}\Delta_\text{int}=1-
\left(\frac{\text{d}\sigma_\text{TT}}{\text{d}\mathcal{O}}
+\frac{\text{d}\sigma_\text{TL}}{\text{d}\mathcal{O}}
+\frac{\text{d}\sigma_\text{LT}}{\text{d}\mathcal{O}}
+\frac{\text{d}\sigma_\text{LL}}{\text{d}\mathcal{O}}\right)
\left(\frac{\text{d}\sigma^\text{DPA}_\text{UU}}{\text{d}\mathcal{O}}\right)^{-1}.
\label{Deltaint}
\end{align}
Because $\text{d}\sigma_\text{TT}$ contains all interferences between the different transverse polarisation
states, $\Delta_\text{int}$ quantifies only the effects of interference terms between
transversely and longitudinally polarised bosons (\refta{tab:sigmaPol}). 
In a separate calculation we have checked
that the interference terms among the different transverse
polarisation states are also small (\refta{tab:sigmaPol_Trans}). In
  the SCET results for transversely polarised bosons in the following
  sections, all interferences are omitted.
Note that $\Delta_\text{int}$ is part of $\Delta_\text{pol}$, and the additional contributions to $\Delta_\text{pol}$ are given by the mixed polarised contributions $\text{d}\sigma_\text{LT/TL}$.

\begin{table}
\centering
\renewcommand{\arraystretch}{1.3}
\begin{tabular}{|l|c|r|}
\hline
Pol.& $\sigma^\text{ZZ}/\text{fb}$ & \multicolumn{1}{c|}{$\delta$}\\
\hline
UU&$1.2274(3)\cdot10^{-2}$&100\%\\%
TT&$1.2249(11)\cdot10^{-2}$&99.8\%\\%
TL+LT&$1.419(3)\cdot10^{-5}$&0.11\%\\%
LL&$3.533(4)\cdot10^{-8}$&$\mathcal{O}(10^{-4})$\%\\%
$\sum$&$1.2264(11)\cdot10^{-2}$&99.9\%\\%
\hline
\end{tabular}
\quad
\begin{tabular}{|l|c|r|}
\hline
Pol. & $\sigma^\text{WW}/\text{fb}$ & \multicolumn{1}{c|}{$\delta$}\\
\hline
UU&$1.456(4)$&100\%\\%
TT&$1.418(4)$&97.4\%\\%
TL+LT&$1.5013(10)\cdot10^{-3}$&0.1\%\\%
LL&$3.757(3)\cdot10^{-2}$&2.5\%\\%
$\sum$&$1.457(4)$&100.1\%\\%
\hline
\end{tabular}
\caption{Integrated fiducial cross sections for $\eezzllll$ (left) and $\eewwllll$ (right) with definite polarisation states. The sum $\Sigma$ includes TT, TL+LT, and LL. All percentages $\delta$ are given with respect to the unpolarised result.}
\label{tab:sigmaPol}
\end{table}
\begin{table}
\centering
\renewcommand{\arraystretch}{1.3}
\begin{tabular}{|l|c|r|}
\hline
Pol. & $\sigma^\text{ZZ}/\text{fb}$ & \multicolumn{1}{c|}{$\delta$}\\
\hline
TT&$1.2249(11)\cdot10^{-2}$&100\%\\%
$ ++ $ & $7.430(8)\cdot10^{-9}$&$\mathcal{O}(10^{-5})$\%\\%
$ +- $ & $6.172(6)\cdot10^{-3}$&50.4\%\\%
$ -+ $ & $6.171(6)\cdot10^{-3}$&50.4\%\\%
$ -- $ & $7.435(8)\cdot10^{-9}$&$\mathcal{O}(10^{-5})$\%\\%
$\sum$&$1.2343(8)\cdot10^{-2}$&100.8\%\\%
\hline
\end{tabular}
\quad
\begin{tabular}{|l|c|r|}
\hline
Pol. & $\sigma^\text{WW}/\text{fb}$ & \multicolumn{1}{c|}{$\delta$}\\
\hline
TT&$1.418(3)$&100\%\\%
$++$&$\sim 5\cdot 10^{-7}$&$\mathcal{O}(10^{-5})$\%\\%
$+-$&$0.06933(10)$&4.9\%\\%
$-+$&$1.349(3)$&95.1\%\\%
$--$&$\sim 5\cdot 10^{-7}$&$\mathcal{O}(10^{-5})$\%\\%
$\sum$&$1.418(4)$&100\%\\%
\hline
\end{tabular}
\caption{Integrated fiducial cross sections for $\eezzllll$ (left) and $\eewwllll$ (right) with definite polarisation states. The sum $\Sigma$ includes $++$, $+-$, $-+$, and $--$. All percentages $\delta$ are given with respect to the TT value.}
\label{tab:sigmaPol_Trans}
\end{table}
The integrated cross sections for all possible polarisation states are
collected in Tables~\ref{tab:sigmaPol} and \ref{tab:sigmaPol_Trans}. The mass-suppressed contributions (mixed and longitudinal for ZZ~production and mixed polarisations for $\text{W}^+\text{W}^-$~production) can safely be neglected. The same holds for the interference terms, since the sum of the polarisation states reproduces the unpolarised cross section within 0.1\%.
Both the contributions from mixed transverse/longitudinal
configurations and the interference between transverse and
longitudinal polarisation states account for less than 1\% of the
unpolarised cross section. We note that this holds almost on the whole
phase space. 

\begin{figure}
\centering
\includegraphics[width=0.45\textwidth]{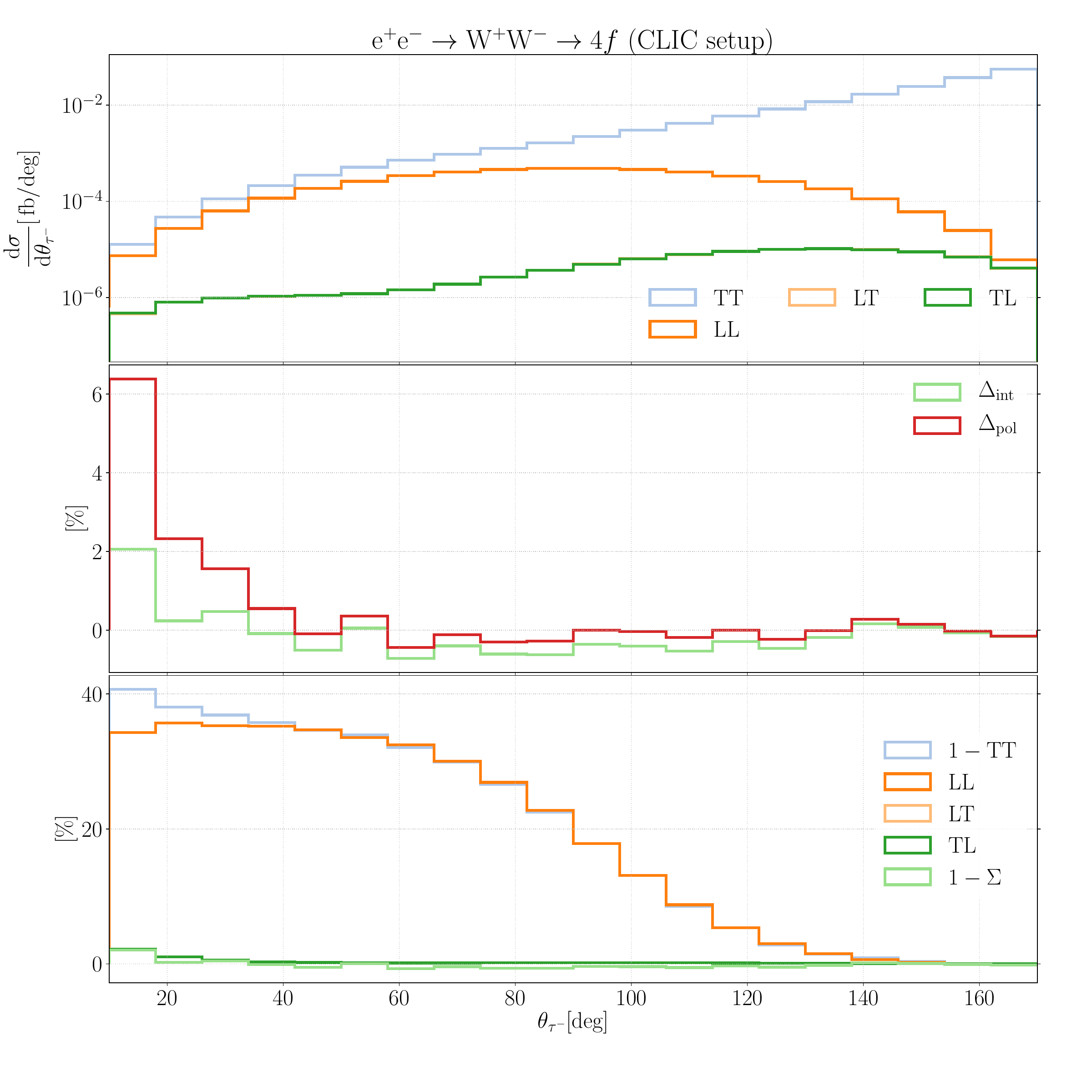}
\includegraphics[width=0.45\textwidth]{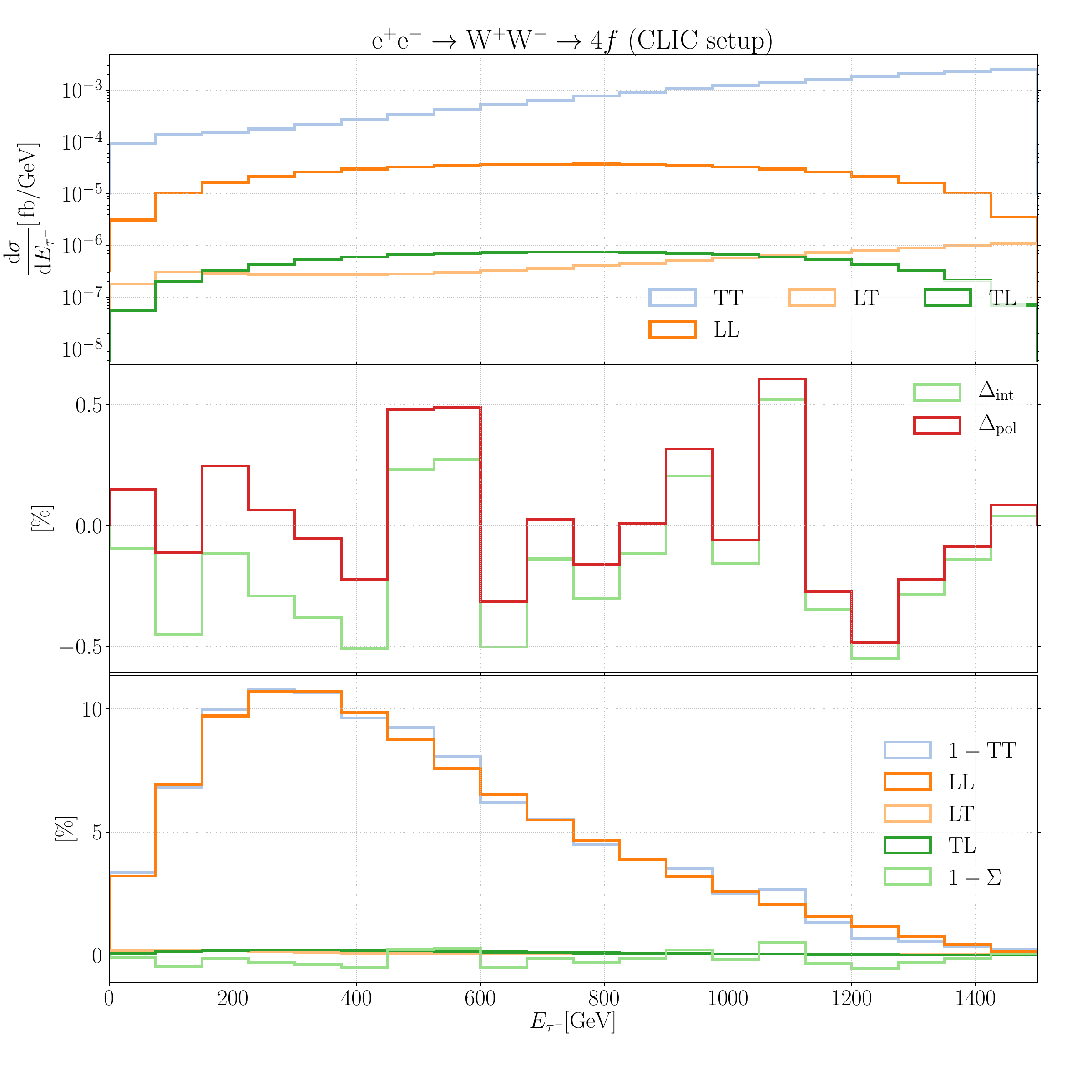}
\caption{Polarised differential cross sections in the $\tau^-$
  production angle and energy for $\eewwllll$. The curves for TL and
  LT are sometimes not to distinguish. The quantities $\Delta_\text{int}$ and $\Delta_\text{pol}$,
defined in (\ref{Deltapol}) and (\ref{Deltaint}), measure the deviation owing to interference contributions and mixed-polarisation contributions, respectively.}
\label{Fig:PolFig_eeWW}
\end{figure}

Differential distributions in the $\tau$
production angle and energy for $\eewwllll$ can be found in
Fig.~\ref{Fig:PolFig_eeWW}. While the upper panels contain the different
polarised cross sections, the middle panels show $\Delta_\text{pol}$ and
$\Delta_\text{int}$, and the lower panels the fraction of the
polarised cross sections with respect to the unpolarised one.
There is an increase of the mixed-polarisation contributions in the
backward region (large antimuon angle, small tau production
angle).
Here the mixed contributions, i.e.\ the difference between  $\Delta_\text{pol}$ and $\Delta_{\text{int}}$ account for up to 4\%, and the total deviation is dominated by the mixed-polarised contributions. Since the cross section is smaller by up to three orders of magnitude, the influence on the integrated cross section is still negligble. In the energy distribution the mixed polarised and interference contributions nowhere exceed $1.5\%$. 

\subsubsection{$\text{SCET}_\text{EW}$ vs.\ fixed-order}
\label{Sec:ValiRes}
Next we check the validity of the $\text{SCET}_\text{EW}$ approximation ($s,|t|,|u|\gg M_\text{W}^2$). 
In order to analyse the quality of this assumption, we first consider unresummed $\text{SCET}_\text{EW}$, meaning that the exponentiated amplitude is expanded to first order in $\alpha$. In this approximation the $\text{SCET}_\text{EW}$~results agree with the fixed-order one-loop results up to powers of $M^2/s_{ij}$ with $M$ being any of the EW mass scales and $s_{ij}\in\{s,t,u\}$.  

\begin{figure}
\centering
\includegraphics[width=0.475\textwidth]{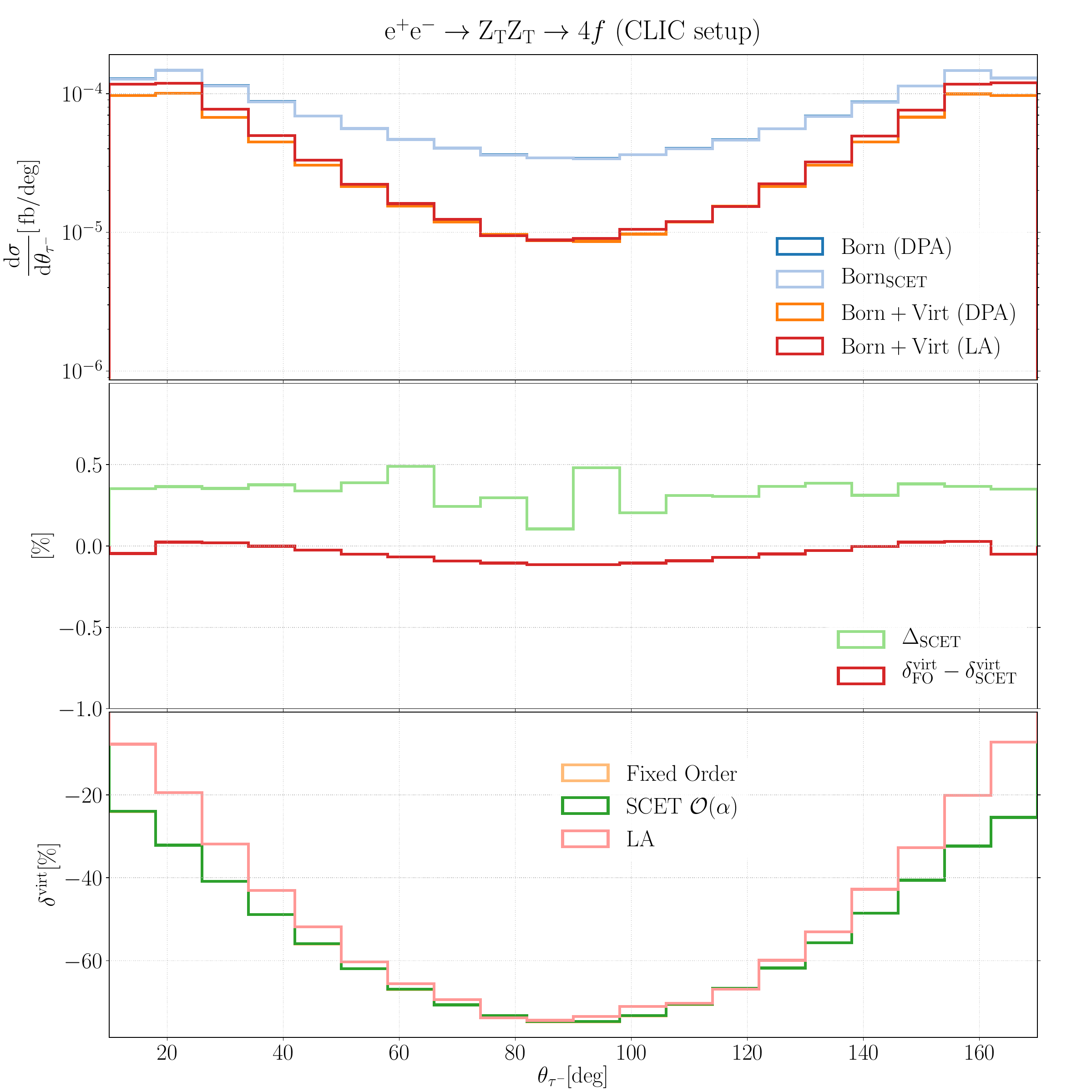}
\includegraphics[width=0.475\textwidth]{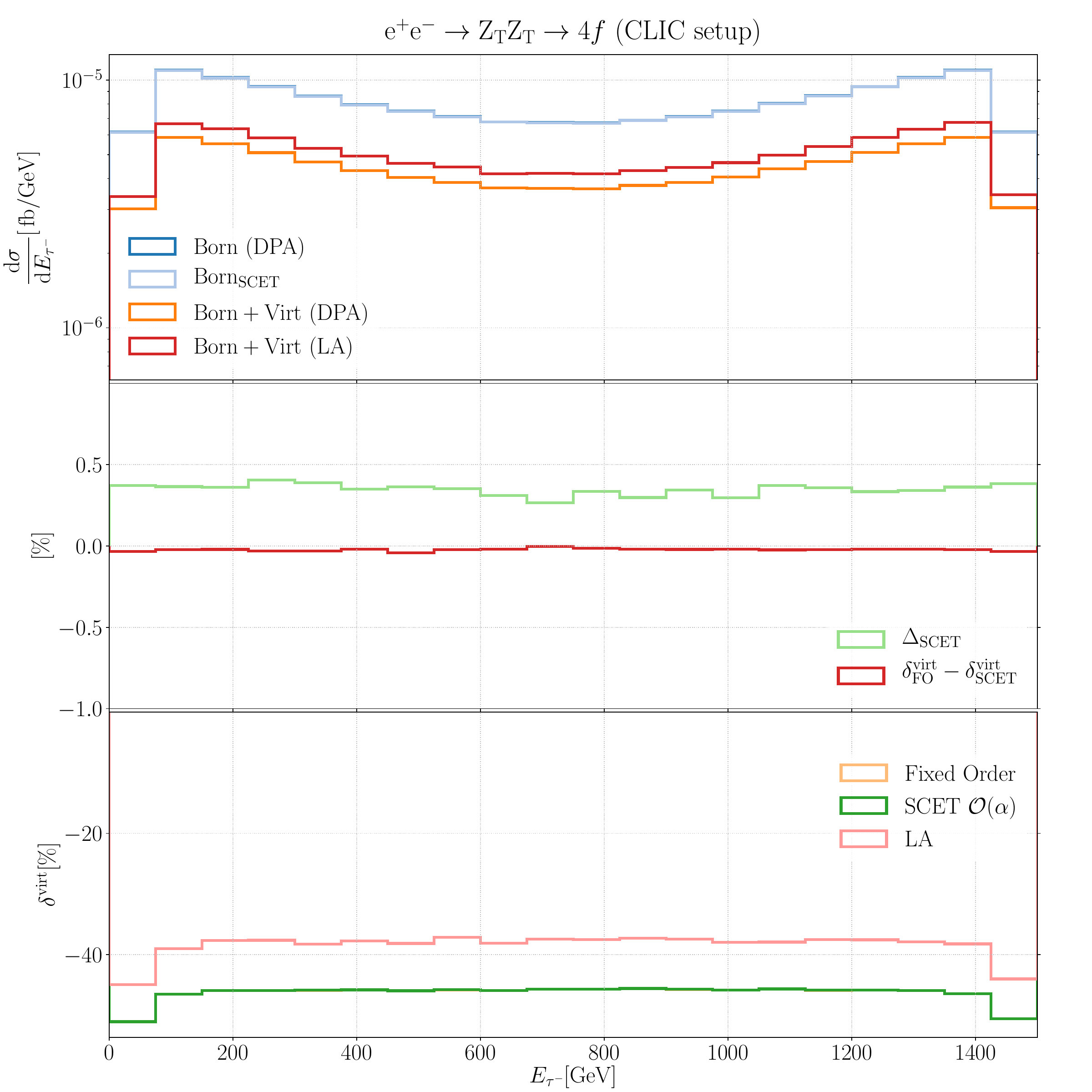}
\caption{Virtual corrections to the $\tau$-production-angle and $\tau$-energy
  distributions for $\eezzllll$ calculated in conventional fixed-order perturbation theory within the DPA compared to the LA and the first-order expansion of the SCET results in~$\alpha$. The orange curves in the bottom panels are hidden behind the green ones.}
\label{Fig:eeZZVali}
\end{figure}%
\begin{figure}
\centering
\includegraphics[width=0.475\textwidth]{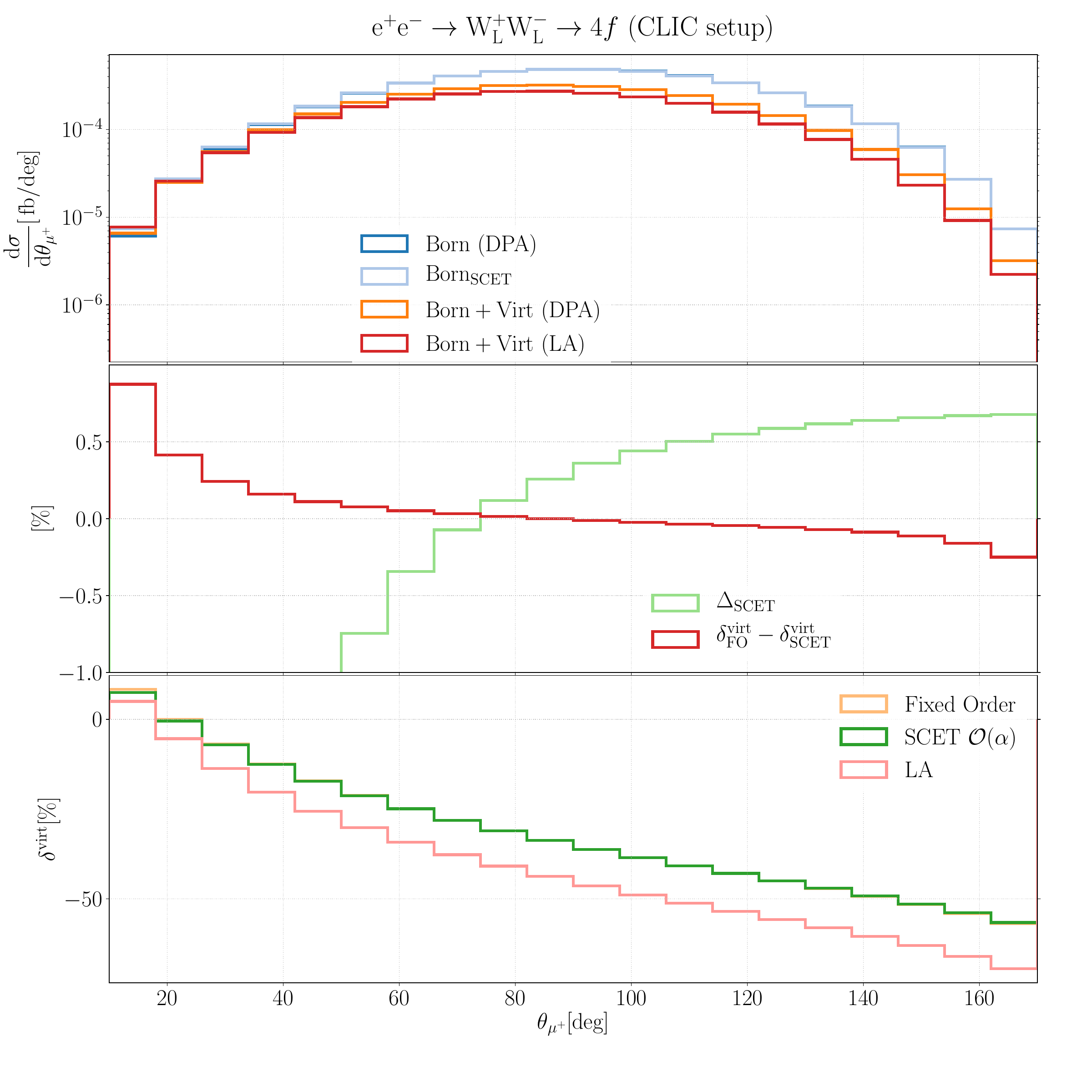}
\includegraphics[width=0.475\textwidth]{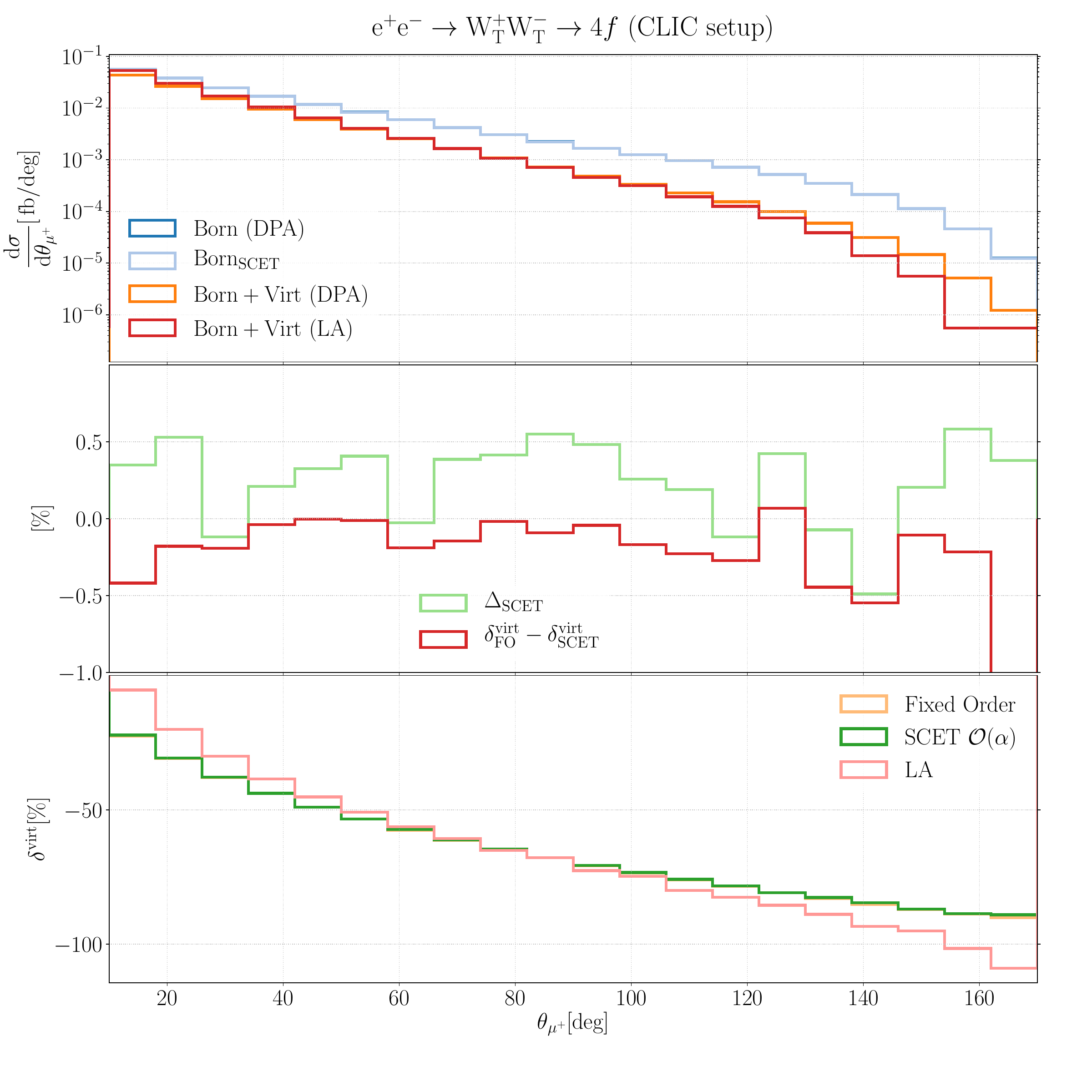}\\
\caption{Virtual corrections to muon-production-angle distribution
  for $\eewwllll$ with longitudinally (left) and transversely (right) polarised W~bosons calculated in conventional fixed-order perturbation theory within the DPA compared to the LA and the first-order expansion of the SCET results in $\alpha$.}
\label{Fig:eeWWVali}
\end{figure}%
We organise the plots in Figs.~\ref{Fig:eeZZVali} and
\ref{Fig:eeWWVali} as follows:
\begin{itemize}
\item The upper panels show the LO differential cross section
  both in fixed order and using $\text{SCET}_\text{EW}$ on the HE phase space (\ref{SudPS}).
Moreover, the sum of LO and IR-finite virtual corrections is
displayed in fixed order and using the logarithmic approximation (LA).
To estimate the error of the $\text{SCET}_\text{EW}$ approximation alone, 
all fixed-order results are computed in DPA.
\item The middle panels demonstrate the quality of the high-energy approximation, showing the quantities
\begin{align}
\Delta_\text{SCET}&=1-\frac{\text{d}\sigma^\text{Born, SCET}/\text{d}\mathcal{O}}{\text{d}\sigma^\text{Born, FO}/\text{d}\mathcal{O}},\nonumber\\\delta^\text{virt}_\text{FO}-\delta^\text{virt}_\text{SCET}&=\frac{\text{d}\sigma^\text{virt, FO, fac}/\text{d}\mathcal{O}}{\text{d}\sigma^\text{Born, FO}/\text{d}\mathcal{O}}-\frac{\text{d}\sigma^\text{virt, SCET}/\text{d}\mathcal{O}}{\text{d}\sigma^\text{Born, SCET}/\text{d}\mathcal{O}}\label{DeltaSCET},
\end{align}
with $\text{d}\sigma^\text{virt, FO, fac}$ denoting the factorisable virtual corrections in DPA.
The quantities in (\ref{DeltaSCET}) quantify the validity of the $\text{SCET}_\text{EW}$ approximation at tree-level and one-loop level, respectively. Note that both $\text{d}\sigma^\text{Born, SCET}$ and $\text{d}\sigma^\text{virt, SCET}$ are evaluated on the HE phase space defined by (\ref{SudPS}).
\item The lower panels show the relative virtual corrections
  calculated in fixed order, using the
  LA, and using $\text{SCET}_\text{EW}$ on the HE
  phase space.
\end{itemize}

In the $\tau$-energy and $\tau$-production-angle distributions in ZZ production
(Fig.~\ref{Fig:eeZZVali}), the deviation between the fixed-order result and
$\text{SCET}_\text{EW}$ approximation, parameterised as in (\ref{DeltaSCET}), is about $0.4\%$ at Born level and roughly constant over both distributions. The accuracy of the relative virtual corrections is even better: $\delta^\text{virt}_\text{FO}-\delta^\text{virt}_\text{SCET}$ is $\lesssim 0.1\%$ on the whole fiducial phase space. The LA describes the full result well only in the central region $50^\circ\leq\theta_\tau\leq140^\circ$. Outside this region, the omitted $\mathcal{O}(\alpha)$ terms contribute by up to 15\% with respect to LO. 

The results for the distribution in the muon production angle in $\text{W}^+\text{W}^-$ production are displayed in
Fig.~\ref{Fig:eeWWVali} for longitudinal and transverse polarisations
separately, both as a consistency check and in order to spot possible
differences: The unpolarised results are qualitatively well described
by the purely transverse contributions in all cases.  
In the longitudinal
case, $\Delta_\text{SCET}$ shows an asymmetric behaviour, ranging from
$\sim +0.7\%$ in the forward to $-5\%$ in backward region (not
visible in the plot). The deviation in the virtual corrections,
$\delta^\text{virt}_\text{FO}-\delta^\text{virt}_\text{SCET}$
grows from $-0.2\%$ in the forward region to only 0.8\% in
the backward region. In the central region which dominates the cross
section both quantities are close to 0. Together with the cancellation
of positive and negative deviations this yields a value below $0.1\%$ for
the fiducial cross section. In the transverse case both
$\Delta_\text{SCET}$ and
$\delta^\text{virt}_\text{FO}-\delta^\text{virt}_\text{SCET}$ vary
between $-0.5\%$ and $+0.5\%$, except for the last bin in the backward
region, where the cross section is suppressed.
For transverse W-pair production the LA is a reasonable approximation in the central region
(similar as in the ZZ case), while for longitudinal W-boson production 
the non-logarithmic $\mathcal{O}(\alpha)$ terms contribute more than ten percent
over most of the distribution.

All in all the deviations between fixed-order and SCET results are at
the level of one percent and hence in the range expected for
power-suppressed corrections, which can safely be neglected in the
considered setups. An exception is given by 
the purely longitudinally polarised $\text{W}^+\text{W}^-$ production in the backward region, which is, however, phenomenologically not relevant.

\subsubsection{Individual $\text{SCET}_\text{EW}$ contributions }
\label{Sec:IndiRes}

\begin{figure}
\centering
\includegraphics[width=0.475\textwidth]{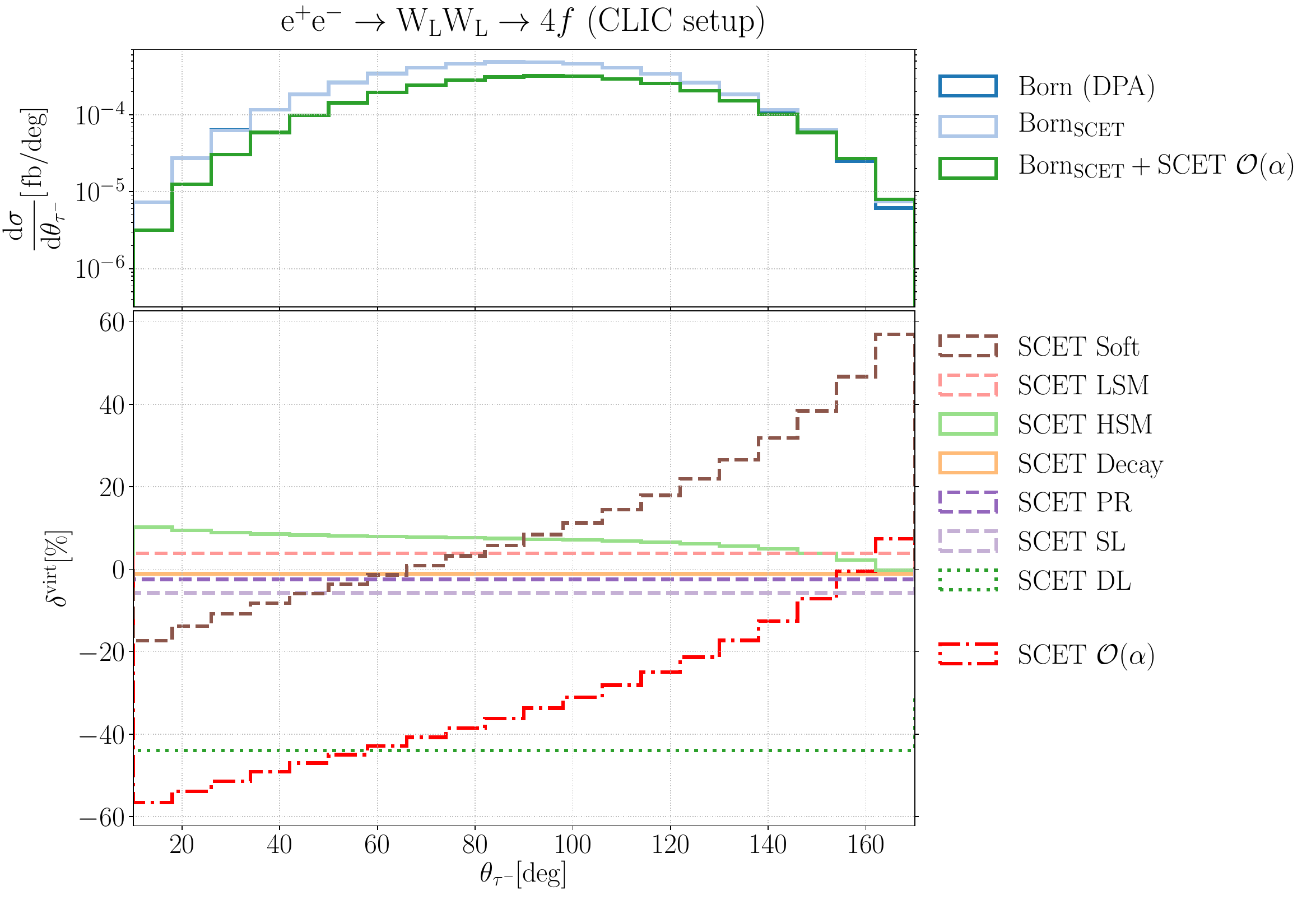}
\includegraphics[width=0.475\textwidth]{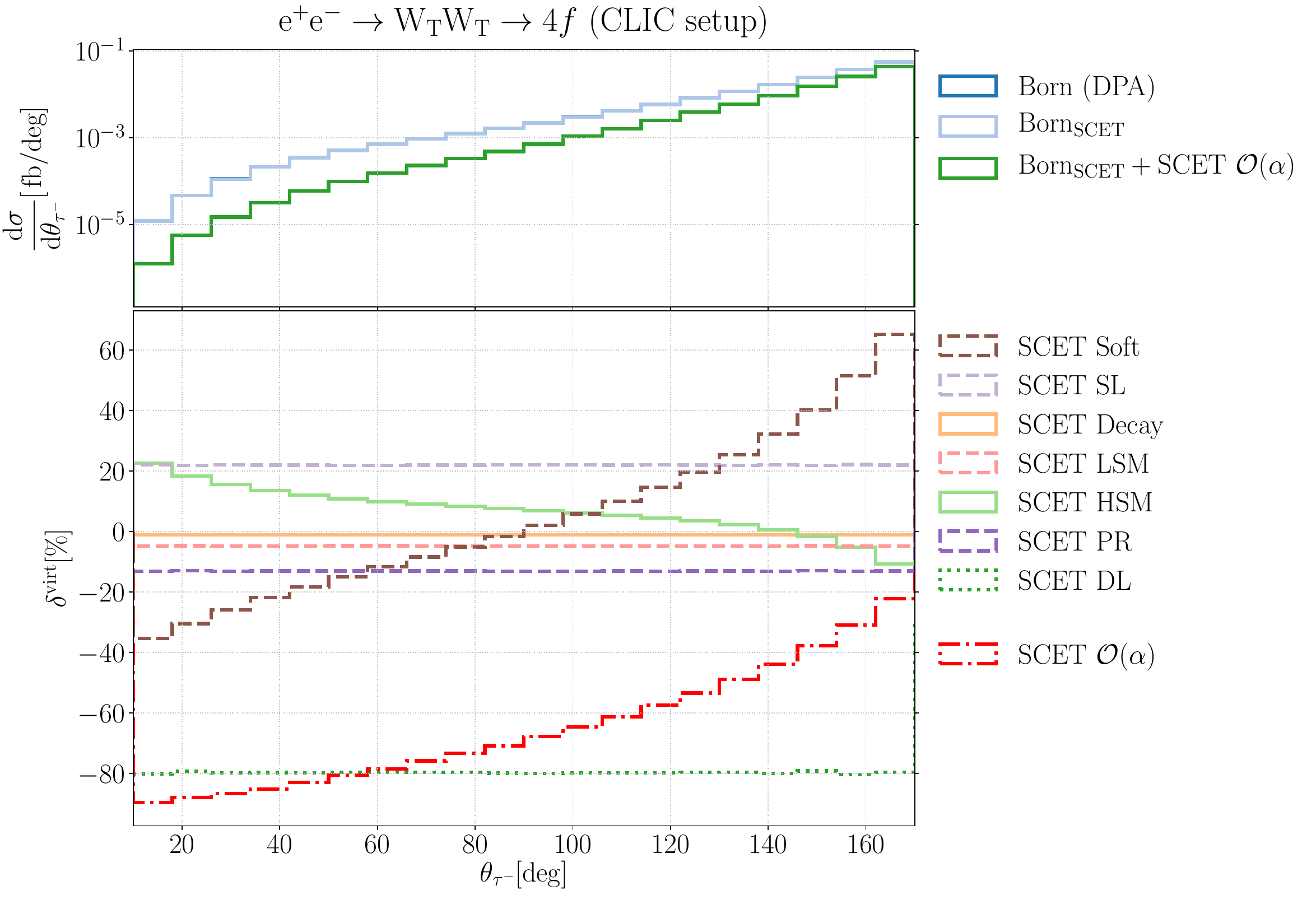}\\
\includegraphics[width=0.475\textwidth]{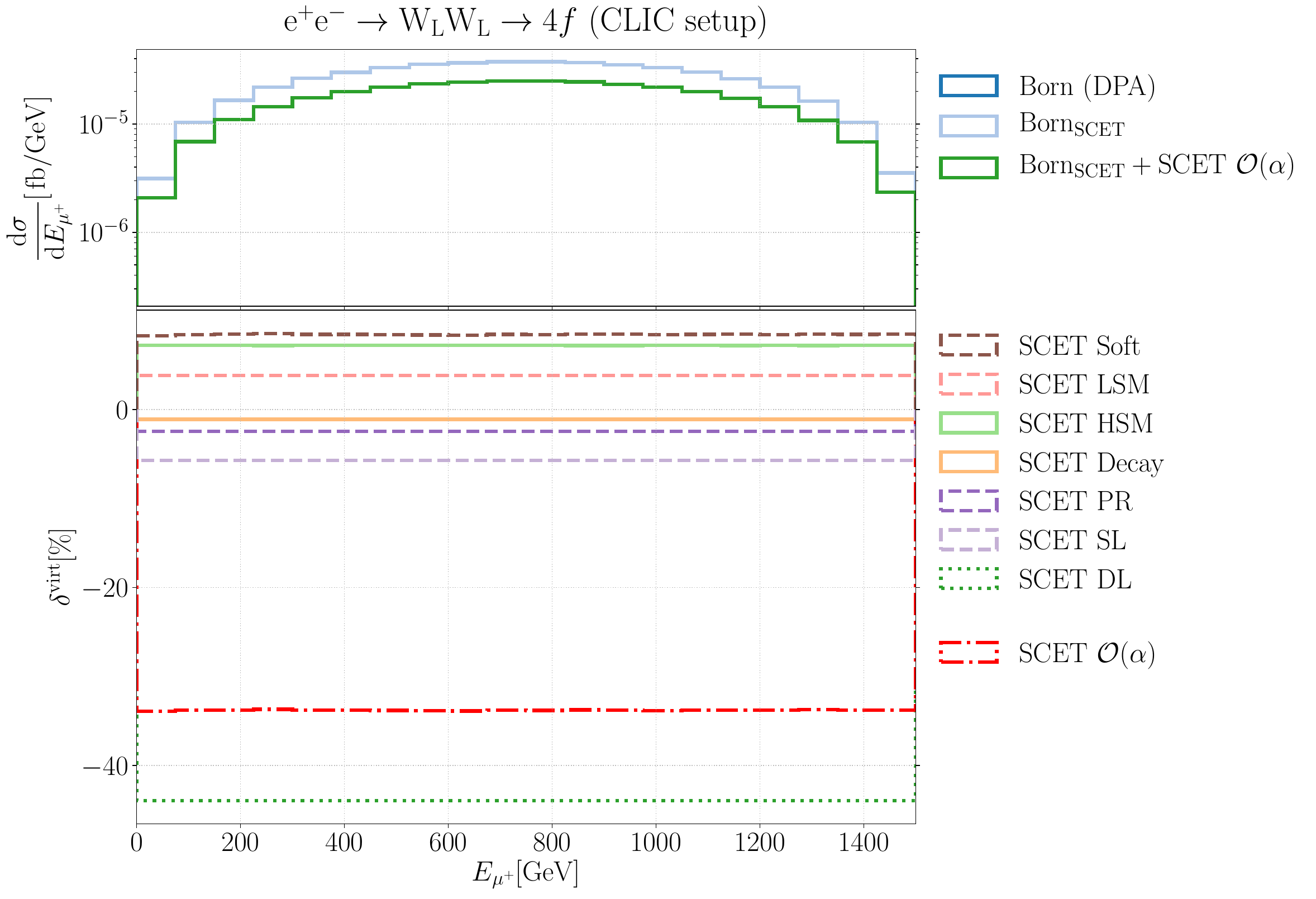}
\includegraphics[width=0.475\textwidth]{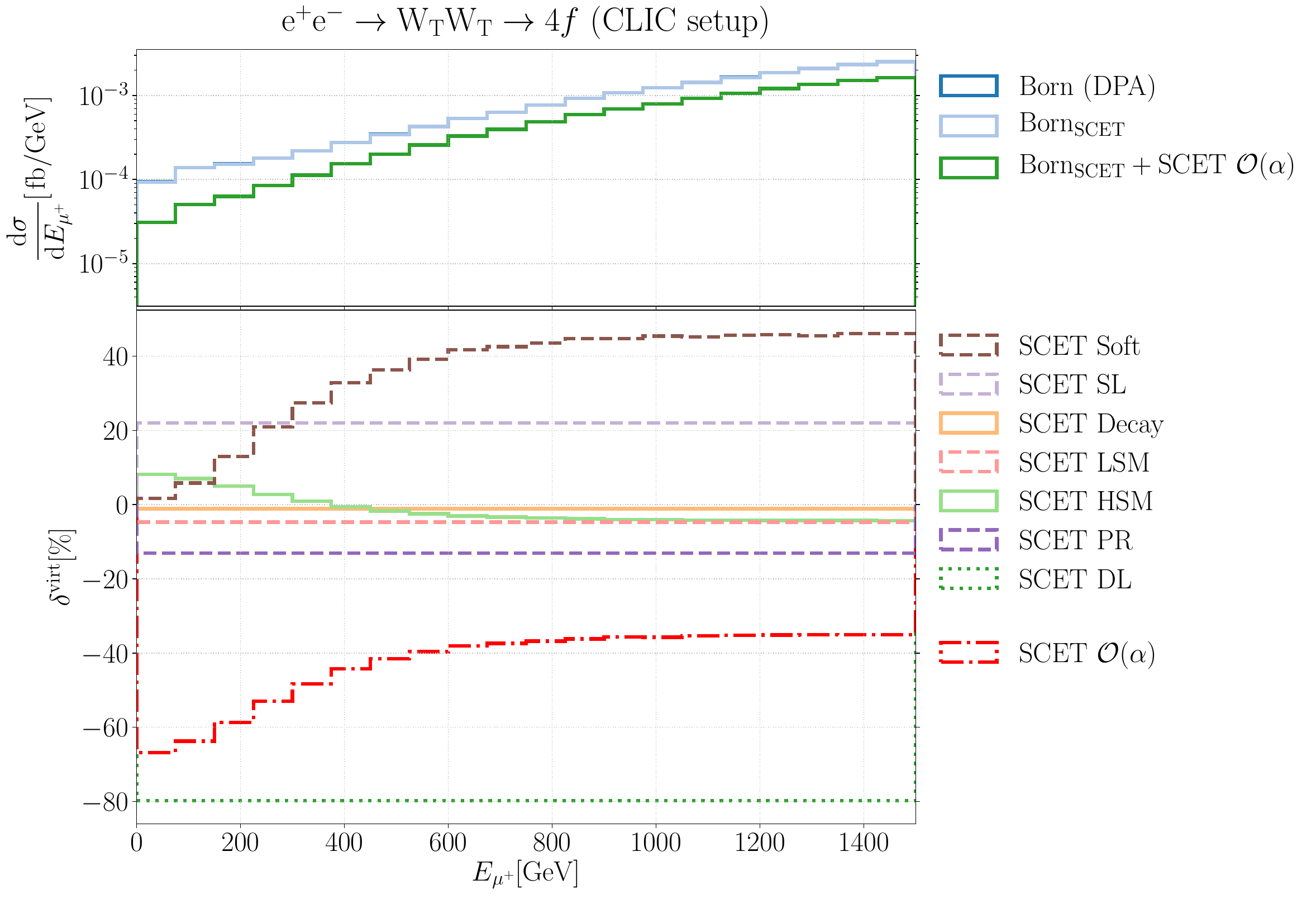}\\
\caption{Individual SCET contributions to the $\tau$-production-angle
  and $\tau$-energy distributions in $\eewwllll$ with longitudinally (left) and transversely (right) polarised W~bosons. The meaning of the abbreviations is explained in the text.}
\label{Fig:eeWWContri}
\end{figure}

\begin{figure}
\centering
\includegraphics[width=0.45\textwidth]{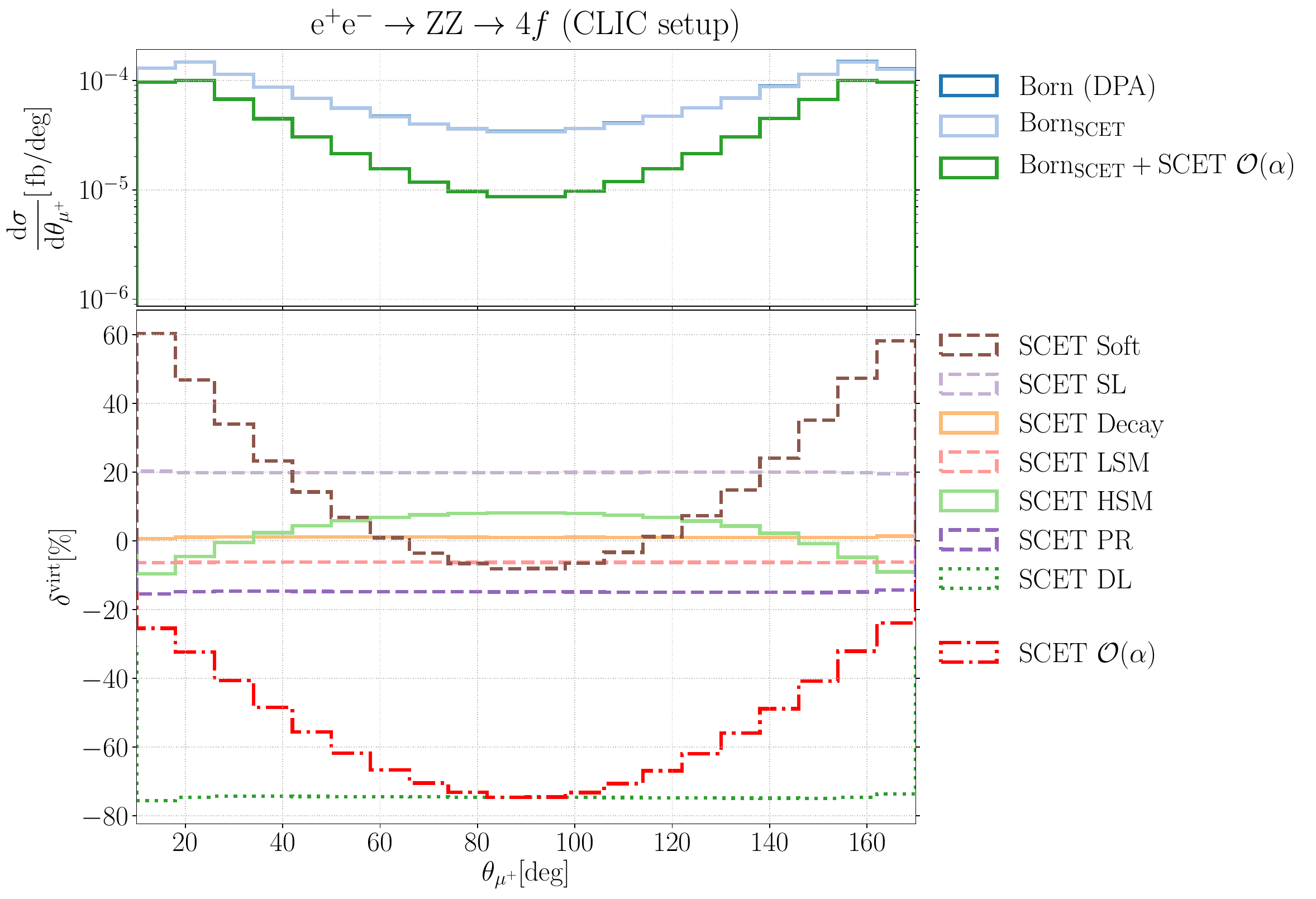}
\includegraphics[width=0.45\textwidth]{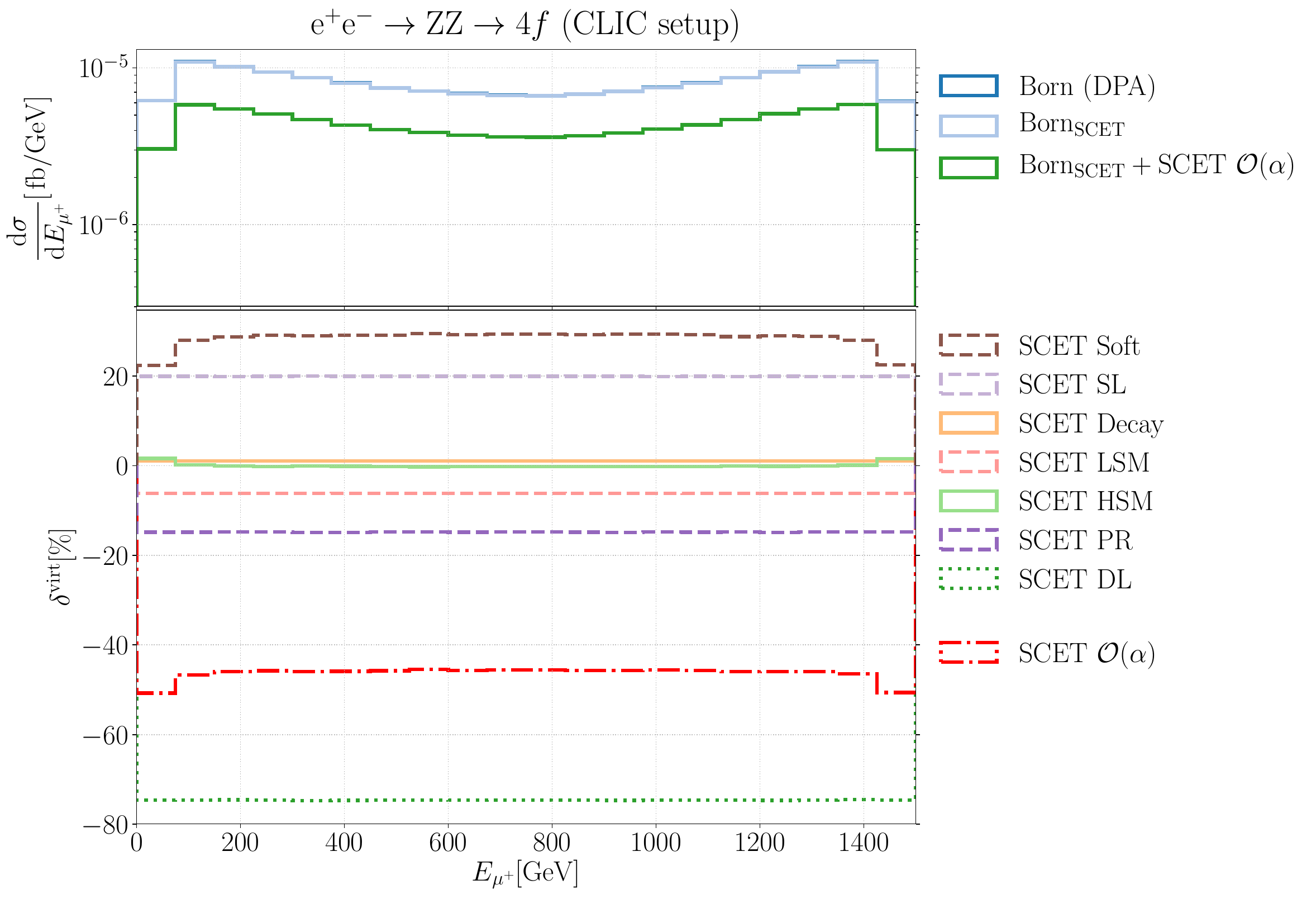}
\caption{Individual SCET contributions to the distributions in the antimuon production angle and antimuon energy in $\eezzllll$. The meaning of the abbreviations is explained in the text.}
\label{Fig:eeZZContri}
\end{figure}

In Figs.~\ref{Fig:eeWWContri} and \ref{Fig:eeZZContri} we demonstrate
exemplarily the role of the individual contributions entering the
results in Sec.~\ref{Sec:ValiRes}.
The different curves are labelled as follows:
\begin{itemize}
\item DL: Double-logarithmic contributions from the collinear anomalous dimension $\gamma_\text{C}$.
\item SL: Single-logarithmic contributions from $\gamma_\text{C}$.
\item PR: Corrections associated with the renormalisation of $\alpha$ and $\theta_\text{w}$. Both logarithmic and finite contributions are included.
\item Soft: Angular-dependent single logarithms from $\boldsymbol{\gamma}_\text{S}$.
\item HSM: High-scale matching coefficient: The $\mathcal{O}(\alpha)$~corrections evaluated in the SySM.
\item LSM: Low-scale corrections: The logarithmic and the finite part of $D_\text{C}$ to $\mathcal{O}(\alpha)$.
\item Decay: Corrections associated with the W- or Z-boson decay.
\item The sum of all is denoted as $\text{SCET}$ $\mathcal{O}(\alpha)$.
\end{itemize}
Contributions involving single logarithms are drawn with
    dashed lines, those with double-logarithms
are dotted, while constant $\mathcal{O}(\alpha)$ contributions are solid and the sum is
dash--dotted.
For the definitions of the quantities $D_\text{C}$, $\gamma_\text{C}$, and $\boldsymbol{\gamma}_\text{S}$ we refer to Sec.~\ref{Sec:Ingredients}.
It should be stressed that the distinction of these contributions is only possible if the $\text{SCET}_\text{EW}$ amplitude is expanded in $\alpha$. Otherwise the matrix structure of the anomalous dimension mixes with the high-scale matching coefficients producing terms that can not unambiguously be identified with one of the above categories. We present these results in order to give a rough estimate of the respective effects. 

Conceivably the DL contributions are by far the dominant ones, followed by the SL, Soft, and PR ones. The most important qualitative difference between the two sample processes is the sign of the SL contribution, which is positive for transverse gauge bosons and fermions. When longitudinal gauge bosons are involved, the top-mass enhanced last term in (\ref{gammaCPhi}), which comes with a different sign, dominates the SL contribution and renders it negative (see the left distribution of Fig.~\ref{Fig:eeWWContri}).

Many quantities are constant or almost constant over the whole phase
space: The phase-space dependence, i.e.\ the shape of the
distributions is mostly determined by the soft and the high-scale
matching contributions. For $\text{W}^+\text{W}^-$ production all
other contributions are completely flat for both polarisation
states. In the ZZ plots the other contributions have a slight angular
dependence owing to the different corrections to the subamplitudes
associated with left- and right-handed electrons in the initial
state. Also the decay corrections are phase-space dependent resulting from different corrections depending on the helicities of the final-state leptons. However, both effects are small compared to the variations of Soft and HSM contributions. In this context it is worth mentioning that one can observe a partial cancellation between the HSM and 
angular-dependent (Soft) contributions, which can be explained by the fact that within the $\text{SCET}_\text{EW}$ formalism the angular-dependent logarithms in the total corrections are split according to (we pick the Mandelstam variable $t$ for definiteness)
\begin{align}\label{eq:anglogsplit}
\frac 12 \log^2\left(\frac{-t}{M_\text{W/Z}^2}\right)\rightarrow \underbrace{\frac 12 \log^2\left(\frac{-t}{s}\right)}_{\in\text{HSM}}+\underbrace{\log\left(\frac{-t}{s}\right)\log\left(\frac{s}{M_\text{W/Z}^2}\right)}_{\in\boldsymbol{\gamma}_\text{S}}+\underbrace{\frac 12 \log^2\left(\frac{s}{M_\text{W/Z}^2}\right)}_{\in\gamma_\text{C}}.
\end{align}
The l.h.s.\ is proportional to a single contribution obtained in LA in fixed-order (see Ref.~\cite{DennerPozz1}). The first term on the r.h.s.\ contains no low-energy information and is hence part of the high-scale matching coefficients, while the second and third terms are part of the anomalous dimension. If $-t$ becomes small with respect to $s$, the first term is the dominant one in the HSM and gives a positive contribution, while the second one gives a negative contribution. Therefore in the small-$|t|$ tails the HSM- and $\boldsymbol{\gamma}_\text{S}$-related contributions are always of opposite sign.%
\footnote{We note that the terms involving
    $\log^2\left({-t}/{s}\right)$ on the r.h.s.\ of
    \refeq{eq:anglogsplit} have been used in \citeres{Pagani_Zaro,Lindert:2023fcu} to improve the
    quality of the logarithmic approximation for many processes.}

\subsubsection{Resummed results}
\label{Sec:ResumRes}
\begin{table}
\centering
\makegapedcells
\begin{tabular}{|l|c|r|}
\hline
Accuracy & $\sigma^\text{WW}/\text{fb}$ & $\delta_\text{EW}$\\
\hline
Fixed order&1.519(9)&$-13.7\%$\\%
SCET $\mathcal{O}(\alpha)$&1.524(9)&$-13.4\%$\\%
LL+NLO&1.901(9)&8.0\%\\%
$\text{NLL}_\text{FO}$+NLO&1.415(9)&$-19.6\%$\\%
NLL+NLO&1.563(9)&$-11.2\%$\\%
LL+NLO+running&2.044(9)&16.1\%\\%
\hline
\end{tabular}
\quad
\begin{tabular}{|l|c|r|}
\hline
Accuracy & $\sigma^\text{ZZ}/\text{fb}$ & $\delta_\text{EW}$\\
\hline
Fixed order&0.00958(5)&$-26.6\%$\\%
SCET $\mathcal{O}(\alpha)$&0.00960(5)&$-26.4\%$\\%
LL+NLO&0.01262(5)&$-3.3\%$\\%
$\text{NLL}_\text{FO}$+NLO&0.00952(5)&$-27.0\%$\\%
NLL+NLO&0.01127(5)&$-13.6\%$\\%
\hline
\end{tabular}
\caption{Integrated fiducial cross sections for $\eewwllll$ (left) and $\eezzllll$ (right) with the virtual corrections replaced by the respective $\text{SCET}_\text{EW}$-resummed results. The contributions included in the (N)LL+NLO calculations are given in Sec.~\ref{Sec:LogCount}.}
\label{tab:sigmaInt}
\end{table}
In this section, we present our main results, i.e.\ the
$\text{SCET}_\text{EW}$ results including a resummation of the Sudakov
logarithms. Following the insights of the analysis of the DPA in
Sec.~\ref{Sec:DPARes} we apply the DPA only to the virtual corrections
and employ the rescaling (\ref{virtrescale}) afterwards. All other
contributions are computed fully off shell.  In the case of
$\text{W}^+\text{W}^-$ production (l.h.s.\ of Table~\ref{tab:sigmaInt})
the LL resummation shifts the cross section by 21.4\% with respect to
the LO cross section, which is obtained as the difference between
$\text{SCET}$ $\mathcal{O}(\alpha)$ and LL+NLO. The magnitude of this
effect can be estimated from the size of the DL corrections to the
transverse polarisations (80\%) as $\exp(-0.8)-1+0.8\approx25\%$.
The difference between LL+NLO and NLL+NLO is about $-19\%$ with respect to Born.
This quantity serves as a measure for the effect of the NLL resummation alone (without the effect of the LL resummation). 
To determine the effect of the most important terms of the NLL
resummation, we can consider the difference between LL+NLO and $\text{NLL}_\text{FO}$+NLO, which is about $-28\%$. These schemes differ by the terms given in the second column of (\ref{FOLogCount}) (excluding the uppermost one of course), of which the dominant one is the $\alpha^2\mathsf{L}^3$ term. It turns out that these terms have an even larger impact 
than the ones in the first column (again excluding the first one). The remaining $9\%$ are then given by the resummed single logarithms, i.e.\ the second column of (\ref{EFTLogCount}).  

For ZZ-pair production one has a similar picture: The LL resummation accounts for a $\sim 23\%$ shift. Going to $\text{NLL}_\text{FO}$ slightly overcompensates this effect: The cross
section is shifted by $-24\%$. And the remaining NLL terms produce another positive shift of $\sim14.5\%$.

It can therefore be concluded that an accurate theoretical prediction
for this collider setup should include at least NLL-resummed EW
corrections. Also the influence of higher-order terms in the
logarithm~counting as well as a resummation of the logarithms
associated with coupling renormalisation should be analysed.

\begin{figure}
\centering
\includegraphics[width=0.475\textwidth]{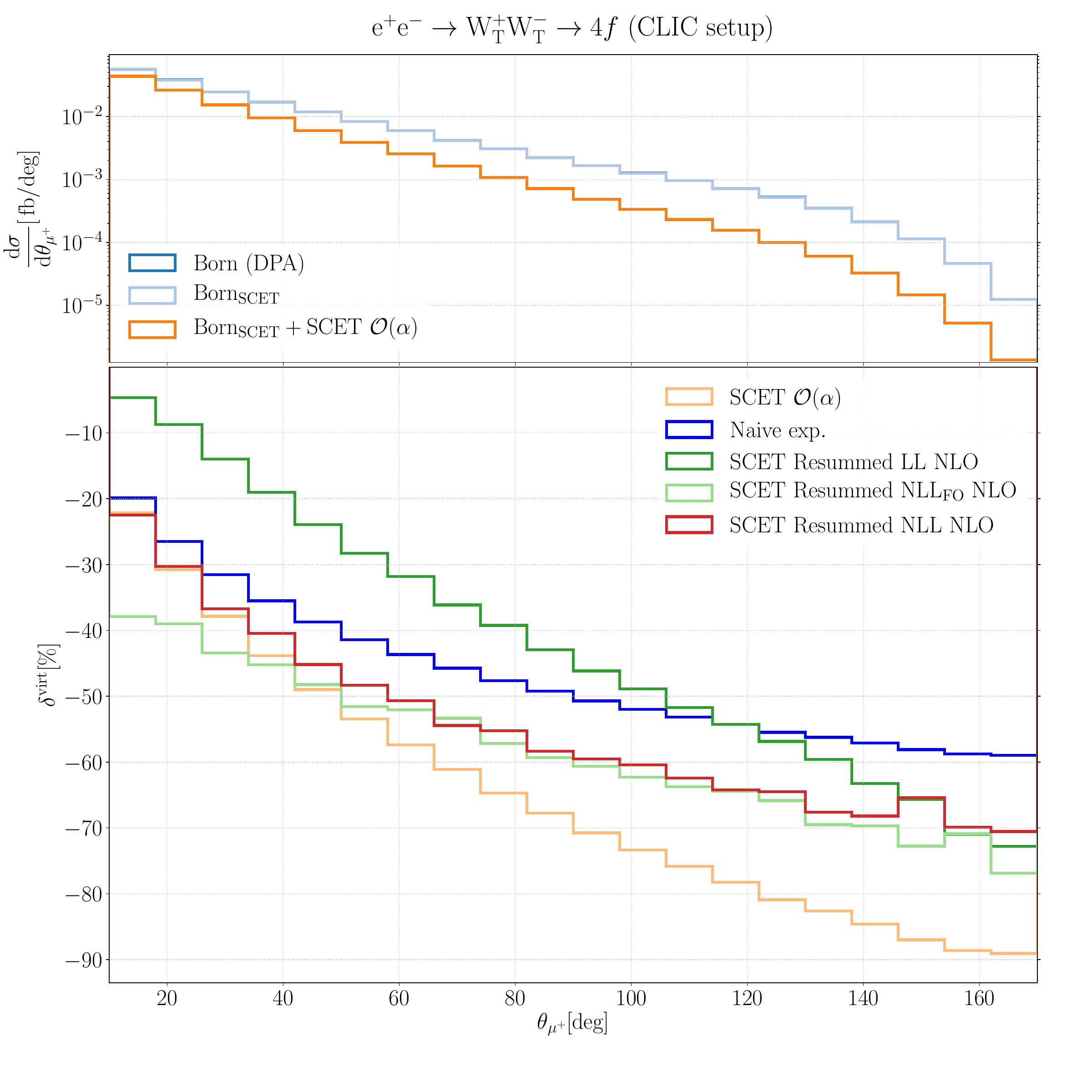}
\includegraphics[width=0.475\textwidth]{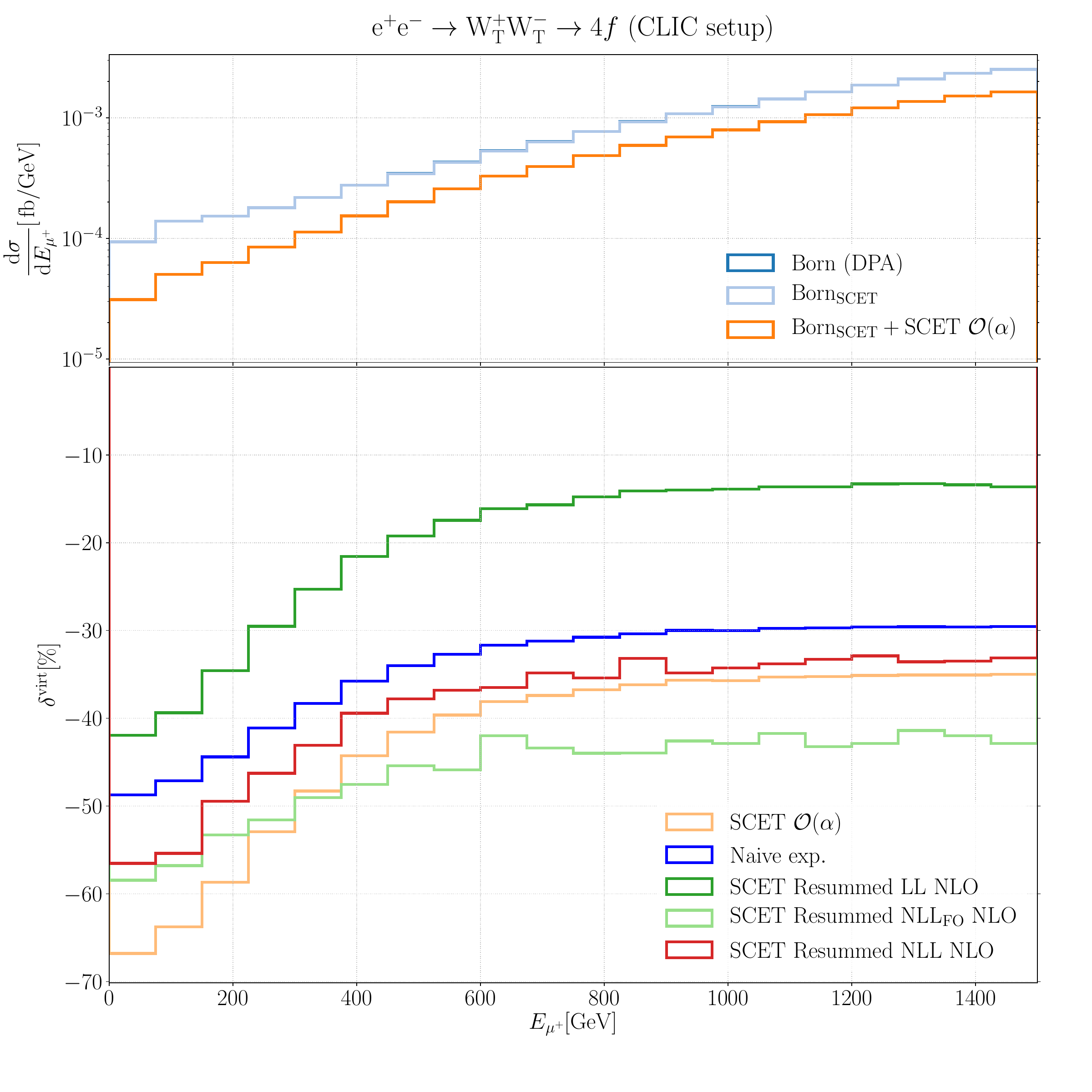}
\caption{Effect of the $\text{SCET}_\text{EW}$ resummation in $\eewwllll$ with transversely polarised W~bosons differential in
the muon production angle and energy. The meaning of the abbreviations is explained in the text.}
\label{Fig:eeWWResumTrans}
\end{figure}

\begin{figure}
\centering
\includegraphics[width=0.475\textwidth]{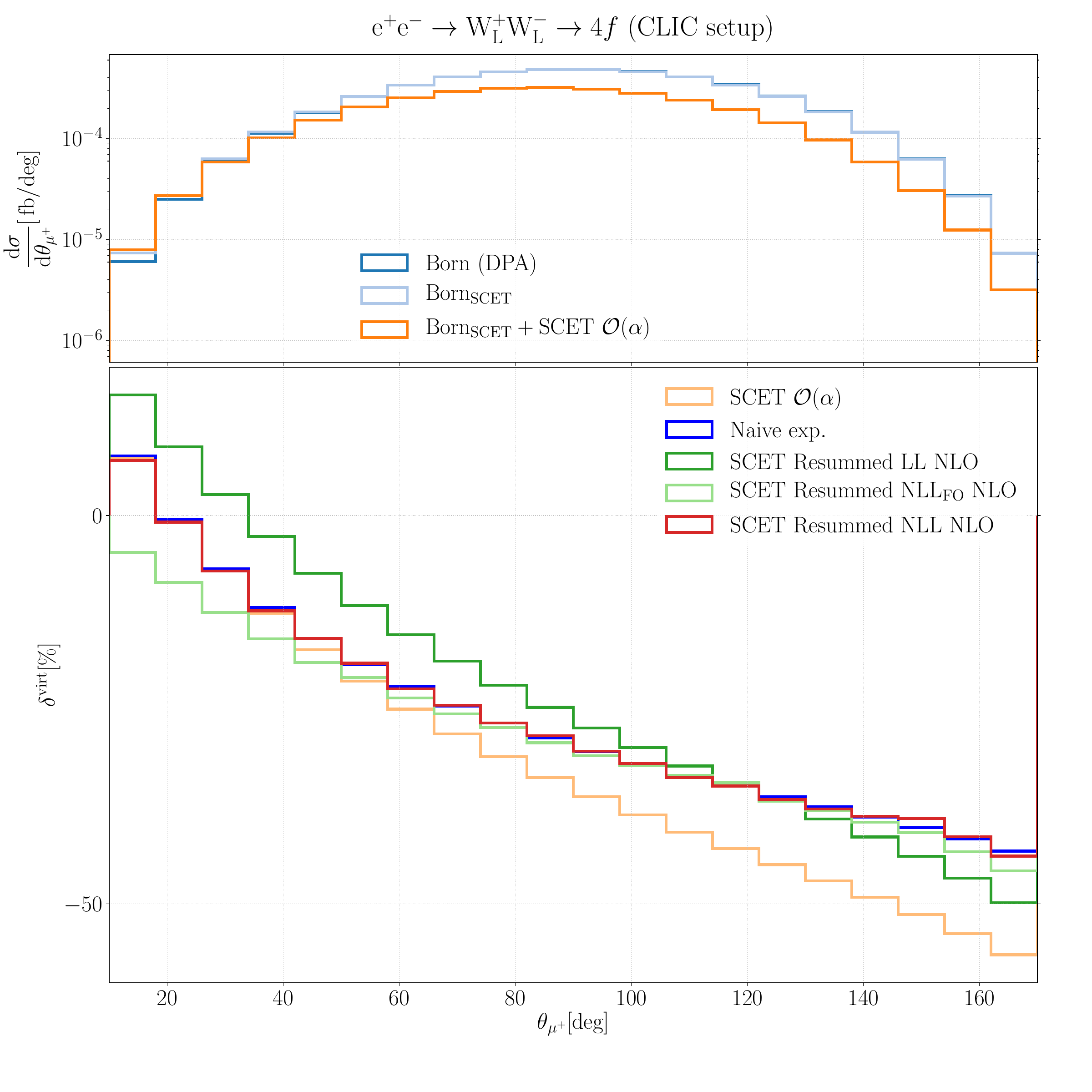}
\includegraphics[width=0.475\textwidth]{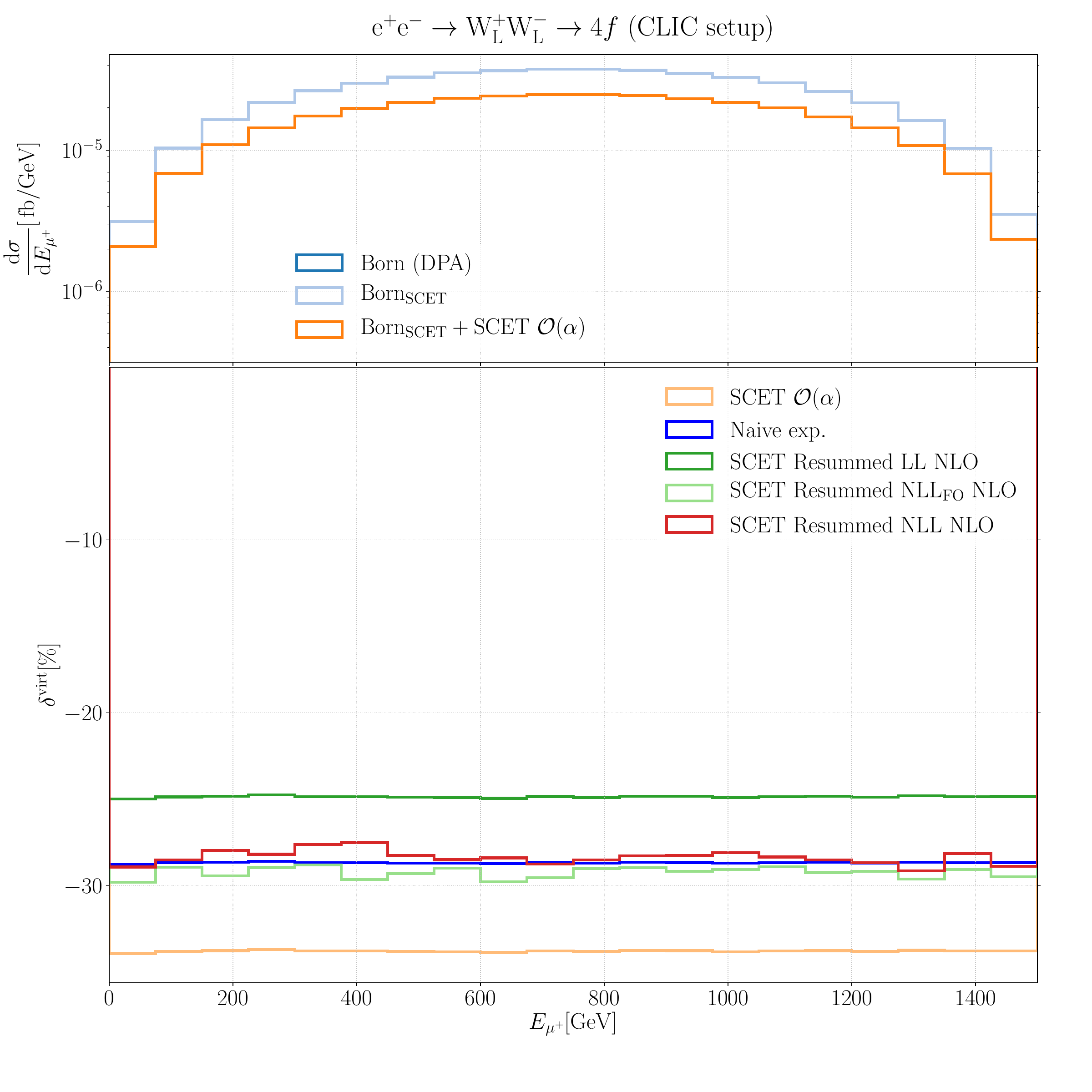}
\caption{Effect of the $\text{SCET}_\text{EW}$ resummation in
  $\eewwllll$  with longitudinally polarised W~bosons differential in
the muon production angle and energy. The meaning of the abbreviations is explained in the text.}
\label{Fig:eeWWResumLong} 
\end{figure}

Differential results in the muon production angle and energy
for transverse and longitudinal W-boson pair production are
shown in Figs.~\ref{Fig:eeWWResumTrans} and \ref{Fig:eeWWResumLong}.
While the upper panels show the distributions for the leading order in
ordinary perturbation theory and SCET as well as the expanded
$\mathcal{O}(\alpha)$ results in SCET, the lower panels show the
corrections relative to the leading order for the expanded SCET result
and different resummed results as defined in (\ref{LogCountings}). 
In addition we show the curves from the ``naive'' exponentiation
\begin{align}
\delta_\text{FO}^\text{virt}\rightarrow \exp(\delta_\text{FO}^\text{virt}),
\end{align}
which has occasionally been employed such as, for instance, in \citere{Sherpa_Sudakov}.

The effect of the LL resummation is (almost) constant over phase space. This is (roughly) the deviation between the $\mathcal{O}(\alpha)$ (orange) curve and the dark green curve, up to $\mathcal{O}(\alpha^2\mathsf{L})$-contributions from the product of one-loop matching and $\mathcal{O}(\alpha)$ soft anomalous dimension,
 see (\ref{LogCountings}) and the footnote underneath. 
The effects of the NLL and $\text{NLL}_\text{FO}$ terms are strongest
in the forward region, where the angular-dependent logarithms are
large (see the ``SCET Soft" curves in Fig.~\ref{Fig:eeWWContri}). Here
the total negative shift due to the $\text{NLL}_\text{FO}$ terms (the
difference between the dark green and the pale green curve, ${\mathcal{O}}(\alpha^2\mathsf{L}^3)$) grows up to $\sim 20\%$ in the longitudinal case and exceeds $30\%$ in the transverse case. The larger impact in the transverse case is explained both by the larger values of the double logarithms (owing to larger EW Casimir for transverse W bosons) and because of the fact that the single-logarithmic corrections resulting from the Yukawa-coupling term in the 
Goldstone-boson anomalous dimension (\ref{gammaCPhi}) contributes with a negative sign to the SL corrections, leading to a smaller value for the SL corrections in the longitudinal case. Towards the backward region the magnitude of the shifts shrinks until it eventually changes the sign and becomes positive for small scattering angles ($\theta_{\mu^+}\gtrsim150^\circ$). This region,
however, contributes little to the total cross section. 
The NLL terms (difference between pale green and red curve) show a
similar behaviour and range between $+15\%$ and about $0\%$ for both
polarisation states. Since these terms are dominated by the
exponential of the soft anomalous dimension their dependence on the
polarisation state is weaker: the angular-dependent logarithms
(Soft) have a similar
magnitude ($\sim60\%$) for both polarisation states (see Fig.~\ref{Fig:eeWWContri}).

In the muon-energy distribution all resummation effects are more or less flat in the longitudinal case. In the transverse case the effects of both the $\text{NLL}_\text{FO}$ and NLL terms are largest in the high-energy region and decrease towards lower energies. 

The quality of the naive exponentiation prescription is
    different for the different polarisation states: Because of the small
single-logarithmic corrections, the corrections in the longitudinal case are strongly dominated by the DL corrections
and the naive exponentiation prescription works well in the sense that it reproduces
the NLL+NLO result within $1\%$. In the dominating transverse case the difference between the NLL+NLO
result and the naive prescription is significant, reaching more than $10\%$ with respect to Born
in the backward region. In the small-energy regime and in the forward region, where the cross section
is large, the difference amounts to less than $5\%$.

\begin{figure}
\centering
\includegraphics[width=0.45\textwidth]{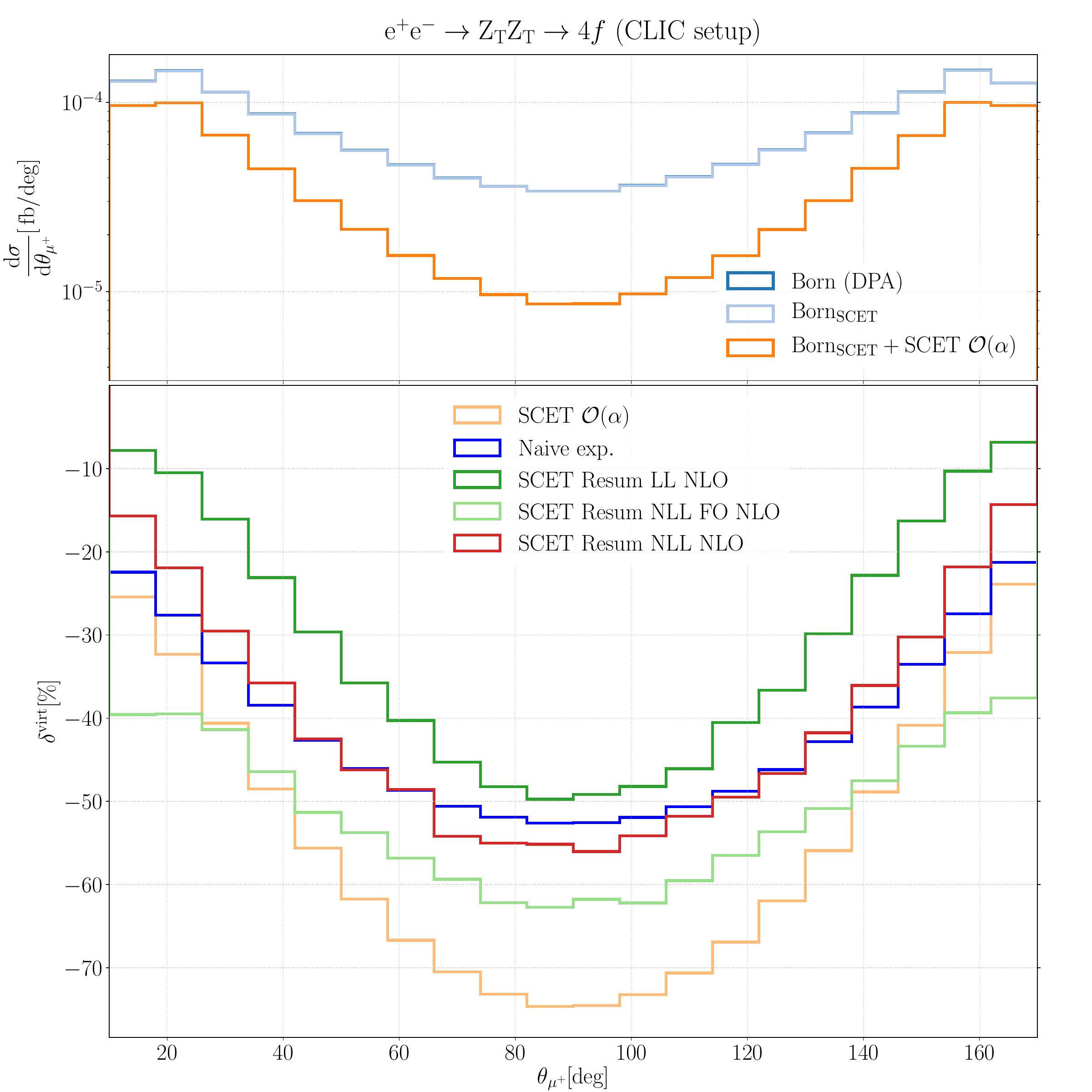}
\includegraphics[width=0.45\textwidth]{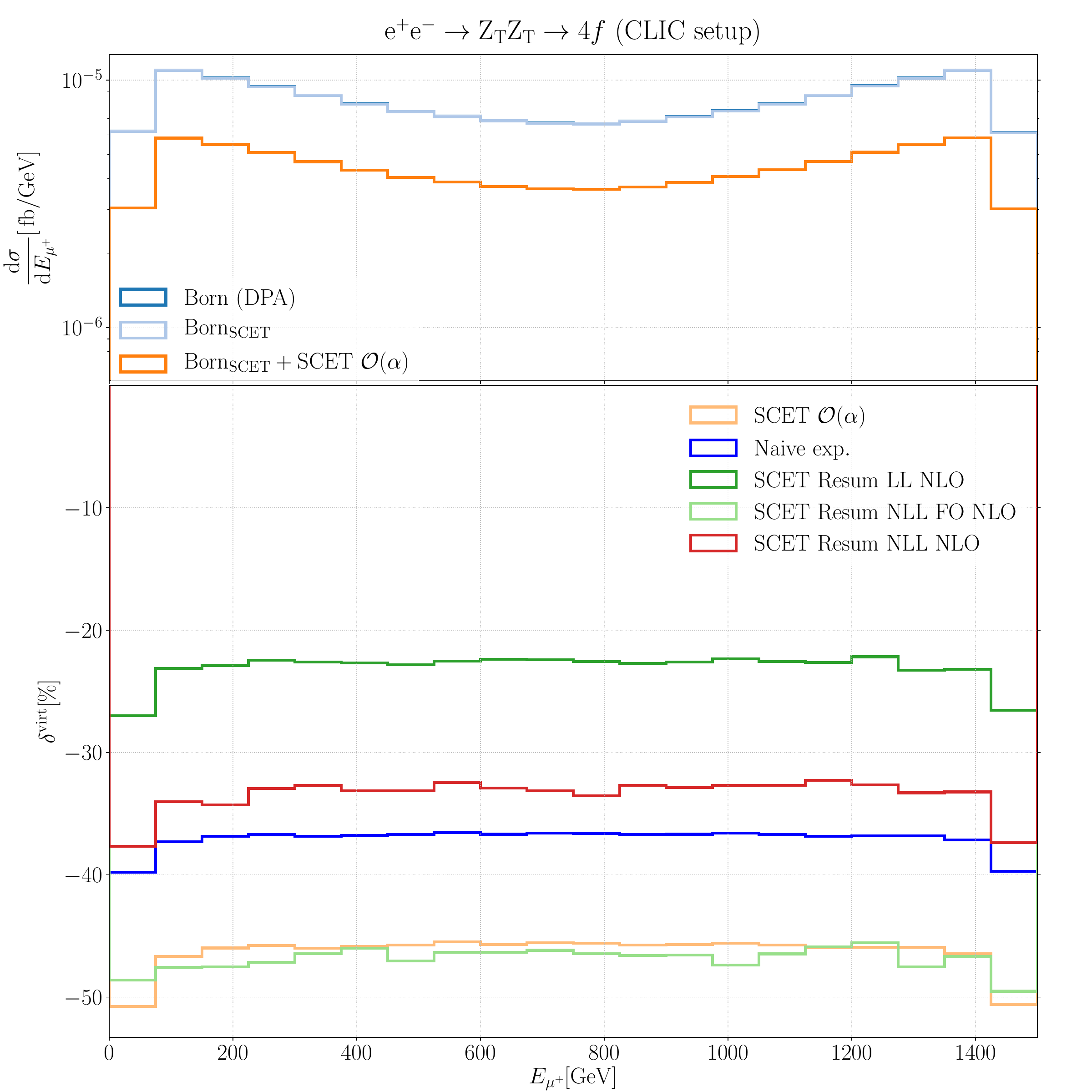}
\caption{Effect of the $\text{SCET}_\text{EW}$ resummation $\eezzllll$ differential in
the antimuon production angle and energy. The meaning of the abbreviations is explained in the text.}
\label{Fig:eeZZResum}
\end{figure}

In Fig.~\ref{Fig:eeZZResum} we show the analogous distributions for ZZ
production. 
For the antimuon-energy distribution the curves for the relative corrections are approximatively parallel, which implies that the resummation effects are more or less constant. In the angular distribution for the antimuon the $\text{NLL}_\text{FO}$ and NLL
terms have larger effects in the forward and backward region: The shift between the green curves reaches 30\% in the first and last bin, while in the central region it is about 12\%. The shift between pale green and red curve is about $25\%$ in the forward and backward region and about 7\% in the central region. 
At the end this implies that going from LL to NLL (from dark green to
red) yields a roughly constant shift of about 5\%. 

The naive exponentiation describes the exponentiation well for moderate lepton production angles ($40^\circ\lesssim\theta_{\mu^+}\lesssim70^\circ$), while both in the
central region and in the forward and backward regimes it introduces an error of $3$--$5\%$. Over the lepton energy,
the error is more or less uniformly distributed, accounting for $\sim 4\%$.

\begin{figure}
\centering
\includegraphics[width=0.45\textwidth]{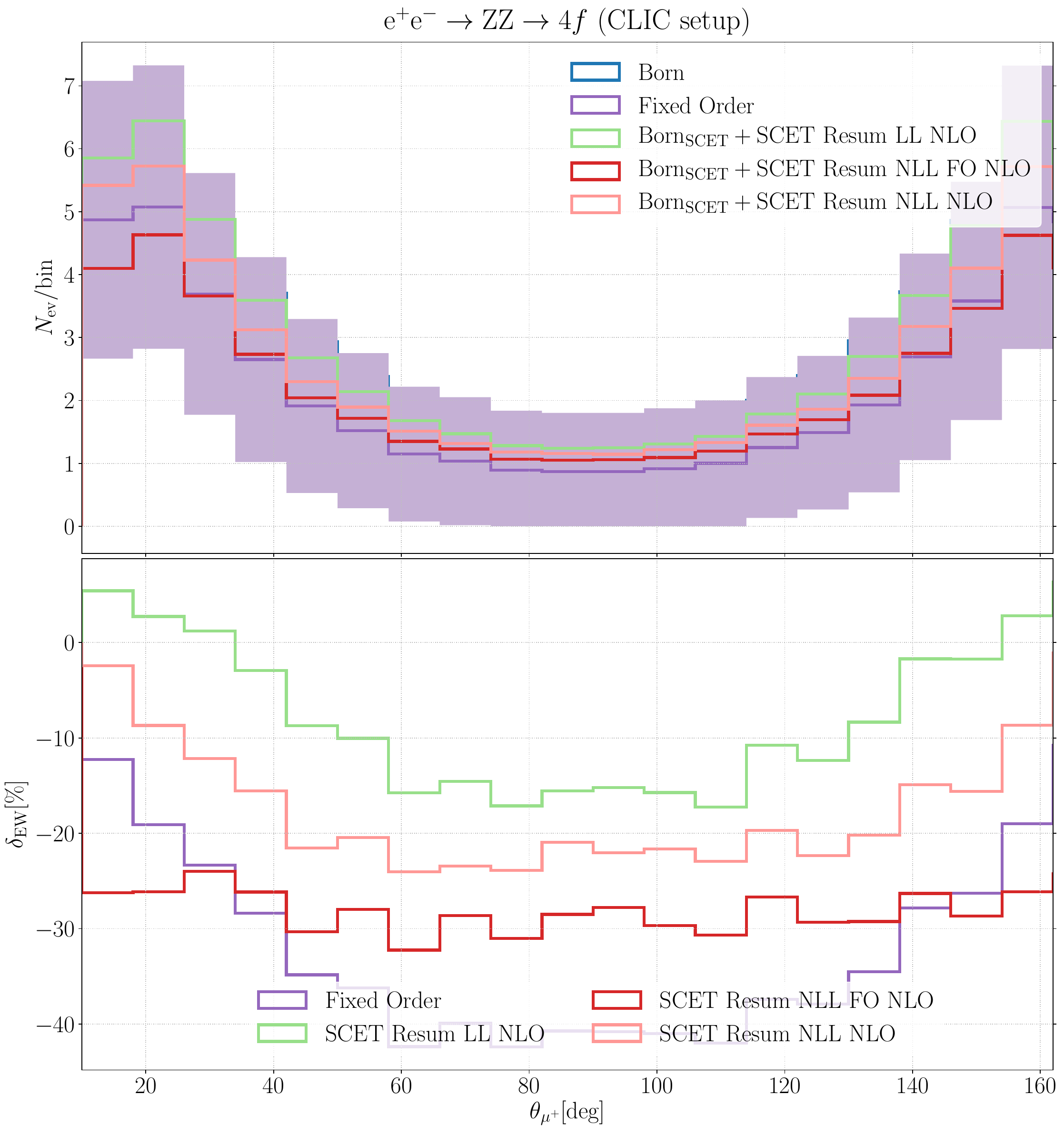}
\includegraphics[width=0.45\textwidth]{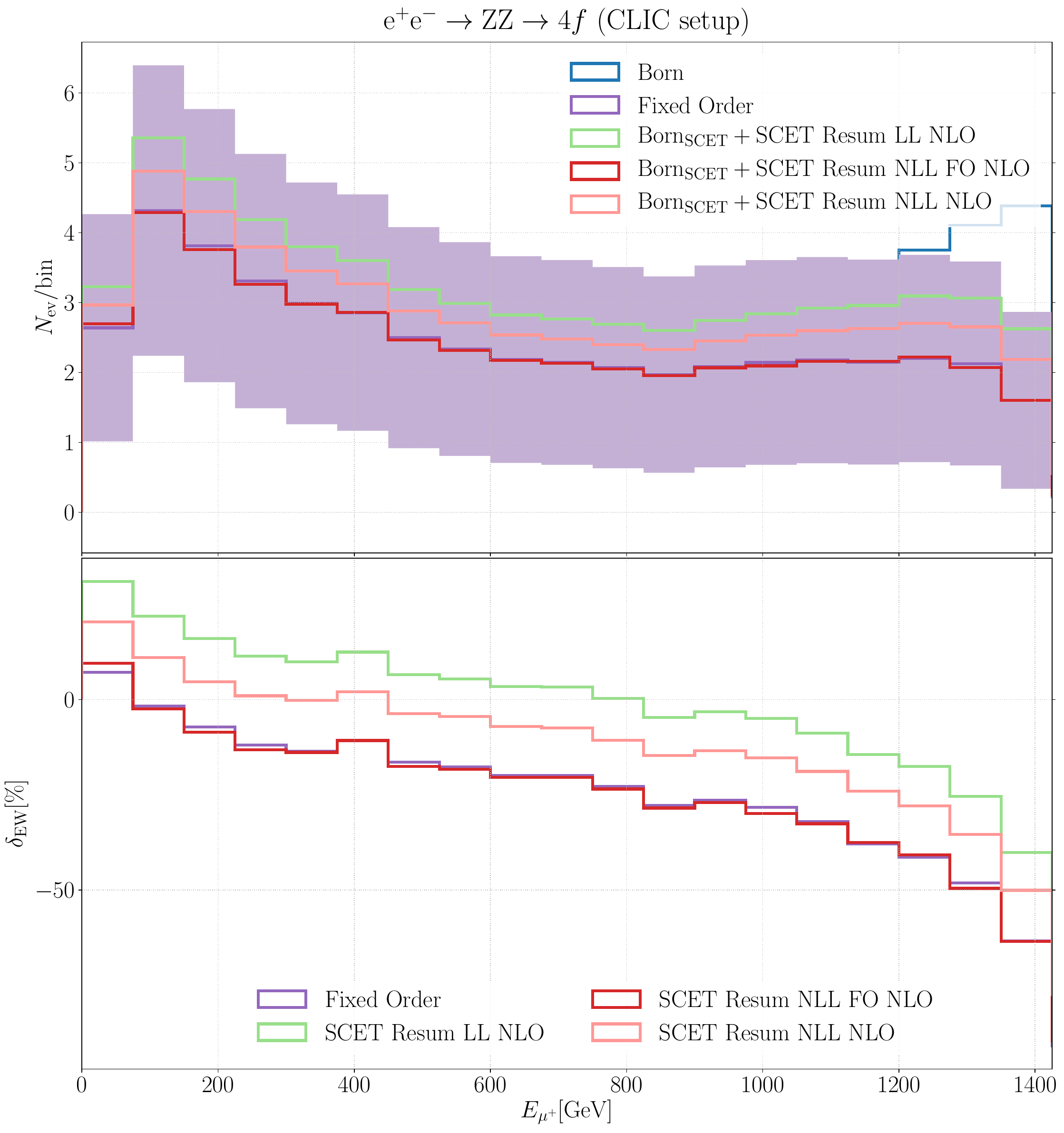}
\caption{Differential distributions in the antimuon production angle
  and energy for $\eezzllll$ with the error on the counting rates
  shaded around the purple curves. The results include real, virtual, and integrated dipole contributions. The
upper panels show the expected event numbers per bin. The various
curves differ in the treatment of the virtual corrections only.}
\label{Fig:eeZZResum_Shade}
\end{figure}
\begin{figure}
\centering
\includegraphics[width=0.45\textwidth]{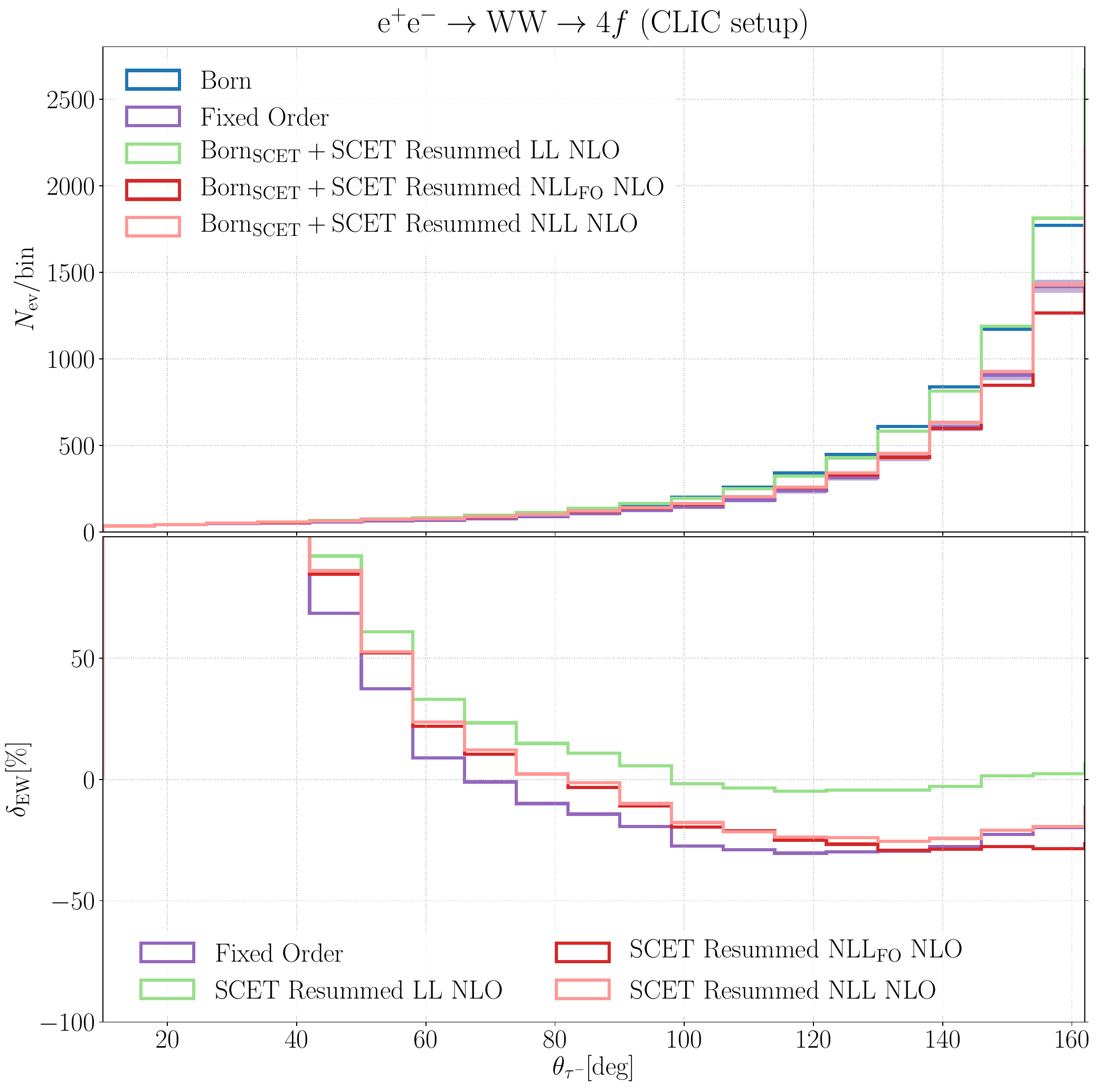}
\includegraphics[width=0.45\textwidth]{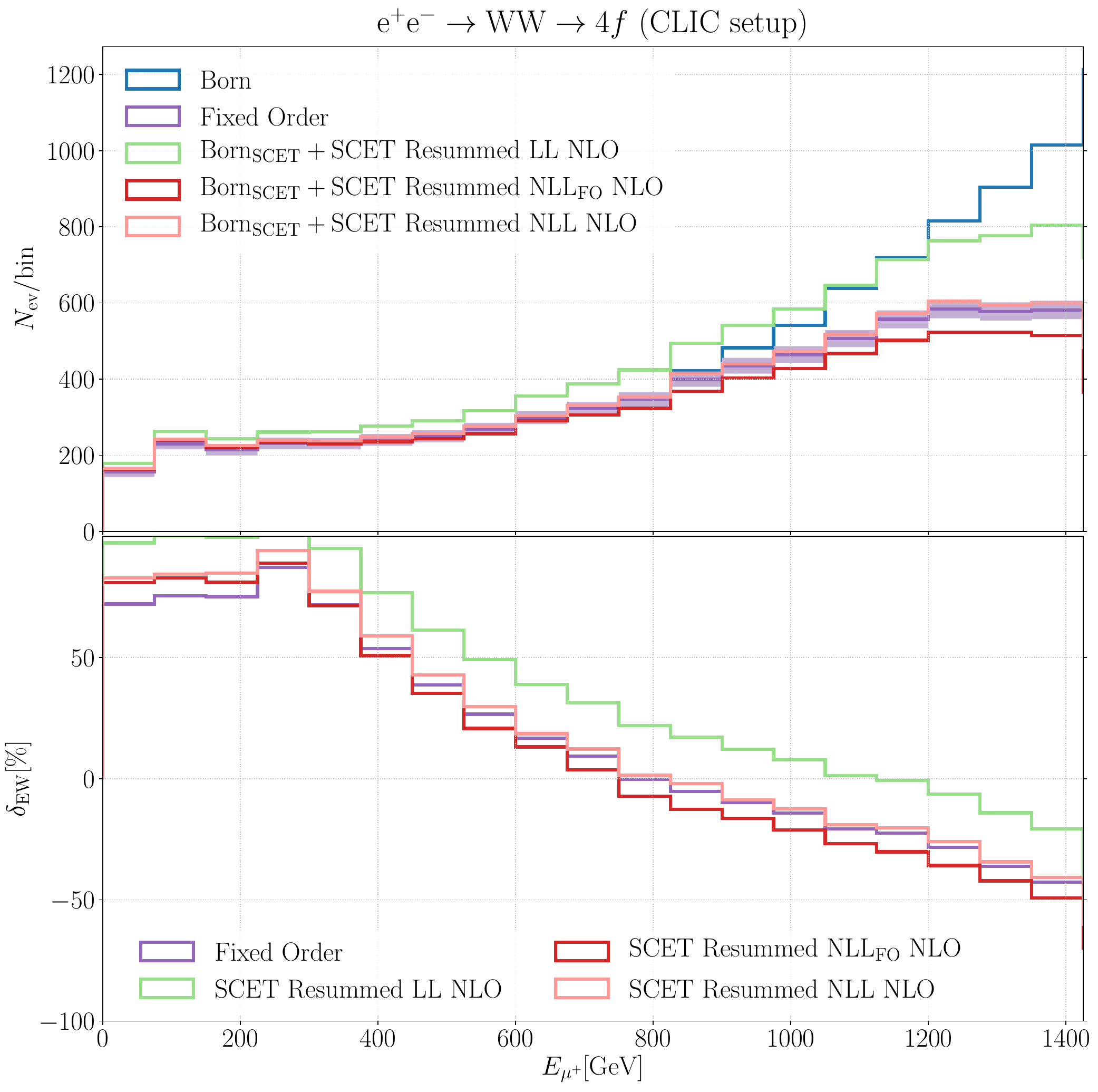}
\includegraphics[width=0.45\textwidth]{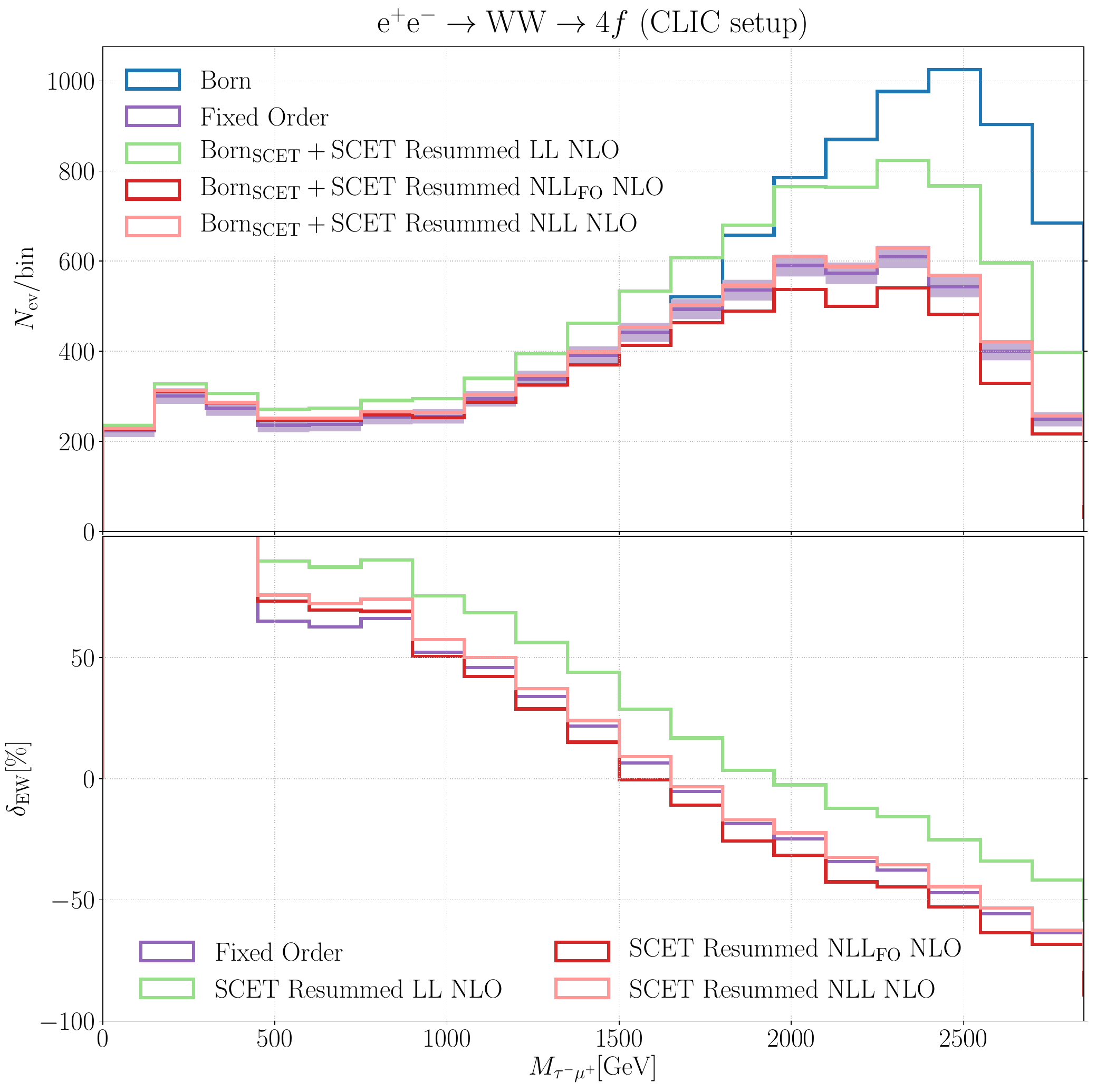}
\caption{Differential distributions in the tau production angle, the
  antimuon energy, and the tau--antimuon invariant mass. Same layout
  as in Fig.~\ref{Fig:eeZZResum_Shade}.}
\label{Fig:eeWWResum_Shade}
\end{figure}
Figures~\ref{Fig:eeZZResum_Shade} and \ref{Fig:eeWWResum_Shade} show
the same differential distributions as Figs.~\ref{Fig:eeWWResumTrans},
\ref{Fig:eeWWResumLong}, and \ref{Fig:eeZZResum}, but with the full
NLO cross sections summed over the polarisations and including
contributions of real corrections, virtual (factorisable and
non-factorisable) corrections, and integrated dipoles, converted to expected
event rates at CLIC assuming an integrated luminosity of~\cite{CLIC_Lumi} 
\begin{align}
\mathcal{L}^\text{CLIC}_\text{int}=5\,\text{ab}^{-1}.
\end{align}
Figure~\ref{Fig:eeZZResum_Shade} confirms that the anticipated
statistics of the considered ZZ decay channel is rather low, owing to
the purely leptonic final state: The cross section of $\sim 10\,\text{ab}$ is expected to yield around 50 events in total, rendering a measurement at the differential level impossible. One should, however, notice that the $\text{SCET}_\text{EW}$ results can be used in the same way for the more prominent decay channels with expected event rates being larger by a factor 10--100. Besides that the radiative corrections are dominated by real-radiation effects in the low lepton-energy regime (see right plot), where including the real corrections renders the total 
NLO EW corrections positive. In the angular distribution on
the l.h.s.\ the effects of the real corrections appear to be strongest in the central region, thus flattening the curves of the relative corrections compared to Fig.~\ref{Fig:eeZZResum}. 

In the $\text{W}^+\text{W}^-$ case (Fig.~\ref{Fig:eeWWResum_Shade}),
in contrast, the expected statistics is sufficient on a wide range of
phase space. In regions with particularly large cross sections, such
as the forward direction or the high-energy and high-invariant-mass
regions both the effects of the LL and $\text{NLL}_\text{FO}$
contributions clearly exceed the numerical uncertainties, suggesting
that the NLL- and NLO-effects may be visible within this experimental
setup. At the end the NLL results incidently lie within the shaded bands again. 
In the backward region (small $\theta_{\tau^-}$) we observe that, while the total cross section is suppressed, the relative EW corrections reach high positive values, because they are, like in the ZZ case dominated by real-radiation effects. This can be explained by the fact that 
the additional photon opens up kinematically suppressed phase-space regions. A similar effect can be observed in the region of small muon energies and small dilepton invariant masses. 

\subsubsection{Effect of the running couplings}
In this section, we investigate the impact of the parameter-logarithm resummation, as compared to the LL+NLO scheme (see Sec.~\ref{Sec:LogCount}).
Recapitulate that the included terms are given by
(\ref{RunningDef}). In addition we fix $\alpha$ by its value in the
$\overline{\text{MS}}$ scheme at $M_\text{Z}$ (see Sec.~10 of Ref.~\cite{PDG})
\begin{equation}
\alpha(M_\text{Z})=\frac{1}{127.952}
\end{equation}
at the low scale. From the input for $c_\text{w}$ and $s_\text{w}$,
\begin{align}
c_\text{w}^2(M_\text{Z})=\frac{M_\text{W}^2}{M_\text{Z}^2},\qquad s_\text{w}^2(M_\text{Z})=1-c_\text{w}^2(M_\text{Z}),
\end{align}
we infer the values for the U(1) and SU(2) gauge couplings at $M_\text{Z}$:
\begin{align}
g_1(M_\text{Z})=\frac{\sqrt{4\pi\alpha(M_\text{Z})}}{c_\text{w}(M_\text{Z})},\qquad g_2(M_\text{Z})=\frac{\sqrt{4\pi\alpha(M_\text{Z})}}{s_\text{w}(M_\text{Z})}.
\end{align}
For the high-scale contributions the one-loop RGE solutions for the EW gauge couplings are used as input parameters:
\begin{align}
g^2_{1/2}(\mu_\text{h})=\frac{g^2_{1/2}(M_\text{Z})}{1-\beta_{0,{1/2}}\frac{g_{1/2}^2(M_\text{Z})}{8\pi^2}\log\frac{\mu_\text{h}}{M_\text{Z}}},
\end{align}
with $\beta_{0,1/2}$ as defined in (\ref{beta0Def}).

The difference between the two setups at the integrated level is about
$8\%$ with respect to the Born for $\text{W}^+\text{W}^-$ production,
see rows 4 and 7 on the left of Table~\ref{tab:sigmaInt}. 
The major part of this effect is explained by the
$\alpha^2\mathsf{L}^3$ contributions associated with the running of
the couplings,
which are not present in our definition of LL+NLO. While this effect is conceivably smaller than
the difference between LL and $\text{NLL}_\text{FO}$, which is also caused by $\alpha^2\mathsf{L}^3$ effects,
it is not negligible and a should be taken into account. 

\begin{figure}
\centering
\includegraphics[width=0.45\textwidth]{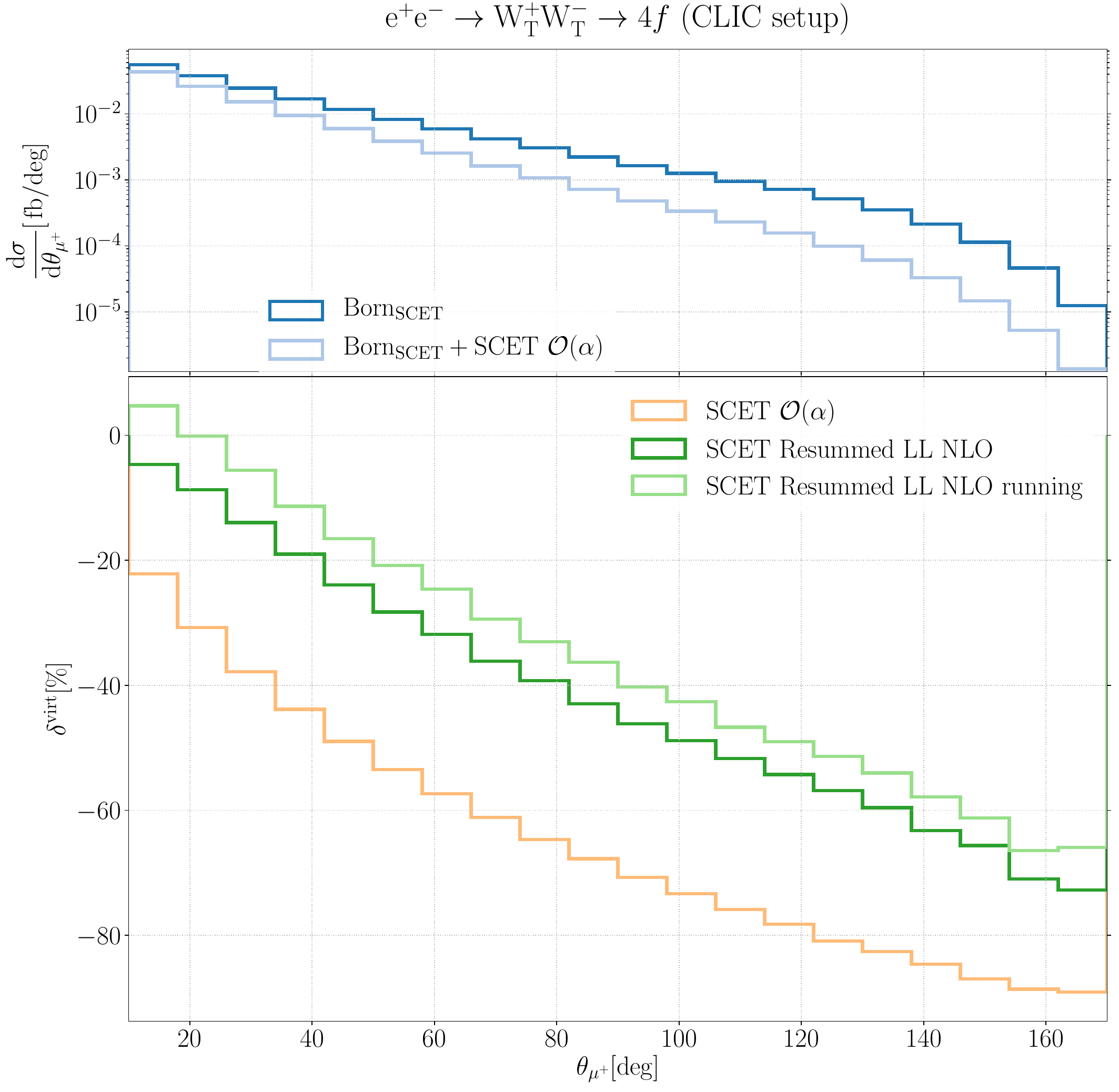}
\includegraphics[width=0.45\textwidth]{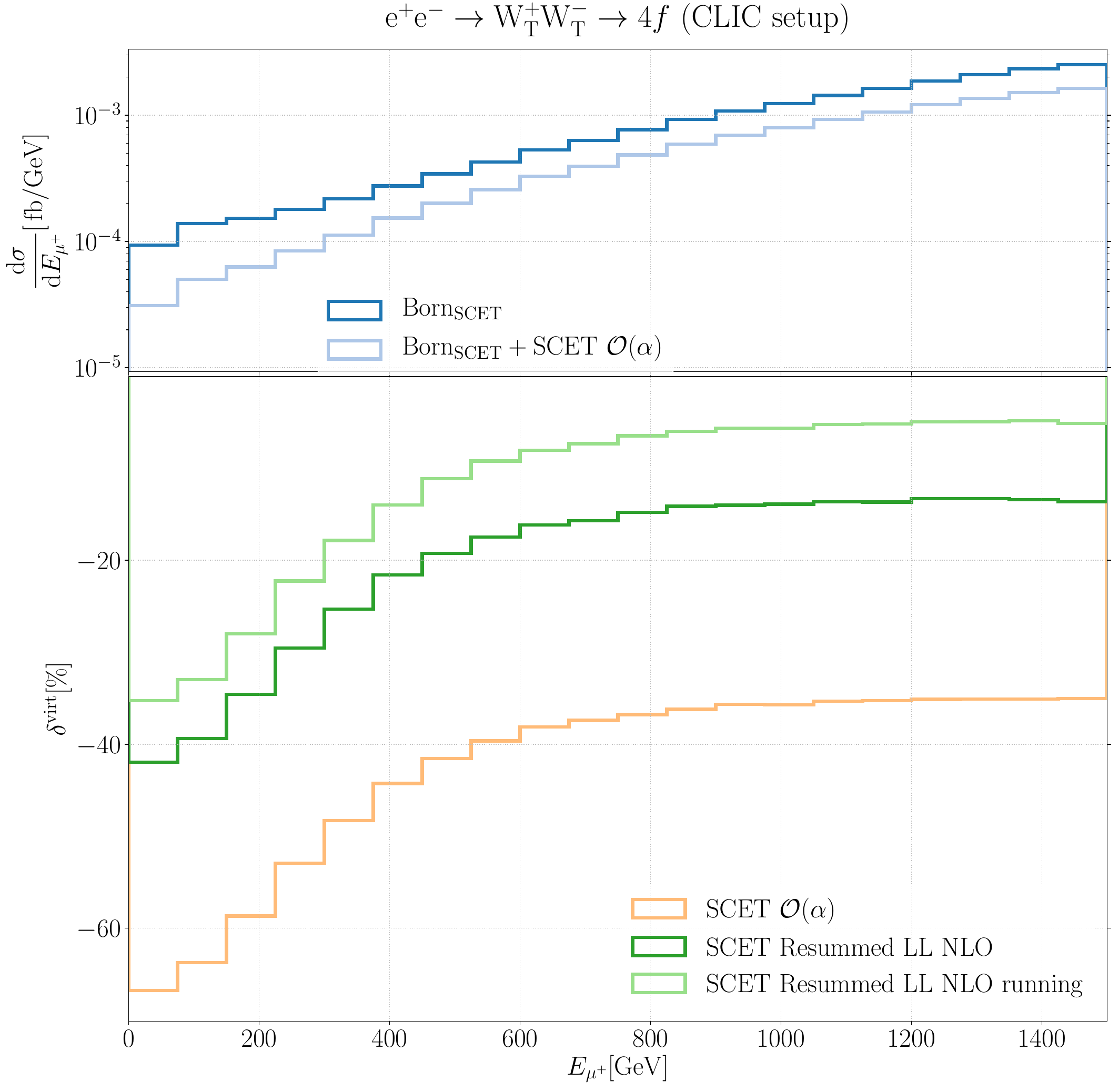}
\caption{Effect of the running EW gauge couplings in 
  $\text{e}^+\text{e}^-\rightarrow\text{W}_\text{T}^+\text{W}_\text{T}^-\rightarrow
  \mu^+\nu_\mu\bar{\nu}_\tau\tau^-$ differential in the muon
  production angle and energy.}
\label{Fig:eeWWRunningTrans}
\end{figure}
\begin{figure}
\centering
\includegraphics[width=0.475\textwidth]{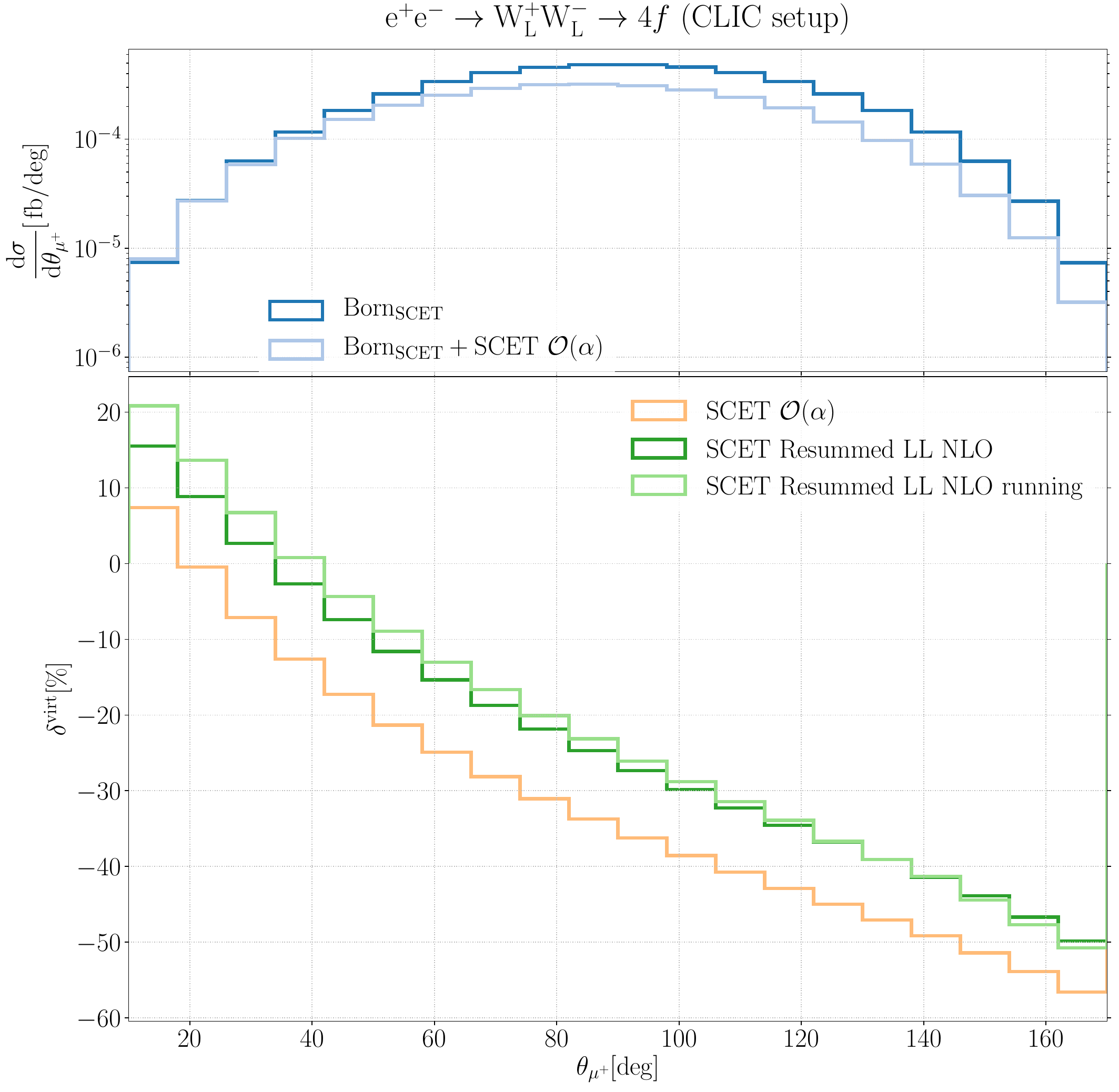}
\includegraphics[width=0.475\textwidth]{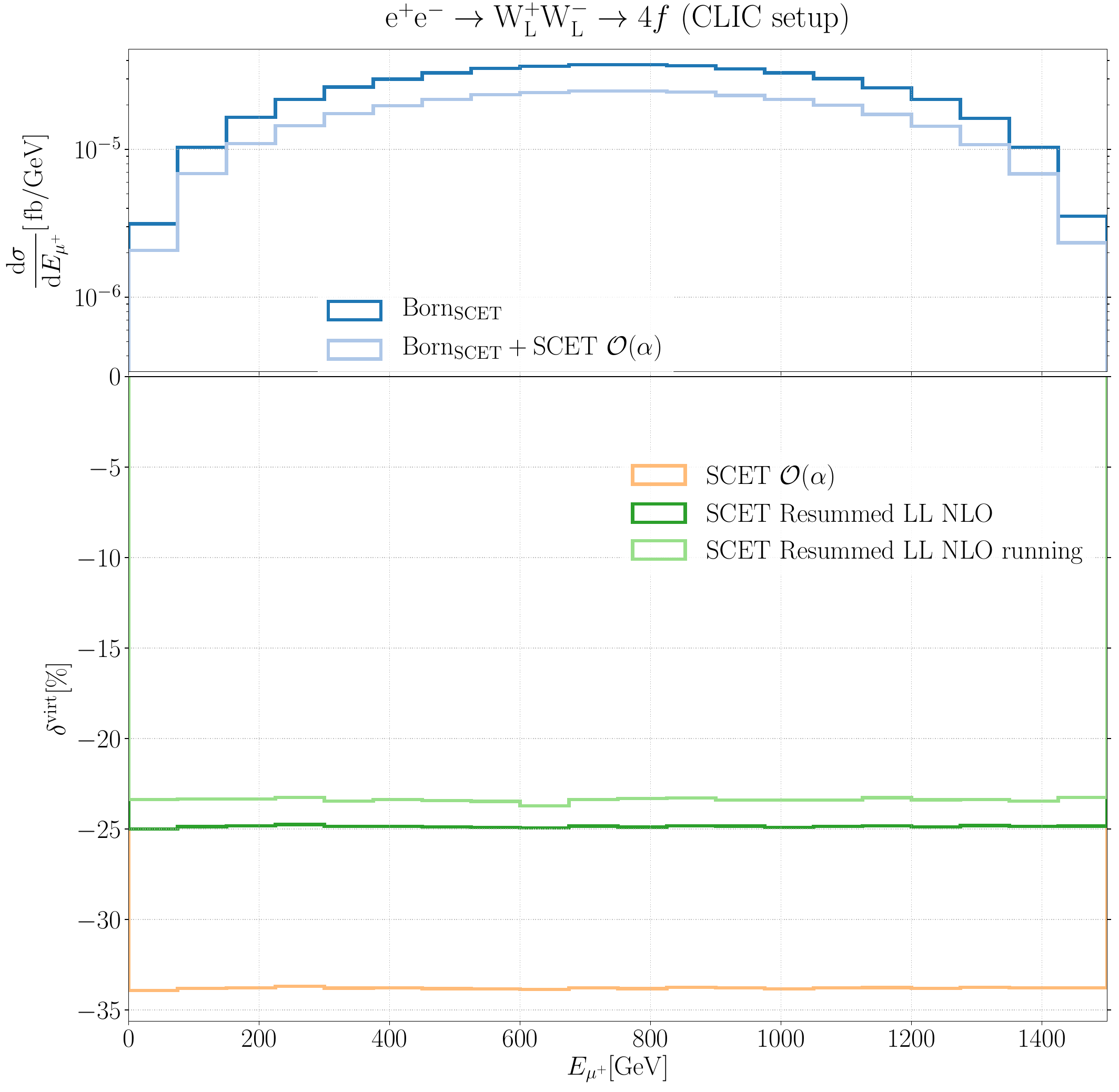}
\caption{Effect of the running EW gauge couplings in 
  $\text{e}^+\text{e}^-\rightarrow\text{W}_\text{L}^+\text{W}_\text{L}^-\rightarrow
  \mu^+\nu_\mu\bar{\nu}_\tau\tau^-$ differential in the muon
  production angle and energy.}
\label{Fig:eeWWRunningLong}
\end{figure}
The differential results are shown in Figs.~\ref{Fig:eeWWRunningTrans} and \ref{Fig:eeWWRunningLong}
 for the production of transverse and longitudinal W bosons, respectively. The phase-space variation of the difference is at the level of few percent.
While the aforementioned $\alpha^2\mathsf{L}^3$ terms are constant, a small dependence is explained by the fact,
that the dominant phase-space dependent contributions, namely the angular-dependent logarithms, are calculated with a different value of $\alpha$. 

\subsection{Results for the FCC--hh collider}
High-energy lepton collisions are a natural environment to study the resummation of EW Sudakov logarithms,
because they are expected to yield the bulk of their events in the high-energy region.
While future hadron colliders potentially achieve much higher energies, most produced events actually happen at
lower energies owing to the PDF fall-off with energy. Many searches for new physics will, however, focus on precision measurements
in the high-energy tails, where EW Sudakov logarithms become large.  

We study the resummation of these logarithms considering a setup inspired by the FCC-hh project, i.e.\ a proton--proton collider operating at a CMS energy of
\begin{align}
\sqrt{s}=100\,\text{TeV}.
\end{align}
We investigate the four diboson production processes:
\begin{align}
&\ppwzllll,\\
&\ppzzllll,\\
&\ppwwllll,\\
&\aawwllll.
\end{align}
We consider the photonic channel of $\text{W}^+\text{W}^-$
production as a distinct process, although it is
actually contained in the process $\ppwwllll$.
This approach is motivated by several facts:
\begin{itemize}
\item The Sudakov corrections are particularly large for this process, since it involves four gauge bosons with large EW Casimir invariants.
\item It is a vector-boson-scattering process: Since the leading EW
  logarithms depend only on the external particles, results obtained
  for $\gamma\gamma\rightarrow \mathrm{W}^+\mathrm{W}^-$ might hint at
  what is to be expected for more complicated scattering processes,
  such as $\mathrm{W}^+\mathrm{W}^+$ scattering. In addition,
  vector-boson-scattering processes have not yet been considered in the context of $\text{SCET}_\text{EW}$ or IR evolution equations. We point out that especially the high-scale matching is much more involved than in the other diboson processes, making it particularly well-suited for an automated computation.
\item In elastic pp scattering the process can actually be
  distinguished experimentally
  \cite{CMS_AA_1,CMS_AA_2,ATLAS_AA,ATLAS_AA_new} from the
  quark-induced channels: if the protons remain intact, the production
  of a colour singlet final state such as two EW gauge bosons can only
  be mediated by colour-singlet exchange, the quark-induced production
  is therefore prohibited by colour conservation. 
In the following, we do, however, consider only the production from inelastic scattering, using the photon PDF set from Ref.~\cite{NNPDF31}.
\end{itemize}
Note also that in the following the process $\ppwwllll$ refers only to the quark-induced partonic channels.

\subsubsection*{Event Selection}

We apply the recombination (\ref{recoco}) and use the standard cut set
\begin{align}
p_{\text{T},\ell}>20\,\text{GeV},\qquad -2.5<y_\ell<2.5,\qquad M_{\text{inv},\ell\ell'}>10\,\text{GeV},\label{standardcuts_FCC}
\end{align}
which is the same as in the CLIC setup [see (\ref{standardcuts_CLIC})], with the angular cut replaced by a rapidity cut, because 
at a hadron collider the partonic interactions are boosted with respect to the laboratory frame. 
In addition we use Z-window cuts,
\begin{align}
81\,\text{GeV}<M_{\mu^+\mu^-}<101 \,\text{GeV},\qquad
81\,\text{GeV}<M_{\text{e}^+\text{e}^-}<101 \,\text{GeV},\label{ZWindow_FCC}
\end{align}
and the condition (\ref{stu}) as a technical cut whenever $\text{SCET}_\text{EW}$ is applied. The Mandelstam variables $s,t,u$ in (\ref{stu}) refer to the sub-processes
\begin{align}
\bar{q}q/\gamma\gamma\rightarrow V V'.
\end{align}
\subsubsection{Integrated results and DPA analysis}
\label{Sec:DPAFCC}
\begin{figure}
\centering
\includegraphics[width=0.45\textwidth]{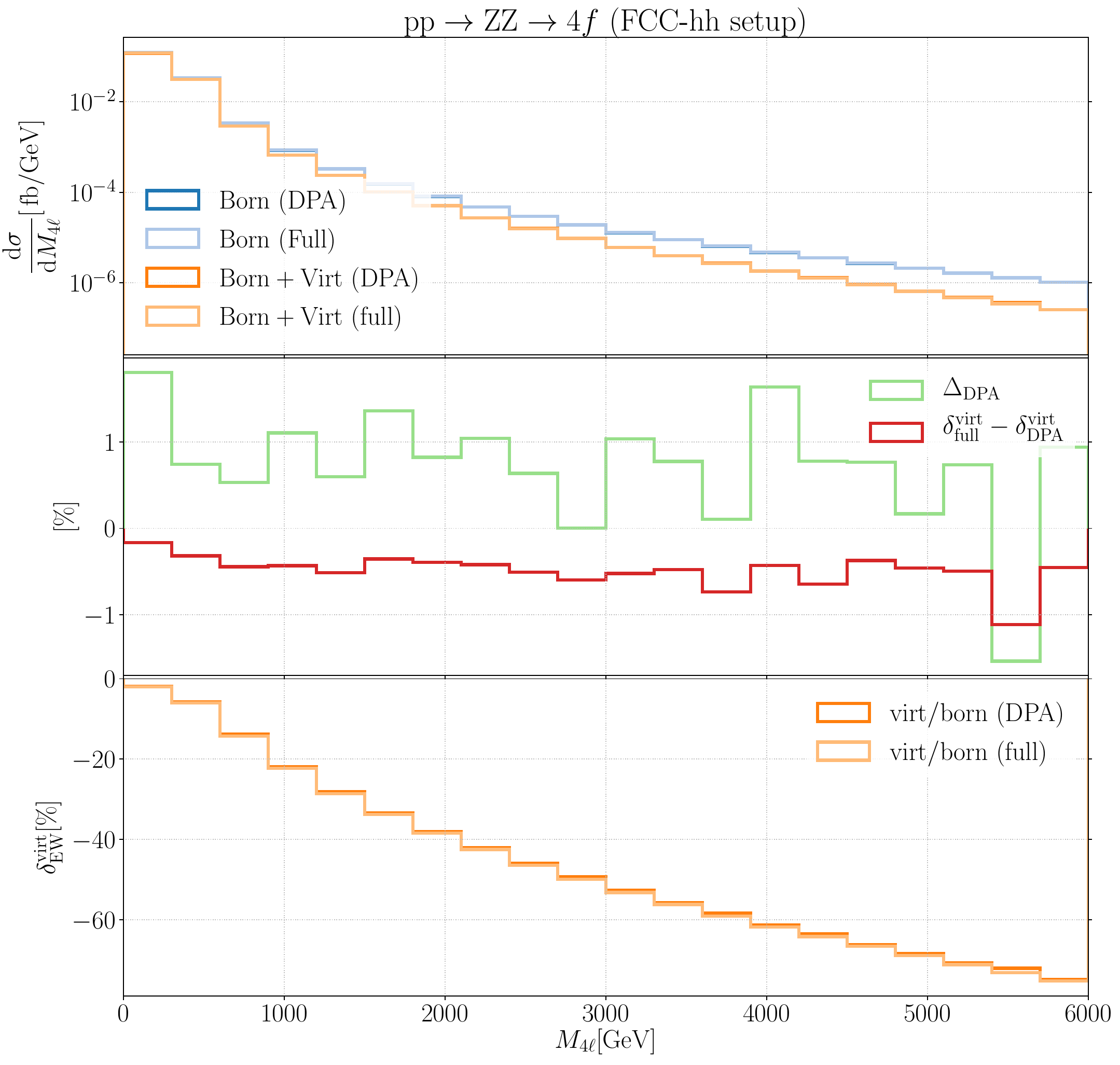}
\includegraphics[width=0.45\textwidth]{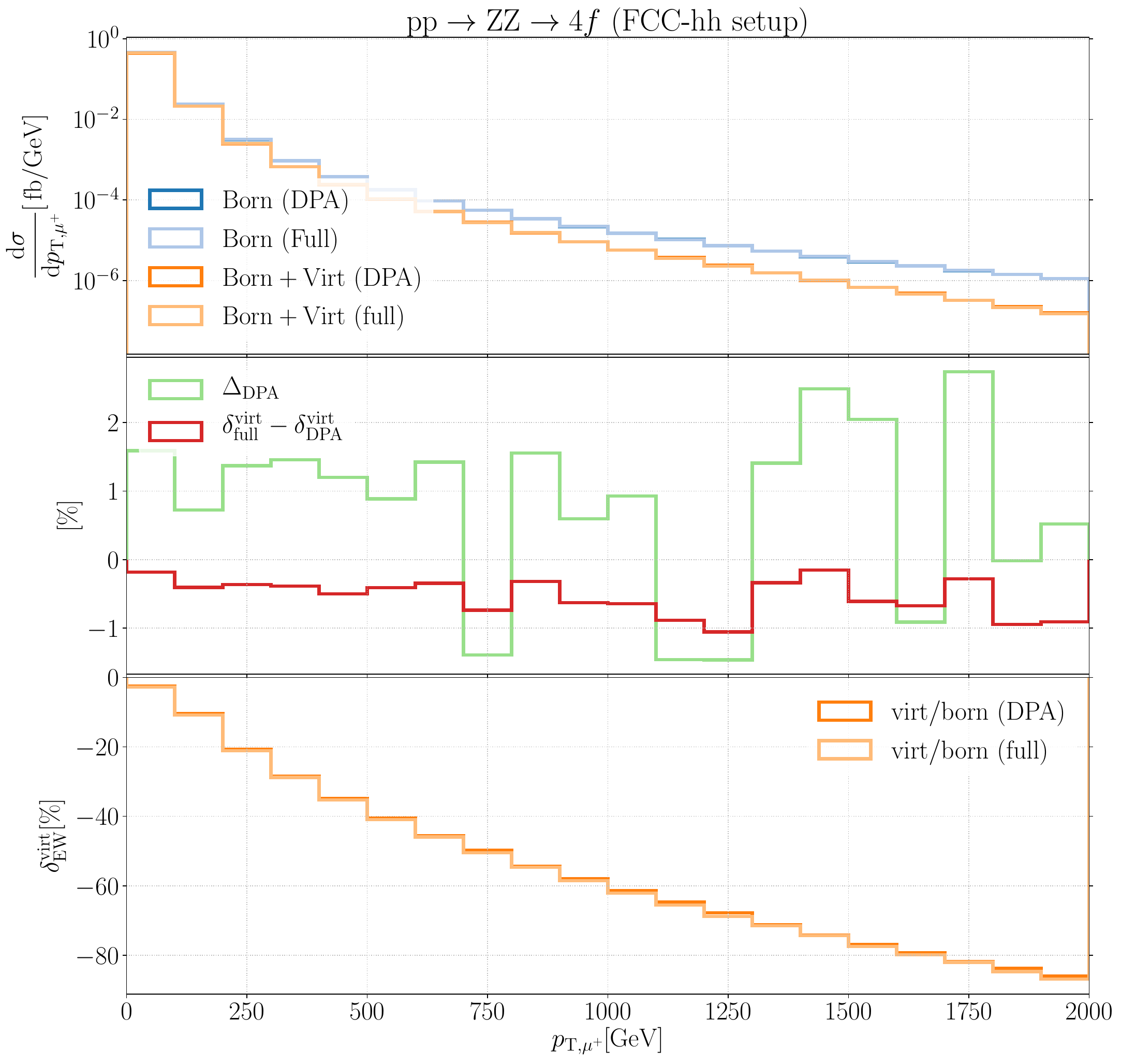}
\caption{Comparison between DPA and full off-shell calculation for
  $\ppzzllll$: Differential distributions in the four-lepton invariant
  mass and the antimuon transverse momentum. The upper panels show the differential cross sections, the middle ones the deviations owing to the DPA with $\Delta_\text{DPA}$ defined in ($\ref{dDPA}$), and the lower ones the relative virtual EW corrections. }
\label{Fig:DPA_ZZ}
\end{figure}

\begin{figure}
\centering
\includegraphics[width=0.45\textwidth]{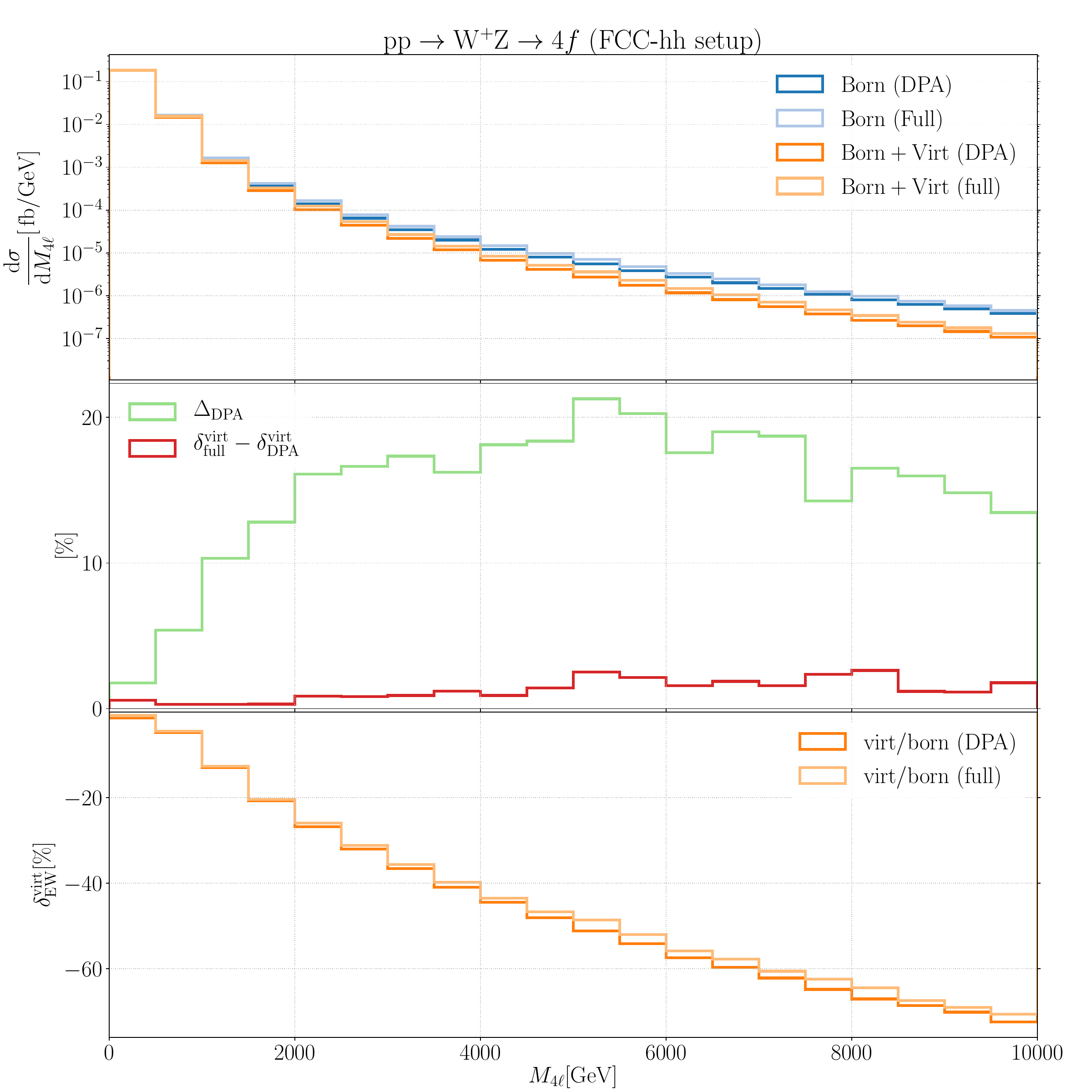}
\includegraphics[width=0.45\textwidth]{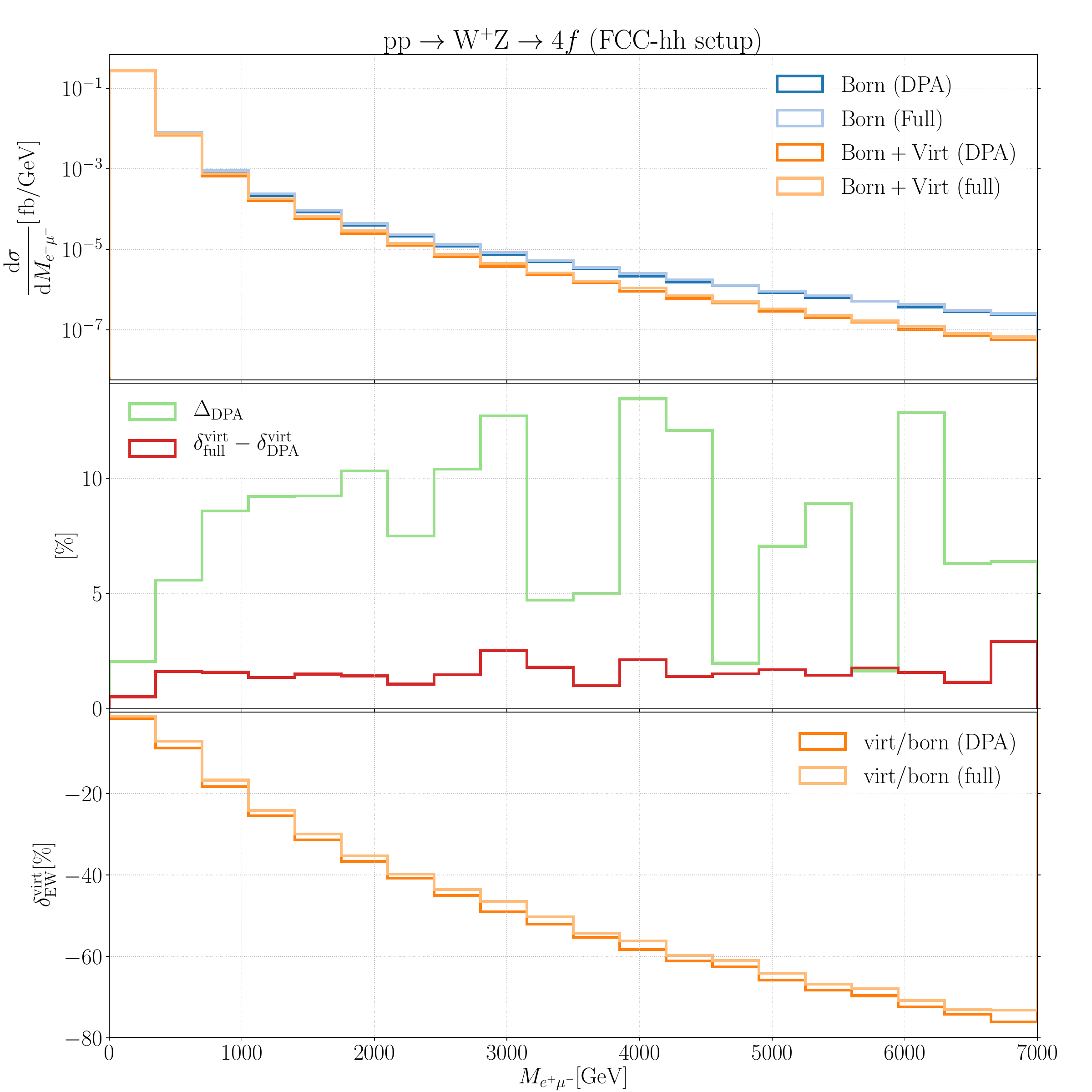}
\includegraphics[width=0.45\textwidth]{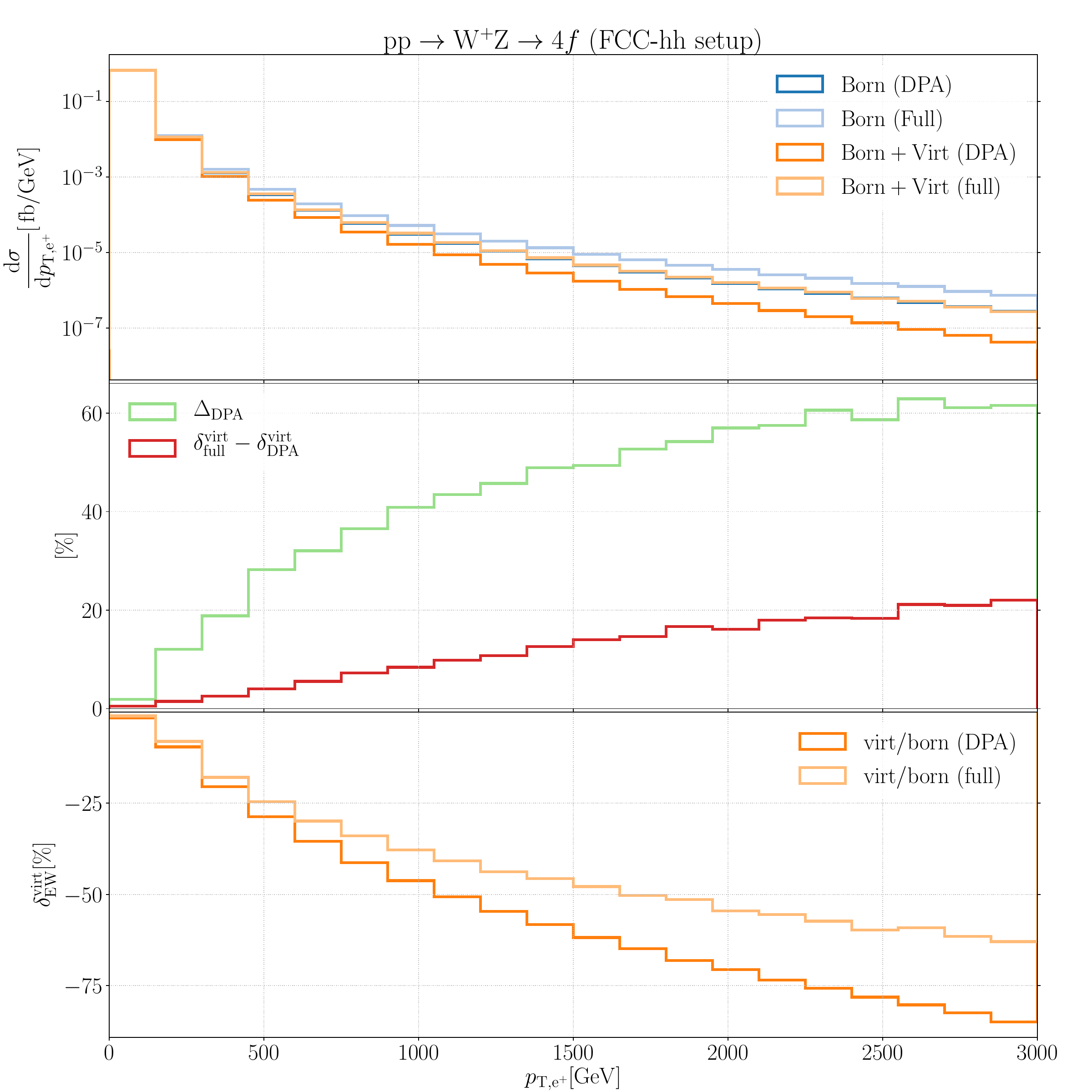}
\includegraphics[width=0.45\textwidth]{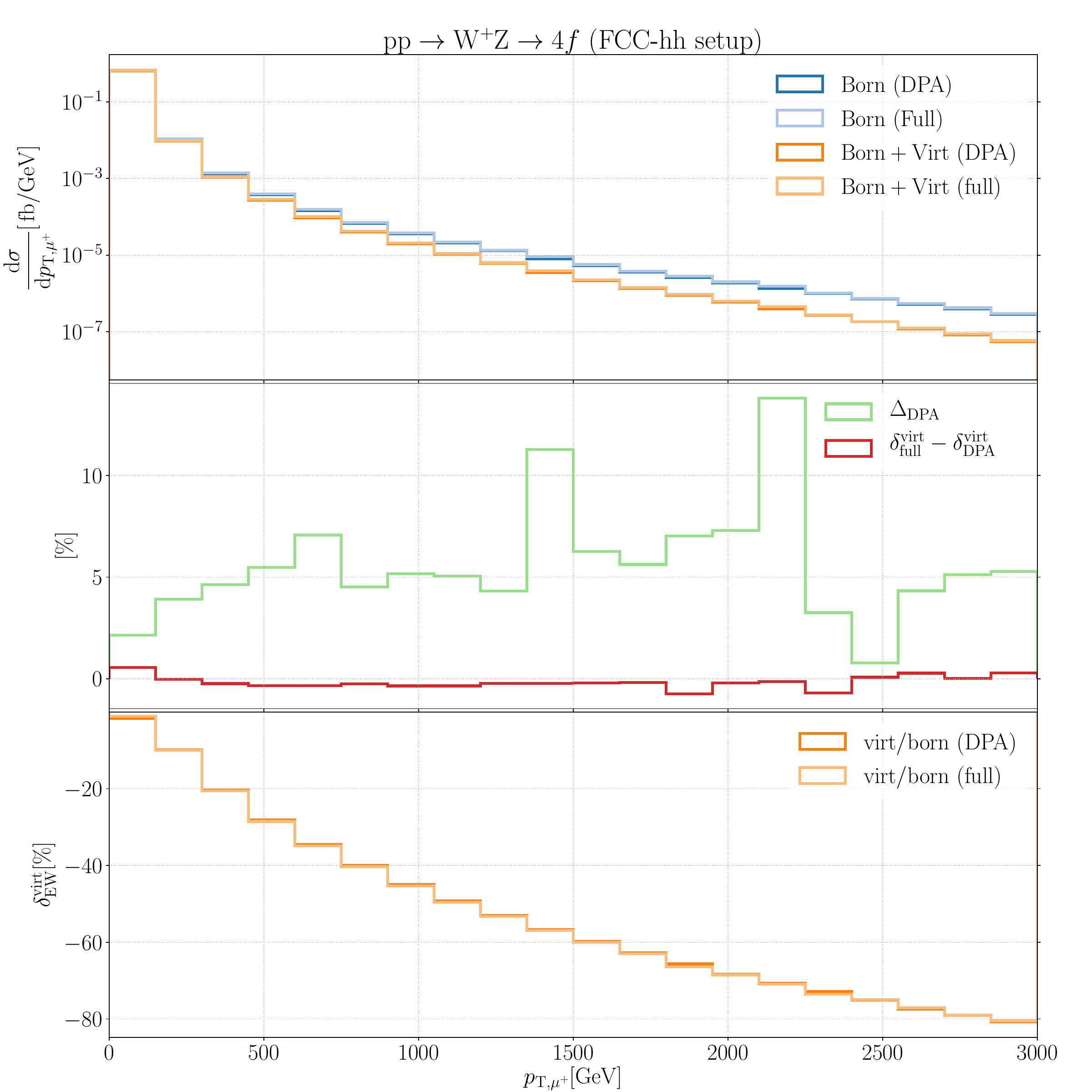}
\caption{Comparison between DPA and full off-shell calculation for
  $\ppwzllll$: Differential distributions in the four-lepton invariant
  mass, the $\Pe^+\mu^-$ invariant mass, and the charged-lepton 
  transverse momenta. The plots are organised as in Fig.~\ref{Fig:DPA_ZZ}.}
\label{Fig:DPA_WZ}
\end{figure}

\begin{figure}
\centering
\includegraphics[width=0.45\textwidth]{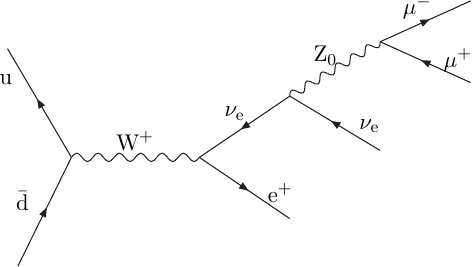}
\hspace{10pt}
\includegraphics[width=0.9\textwidth]{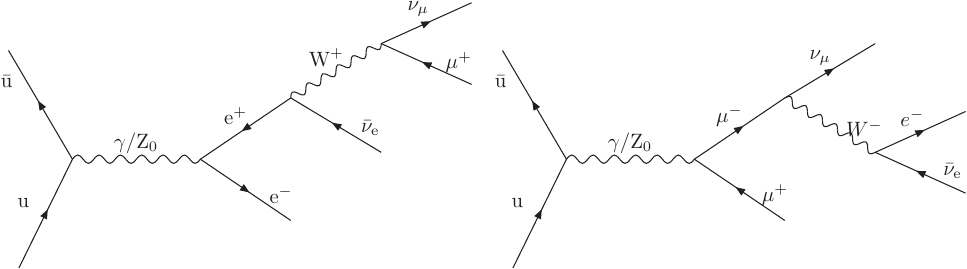}\\
\caption{Singly-resonant diagrams that spoil the validity of the DPA
  in $\text{W}^+\text{Z}$ production (top) and $\text{W}^+\text{W}^-$
  production (bottom) for high lepton transverse momenta. }
\label{Fig:DomChan_WZ}
\end{figure}

In this section, we inspect the quality of the DPA in the high-energy tails in diboson production in high-energy proton--proton collisions. 
The plots are organised as in Sec.~\ref{Sec:DPARes}.
\paragraph{ZZ production}
For ZZ production we obtain the fiducial cross sections (all errors are MC integration errors)
\begin{equation}
\sigma_\text{full}=48.285(14)\,\text{fb},\qquad
\sigma_\text{DPA}= 47.54(4)\,\text{fb}
\end{equation}
in the DPA and fully off shell, respectively. This yields a deviation of $\Delta_\text{DPA}=1.54(8)\%$ on the integrated cross section. The relative virtual corrections read
\begin{align}
\delta^\text{virt}_\text{full}=-3.381(8)\%,\qquad
\delta^\text{virt}_\text{DPA}=-3.16(3)\%,\qquad
\delta^\text{virt}_\text{full}-\delta^\text{virt}_\text{DPA}=-0.22(3)\%.
\end{align}
The comparison of the differential cross section for ZZ production
between the full off-shell and the DPA computation differential in the
four-lepton invariant mass and the antimuon transverse momentum can be found in Fig.~\ref{Fig:DPA_ZZ}. 
Note that this and the following plots contain large fluctuations 
owing to low statistics in the high-energy tails. This is especially the case for quantities 
like $\Delta_\text{DPA}$ and $\delta^\text{virt}_\text{full}-\delta^\text{virt}_\text{DPA}$ that suffer from cancellations.
In the high-energy and high-transverse-momentum tails
$\Delta_\text{DPA}$ is about 1\%, while the difference of the relative
corrections ranges between $0\%$ and $-1\%$. Similar to the case of $\text{e}^+\text{e}^-$ collisions, the non doubly-resonant contributions are 
effectively suppressed by the Z-window cut. 
\paragraph{$W^+Z$ production}
The fiducial cross sections for $\text{W}^+\text{Z}$ production in DPA and fully off shell read
\begin{align}
\sigma_\text{full}=102.71(5)\,\text{fb},\qquad\sigma_\text{DPA}=100.5(2)\,\text{fb},\qquad\Delta_\text{DPA}=2.2(2)\%,
\end{align}
and the relative virtual corrections are given by
\begin{align}
\delta^\text{virt}_\text{full}=-1.313(6)\%,\qquad\delta^\text{virt}_\text{DPA}=-1.19(3)\%,\qquad\delta^\text{virt}_\text{full}-\delta^\text{virt}_\text{DPA}=-0.12(3)\%.
\end{align}
Differential results for the DPA comparison in $\text{W}^+\text{Z}$ production are shown in Fig.~\ref{Fig:DPA_WZ}.
Compared to the ZZ case we note a worse agreement of the Born cross
sections in the tails: $\Delta_\text{DPA}$ grows up to 10--20\% in the
high-invariant-mass tails. Even stronger deviations can be found in
the high-$p_{\text{T},\text{e}^+}$ tail (note that the positron is the
decay product of the $\text{W}^{+}$ boson). Here the DPA cross section
is not even close to the full result ($\Delta_\text{DPA}>50\%$). This
is because this region is dominated by contributions originating from
diagrams depicted in Fig.~\ref{Fig:DomChan_WZ}: Although they are not
doubly resonant, they are strongly enhanced for high positron
transverse momenta, because the positron recoils against two or three final-state particles.
This mechanism is described in more detail in Ref.~\cite{WpWm_NLOEW_offshell}.
\paragraph{$W^+W^-$ production from quarks}
For $\text{W}$-boson pair production we obtain the fiducial LO cross sections
\begin{align}
\sigma_\text{full}=2540(3)\,\text{fb},\qquad\sigma_\text{DPA}=2487(2)\,\text{fb},\qquad\Delta_\text{DPA}=2.09(15)\%,
\end{align}
and the relative corrections
\begin{align}
\qquad\delta^\text{virt}_\text{full}=-0.759(5)\%,\qquad\delta^\text{virt}_\text{DPA}=-0.617(2)\%,\qquad\delta^\text{virt}_\text{full}-\delta^\text{virt}_\text{DPA}=-0.14(6)\%.
\end{align}
\begin{figure}
\centering
\includegraphics[width=0.45\textwidth]{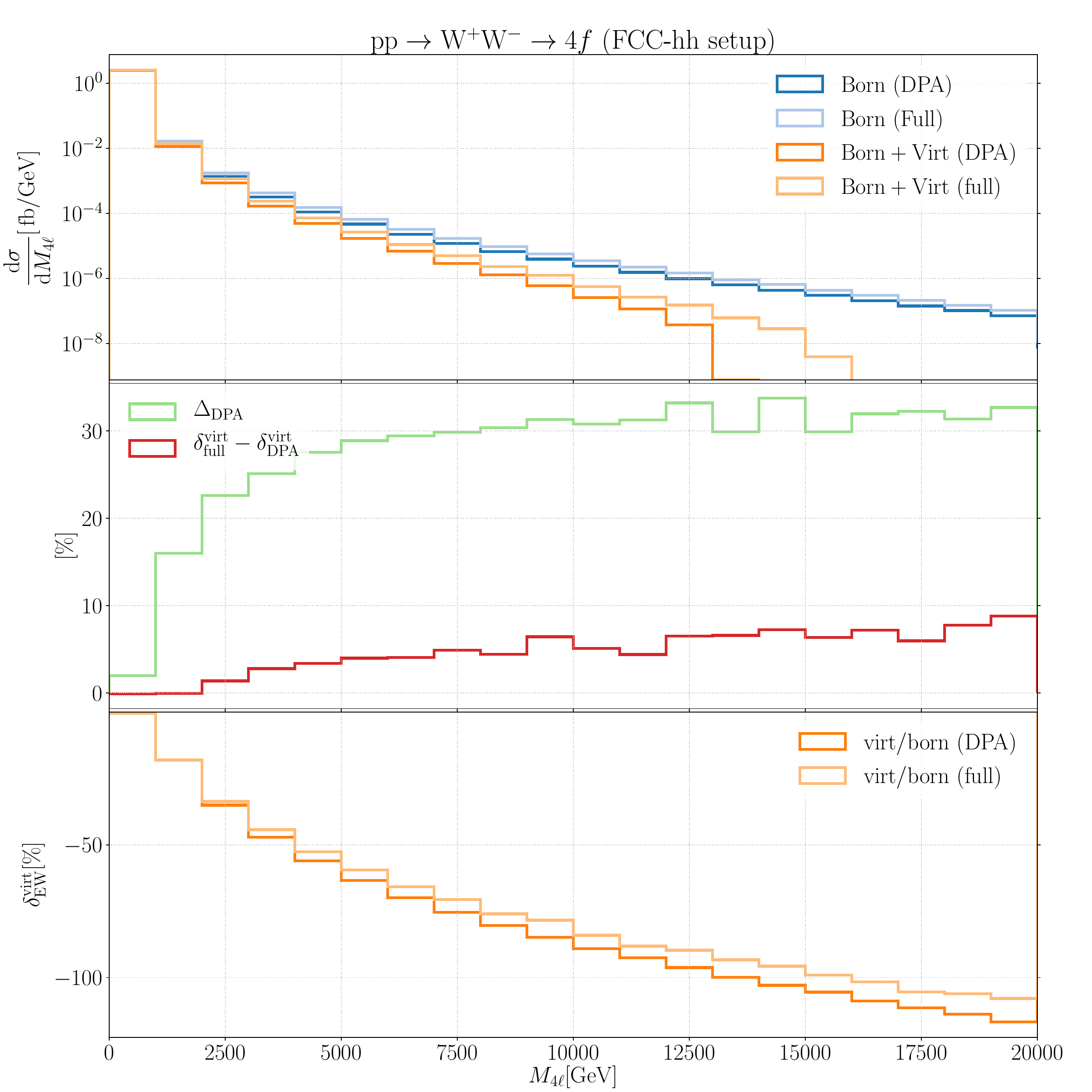}
\includegraphics[width=0.45\textwidth]{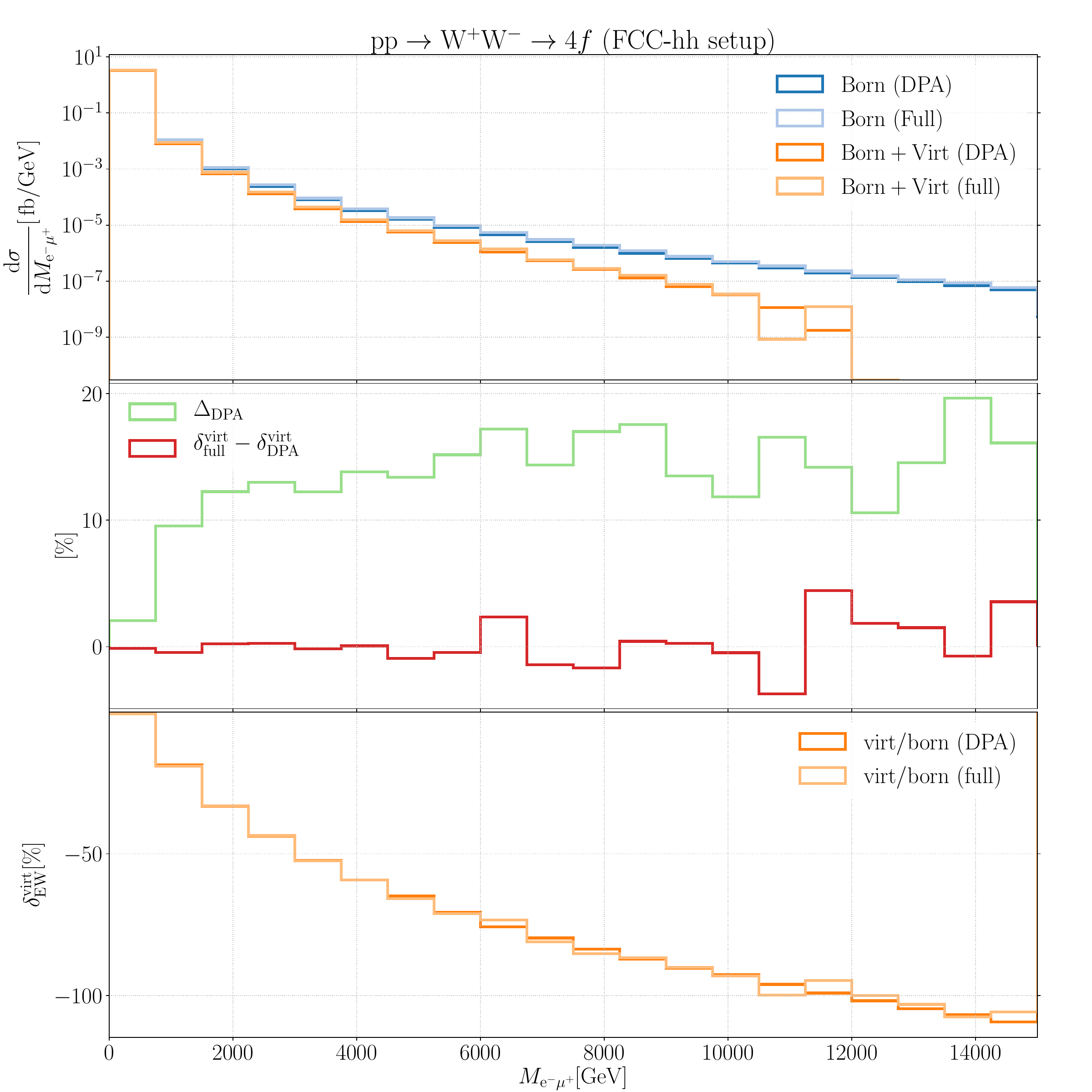}
\includegraphics[width=0.45\textwidth]{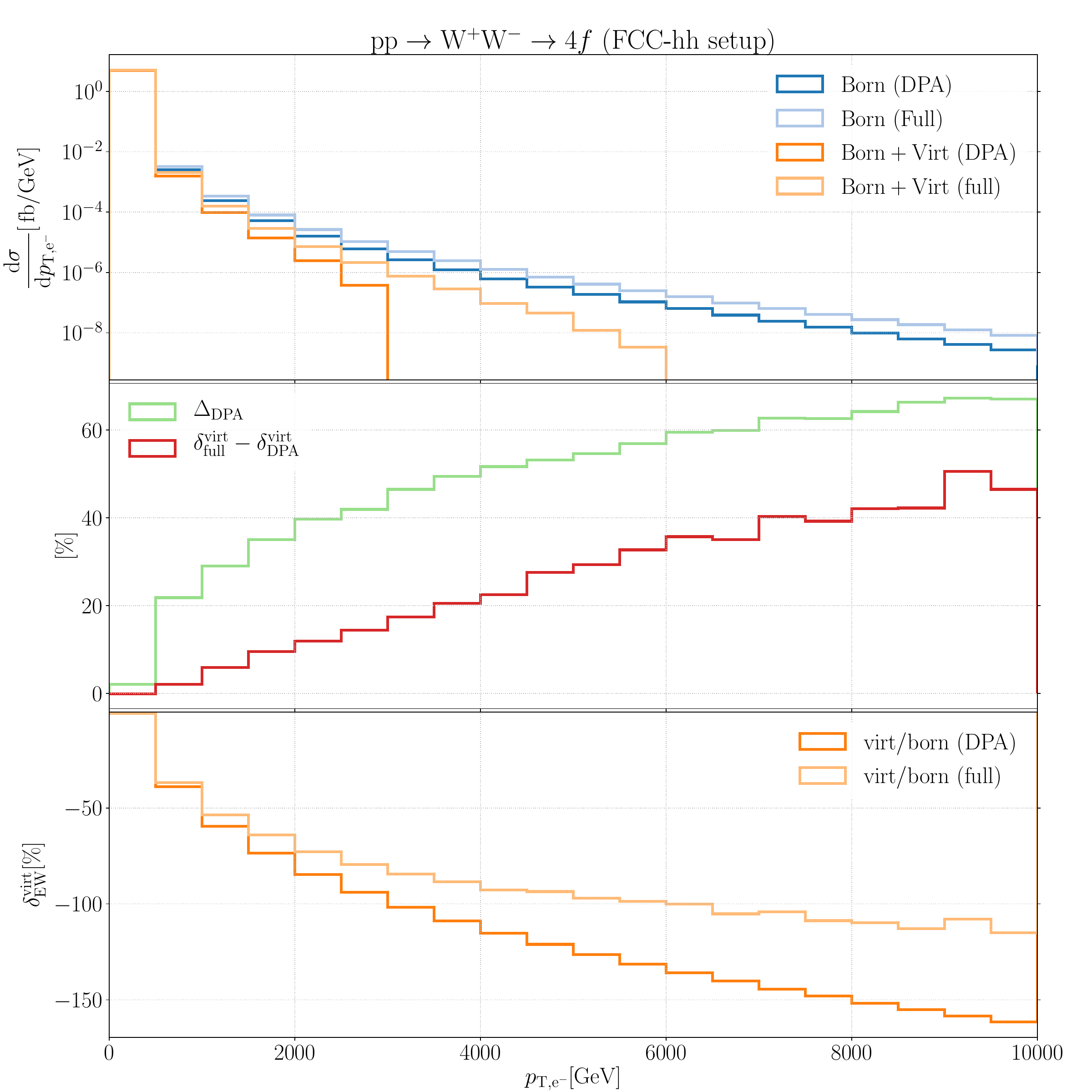}
\includegraphics[width=0.45\textwidth]{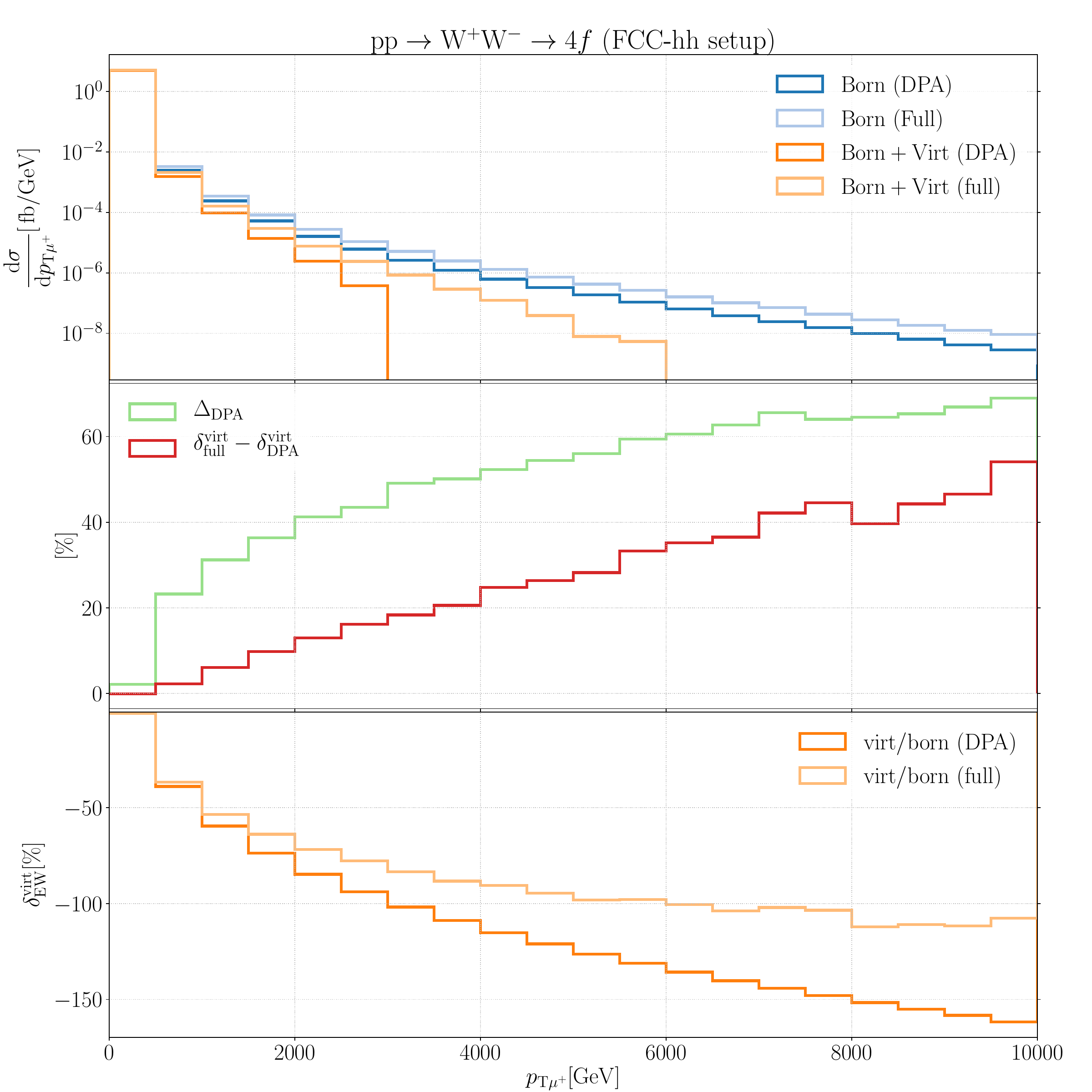}
\caption{Comparison between DPA and full off-shell calculation for
  $\ppwwllll$: Differential distributions in the four-lepton invariant
  mass, the $\Pe^-\mu^+$ invariant mass, and the charged-lepton
  transverse momenta. The plots are organised as in Fig.~\ref{Fig:DPA_ZZ}.}
\label{Fig:DPA_WW}
\end{figure}%
In Fig.~\ref{Fig:DPA_WW} we show several distributions in energy-like phase-space variables. 
In the tails we can observe a similar behaviour as in the
$\text{W}^+\text{Z}$ case. The diagrams drawn in the second row of
Fig.~\ref{Fig:DomChan_WZ} give sizeable contributions for high invariant mass and especially
high~$p_{\text{T},\ell}$. In all plots $\Delta_\text{DPA}$ grows towards the
tail, reaching $\sim20\%$ for high lepton-pair invariant masses, $\sim
30\%$ for high four-lepton invariant masses, and more than $60\%$ for high
lepton transverse momenta. 

\begin{figure}
\centering
\includegraphics[width=0.45\textwidth]{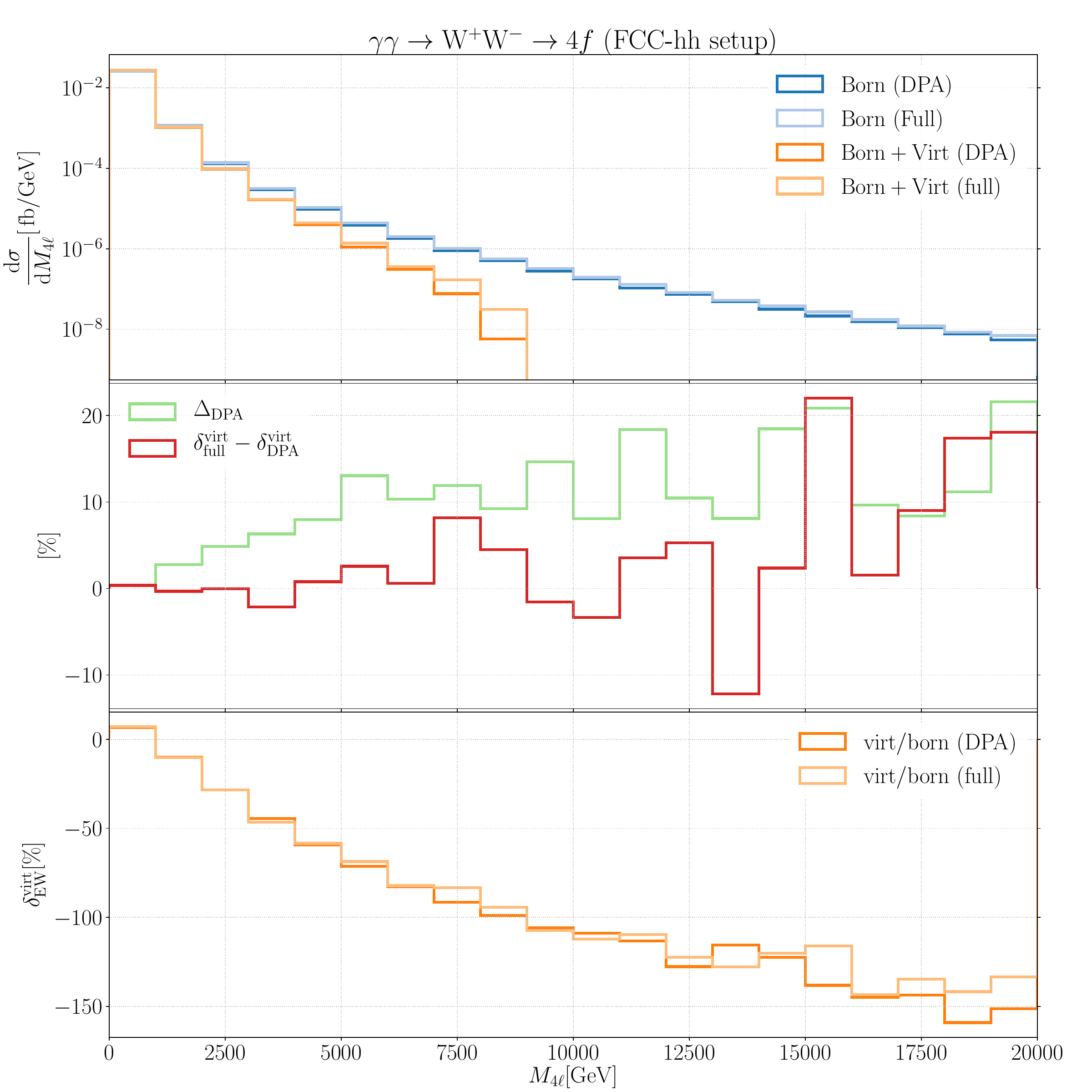}
\includegraphics[width=0.45\textwidth]{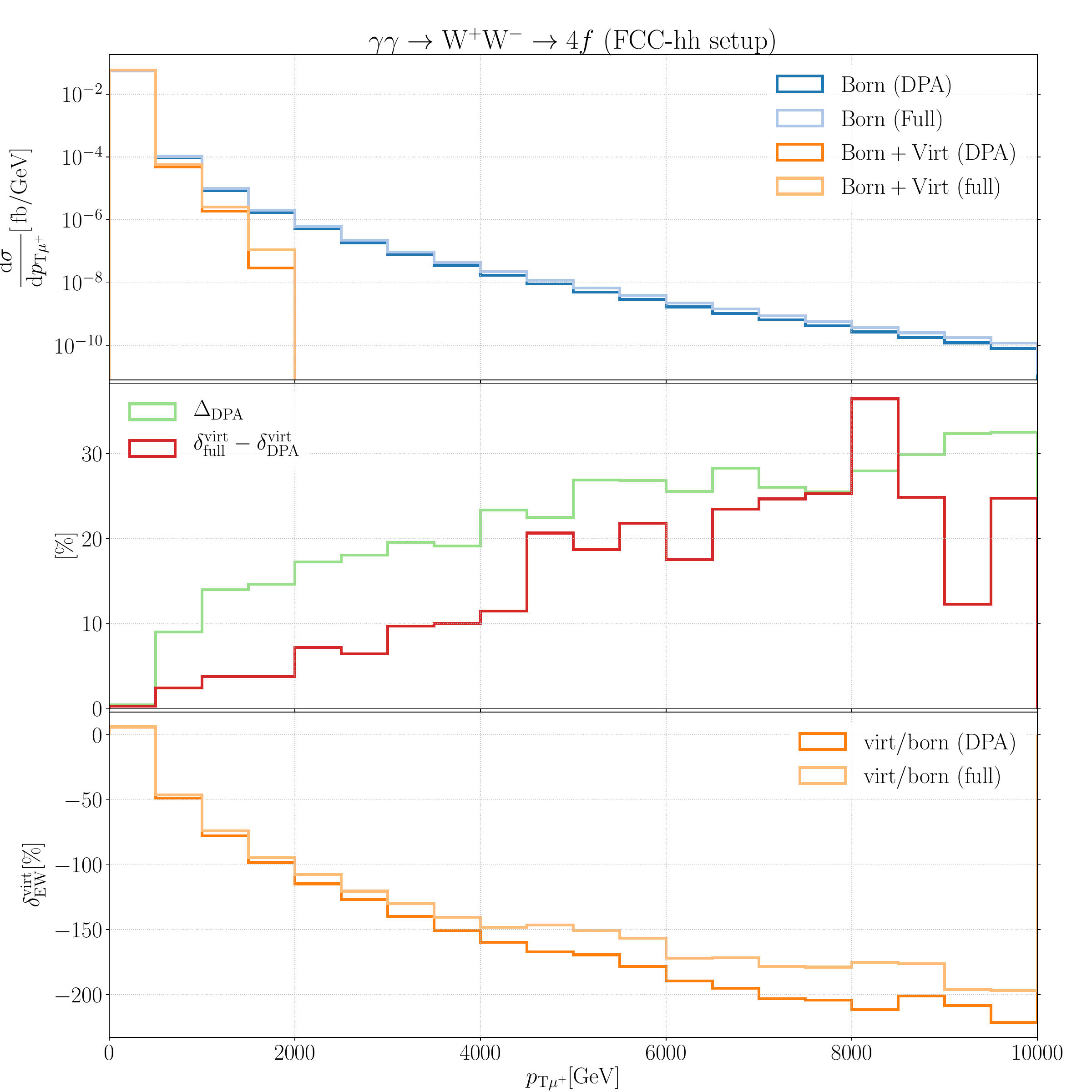}
\caption{Comparison between DPA and full off-shell calculation for
  $\aawwllll$: Differential distributions in the four-lepton invariant
  mass and the muon transverse momentum. The plots are organised as in
  Fig.~\ref{Fig:DPA_ZZ}.}
\label{Fig:DPA_AA}
\end{figure}

Meanwhile, in the high-$M_{4\ell}$ and high-$M_{\Pe^-\mu^+}$ tails the
difference of the relative correction remains below $10\%$. In the
high-$M_{\Pe^-\mu^+}$ tail, $\delta^\text{virt}_\text{full}-\delta_\text{DPA}^\text{virt}$ fluctuates
in the range of $1$--$5\%$, while in the high-$M_{4\ell}$ tail there is a conceivable systematic growth up to $10\%$. 
The fact that $\Delta_\text{DPA}$, which measures the deviation of the Born cross sections is between $20\%$ and $30\%$ indicates strong contributions of non-doubly resonant diagrams,
which, however, are less relevant for the relative virtual corrections. In the high-$p_{\text{T},\ell}$ tails, however, $\delta^\text{virt}_\text{full}-\delta_\text{DPA}^\text{virt}$ grows up to $50\%$.
We thus have to conclude that in the high-$p_{\text{T}\ell}$ tails the DPA is not applicable anymore, while in the high-$M_{4\ell}$ tail its accuracy is limited to the level of 5--10\%. 
In the high-$M_{\Pe^-\mu^+}$ tail, the DPA shows a reasonable
precision of less than $5\%$. 
\paragraph{${W}^+{W}^-$ production from photons}
For photon-induced W-boson pair production we obtain the Born cross section
\begin{align}
\sigma_\text{full}=27.33(2)\,\text{fb},\qquad\sigma_\text{DPA}=27.200(16)\,\text{fb},\qquad\Delta_\text{DPA}=0.47(8)\%,
\end{align}
and the relative corrections
\begin{align}
\delta^\text{virt}_\text{full}=6.037(12)\%,\qquad
\delta^\text{virt}_\text{DPA}=5.745(11)\%,\qquad
\delta^\text{virt}_\text{full}-\delta^\text{virt}_\text{DPA}=-0.219(16)\%.
\end{align}
Differential distributions for the respective photon-induced channel
can be found in Fig.~\ref{Fig:DPA_AA}. 
In the high-lepton-$p_\text{T}$ tail of the distributions a similar
behaviour as in the quark-induced case can be observed. The deviations
are, however, not as strong, which can explained by the fact, that
there are no singly-resonant $s$-channel topologies for this
process. The deviation $\Delta_\text{DPA}$ still reaches more than
$30\%$, and in contrast to the quark-induced processes,
$\delta^\text{virt}_\text{full}-\delta^\text{virt}_\text{DPA}$ has a
similar order of magnitude.  In the high-$M_{4\ell}$ tail
$\Delta_\text{DPA}$ varies between $10\%$ and $20\%$. The difference between the relative virtual corrections in DPA and full off~shell is about $1\%$ up to $M_{4\ell}\approx 7\,\text{TeV}$.

\subsubsection{$\text{SCET}_\text{EW}$ vs.\ fixed order}
The validity of $\text{SCET}_\text{EW}$ results is restricted to the region $s,|t|,|u|\gg M_\text{W}^2$.
However, we have to expect large contributions from phase-space regions, where the condition (\ref{stu}) is not satisfied: In the high-energy limit the Born matrix elements for the dominant $\bar{q}q$ and $q\bar{q}$ production channels of transversely (T) and longitudinally (L) polarised gauge bosons can be decomposed as \cite{DennerPozz1}
\begin{equation}
\mathcal{M}_\text{T}=\frac{\mathcal{M}_t}{t}+\frac{\mathcal{M}_u}{u},\qquad\mathcal{M}_\text{L}=\frac{\mathcal{M}_s}{s}
\end{equation}
with $\mathcal{M}_s$, $\mathcal{M}_t$, $\mathcal{M}_u$ being analytical in $s$, $t$, $u$, and all masses. This form holds for all diboson production processes from fermions.

For $\gamma\gamma\rightarrow\text{W}^+\text{W}^-$, the amplitudes have the form \cite{Accomando_VBS}
\begin{align}
\mathcal{M}_\text{T}=\frac{\mathcal{M}_{s/u}s}{u}+\frac{\mathcal{M}_{t^2/us}t^2}{us}+\frac{\mathcal{M}_{u/s}u}{s}+\frac{\mathcal{M}_{s/t}s}{t}+\frac{\mathcal{M}_{t/s}t}{s}+\frac{\mathcal{M}_{u^2/ts}u^2}{ts}
\end{align}
in the high-energy limit for transverse gauge bosons and
\begin{align}
\mathcal{M}_\text{L}=\frac{\mathcal{M}_1u}{s}+\frac{\mathcal{M}_2t}{s}
\end{align}
in the longitudinal case. Since the smallest invariant is relevant for
estimating the quality of the $\text{SCET}_\text{EW}$ assumption, we
 investigate the behaviour for $|t|,|u|\ll s$. In the fermionic case
squaring the matrix element yields another factor of the invariant, making the contribution of the squared matrix element behave as
\begin{align}
\left|\frac{\mathcal{M}_t}{t}\right|^2_{|t|\ll s}\sim\frac st
\end{align}
and similar for $\mathcal{M}_u$.
In the photon-induced case this effect is absent, and the squared matrix element scales as
\begin{equation}
\left|\frac{\mathcal{M}_{s/t}s}{t}\right|^2_{|t|\ll s}\sim\frac {s^2}{t^2},\label{s2t2}
\end{equation}
and analogously for the other matrix elements.

In any case, the total cross section is expected to be dominated by regions, in which at least one of the Mandelstam variables is small [of $\mathcal{O}(M_\text{W}^2)$].
Therefore, we do not discuss the influence of the resummation on integrated cross sections and focus on differential distributions in energy-like observables and study the behaviour in the tails only.
In order to analyse the quality of this approximation, we first consider unresummed $\text{SCET}_\text{EW}$, meaning that the exponentiated amplitude is expanded to first order in $\alpha$. In this approximation the $\text{SCET}_\text{EW}$~results agree with the fixed-order one-loop results up to powers of $M^2/s_{ij}$ with $M$ being any of the EW mass scales and $s_{ij}\in\{s,t,u\}$.  

\begin{figure}
\centering
\includegraphics[width=0.45\textwidth]{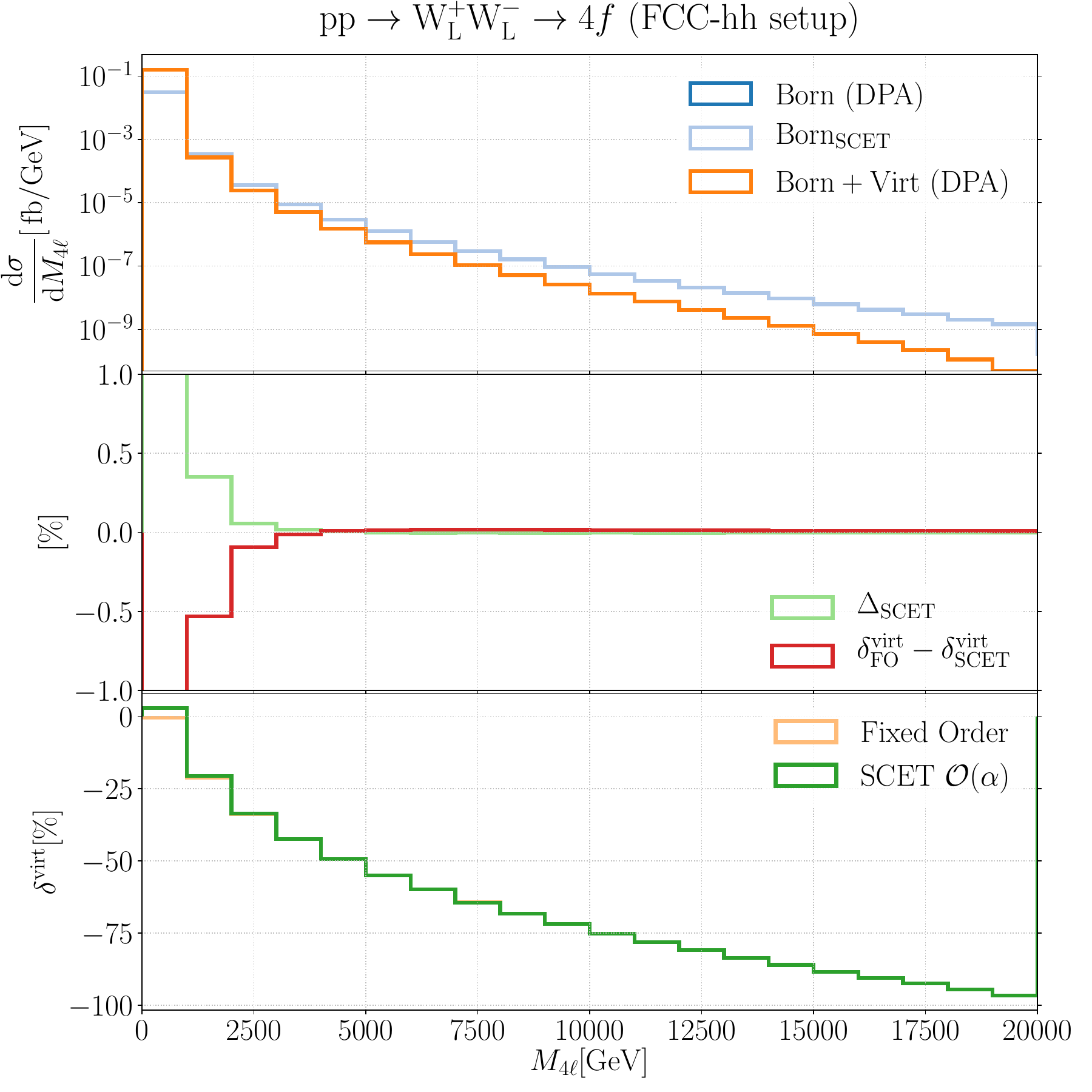}
\includegraphics[width=0.45\textwidth]{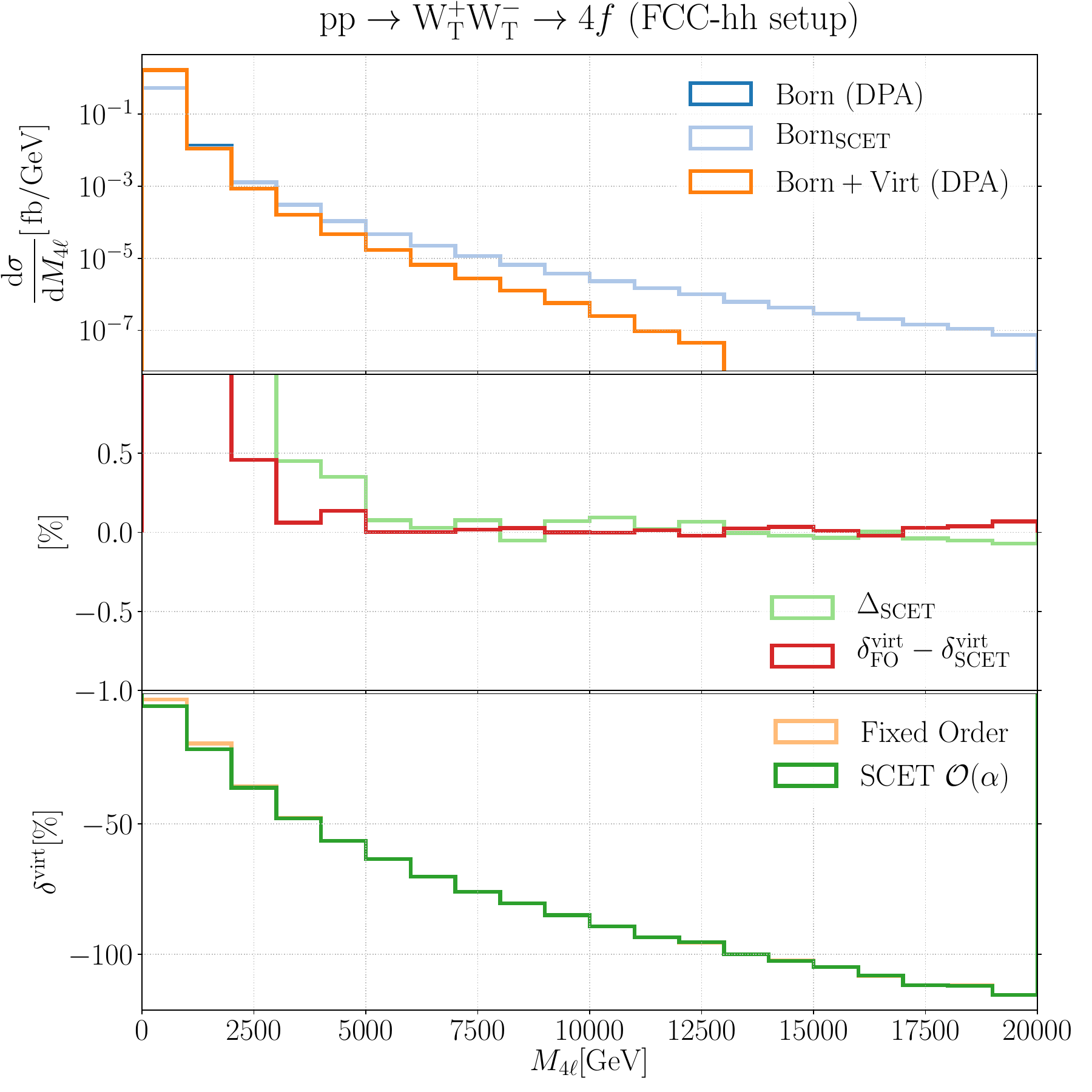}\\
\includegraphics[width=0.45\textwidth]{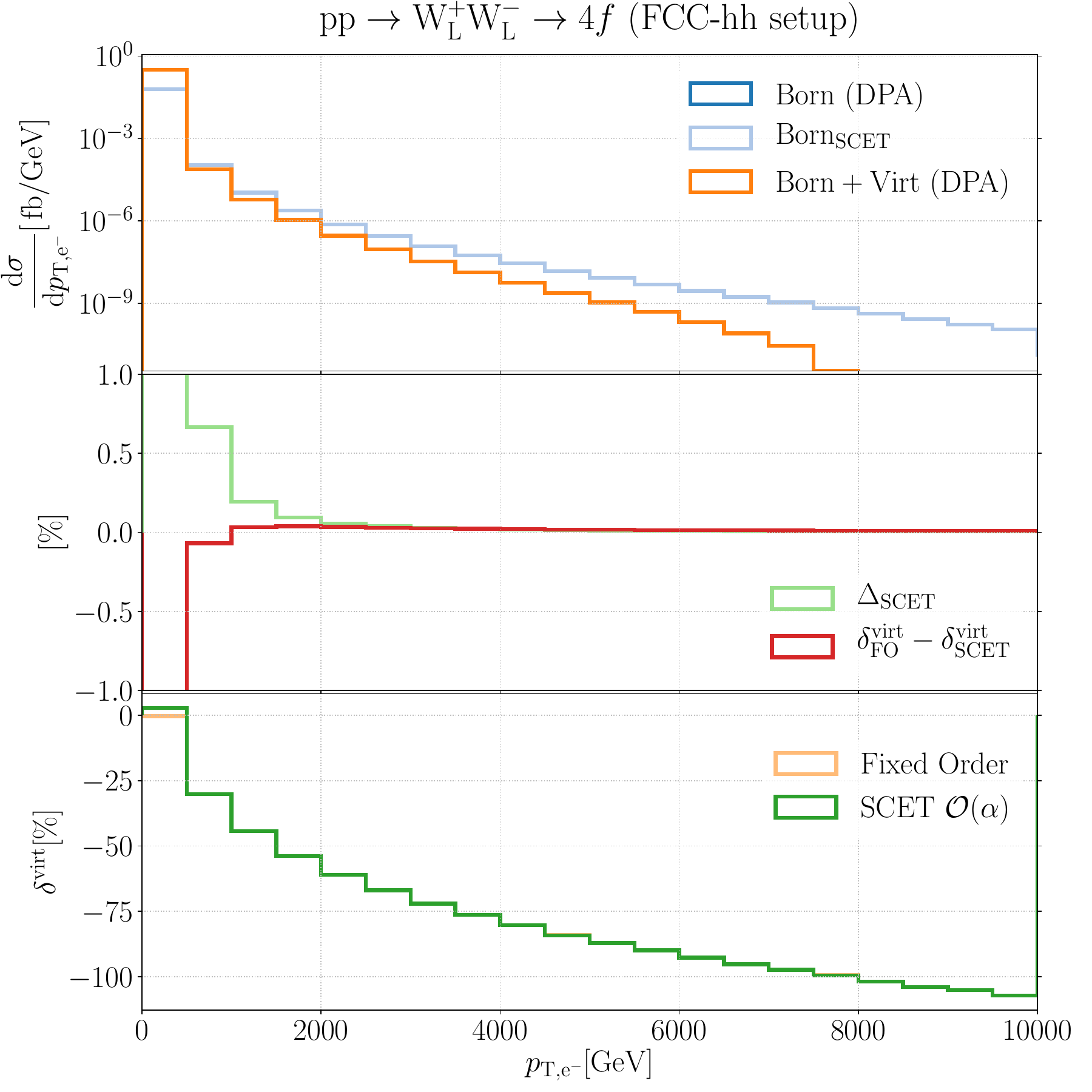}
\includegraphics[width=0.45\textwidth]{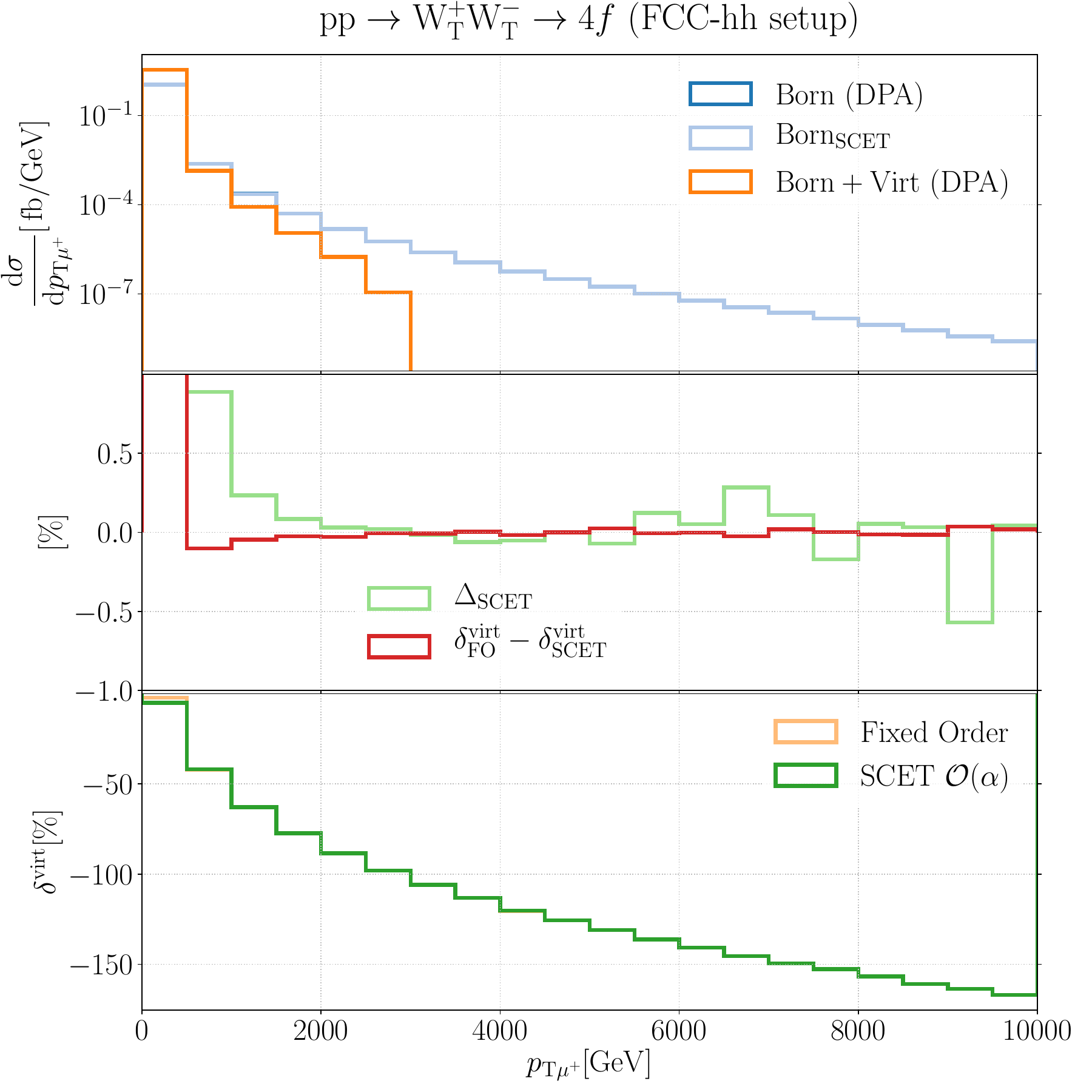}
\caption{Virtual corrections to longitudinal (left) and transverse
  (right) $\text{W}^+\text{W}^-$ production in $\ppwwllll$ calculated
  in conventional fixed-order perturbation theory compared to the
  first-order expansion of the SCET results in $\alpha$. 
}
\label{Fig:WWVali}
\end{figure}

\begin{figure}
\centering
\includegraphics[width=0.45\textwidth]{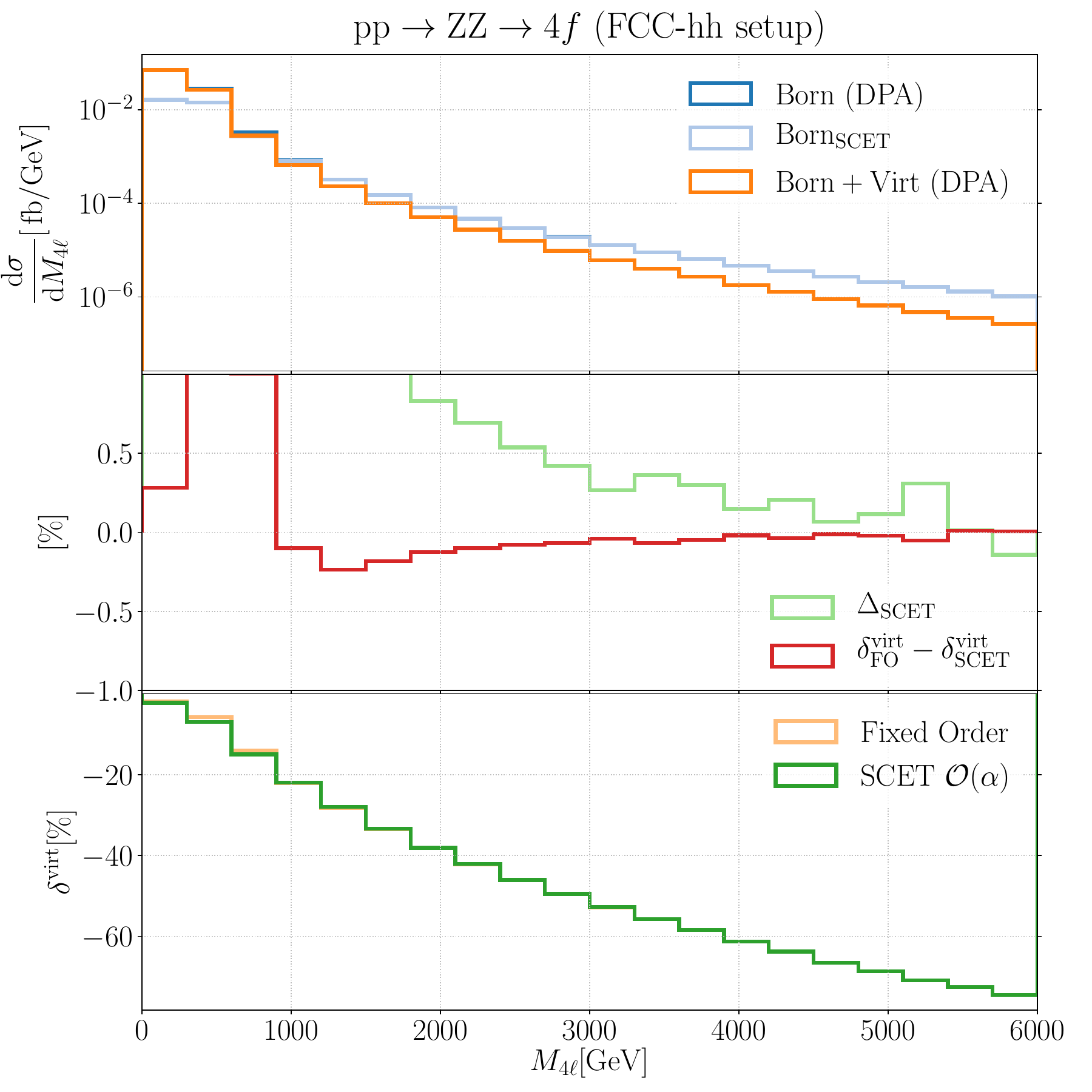}
\includegraphics[width=0.45\textwidth]{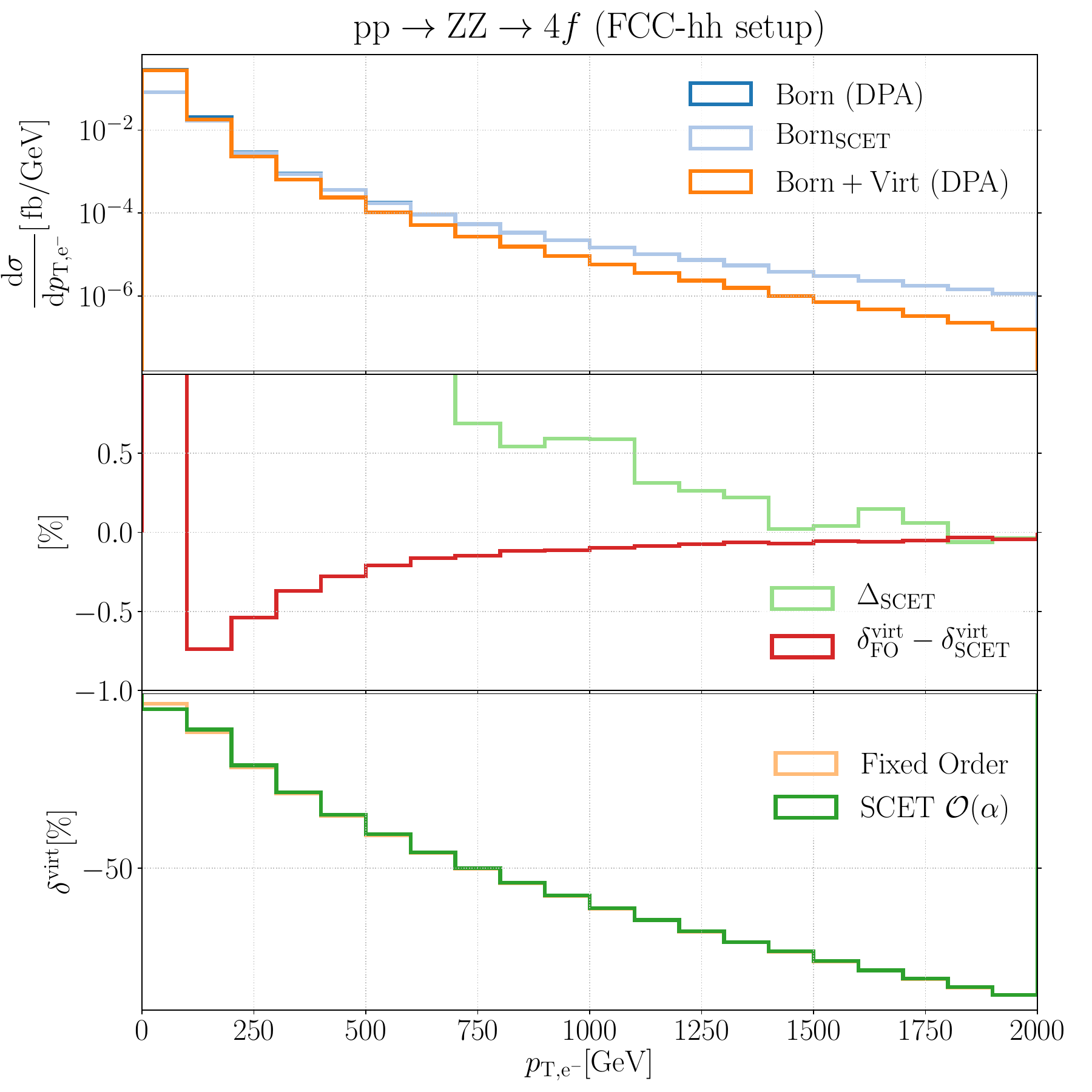}
\caption{Virtual corrections to $\ppzzllll$ calculated in conventional
  fixed-order perturbation theory compared to the first-order
  expansion of the SCET results in $\alpha$. 
}
\label{Fig:ZZVali}
\end{figure}

\begin{figure}
\centering
\includegraphics[width=0.45\textwidth]{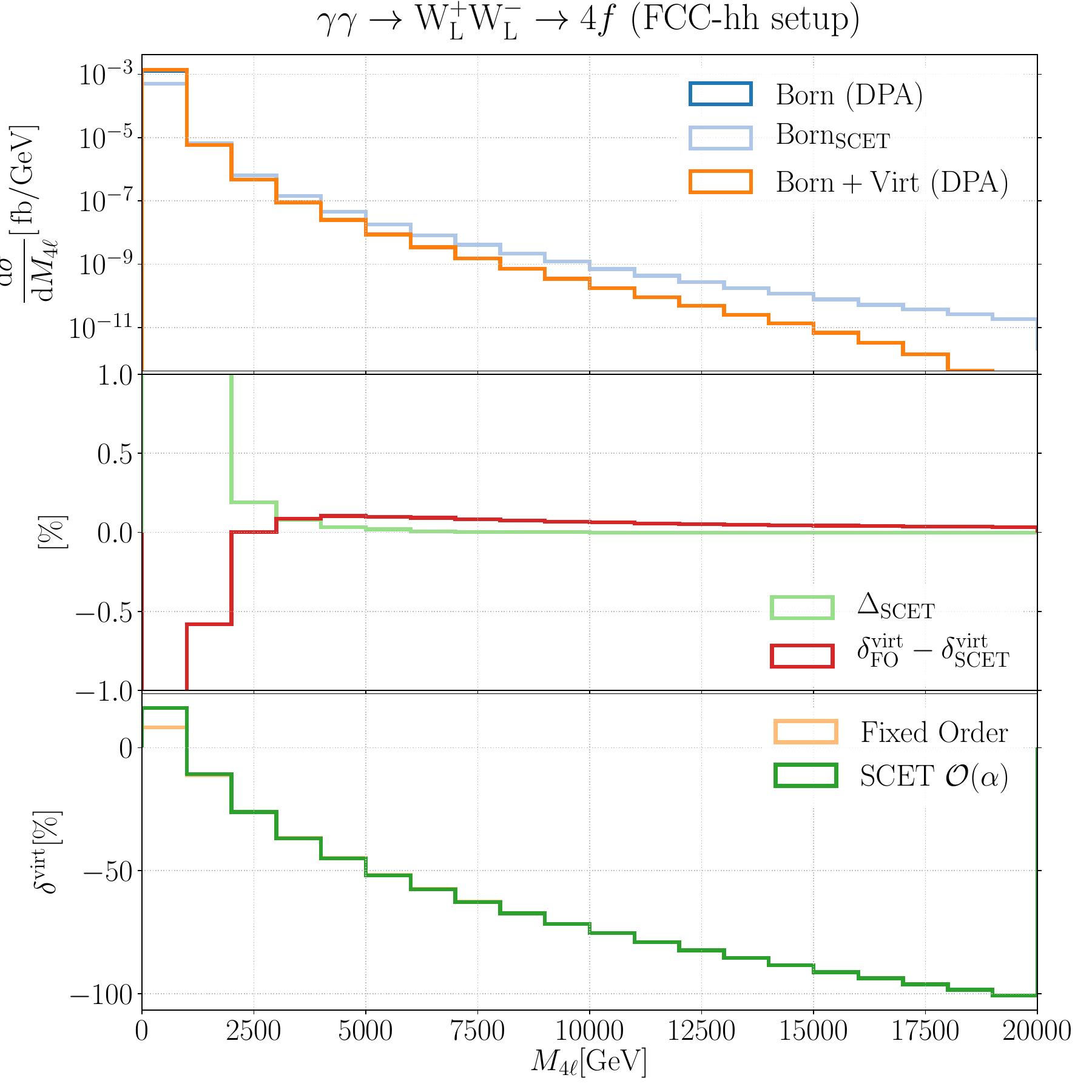}
\includegraphics[width=0.45\textwidth]{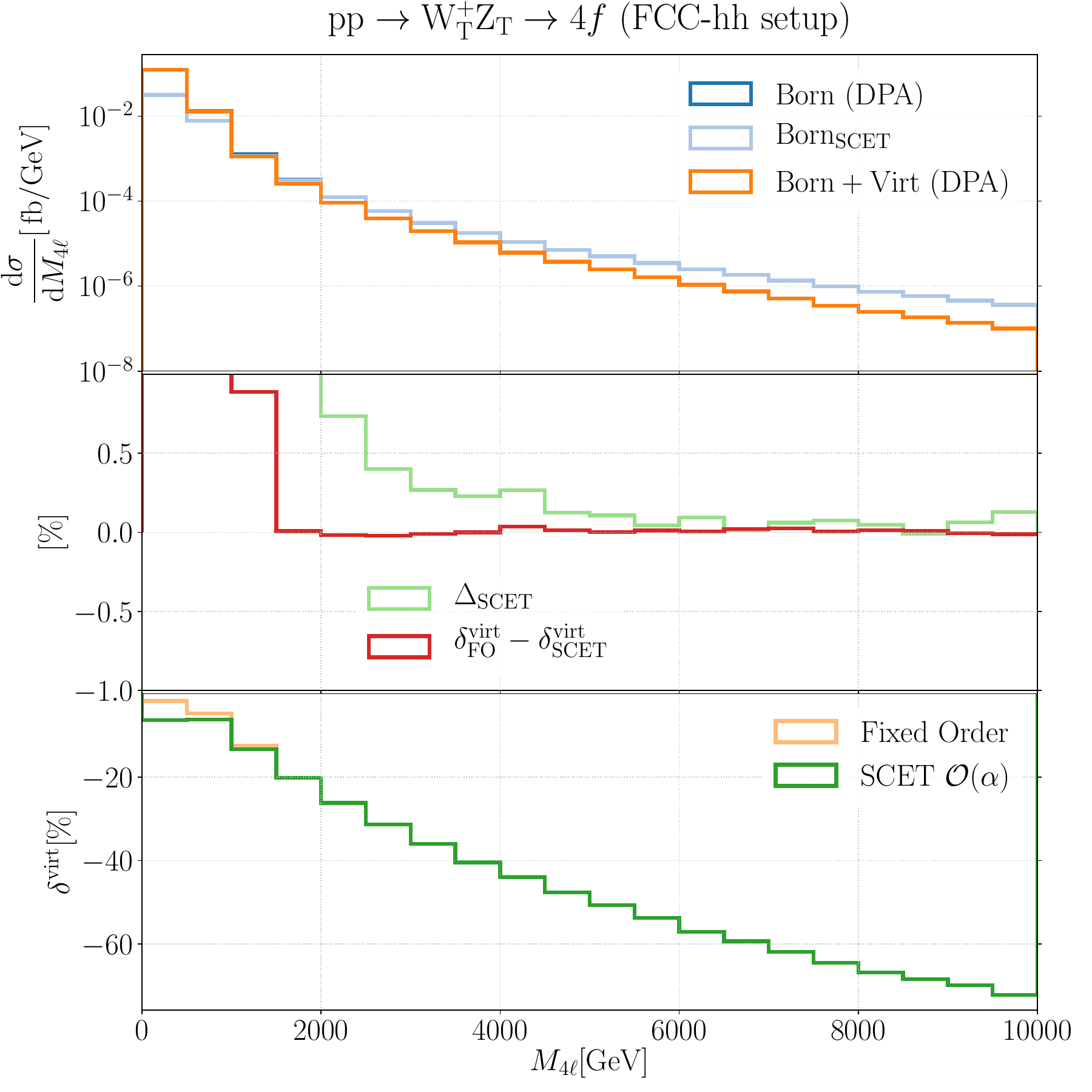}\\
\includegraphics[width=0.45\textwidth]{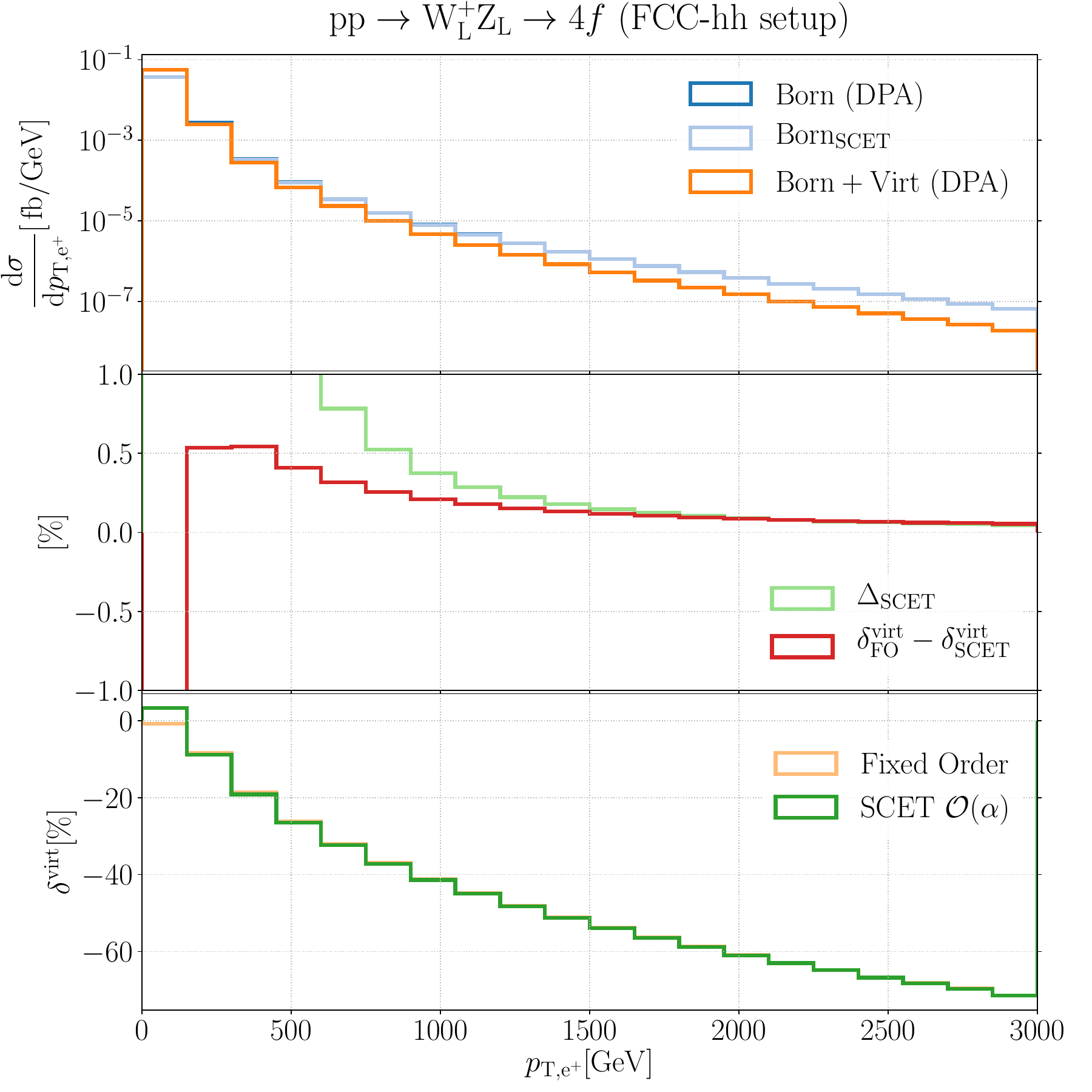}
\includegraphics[width=0.45\textwidth]{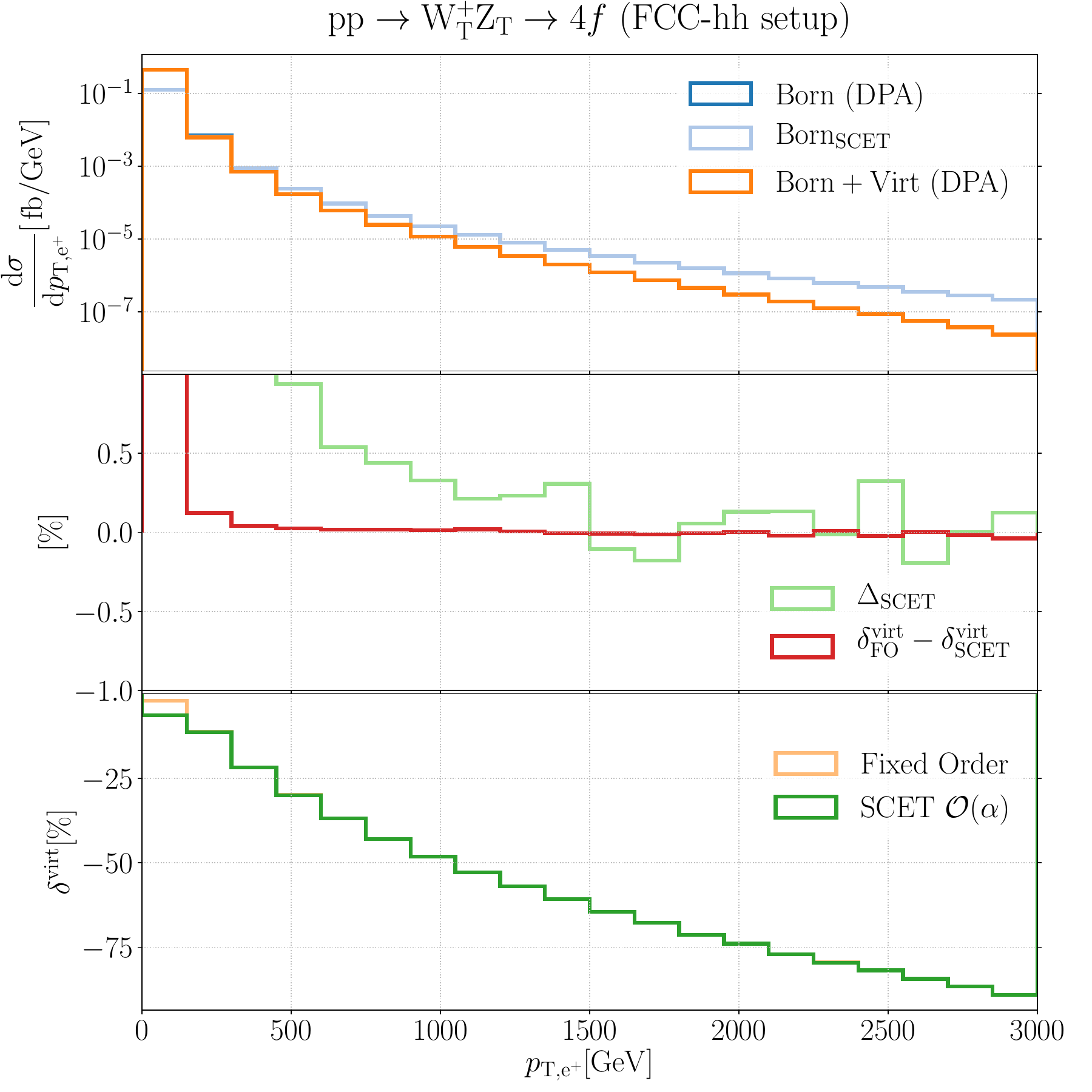}
\caption{Virtual corrections to longitudinal (left) and transverse
  (right) $\text{W}^+$Z production in $\ppwzllll$ calculated in
  conventional fixed-order perturbation theory compared to the
  first-order expansion of the SCET results in $\alpha$. 
}
\label{Fig:WZVali}
\end{figure}

\begin{figure}
\centering
\includegraphics[width=0.45\textwidth]{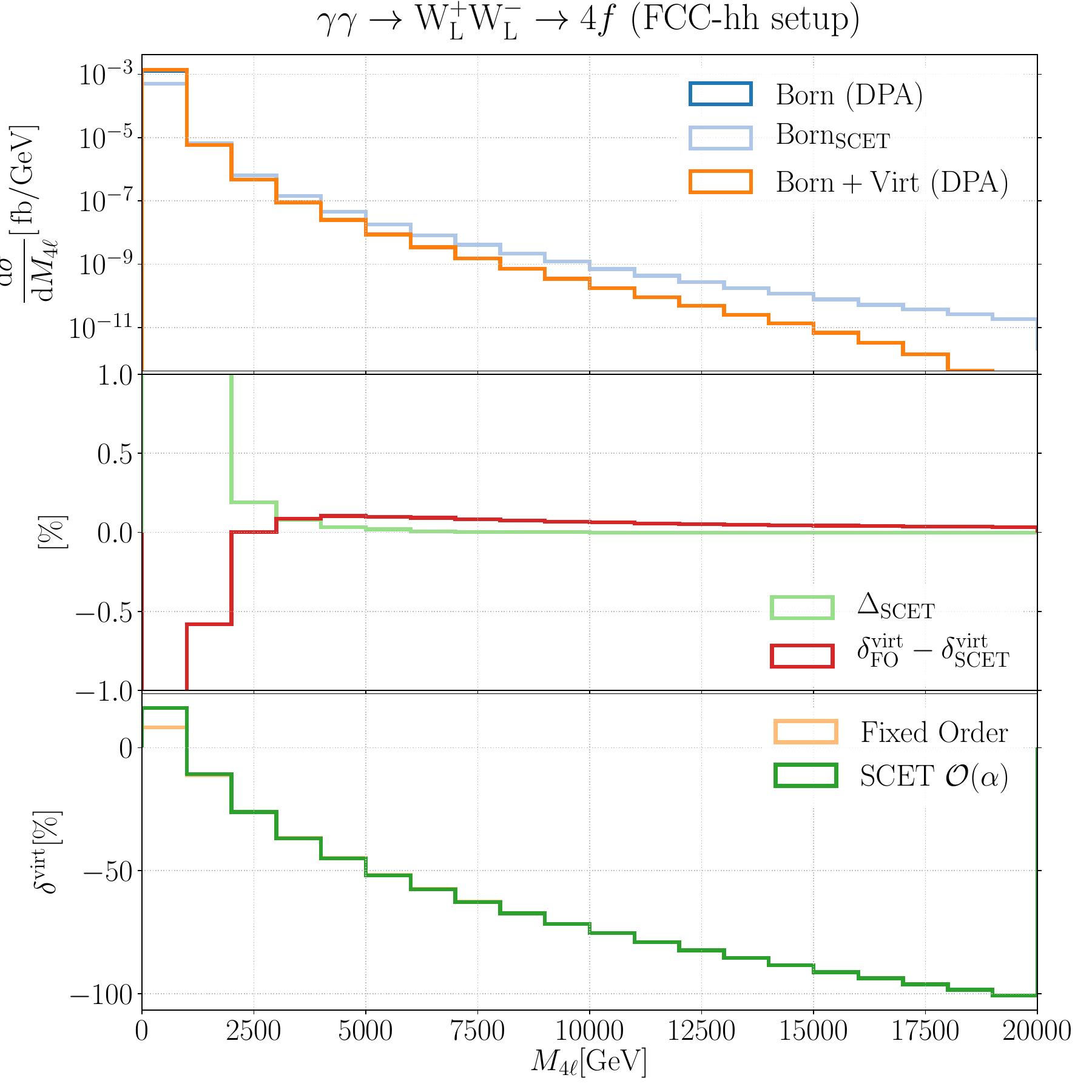}
\includegraphics[width=0.45\textwidth]{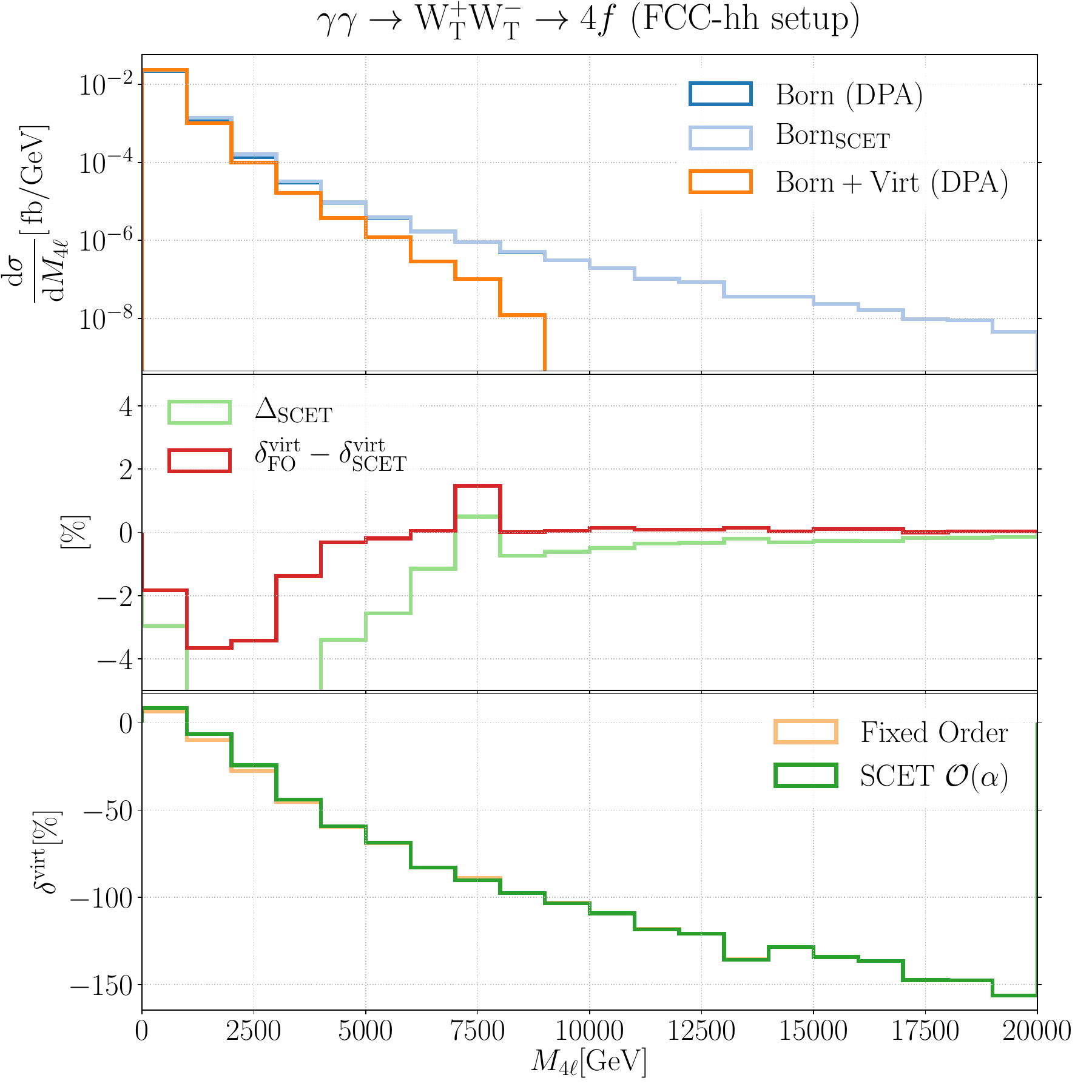}\\
\includegraphics[width=0.45\textwidth]{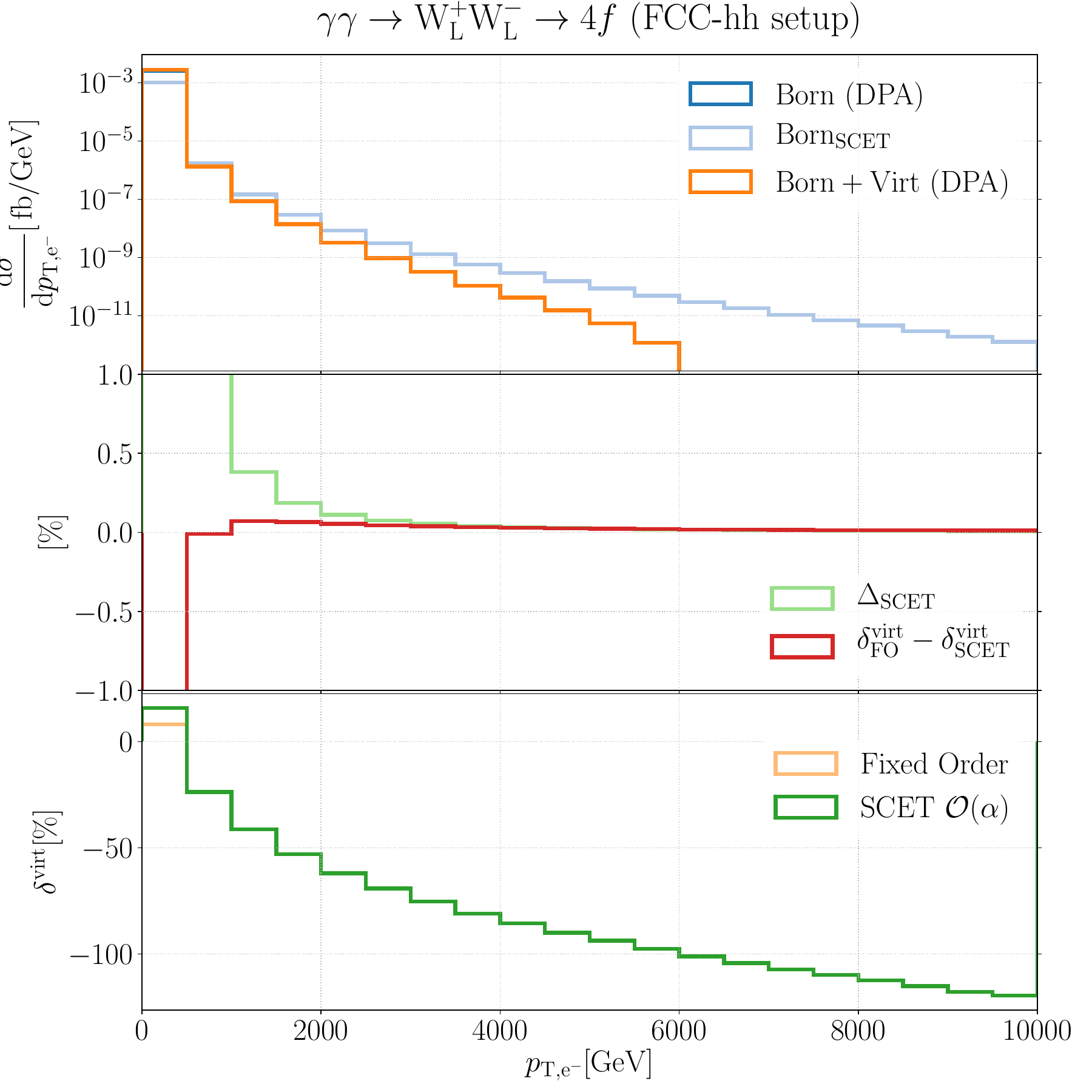}
\includegraphics[width=0.45\textwidth]{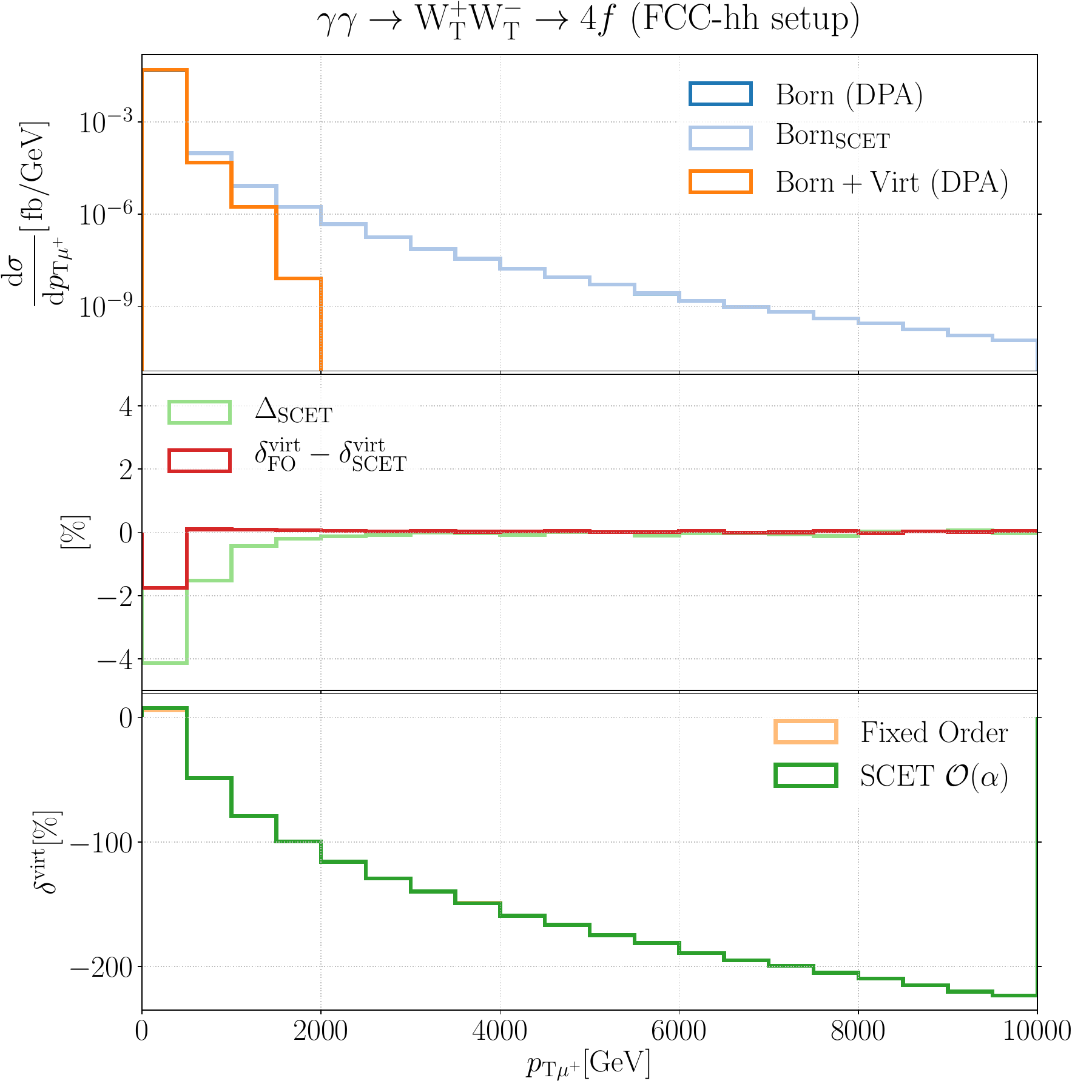}
\caption{Virtual corrections to longitudinal (left) and transverse
  (right) $\aawwllll$~production calculated in conventional
  fixed-order perturbation theory compared to the first-order
  expansion of the SCET results in $\alpha$. 
}
\label{Fig:AAVali}
\end{figure}

In Figs.~\ref{Fig:WWVali}, \ref{Fig:ZZVali}, \ref{Fig:WZVali}, and
\ref{Fig:AAVali} we show differential distributions in the lepton
transverse momenta and the four-lepton invariant masses 
for the four diboson processes under consideration.
The panels are organised as described in the beginning of Sec.~\ref{Sec:ValiRes}.

In the first few bins that dominate the cross section the $\text{SCET}_\text{EW}$ results are smaller by a factor of 2--5 than the fixed-order ones because of the technical cut (\ref{stu}):
The $\text{SCET}_\text{EW}$ matrix elements thus are not appropriate.
Towards the high-energy tails the results converge against each other,
up to sub-percent accuracy from $M_{4\ell}\approx1000\,\text{GeV}$ and
$p_{\text{T},\ell}\approx 500\,\text{GeV}$ depending on the process
and the polarisation state.\footnote{Note that the plots have
  different ranges: For larger cross~sections a longer high-energy tail is shown. Thus, $M_{4\ell}\approx 1000\,\text{GeV}$ is not at the same point in each plot!} In the tails, i.e.\ at energy scales of several TeV the deviation becomes of $\mathcal{O}(10^{-4})$, which is the expected order of magnitude for power corrections. 

In general, in the results for the transverse polarisations the
agreement is worse, since they are dominated by the small-$|t|$ or
small-$|u|$ regime, as described above. We also recognise that the
results for $\gamma\gamma\rightarrow\text{W}^+\text{W}^-$, shown on
the r.h.s.\ of Fig.~\ref{Fig:AAVali}, exhibit a slower
convergence in the high-energy tails. This is due to the quadratic $t$
dependence of the tree-level amplitudes, see (\ref{s2t2}).  We note,
however, that even in regions where the Born results differ by a few
percent (where the green curve is outside the range of the middle
panels), the difference in the relative corrections is already of the
order of 0.1\% (see for instance $\gamma\gamma$-induced transverse WW
production in Fig.~\ref{Fig:AAVali}). This indicates that the error owing
to small-$|t|$ or small-$|u|$ events is strongly reduced for the relative virtual corrections. 

For $\gamma\gamma\rightarrow
\text{W}_\text{T}\text{W}_\text{T}$, which is dominated by 
W production in the forward/backward direction, 
$\Delta_\text{SCET}$ reaches the subpercent level at $M_{4\ell}\approx
8\,\text{TeV}$, and
$\delta_\text{FO}^\text{virt}-\delta_\text{SCET}^\text{virt}$ at about
$M_{4\ell}\approx 5\TeV$. In these regions, however, almost no
statistics is to be expected, since the differential cross section is
already below $10^{-6}\,\text{fb}/\text{GeV}$. Even if one would
consider an overflow bin from $M_{4\ell}=10\TeV$ to infinity the cross
section in that bin would not exceed $1\,\text{ab}$. In the
high-$p_{\text{T},\ell}$ tails, events with low $|t|$ or $|u|$ are suppressed: because the gauge bosons are produced with high energy, their decay products are preferably radiated in the forward direction. A high lepton~$p_\text{T}$ is thus likely to result from the decay of a gauge boson with high $p_\text{T}$. Thus in the high-$p_{\text{T},\ell}$ tails the convergence looks better, reaching the subpercent level between 1 and $2\,\text{TeV}$. However, for the same reason the cross section falls off very fast in these distributions, and only for $p_{\text{T},\ell}\lesssim2\TeV$ a measurable cross section is to be expected. Thus, the window in which $\text{SCET}_\text{EW}$ can be applied and at the same time sufficient statistics can be expected is rather narrow for this process.

\subsubsection{Resummed results}
\begin{figure}
\centering
\includegraphics[width=0.45\textwidth]{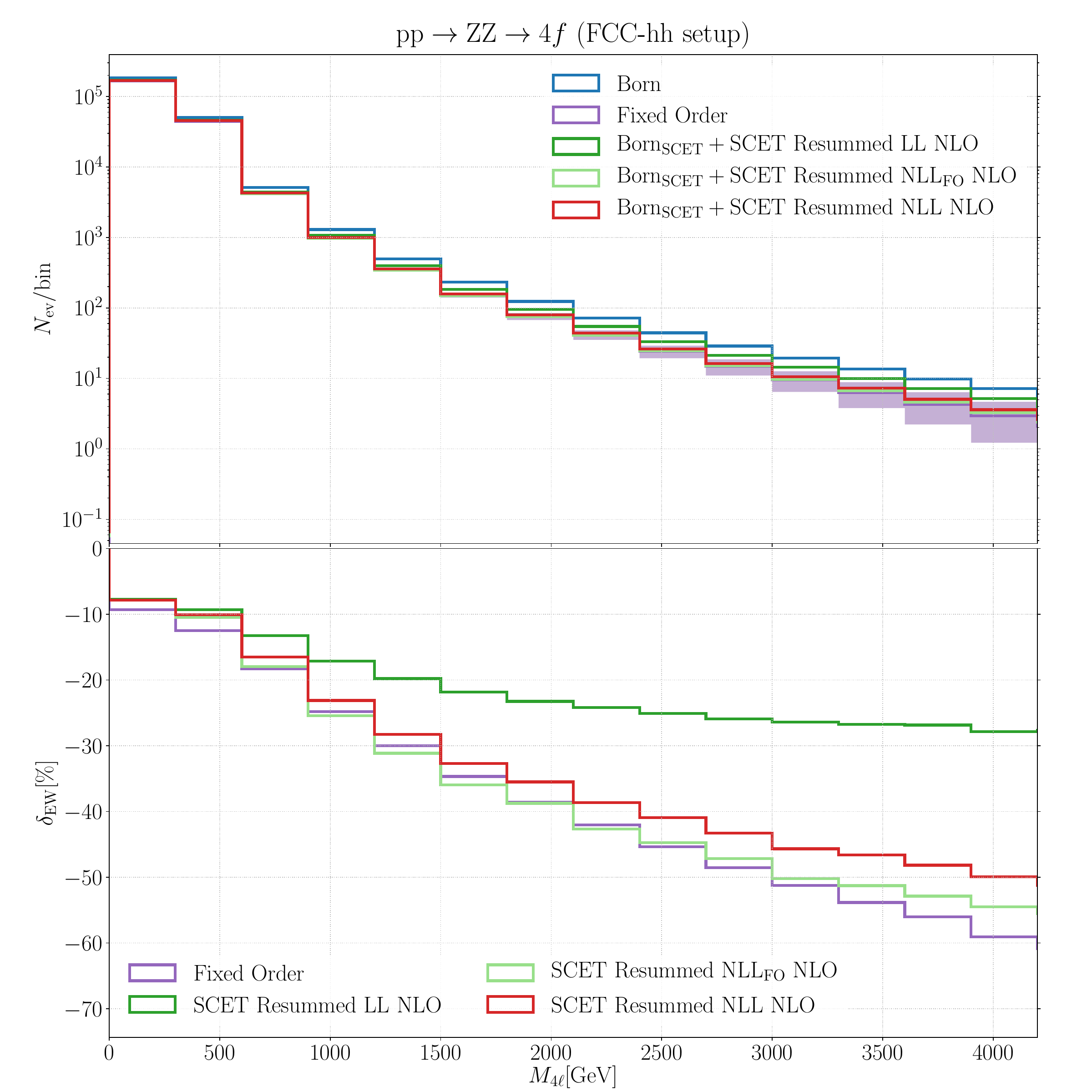}
\includegraphics[width=0.45\textwidth]{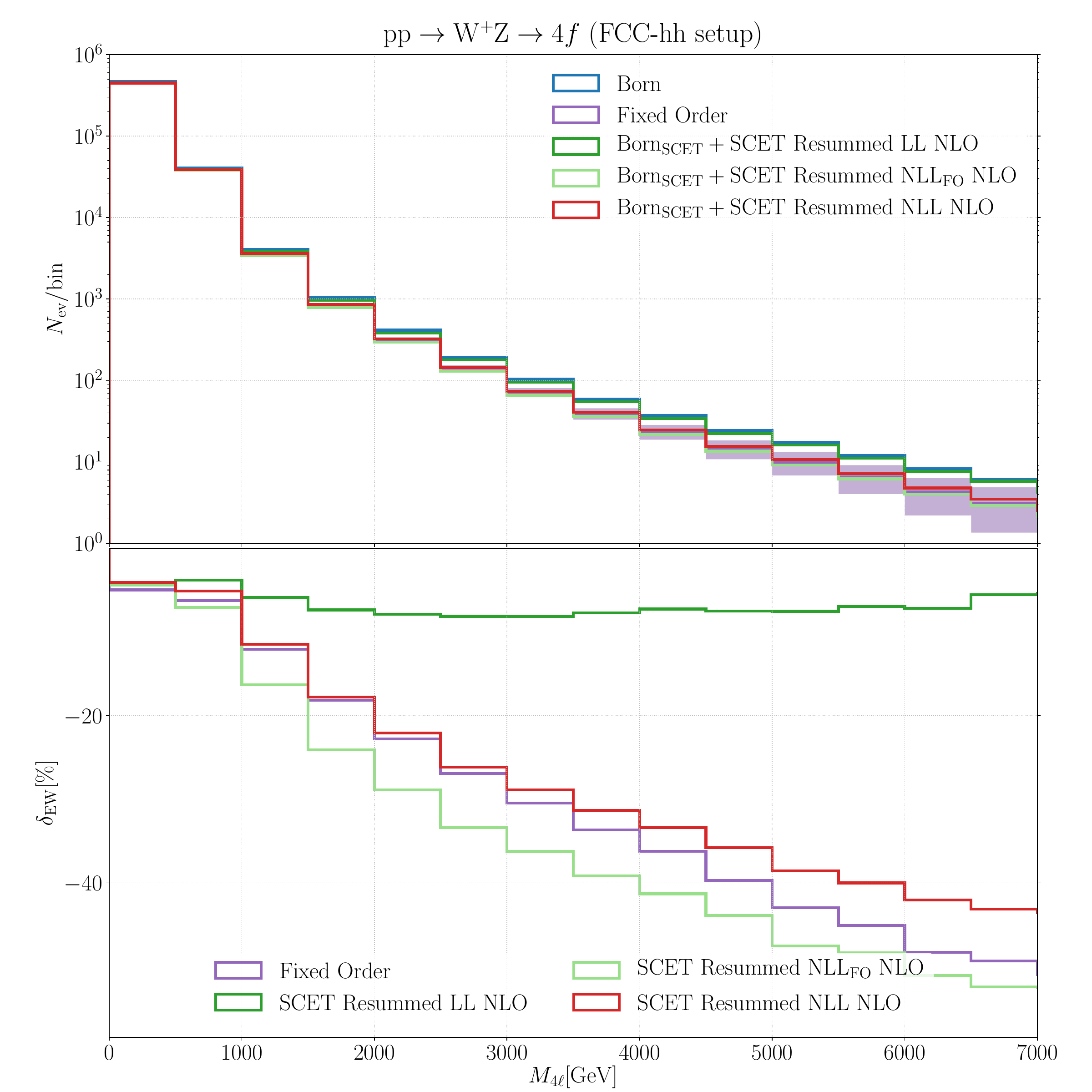}\\
\includegraphics[width=0.45\textwidth]{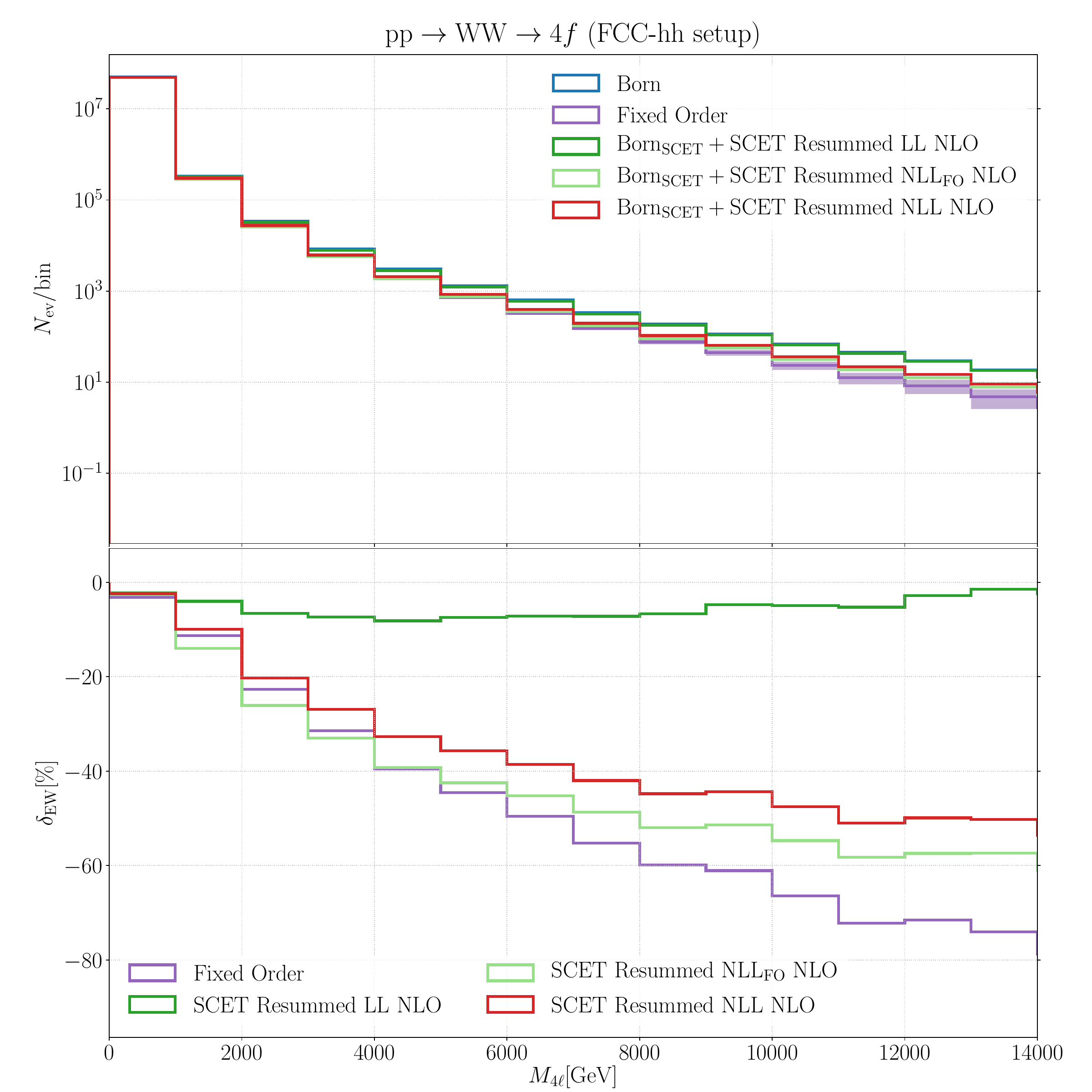}
\includegraphics[width=0.45\textwidth]{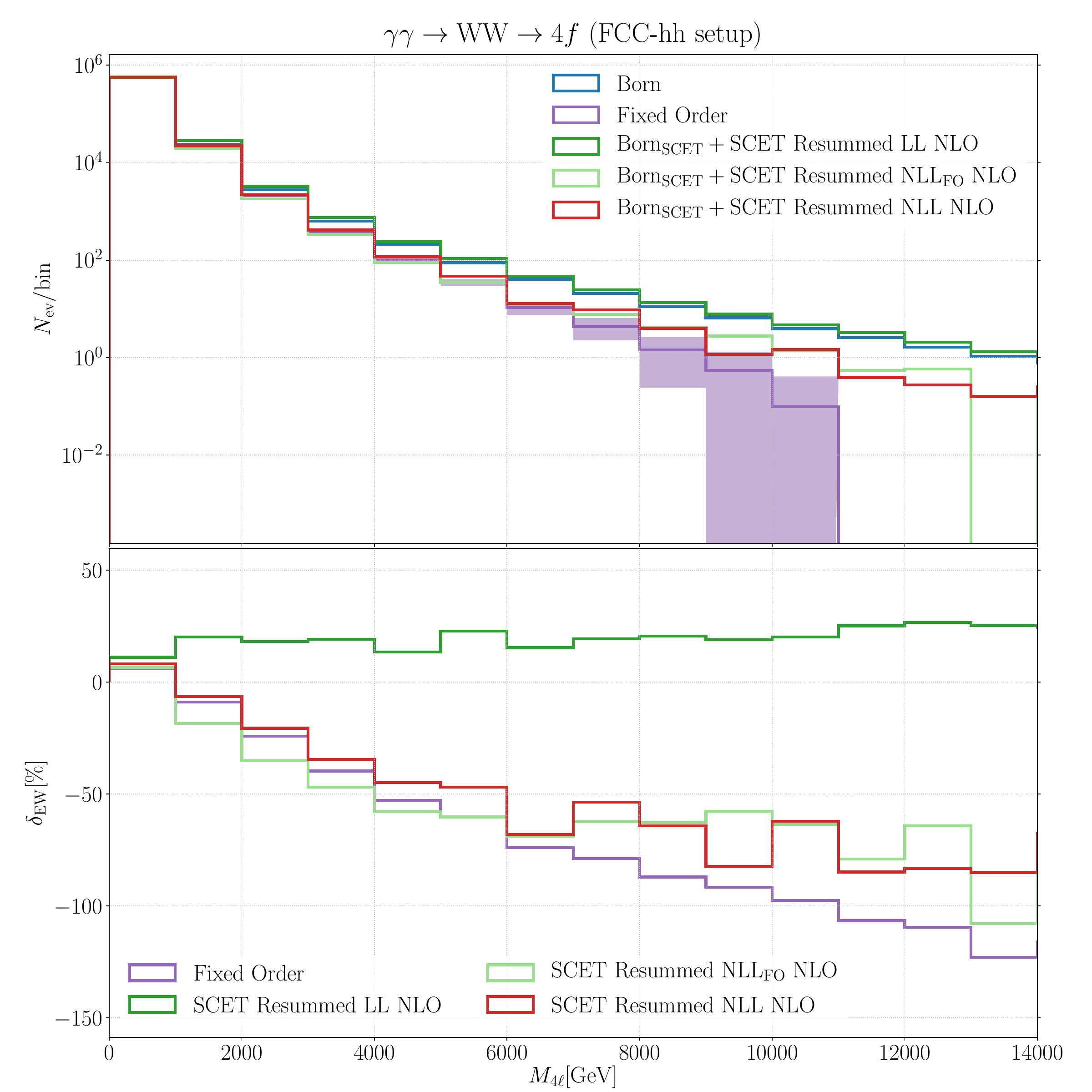}
\caption{Differential
  distributions in the four-lepton invariant masses for unpolarised
  $\ppzzllll$, $\ppwzllll$, $\ppwwllll$, and $\aawwllll$, production with the error on the
  counting rates shaded around the purple curves. The results include
  real, virtual, and integrated dipole contributions. The upper panels
  show the expected event numbers per bin assuming an integrated
  luminosity of $\mathcal{L}=20\,\text{ab}^{-1}$. The various curves differ
  in the treatment of the virtual corrections only.  }
\label{Fig:FCChhResum}
\end{figure}
At this point we can discuss the impact of the resummation taking into
account the expected counting rates at the FCC--hh. In
Fig.~\ref{Fig:FCChhResum} we show some high-energy tails with the
respective counting rates instead of cross sections, including the
statistical error on the fixed-order rate as a band for demonstration. The counting rates are obtained from the differential cross sections and the assumed FCC--hh luminosity of~\cite{FCC_hh_lumi} 
\begin{align}
\mathcal{L}_\text{FCC-hh}=20\,\text{ab}^{-1}.
\end{align}
The results are summed over the polarisations and include besides the
factorisable virtual corrections also real corrections, integrated-dipole
contributions and non-factorisable virtual corrections.

For $\text{ZZ}$ production (upper left plot in Fig.~\ref{Fig:FCChhResum}) we see that in the window of $1.5\,\text{TeV}<M_{4\ell}<3\,\text{TeV}$ the impact of the LL resummation exceeds the statistical error on the counting rate within the given binning, indicating that a resummation of the leading logarithms is indeed necessary at these energies. 
Note that Fig.~\ref{Fig:ZZVali} shows that the
error due to the $\text{SCET}_\text{EW}$ assumption in this
phase-space region amounts to
\mbox{$\delta_\text{FO}^\text{virt}-\delta_\text{SCET}^\text{virt}<0.5\%$}.
The effect of the LL resummation varies between 15\% and 30\% in this range, the NLL effect is of the same order of magnitude, but with a different sign, 
such that the NLL+NLO results do not differ by more than 10\% from the fixed-order result in the respective range, 
which is not significant in view of the experimental error. One should again note that this cancellation is accidental and should not lead to the conclusion
that the resummation effects in general are small or even negligible.

In $\text{W}^+\text{Z}$ and $\text{W}^+\text{W}^-$ production (upper right and lower left plot in Fig.~\ref{Fig:FCChhResum}), which has by far the highest cross section of the diboson processes, the significance of the LL resummation, which shifts the cross section by more than 50\% in the tails, is clearer. 
The effect of the LL resummation significantly exceeds the statistical uncertainty
on the counting rates up to an invariant mass of $M_{4\ell}\approx
6\,\text{TeV}$ and $M_{4\ell}\approx 12\,\text{TeV}$ for
$\text{W}^+\text{Z}$ and $\text{W}^+\text{W}^-$ production, respectively.
Including the $\text{NLL}_\text{FO}$ resummation again has a similarly large effect 
with an opposite sign. The NLL-resummed result shifts the differential cross section
by 5--10\% and differs from the fixed-order result
by a few percent with respect to LO.
Therefore, $\text{SCET}_\text{EW}$ resummation of both the LL and NLL
contributions should be included if a precise comparison is aimed for.

In the photon-induced case (lower right plot in Fig.~\ref{Fig:FCChhResum}) the effect of the resummation, normalised to the LO cross section, is largest because of the large Casimir invariant of the W boson. 
The expected event rate is, similar to the $\text{W}^+\text{Z}$ and ZZ case, only significant up to $M_{4\ell}\sim 6\,\text{TeV}$ 
(note further that the photon PDFs also introduce higher uncertainties in particular at large momentum fractions). 
The qualitative behaviour of the resummed curves is similar to the
quark-induced cases: for $M_{4\ell}\sim 6\,\text{TeV}$
the LL resummation increases the cross section by almost $100\%$, but the inclusion of the $\text{NLL}_\text{FO}$ and NLL undoes the bulk of the effect such that the difference between
the fixed-order and NLL-resummed results is at the level of $10\%$. 

\section{Conclusions}

At high energies, electroweak corrections are dominated by large
logarithms that have to be resummed for decent predictions. Soft-collinear 
effective theory ($\text{SCET}_\text{EW}$) has been proposed as an appropriate
framework for this task.  We have implemented $\text{SCET}_\text{EW}$
into a Monte Carlo integration code and applied it to vector-boson
pair production including the decays of the vector bosons. The
application of $\text{SCET}_\text{EW}$ to such a process requires the
use of a number of approximations, including the Double-Pole
Approximation (DPA), the use of polarised cross sections and the
approximation of large kinematical invariants.

We have considered two different future collider setups inspired by
the CLIC and FCC--hh projects for diboson production processes. Within
these setups we have investigated the accuracy of the different
approximations and the effects of the resummation of the LL and NLL
electroweak logarithms for both integrated cross sections and
differential distributions of all diboson production processes.

In the considered CLIC setup ($\Pe^+\Pe^-$ collisions at $3\TeV$) the errors owing to the
$\text{SCET}_\text{EW}$ assumption, $s,|t|,|u|\gg M_\text{W}^2$, are below
$0.5\%$ on the level of integrated cross sections. The resummation of
the Sudakov double logarithms ($\alpha\mathsf{L}^2$) shifts the cross sections by about
$+23\%$ in ZZ production and by about $+21\%$ in
$\text{W}^+\text{W}^-$ production in $\text{e}^+\text{e}^-$ collisions
(all percentages with respect to the Born cross section).
The resummation of the next-to-leading logarithms accounts for $-10\%$
for ZZ and $-19\%$ for $\text{W}^+\text{W}^-$ with the
major effects arising from the $\alpha^2\mathsf{L}^3$ contributions.
The matching corrections account for 5--10\% and are relevant
if a high accuracy is aimed for.

In the FCC--hh setup ($\Pp\Pp$ collisions at $100\TeV$) the
diboson-production cross sections are dominated by phase-space
regions in which the $\text{SCET}_\text{EW}$ assumption does not hold.
This is especially drastic for photon-induced W-boson pair production.
In the high-energy tails of distributions in processes with a high
cross section, such as $\text{W}^+\text{W}^-$ production, there is
nevertheless a window, in which $\text{SCET}_\text{EW}$ is applicable
(with a subpercent error) and the effect of the resummation is
significant. Depending on the process, this window is typically within
an energy range of $3$--$12\,\text{TeV}$.

In both setups the DPA, applied only to the virtual corrections, has a
limited accuracy for some distributions in processes involving
external W bosons. In the CLIC setup the error owing to the
DPA does not exceed 1\% in the regions that dominate the cross
section.  In the backward region, where the cross section is
suppressed, we find discrepancies of up to 40\% indicating a failure
of the DPA.  In the FCC--hh setup for $\text{W}^+\text{W}^-$ and
$\text{W}^+\text{Z}$ production, however, the DPA error reaches 10\%
for large invariant masses and up to 40\% for larger transverse
momenta of the leptons.

Both, the errors owing to the $\text{SCET}_\text{EW}$ assumption and
the use of the DPA limit the accuracy of our results and contribute to
their theoretical uncertainty. Further sources of theoretical
uncertainty are the size of the higher-order corrections that are not
included. The missing single logarithmic terms of
$\mathcal{O}(\alpha^2\mathsf{L}^2)$ can for instance be estimated upon
exponentiating the $\mathcal{O}(\alpha\mathsf{L})$ corrections. While
it is difficult to come up with a simple recipe for the theoretical
uncertainty of our results, a combination of the mentioned sources of
errors suggests that an accuracy at the level of few percent is
feasible in relevant phase-space regions.

While the resummation of large EW logarithms is a must at future
high-energy colliders, the application of the $\text{SCET}_\text{EW}$
formalism to realistic diboson processes is nontrivial and requires a
number of approximations that need to be carefully checked.  

\section*{Acknowledgements}
The authors are grateful to Jean-Nicolas Lang for great support and advice in using {\sc{Reptil}}
and Jean-Nicholas Lang and Sandro Uccirati for maintaining \recola.
This work is supported by the German Research Foundation
\looseness -1
(DFG) under reference number DFG 623/6-1.

\appendix
\section{Low-scale corrections for the Standard Model}
\label{Sec:SCFunctions}
In this appendix we collect the results for the $\text{SCET}_\text{EW}$ one-loop low-scale corrections in the broken phase. 
\subsubsection*{Extraction of the IR divergences}
The soft and collinear functions have first been published in Ref.~\cite{SCET_SM} without any IR-divergent contributions. 
Because we use the results in the context of a fixed-order calculation we reintroduce the IR divergences.
To this end we employ two different strategies for the case of photonic corrections to massless external fermions and W bosons, respectively. 
\paragraph{External W bosons}
In the case of
W bosons the IR poles can be calculated within boosted-Heavy-Quark Effective Theory (bHQET), following Ref.~\cite{CGKM2}, where the bHQET corrections for the case of a gauge-boson exchange 
between two heavy quarks with masses $M_1$, $M_2$, have been computed. Note that one heavy quark is considered to be incoming and one outgoing. 
Assuming the gauge-boson mass $\lambda$ much
smaller than the heavy quark masses $M_1$, $M_2$, 
\begin{align}
\lambda^2\ll M_1^2\sim M_2^2\ll 2 p_1\cdot p_2,
\end{align}
the result reads (see Eq.~(81) of Ref.~\cite{CGKM2} and the explanations below)
\begin{align}
-\frac{\alpha}{2\pi}\log\frac{ 2 p_1\cdot p_2}{M_1M_2}\left(\frac{1}{\varepsilon_\text{UV}}+\log\frac{\mu_\text{UV}^2}{\lambda^2}\right).
\end{align}
Interpreting $\lambda$ as the photon mass and making use of the fact that the IR divergences are of pure soft origin, we can translate the result to dimensional regularisation
using
\begin{align}
\log{\lambda^2}\Rightarrow(4\pi)^{\varepsilon_\text{IR}}\Gamma(1+\varepsilon_\text{IR})\frac{1}{\varepsilon_\text{IR}}+\log{\mu_\text{IR}^2},
\label{RegularisationTranslation}
\end{align}
with regularisation parameter $\varepsilon_\text{IR}=(4-D)/2$ and
corresponding mass scale $\mu_\text{IR}$.
After these steps the UV-finite photonic radiative corrections to an $n$-particle amplitude with massive external legs (the case of massless fermions is treated below) are obtained as a sum 
over pairs,
\begin{align}
\boldsymbol{D}^\gamma=\sum_{\left<ij\right>}\frac{\alpha}{2\pi}\sigma_i\sigma_j Q_iQ_j\log\frac{-s_{ij}-\text{i}0}{M_iM_j}\left(-\frac{1}{\varepsilon_\text{IR}}+\log\frac{\mu_\text{UV}^2}{\mu_\text{IR}^2}\right)\mathbb{1},
\end{align}
where $\sigma_i=1$ for incoming particles and outgoing antiparticles while
$\sigma_i=-1$ for incoming antiparticles and outgoing particles.
Using charge conservation as well as the decomposition
$s_{ij}=\eta_{ij}(\bar{n}_i\cdot p_i)(\bar{n}_j\cdot
p_j)(n_i\cdot n_j)/2$, which holds in the high-energy limit, the
photonic one-particle contributions for an $n_i$-collinear W boson read 
\begin{align}
D_\text{C}^{\text{W},\gamma}(\mu)=\frac{\alpha}{2\pi}\log\frac{\bar{n}_i\cdot p_i}{M_\text{W}}\left(\frac{1}{\varepsilon_\text{IR}}-\log\frac{\mu_\text{UV}^2}{\mu_\text{IR}^2}\right),
\end{align}
which we add to the W-boson low-scale corrections in order to retain the correct IR divergences.
The two-particle contributions are mass independent and read
\begin{align}
\boldsymbol{D}_\text{S}^\gamma=-\frac{\alpha}{2\pi}\sigma_iQ_i\sigma_jQ_j\left(\frac{1}{\varepsilon_\text{IR}}-\log\frac{\mu_\text{UV}^2}{\mu_\text{IR}^2}\right)\log\left(-\eta_{ij}\frac{n_i\cdot n_j}{2}-\text{i}0\right)\mathbb{1}\label{DSphot}.
\end{align}
\paragraph{Massless external fermions}
In the case of massless fermions the photonic $\text{SCET}_\text{EW}$ corrections are scaleless and can hence be written as
\begin{align}
\left(\frac{\mu_\text{UV}^2}{\mu_\text{IR}^2}\right)^\varepsilon\sum_{k=0}^2\left(\frac{C_k}{\varepsilon_\text{UV}^k}-\frac{C_k}{\varepsilon_\text{IR}^k}\right)
\label{Ckepsk}
\end{align}
with constants $C_k$.
The coefficients of the UV poles and the dependence on the UV scale can be extracted using mass regularisation. Within this scheme the one-particle contributions read (see e.g.\ Ref.~\cite{SCET_wo_regulator})
\begin{align}
D_\text{C,MR}^{f,\gamma,\text{bare}}(\mu)&=\frac{\alpha Q_f^2}{4\pi}\left(\frac{c_{\varepsilon_\text{UV}}}{\varepsilon^2_\text{UV}}-\frac{c_{\varepsilon_\text{UV}}}{\varepsilon_\text{UV}}\left(2\log\frac{\bar{n}_i\cdot p_i}{\mu_\text{UV}}-\frac32\right)
+2\log\frac{\lambda^2}{\mu_\text{UV}^2}\log\frac{\bar{n}_i\cdot p_i}{\mu_\text{UV}}\right.\nonumber\\
&\left.\hspace{40pt}
-\frac12\log^2\frac{\lambda^2}{\mu_\text{UV}^2}
-\frac32\log\frac{\lambda^2}{\mu_\text{UV}^2}
-\frac{\pi^2}{2}+\frac94\right)
\label{Dbare}
\end{align}
with the normalisation factor 
\begin{align}
c_{\varepsilon_\text{UV}}=\Gamma(1+\varepsilon_\text{UV})(4\pi)^{\varepsilon_\text{UV}}\label{cvareps},
\end{align}
which is absorbed into the standard divergence following the conventions of Refs.~\cite{Recola1,CollierMan}.
Equation~(\ref{Dbare}) can, for instance, be obtained by the respective Z-boson contributions with $M_\text{Z}$ substituted by $\lambda$. 
In dimensional regularisation the finite parts have to be zero, and
the $\mu_\text{UV}$ dependence has to be compensated by a dependence
on the IR scale. Using (\ref{Ckepsk}) and setting the UV poles to 0 (because the
low-scale corrections are given by the UV-finite part of the $\text{SCET}_\text{EW}$ diagrams) we obtain
\begin{align}
D_\text{C,DR}^{f,\gamma}(\mu)
&=\frac{\alpha Q_f^2}{4\pi}\left(-\frac{c_{\varepsilon_\text{IR}}}{\varepsilon^2_\text{IR}}+\frac{c_{\varepsilon_\text{IR}}}{\varepsilon_\text{IR}}\left(2\log\frac{\bar{n}_i\cdot p_i}{\mu_\text{IR}}-\frac32\right)
\right.\nonumber\\
&\left.\hspace{30pt}{}+2\log\frac{\mu_\text{IR}^2}{\mu_\text{UV}^2}\log\frac{\bar{n}_i\cdot p_i}{\mu_\text{UV}}-\frac12\log^2\frac{\mu_\text{IR}^2}{\mu_\text{UV}^2}-\frac32\log\frac{\mu^2_\text{IR}}{\mu^2_\text{UV}}\right)
\end{align}
with $c_{\varepsilon_\text{IR}}$ being defined in analogy to (\ref{cvareps}).

The two-particle contributions are treated in the same way. In mass regularisation they read
\begin{align}
\boldsymbol{D}^\text{bare}_\text{S,MR}=\sum_{\left<ij\right>}\frac{\alpha}{2\pi}\sigma_iQ_i\sigma_jQ_j\mathbb{1}\left(\frac{c_{\varepsilon_\text{UV}}}{\varepsilon_\text{UV}}+\log\frac{\mu_\text{UV}^2}{\lambda^2}\right)\log\left(-\eta_{ij}\frac{n_i\cdot n_j}{2}-\text{i}0\right),
\end{align}
and their UV-finite part translates to dimensional regularisation to
obtain again (\ref{DSphot}). These results thus imply that the IR-divergent parts of the two-particle low-scale corrections
are given as in (\ref{DForm}), regardless of the external masses. 
\subsubsection*{Final formulae: Fermions}
For massless fermions with helicity $\kappa$ the collinear part of the low-scale corrections yields
\begin{align}
D^{f_\kappa}_\text{C}=\frac{\alpha}{4\pi}\left(\left(I_{\bar{f}f}^{\text{Z},\kappa}\right)^2D_\text{Z}(\mu_\text{l})+\delta_{\kappa\text{L}}\frac{1}{2s_\text{w}^2}D_\text{W}(\mu_\text{l})+Q_f^2D_\gamma(\mu_\text{l})\right)
\end{align}
with the auxiliary functions (in the following $\varepsilon$ always refers to $\varepsilon_\text{IR}$)
\begin{align}
D_\text{W/Z}(\mu_\text{l})&=2\log\frac{M^2_\text{W/Z}}{\mu_\text{l}^2}\log\frac{\bar{n}\cdot p}{\mu_\text{l}}-\frac 12\log^2\frac{M^2_\text{W/Z}}{\mu^2_\text{l}}-\frac 32 \log\frac{M^2_\text{W/Z}}{\mu^2_\text{l}}-\frac{\pi^2}{2}+\frac 94\nonumber,\\
D_\gamma(\mu_\text{l})&=-\frac{c_\varepsilon}{\varepsilon^2}-\frac{c_\varepsilon}{\varepsilon}\left(\frac 32-2\log\frac{\bar{n}\cdot p}{\mu_\text{IR}}\right)+2\log\frac{\mu_\text{IR}^2}{\mu_\text{l}^2}\log\frac{\bar{n}\cdot p}{\mu_\text{l}}-\frac 12\log^2\frac{\mu_\text{IR}^2}{\mu^2_\text{l}}-\frac 32 \log\frac{\mu^2_\text{IR}}{\mu^2_\text{l}}\label{DWZgamma}
\end{align}
and the respective Z-boson couplings
\begin{align}\label{I^Z}
I^{\text{Z},\kappa}_{f\bar{f}}=\frac{I^\kappa_{3,f}-s^2_\text{w}Q_f}{s_\text{w}c_\text{w}}.
\end{align}
Note that the normalisation factor (\ref{cvareps}) leads to differences in the $\pi^2$~terms, which is why in Ref.~\cite{SCET_SM} the term $-\pi^2/2$ in $D_\text{W/Z}$ is replaced by $-5\pi^2/12$. This corresponds to the replacement
\begin{align}
c_\varepsilon\rightarrow c_\varepsilon'=e^{-\gamma_\text{E}\varepsilon}(4\pi)^\varepsilon,\label{cprimeepsilon}
\end{align}
with $\gamma_\text{E}$ denoting the Euler--Mascheroni constant.
The functions in Eq.~(\ref{DWZgamma}) already contain the contribution from the respective field-renormalisation constants.
\subsubsection*{Final formulae: Transverse gauge bosons}
For gauge bosons we introduce the functions
\begin{align}
F_\text{W/Z}(\mu_\text{l})&=2\log\frac{M^2_\text{W/Z}}{\mu_\text{l}^2}\log\frac{\bar{n}\cdot p}{\mu_\text{l}}-\frac 12\log^2\frac{M^2_\text{W/Z}}{\mu^2_\text{l}}-\log\frac{M^2_\text{W/Z}}{\mu^2_\text{l}}-\frac{\pi^2}{2}+1,\nonumber
\\
F_\gamma(\mu_\text{l})&=2\frac{c_{\varepsilon}}{\varepsilon}\log\frac{\bar{n}\cdot p}{M_\text{W}}+2\log\frac{\mu^2_\text{IR}}{\mu_\text{l}^2}\log\frac{\bar{n}\cdot p}{M_\text{W}}+\frac 12\log^2\frac{M_\text{W}^2}{\mu^2_\text{l}}-\log\frac{M_\text{W}^2}{\mu^2_\text{l}}+2,\label{FWZgamma}
\end{align}
as well as the integral
\begin{align}
f_\text{S}(w,z)=\int_0^1\text{d}x\,\frac{2-x}{x}\log\frac{1-x+zx-wx(1-x)}{1-x},
\end{align}
defined in App.~B of Ref.~\cite{SCET_4f}.
The function $F_\gamma$ comprises both the finite
$\text{SCET}_\text{EW}$ corrections related to photon exchange and the
IR poles obtained from bHQET. In our implementation, the poles
are discarded and the IR scale is identified with the low scale, $\mu_\text{IR}=\mu_\text{l}$.
The integral $f_\text{S}$
can be written in terms of the Passarino--Veltman two-point and three-point standard integrals (for their definition see e.g.\ Ref.~\cite{BDJ}):
\begin{align}
f_\text{S}\left(\frac{p^2}{M^2},\frac{m^2}{M^2}\right)&=-\lim_{r\rightarrow \infty}r\left(C_0(p^2,r,p^2,M^2,m^2,m^2)-\left.C_0(0,r,0,0,0,0,0)\right|_{\mu_\text{IR}^2=M^2}\right)\nonumber\\
&\hspace{30pt}+\frac{\pi^2}{2}+B_0(p^2,M^2,m^2)-B_0(0,M^2,0).
\end{align}
Note that also the standard functions (\ref{FWZgamma}) differ from the expressions given in Ref.~\cite{SCET_SM} by a contribution of $\pi^2/12$, which is due to the convention Eq.~(\ref{cprimeepsilon}).
In terms of these functions the low-scale corrections read
\begin{align}
D^{W\to W}_\text{C}(\mu_\text{l})&=\frac{\alpha}{4\pi}\left[\frac{c_\text{w}^2}{s_\text{w}^2}\left(F_Z+f_\text{S}\left(\frac{M_\text{W}^2}{M_\text{Z}^2},\frac{M_\text{W}^2}{M_\text{Z}^2}\right)\right)+\frac{c_\text{w}^2}{s_\text{w}^2}\left(F_W+f_\text{S}\left(1,\frac{M_\text{Z}^2}{M_\text{W}^2}\right)\right)\right]\nonumber\\
&\hspace{40pt}+\frac{\alpha}{4\pi}\left[F_\gamma+F_W+f_\text{S}(1,0)\right]+\left.\frac 12 \delta Z_\text{W}\right|_{\mu_\text{UV}=\mu_\text{l}}
\end{align}
for external W bosons. The low-scale corrections for the Z boson and the photon depend on the subamplitudes and read
\begin{align}
D^{W^3\rightarrow Z}_\text{C}&(\mu_\text{l})=\frac{\alpha}{2\pi s_\text{w}^2}\left(F_\text{W}+f_\text{S}\left(\frac{M^2_\text{Z}}{M_\text{W}^2},1\right)\right)+\left.\frac 12 \delta Z_\text{ZZ}\right|_{\mu_\text{UV}=\mu_\text{l}}+\left.\frac{s_\text{w}}{c_\text{w}} \frac12\delta Z_\text{AZ}\right|_{\mu_\text{UV}=\mu_\text{l}}\nonumber,\\
D^{W^3\rightarrow \gamma}_\text{C}&(\mu_\text{l})=\frac{\alpha}{2\pi s_\text{w}^2}\left(F_\text{W}+f_\text{S}(0,1)\right)+\left.\frac 12 \delta Z_\text{AA}\right|_{\mu_\text{UV}=\mu_\text{l}}+\frac12\left.\frac{c_\text{w}}{s_\text{w}} \delta Z_\text{ZA}\right|_{\mu_\text{UV}=\mu_\text{l}}\nonumber,\\
D^{B\rightarrow Z}_\text{C}&(\mu_\text{l})=\left.\frac 12 \delta Z_\text{ZZ}\right|_{\mu_\text{UV}=\mu_\text{l}}-\left.\frac12\frac{c_\text{w}}{s_\text{w}} \delta Z_\text{AZ}\right|_{\mu_\text{UV}=\mu_\text{l}}\nonumber,\\
D^{B\rightarrow \gamma}_\text{C}&(\mu_\text{l})=\left.\frac 12 \delta Z_\text{AA}\right|_{\mu_\text{UV}=\mu_\text{l}}-\left.\frac12\frac{s_\text{w}}{c_\text{w}} \delta Z_\text{ZA}\right|_{\mu_\text{UV}=\mu_\text{l}}.
\end{align}
Note that we use the definitions of the $\gamma/\text{Z}$-mixing field-renormalisation constants as in Ref.~\cite{DennerHabil}, which differ from the ones given in Ref.~\cite{SCET_SM} by a factor 2. 
\subsubsection*{Final formulae: Longitudinal gauge bosons/scalars}
For the case of longitudinal W bosons the respective functions read
\begin{align}
D^{\phi\to W_\text{L}}_\text{C}(\mu_\text{l})&=\frac{\alpha}{4\pi}\left(\frac{c_\text{w}^2-s_\text{w}^2}{4c_\text{w}^2s_\text{w}^2}\left(F_Z+f_\text{S}\left(\frac{M_\text{W}^2}{M_\text{Z}^2},\frac{M_\text{W}^2}{M_\text{Z}^2}\right)\right)+\frac{1}{4s_\text{w}^2}\left(F_W+f_\text{S}\left(1,\frac{M_\text{Z}^2}{M_\text{W}^2}\right)\right)\right.\nonumber\\
&\hspace{40pt}\left.{}+\frac{1}{4s_\text{w}^2}\left(F_W+f_\text{S}\left(1,\frac{M_\text{H}^2}{M_\text{W}^2}\right)\right)+F_\gamma\right)
+\left. \delta C_\phi\right|_{\mu_\text{UV}=\mu_\text{l}}
\end{align}
with the charged-boson GBET correction factor
\begin{align}
\delta C_\phi=\frac12\delta Z_\text{W}+\frac{\delta M_\text{W}}{M_\text{W}}-\frac{\Sigma^{\text{W}\phi}(M_\text{W}^2)}{M_\text{W}}-\frac{\Sigma_\text{L}^{\text{WW}}(M_\text{W}^2)}{M_\text{W}^2},
\end{align}
[see also Eqs.~(\ref{GBET}),~(\ref{GBET1LOOP})]. Remember that the
Goldstone-boson field remains unrenormalised by convention. Similarly
to the $\gamma$/Z mixing in the transverse case one finds different operator corrections for operators containing the $\phi_2$ field, depending on whether the external state is a longitudinal Z or a Higgs boson.
In the first case we have
\begin{align}
D_\text{C}^{\phi_2\rightarrow Z_\text{L}}&(\mu_\text{l})=\frac{\alpha}{4\pi}\left(\frac{1}{4c_\text{w}^2s_\text{w}^2}\left(F_Z+f_\text{S}\left(1,\frac{M_\text{H}^2}{M_\text{Z}^2}\right)\right)+\frac{1}{2s_\text{w}^2}\left(F_W+f_\text{S}\left(\frac{M_\text{Z}^2}{M_\text{W}^2},1\right)\right)\right)\nonumber\\
&\hspace{40pt}+\left.\delta C_\chi\right|_{\mu_\text{UV}=\mu_\text{l}}
\end{align}
with the neutral-boson GBET correction factor
\begin{align}
\delta C_\chi=\frac 12 \delta Z_\text{ZZ}+\frac{\delta M_\text{Z}}{M_\text{Z}}+\text{i}\frac{\Sigma^{\text{Z}\chi}(M_\text{Z}^2)}{M_\text{Z}}-\frac{\Sigma_\text{L}^{\text{ZZ}}(M_\text{Z}^2)}{M_\text{Z}^2}.
\end{align}
For an external Higgs boson one finds
\begin{align}
D_\text{C}^{\phi_2\rightarrow\eta}&(\mu_\text{l})=\frac{\alpha}{4\pi}\left(\frac{1}{4c_\text{w}^2s_\text{w}^2}\left(F_Z+f_\text{S}\left(\frac{M_\text{H}^2}{M_\text{Z}^2},1\right)\right)+\frac{1}{2s_\text{w}^2}\left(F_W+f_\text{S}\left(\frac{M_\text{H}^2}{M_\text{W}^2},1\right)\right)\right)\nonumber\\
&\hspace{40pt}+\left.\frac 12 \delta Z_\text{H}\right|_{\mu_\text{UV}=\mu_\text{l}}.
\end{align}

The low scale corrections can be calculated in an alternative way by
performing a high-energy expansion of the complete NLO results and subtracting the high-scale
matching corrections as well as the contributions of the anomalous
dimensions. Since all contributions that are not soft and/or collinear
singular are fully contained in the high-scale matching terms, the
subtraction and expansion has to be done only for the soft/and or
collinear contributions. These are precisely the terms that have been
analysed in Refs.~\cite{DennerPozz1,DennerPozz2}. In this approach,
only standard scalar one-loop integrals have to be evaluated and
manipulated. The results are in agreement with those presented above
both for the non-photonic contributions and the photonic contributions
to the low-scale corrections.

\bibliographystyle{JHEPmod}
\bibliography{SCET_bib}
\end{document}

%% file: macros.tex
\def\refeq#1{\mbox{Eq.~(\ref{#1})}}

\def\refta#1{\mbox{Table~\ref{#1}}}

\def\citere#1{\mbox{Ref.~\cite{#1}}}
\def\citeres#1{\mbox{Refs.~\cite{#1}}}

\newcommand{\newc}{\newcommand}
\newc{\beq}{\begin{equation}}
\newc{\eeq}{\end{equation}}
\newc{\bit}{\begin{itemize}}
\newc{\eit}{\end{itemize}}
\newc{\ben}{\begin{enumerate}}
\newc{\een}{\end{enumerate}}
\newc{\bce}{\begin{center}}
\newc{\ece}{\end{center}}
\newc{\bfi}{\begin{figure}}
\newc{\efi}{\end{figure}}

\newcommand{\TeV}{\ensuremath{\,\text{TeV}}\xspace}

\newcommand{\Pq}{\ensuremath{q}\xspace}

\newcommand{\Pp}{\ensuremath{\text{p}}}
\newcommand{\Pe}{\ensuremath{\text{e}}\xspace}

\newcommand{\PW}{\ensuremath{\text{W}}\xspace}
\newcommand{\PZ}{\ensuremath{\text{Z}}\xspace}

\newcommand{\MW}{\ensuremath{M_\PW}\xspace}

\newcommand{\recola}{{\sc Recola}\xspace}
\newcommand{\recolaone}{{\sc Recola1}\xspace}
\newcommand{\recolatwo}{{\sc Recola2}\xspace}

\newcommand{\mocanlo}{{\sc MoCaNLO}\xspace}
\newcommand{\collier}{{\sc Collier}\xspace}

\newcolumntype{.}{D{.}{.}{-1}}
\newcolumntype{d}[1]{D{.}{.}{#1}}
\colorlet{tableoverheadcolor}{gray!37.5}
\colorlet{tableheadcolor}{gray!25}
\colorlet{tablerowcolor}{gray!12.5}

\def\draftdate{\relax}
\def\mda{\relax}
\def\mua{\relax}
\def\mla{\relax}
\def\draft{
\def\thtystars{******************************}
\def\sixtystars{\thtystars\thtystars}
\typeout{}
\typeout{\sixtystars**}
\typeout{* Draft mode!
         For final version remove \protect\draft\space in source file *}
\typeout{\sixtystars**}
\typeout{}
\def\draftdate{\today}
\def\mua{\marginpar[\boldmath\hfil$\uparrow$]%
                   {\boldmath$\uparrow$\hfil}\color{black}%
                    \typeout{marginpar: $\uparrow$}\ignorespaces}
\def\mda{\color{red}\marginpar[\boldmath\hfil$\downarrow$]%
                   {\boldmath$\downarrow$\hfil}%
                    \typeout{marginpar: $\downarrow$}\ignorespaces}
\def\mla{\marginpar[\boldmath\hfil$\rightarrow$]%
                   {\boldmath$\leftarrow $\hfil}%
                    \typeout{marginpar: $\leftrightarrow$}\ignorespaces}
\def\Mua{\marginpar[\boldmath\hfil$\Uparrow$]%
                   {\boldmath$\Uparrow$\hfil}\color{black}%
                    \typeout{marginpar: $\uparrow$}\ignorespaces}
\def\Mda{\color{red}\marginpar[\boldmath\hfil$\Downarrow$]%
                   {\boldmath$\Downarrow$\hfil}%
                    \typeout{marginpar: $\downarrow$}\ignorespaces}
\def\Mla{\marginpar[\boldmath\hfil\textcolor{red}{$\Rightarrow$}]%
                   {\boldmath\textcolor{red}{$\Leftarrow $}\hfil}%
                    \typeout{marginpar: $\leftrightarrow$}\ignorespaces}
\overfullrule 5pt
\oddsidemargin 15mm
\marginparwidth 29mm
}

